\documentclass[a4paper,11pt]{article}
\pdfoutput=1 

\usepackage{jheppub} 
\usepackage{soul}
\usepackage[T1]{fontenc} 
\usepackage{tikz}
\usepackage{pifont}
\usepackage{amsmath} 
\usepackage{mathrsfs} 
\usepackage[scr=boondox]{mathalpha}
\usepackage{xfp} 
\usepackage[outline]{contour} 
\usetikzlibrary{decorations.markings,decorations.pathmorphing}
\usetikzlibrary{angles,quotes} 
\usetikzlibrary{arrows.meta} 
\usetikzlibrary{external}
\contourlength{1.4pt}

\usepackage{braket}
\usepackage{bbm}

\usepackage{amsmath, amssymb, amsfonts, amsthm, latexsym, epsfig, mathrsfs, xcolor, slashed, braket, cancel}
\usepackage[overload]{textcase}

\usepackage{cleveref}
\usepackage{soul}

\usepackage{yfonts}

\newcommand{\Tr}{\textrm{Tr}}
\tikzset{>=latex} 
\colorlet{myred}{red!80!black}
\colorlet{myblue}{blue!80!black}
\colorlet{mygreen}{green!80!black}
\colorlet{mydarkred}{red!50!black}
\colorlet{mydarkblue}{blue!50!black}
\colorlet{mylightblue}{mydarkblue!6}
\colorlet{myvlightblue}{mydarkblue!3}
\colorlet{mypurple}{blue!40!red!80!black}
\colorlet{mydarkpurple}{blue!40!red!50!black}
\colorlet{mylightpurple}{mydarkpurple!80!red!6}
\colorlet{myorange}{orange!40!yellow!95!black}
\tikzstyle{cone}=[mydarkblue,line width=0.2,top color=blue!60!black!30,
 bottom color=blue!60!black!50!red!30,shading angle=60,fill opacity=0.9]
\tikzstyle{cone back}=[mydarkblue,line width=0.1,dash pattern=on 1pt off 1pt]
\tikzstyle{world line}=[myblue!60,line width=0.4]
\tikzstyle{world line t}=[mypurple!60,line width=0.4]
\tikzstyle{particle}=[mygreen,line width=0.5]
\tikzstyle{photon}=[-{Latex[length=4,width=3]},myorange,line width=0.4,decorate,
 decoration={snake,amplitude=0.9,segment length=4,post length=3.8}]
\tikzstyle{singularity}=[myred,line width=0.6,decorate,
 decoration={zigzag,amplitude=2,segment length=6.17}]
\tikzset{declare function={%
 penrose(\x,\c) = {\fpeval{2/pi*atan( (sqrt((1+tan(\x)^2)^2+4*\c*\c*tan(\x)^2)-1-tan(\x)^2) /(2*\c*tan(\x)^2) )}};%
 penroseu(\x,\t) = {\fpeval{atan(\x+\t)/pi+atan(\x-\t)/pi}};%
 penrosev(\x,\t) = {\fpeval{atan(\x+\t)/pi-atan(\x-\t)/pi}};%
 kruskal(\x,\c) = {\fpeval{asin( \c*sin(2*\x) )*2/pi}};
}}

\newtheorem{theorem}{Theorem}

\crefname{section}{sec.}{sec.}
\crefname{appendix}{Appendix}{Appendices}
\crefname{table}{Table}{Tables}

\crefname{definition}{Def.}{Defs.}
\crefname{prop}{Prop.}{Props.}
\crefname{lemma}{Lemma}{Lemmas}
\crefname{corollary}{Cor.}{Cors.}
\crefname{thm}{Theorem}{Theorems}
\crefname{remark}{Remark}{Remarks}

\crefname{ass}{Assumptions}{Assumptions}
\crefname{property}{Properties}{Properties}

\newcommand{\be}{\begin{equation}\begin{aligned}}
\newcommand{\ee}{\end{aligned}\end{equation}}

\newcommand{\lb}{\left}
\newcommand{\rb}{\right}

\newcommand{\mc}{\mathcal}

\newcommand{\ms}{\mathscr}
\newcommand{\mf}{\mathfrak}
\newcommand{\bb}{\mathbb}

\usepackage{enumitem}

\setlist[enumerate]{itemsep=2pt, label=(\arabic*), ref=(\arabic*)}

\definecolor{indigo(dye)}{rgb}{0.0, 0.25, 0.42}

\newcommand{\defn}{\mathrel{\mathop:}=} 

\newcommand{\norm}[1]{\lb\Vert\, #1 \,\rb\Vert}		

\newif\ifslow

\newcommand{\op}[1]{\boldsymbol{#1}}

\newcommand{\Alg}{\mathscr{A}}
\newcommand{\conf}{\Omega}

\newcommand{\s}{\omega}

\newcommand{\Hilb}{\mathscr{H}}

\newcommand{\antiHilb}{%
\hspace{4pt} 
 \vbox{%
 \hrule height 0.5pt
 \kern0.25ex
 \hbox{%
 \kern-0.3em
 \ifmmode\Hilb\else\ensuremath{\Hilb}\fi
 \kern0em
 }
 }
}

\newcommand{\Fock}{\mathscr{F}}

\newcommand{\KG}{\textrm{KG}}

\newcommand{\GR}{\textrm{GR}}

\newcommand{\symp}{\mathscr{W}} 

\let\oldsetminus\setminus
\renewcommand{\setminus}{\!\oldsetminus\!} 

\let\oldint\int
\renewcommand{\int}{\oldint\limits}

\let\oldlim\lim
\renewcommand{\lim}{\oldlim\limits}

\renewcommand{\bar}{\overline}

\newcommand{\scri}{\ms I}

\newcommand{\nfrac}[2]{{{}^#1\!\!/\!_#2}}

\newcommand{\half}{\nfrac{1}{2}}

\renewcommand{\Im}{{\rm Im\,}}

\DeclareMathOperator\supp{supp}

\newcommand{\sA}{\mathscr{A}}

\def\le{\left}
\def\ri{\right}

\newcommand\sH{{\ensuremath{{\mathcal H}}}}

\newcommand\sB{{\ensuremath{{\mathcal B}}}}

\newcommand\om{{\ensuremath{{\omega}}}}

\newcommand\sR{{\ensuremath{{\mathcal R}}}}

\newcommand{\vNext}{\mathfrak{A}_{\textrm{ext.}}}

\newcommand{\hatvNext}{\hat{\mathfrak{A}}_{\textrm{ext.}}}

\title{Generalized Black Hole Entropy is von Neumann Entropy}

\author[a,b]{Jonah Kudler-Flam,}
 \author[c]{Samuel Leutheusser}
\author[c]{Gautam Satishchandran}%

 \affiliation[a]{School of Natural Sciences, Institute for Advanced Study, Princeton, NJ 08540, USA}
 \affiliation[b]{Princeton Center for Theoretical Science, Princeton University, Princeton, NJ 08544, USA}
 \affiliation[c]{Princeton Gravity Initiative, Princeton University, Princeton NJ 08544, USA} 

\emailAdd{jkudlerflam@ias.edu}
\emailAdd{sl9535@princeton.edu}
\emailAdd{gautam.satish@princeton.edu}

\abstract{
It has been argued that, while the individual terms in the generalized entropy, $S_{\textrm{gen.}}=A/4G_{\textrm{N}} + S_{\textrm{ext.}},$ are ill-defined in the semiclassical limit, their sum is well-defined if one takes into account perturbative quantum gravitational effects. The first term diverges as $G_{\textrm{N}}\to 0$ and the second diverges due to the infinite entanglement across the horizon which is characteristic of Type III von Neumann algebras. It was recently shown that the von Neumann algebra of observables ``gravitationally dressed'' to the mass of a Schwarzschild-AdS black hole or the energy of an observer in de Sitter spacetime admit a well-defined trace. The algebras are Type II$_{\infty}$ (which does not admit a maximum entropy state) and Type II$_{1}$ (which admits a maximum entropy state) respectively and the von Neumann entropy of ``semiclassical'' states was found to be (up to an additive constant) the generalized entropy. However, these arguments rely on the existence of a stationary ``equilibrium (KMS) state'' and do not apply to, for example, black holes formed from gravitational collapse, Kerr black holes, or black holes in asymptotically de Sitter spacetime. These spacetimes are stationary but not in thermal equilibrium. In this paper, we present a general framework for obtaining the algebra of ``gravitationally dressed'' observables for a linear, Klein-Gordon field on any spacetime with a (bifurcate) Killing horizon. We prove, assuming the existence of a stationary state --- which is not necessarily KMS --- and suitable asymptotic decay of solutions, a ``structure theorem'' that the algebra of ``gravitationally dressed'' observables always contains a Type II factor of observables ``localized'' on the horizon. These assumptions have been rigorously proven in most cases of interest in this paper. Applying our general framework to the algebra of observables in the exterior of an asymptotically flat Kerr black hole where the fields are dressed to the black hole mass and angular momentum we find that the algebra is the product of a Type II$_{\infty}$ algebra on the horizon and a Type I$_{\infty}$ algebra at past null infinity. The full algebra is Type II$_{\infty}$ and the von Neumann entropy of semiclassical states is the generalized entropy. In the case of Schwarzschild-de Sitter, despite the fact that we must introduce an observer, the algebra of observables dressed to the perturbed areas of the black hole and cosmological horizons is the product of Type II$_{\infty}$ algebras on each horizon. The entropy of semiclassical states is given by the sum of the areas of the two horizons as well as the entropy of quantum fields in between the horizons. Our results suggest that in all cases where there exists another ``boundary structure'' (e.g., an asymptotic boundary or another Killing horizon) the algebra of observables is Type II$_{\infty}$ and in the absence of such structures (e.g. de Sitter spacetime) the algebra is Type II$_{1}$.
\par 
}
\begin{document}

\maketitle
\flushbottom

\section{Introduction}
In 1973 Bekenstein proposed that the entropy of a black hole is proportional to its area \cite{bekenstein1973black}. Shortly after this proposal, Hawking \cite{hawking1975particle} showed --- in a model of linear fields interacting only with geometry --- that the theory of quantum fields on curved spacetime predicts that a black hole formed from collapse will emit thermal radiation. This result, together with the celebrated ``first law of black hole mechanics'' \cite{bardeen1973four} implies that the black hole entropy is precisely 
\begin{equation}
\label{eq:Sbhintro}
S_{\textrm{BH}}=\frac{k_{\textrm{B}}c^{3}A}{4G_{\textrm{N}}\hbar},
\end{equation}
where we have included the fundamental constants of nature (we shall henceforth suppress all constants except for $G_{\textrm{N}}$). This results suggests a ``generalized entropy'' defined as the sum of the entropy of the black hole and the entropy of quantum fields in its exterior \cite{bekenstein1974generalized}:
\begin{equation}
\label{eq:Sgenintro}
S_{\textrm{gen.}} = \frac{A}{4G_{\textrm{N}}} + S_{\textrm{vN.}}(\rho_{\s}) \ .
\end{equation}
Here $\rho_{\s}$ is the density matrix of the quantum state $\ket{\s}$ of the field in the exterior of the black hole and
\begin{equation}
\label{eq:Vnentintro}
S_{\textrm{vN.}}(\rho_{\s}) = - \textrm{Tr}(\rho_{\s}\log\rho_{\s}).
\end{equation}
Indeed, it was proven by Wall that \cref{eq:Sgenintro} obeys a generalized second law in the sense that $\Delta S_{\textrm{gen.}}\geq 0$ \cite{2012PhRvD..85j4049W}. In totality, these results suggest a deep connection between quantum field theory in curved spacetime (in the $G_{\textrm{N}}\to 0$ limit), spacetime geometry, thermal effects, and entropy. 

It was soon realized that the thermal effects are due to the presence of a (bifurcate)\footnote{A black hole formed from gravitational collapse will rapidly settle down to a stationary black hole whose event horizon is, to an excellent approximation, a Killing horizon \cite{Hawking:1971vc,Friedrich:1998wq,2009arXiv0902.1173A} and therefore can be extended to have a bifurcate structure \cite{Racz:1995nh}.} ``Killing horizon'' (i.e., a pair of intersecting null surfaces that we shall generally label as $\mc{H}^{+}$ and $\mc{H}^{-}$ and are orthogonal to the Killing field). The presence of such a horizon can lead to the existence of ``thermal equilibrium states'' (also known as ``KMS states'') which are thermal with respect to the horizon Killing vector. Examples of such spacetimes include Minkowski spacetime (where the isometries correspond to Lorentz boosts), the maximally extended Schwarzschild black hole, and de Sitter spacetime. On these spacetimes the thermal ``vacuum state'' is, respectively, the Minkowski vacuum, the ``Hartle-Hawking'' vacuum \cite{Hartle:1976tp,Israel:1976ur,Gibbons_76,Gibbons_78}, and the ``Bunch-Davies vacuum'' (also known as the ``Euclidean vacuum'') \cite{Bunch:1978yq,Chernikov:1968zm,Schomblond_1976}. These states are ``thermal equilibrium states'' in that they are both stationary with respect to the horizon Killing field and thermal throughout the entire spacetime. Such states can also be obtained by ``Euclidean methods'' using path integral techniques \cite{Hartle:1976tp,Gibbons:1977mu}. However, despite their utility, such thermal equilibrium states do not exist in many spacetimes of physical interest such as, for example, black holes formed from gravitational collapse, Kerr black holes, or black holes in asymptotically de Sitter spacetime \cite{Kay_1988}. Nevertheless, as pointed out originally by Unruh \cite{unruh1976notes}, one can define a suitable ``stationary state'' in the exterior and interior of the black hole which is stationary with respect to the horizon Killing field. This stationary state is essentially the state used in Hawking's original calculation \cite{Fredenhagen:1989kr} and has been rigorously constructed for black holes in asymptotically flat and asymptotically de Sitter spacetimes \cite{Dimock:1987hi,2009arXiv0907.1034D,Brum:2014nea,Hollands:2019whz,Klein:2022jtb}.

In contrast to the status of our understanding of the thermal nature of black holes, the nature of the generalized entropy has remained comparatively obscure. While the black hole entropy has been reproduced by a direct microstate counting in string theory \cite{strominger1996microscopic} and by the gravitational path integral \cite{gibbons1977action}, these methods have a restricted domain of validity to either black holes with a high degree of symmetry or to thermal equilibrium states.\footnote{In the gravitational path integral, the Euclidean time circle shrinks to a point at the black hole horizon, obscuring the interpretation as a thermal trace. The contour of integration that should be used in quantum gravity is also unclear and the metrics used in some entropy calculations, such as Kerr \cite{gibbons1977action}, are not ``allowable,'' as proposed in \cite{2021arXiv210510161K, 2021arXiv211106514W}.} Furthermore, in the semiclassical regime (i.e., $G_{\textrm{N}}\to 0$), the individual terms in the generalized entropy \cref{eq:Sgenintro} are ill-defined. The black hole entropy term \cref{eq:Sbhintro} diverges in this limit. Additionally, the von Neumann entropy \cref{eq:Vnentintro} diverges in quantum field theory in curved spacetime due to the ultraviolet entanglement across the horizon. This divergence is a characteristic feature of the algebra of observables of local quantum field theory which, in the language of Murray, von Neumann, and Connes's classification \cite{murray1936rings,connes1973classification}, is a Type III$_{1}$ von Neumann algebra. These algebras have no pure states and no notion of a trace, so the expression \cref{eq:Vnentintro} is undefined. This is in contrast to the (more familiar) Type I algebras --- on which both pure states and density matrices exist --- that arise in global quantum field theory on a complete Cauchy surface. An intermediate case, which will be important for the considerations of this paper, is that of Type II algebras which have no pure states, but for which one can define density matrices. 

It was argued by Susskind and Uglum \cite{1994PhRvD..50.2700S}, that while both terms in \cref{eq:Sgenintro} are individually divergent in the semiclassical regime, their sum is well-defined if one takes into account (perturbative) quantum gravitational effects.\footnote{This notion was sharpened using language similar to the present paper in \cite{Gesteau:2023hbq}.} Recently, building on the work of Leutheusser and Liu \cite{2021arXiv211005497L,2021arXiv211212156L,2022arXiv221213266L} studying ``large $N$'' algebras in AdS/CFT, Witten \cite{2022JHEP...10..008W} together with Chandrasekaran, Penington (CPW) \cite{2022arXiv220910454C} as well as Longo (CLPW) \cite{Chandrasekaran:2022cip} included such perturbative quantum gravity effects by constructing the algebra of (isometry invariant) ``dressed observables'' in the exterior of a Schwarzschild-AdS black hole and in the ``static patch'' of de Sitter. Remarkably, they found that the algebra is generally Type II --- so the trace is well-defined --- and showed that the von Neumann entropy of so-called ``classical-quantum'' states can be expressed (up to a state independent constant) as the generalized entropy.\footnote{Follow-up work using the ``crossed product'' in high-energy physics includes \cite{Jensen:2023yxy,2023arXiv230607323A,2023arXiv230609314K,2023arXiv230814166B,2023arXiv230712481S,2023arXiv230704233A,2023arXiv230411845G,2023arXiv230214747G,2022arXiv220706704G}.} The algebra in de Sitter spacetime contained a maximum entropy state and was, under Murray and von Neumann's subclassification, Type II$_{1}$. The algebra of observables in the exterior of Schwarzschild-AdS has no maximum entropy state and is Type II$_{\infty}$. 

However, as we shall explain below, the constructions of CPW and CLPW crucially relied on the existence of a KMS state and therefore cannot be straightforwardly applied to more general spacetimes with Killing horizons. An important aim of this work is to extend and generalize the work of CPW and CLPW to arbitrary spacetimes with Killing horizons. Our main result is that, under certain assumptions --- which have been rigorously proven in most of the spacetimes considered in this paper --- the algebra of observables in the ``exterior'' of a bifurcate Killing horizon is Type II. Roughly speaking, we find that in all cases where one has a bifurcate Killing horizon together with another ``invariantly defined'' structure (e.g., another Killing horizon or an asymptotic boundary) the algebra is Type II$_{\infty}$.\footnote{In very interesting recent work \cite{Jensen:2023yxy}, Jensen, Sorce, and Speranza argued that, under certain assumptions, general bounded subregions in quantum gravity have associated Type II$_1$ von Neumann algebras while unbounded subregions have associated Type II$_\infty$ algebras. We will not require these assumptions in our work, and indeed, these assumptions will not be generically satisfied in the spacetimes that we consider. In particular, for the region $\mc{R}$ of Schwarzschild-de Sitter considered in \cref{subsec:sds}, we conclude that the algebra $\vNext(\mc{R},\s_{0})$ is Type II$_{\infty}$.} In contrast, in de Sitter spacetime --- where the bifurcate Killing horizon exists in a (spatially) closed universe --- the algebra is Type II$_{1}$. In our analysis, we will restrict consideration, for simplicity, to the case of a free Klein-Gordon field. Our results directly generalize to electromagnetic fields and linearized gravitational perturbations. In \cref{subsec:interactions} we outline the plausible generalizations of our assumptions necessary to consider general interacting quantum fields.

The key technical ingredient in our general analysis of spacetimes $(\mc{M},g)$ with Killing horizons is \cref{thm:horizontheorem} which is a ``structure theorem'' that illustrates how Type II algebras generically arise for ``dressed observables'' in spacetimes with bifurcate Killing horizons. \Cref{thm:horizontheorem} applies to any spacetime satisfying assumptions \ref{assump1}--\ref{assump4}. Assumptions \ref{assump1} and \ref{assump2} are simply that $\mc{M}$ contains a bifurcate Killing horizon and there exists a ``physically relevant'' subregion $\mc{M}_{\textrm{R}}$ (defined in \cref{eq:MRML}) on which the initial value problem is well-defined (e.g., for an asymptotically flat black hole, $\mc{M}_{\textrm{R}}$ is the interior and ``right'' exterior of the black hole). Assumption \ref{assump3} is that there exists an initial data surface for wave propagation on $\mc{M}_{\textrm{R}}$ which consists of the ``past horizon'' $\mc{H}^{-}$ together with possibly another hypersurface $\Sigma^{\prime}$. Furthermore, we assume that the hypersurface can be chosen such that the initial data can be specified independently on $\mc{H}^{-}$ and $\Sigma^{\prime}$. Finally, in assumption \ref{assump4}, we assume that there exists a stationary state in $(\mc{M}_{\textrm{R}},g)$. Assumptions \ref{assump1} and \ref{assump2} will be satisfied in all spacetimes explicitly considered in this paper. Roughly speaking, assumption \ref{assump3} is equivalent to assuming sufficient decay of solutions in the limit to the past boundary $\mc{H}^{-}\cup \Sigma^{\prime}$ where, in all cases of interest, $\Sigma^{\prime}$ will be a boundary ``at infinity'' or another Killing horizon. Finally, assumption \ref{assump4} reflects the expectation that for a black hole formed by gravitational collapse, the state of a quantum field is expected to approach a stationary quantum state after the black hole has “settled down” to a stationary state. Assumptions \ref{assump3} and \ref{assump4} have been rigorously proven for de Sitter spacetime, Schwarzschild black holes, and Kerr-Newmann-de Sitter black holes \cite{Dafermos:2010hd,Luk:2010jfs,Dappiaggi:2017kka,Hintz:2015jkj}. While the non-trivial decay estimates have not been rigorously proven for general asymptotically flat black holes (e.g., Kerr) it is expected that these assumptions will continue to hold in these cases. 

In \cref{sec:dressed}, we show that assumptions \ref{assump1}---\ref{assump4} directly imply the existence of a Type III$_{1}$ von Neumann algebra $\mf{A}(\mc{H}^{-}_{\textrm{R}},\s_{0})$ of ``initial data'' where $\mc{H}^{-}_{\textrm{R}}$ is the portion of the ``past horizon'' to the past of the bifurcation surface (e.g., for a black hole, $\mc{H}^{-}_{\textrm{R}}$ lies in the past of the exterior region) and $\ket{\s_{0}}$ is the (unique) stationary state. Furthermore, the modular Hamiltonian on this algebra is precisely the generator of Killing translations on $\mc{H}^{-}$. We find that the algebra of observables on $\mc{H}^{-}_{\textrm{R}}$ invariant under the Killing isometries is equivalent to the so-called ``crossed product'' of $\mf{A}(\mc{H}^{-}_{\textrm{R}},\s_{0})$ with respect to its modular automorphism group. It directly follows from a structure theorem of Takesaki \cite{takesaki1973duality} that the resultant algebra is Type II. 

Since many of our arguments require a considerable amount of technical machinery, we now provide a sketch of the construction of the algebra of observables for each spacetime considered in the paper. In each case we will consider the algebra of observables in an (invariantly defined) subregion of the spacetime which we will refer to as the ``right wedge'' $\mc{R}$. For a black hole in asymptotically flat or AdS spacetime, $\mc{R}$ will be the exterior of the black hole. In de Sitter spacetime $\mc{R}$ is the so-called ``static patch'' associated to the domain of dependence of an inertial observer. For black holes in asymptotically de Sitter spacetime, the relevant region will be the spacetime in the exterior of the black hole and in the interior of the cosmological horizon. While, in the main body of the paper, we consider the algebra of observables invariant under the full isometry group of $(\mc{R},g)$, we will focus attention, in this section, on the construction of observables invariant under the Killing horizon isometry since this isometry is most relevant for the emergence of a Type II algebra. As we will show in each section, the inclusion of these additional constraints does not change the type of the algebra.

In \cref{sec_stationary} we revisit the construction of the algebra of observables in the region $\mc{R}$ of a Schwarzschild-AdS black hole and de Sitter spacetime considered in \cite{2022arXiv220910454C,Chandrasekaran:2022cip}. As already stated, quantum fields on these spacetimes enjoy the special property of admitting globally KMS states. The main purpose of revisiting this construction is to prove the main results of these references in a manner that generalizes to arbitrary spacetimes with Killing horizons. In both cases, the relevant isometry group is $\bb{R}\times \textrm{SO}(3)$ where $\bb{R}$ is the isometry group that generates the bifurcate Killing horizon. We treat the Klein-Gordon field propagating on this background spacetime as a ``test field'' which is a first-order perturbation of the background geometry. This field gravitates and its total (conserved) energy (with respect to the horizon Killing field $\xi$) is given by 
\begin{equation}
\label{eq:introFxi}
F_{\xi}=\int_{\Sigma}\sqrt{h}d^{3}x~\delta^{2}T_{ab}n^{a}\xi^{b}
\end{equation}
where $n^{a}$ is the normal to the Cauchy surface $\Sigma$, {$\sqrt{h}d^{3}x$ is the proper volume element on $\Sigma$}, the ``$\delta$'' denotes the order of the perturbation, e.g., $\delta^{2}T_{ab}$ is the (second-order) stress tensor of the (first-order) scalar field. On the phase space of the scalar field, $F_{\xi}$ generates translations along orbits of $\xi$. The energy of the scalar field enters Einstein's equation at second-order and can be related to the difference of (second-order) ``gravitational charges.''

In the case of a Schwarzschild black hole in AdS, considered in \cref{subsec:sads_kms}, if we evaluate \cref{eq:introFxi} on the past horizon $\mc{H}^{-}$ of the black hole, we obtain 
\begin{equation}
\label{eq:introTUUA+A-}
-\int_{-\infty}^{\infty}dU\int_{\bb{S}^{2}}d\Omega_{2}~U\delta^{2}T_{UU} = {\frac{\delta^{2}\mc{Q}_{-}}{4G_{\textrm{N}}\beta}-\frac{\delta^{2}\mc{Q}_{+}}{4G_{\textrm{N}}\beta},}
\end{equation}
{where $U$ is an affine parameter of $\mc{H}^{-}$, $\beta$ is the inverse temperature of the black hole and the second-order gravitation charges in \cref{eq:introTUUA+A-} correspond to the limits of local charges on the horizon} 
\begin{equation}
{\delta^{2}\mc{Q}_{\pm}\defn \lim_{U\to \pm \infty}\delta^{2}\mc{Q}_{U}}
\end{equation}
where \cite{Chandrasekaran:2018aop,Hollands:2024vbe}
\begin{equation}
{\delta^{2}\mc{Q}_{U} = \delta^{2}A_{U} - \int_{S_{U}}d\Omega_2~U\delta^{2}\theta }
\end{equation}
{and $\delta^{2}A_{U}$ is the perturbed area of a constant $U$ cross-section $S_{U}$ and $\delta^{2}\theta$ is the perturbed expansion on that cross-section. When the horizon is stationary, the perturbed expansion vanishes and the charge $\delta^{2}\mc{Q}_{U}$ is equivalent to the perturbed area $\delta^{2}A_{U}$.} Furthermore, since the limiting perturbed charges are equal to the right/left perturbed ADM masses at infinity
\begin{equation}
\frac{\delta^{2}\mc{Q}_{-}}{4G_{\textrm{N}}\beta} = \delta^{2}M_{\textrm{R}},\quad \frac{\delta^{2}\mc{Q}_{+}}{4G_{\textrm{N}}\beta} = \delta^{2}M_{\textrm{L}},
\end{equation}
this constraint is identical to the one considered in \cite{2022arXiv220910454C}. The (Type III$_{1}$) algebra $\mf{A}(\mc{R},\s_{0})$ of field observables in $\mc{R}$ can be specified by choosing a vacuum state $\ket{\s_{0}}$ on the spacetime. In this case, the relevant stationary state $\ket{\s_{0}}$ is the Hartle-Hawking state, which exists on this spacetime. Furthermore, since the past horizon $\mc{H}^{-}$ is an initial data surface, the algebra of local observables is equivalent to the algebra $\mf{A}(\mc{H}_{\textrm{R}}^{-},\s_{0})$ of initial data on the past horizon 
\begin{equation}
\label{eq:introbulkbndry}
\mf{A}(\mc{R},\s_{0})\cong \mf{A}(\mc{H}_{\textrm{R}}^{-},\s_{0}) \quad \quad \textrm{(Schwarzschild-AdS)}.
\end{equation}
To consider perturbative gravitational effects, we must also specify a state of the (second-order) gravitational field subject to the constraint \cref{eq:introTUUA+A-}. As noted in \cite{2022arXiv220910454C}, to ``solve'' these constraints it is sufficient to quantize a single, ``non-propagating'' mode of the second-order gravitational field. We quantize\footnote{In all cases considered in this paper the gravitational charges arise from a locally compact, isometry group $G$ and the Hilbert space will be taken to be square-integrable functions $\Hilb_{\textrm{G}}=L^{2}(G)$ with respect to the (unique) left/right invariant Haar measure on $G$.} this mode by considering square-integrable wave functions of the perturbed mass $\delta^{2}M_{\textrm{R}}$ (or equivalently $\delta^{2}\mathcal{Q}_{-}$) in $\bb{R}$ \begin{equation}
\ms{H}_{\textrm{M}}=L^{2}(\bb{R}),
\end{equation}
which also has the representation of a conjugate ``asymptotic time'' operator $t_{\textrm{R}}$.
Therefore, the states on this extended Hilbert space now include {\em intrinsic} $O(G_{\textrm{N}})$ fluctuations of the black hole area. The perturbed mass $\delta^{2}M_{\textrm{L}}$ (or equivalently $\delta^{2}\mathcal{Q}_{+}$) lies in the causal complement of $\mc{R}$, so these charges commute with local observables in $\mc{H}_{\text{R}}^{-}$. Furthermore, the left-hand side of \cref{eq:introTUUA+A-} is the modular Hamiltonian of $\mf{A}(\mc{H}_{\textrm{R}},\s_{0})$. The physical observables on $\mc{H}_{\textrm{R}}^{-}$ are those in which the Klein-Gordon field is {\em dressed to the (perturbed) ADM mass} so that the joint observable commutes with the constraint \cref{eq:introTUUA+A-}. We will refer to the algebra of physical observables on $\mc{H}_{\textrm{R}}^{-}$ as $\vNext(\mc{H}^{-}_{\textrm{R}},\s_{0})$ which, by \cref{eq:introbulkbndry}, directly yields a construction of the algebra of physical observables in $\mc{R}$ 
\begin{equation}
\vNext(\mc{R},\s_{0})\cong \vNext(\mc{H}^{-}_{\textrm{R}},\s_{0}) \quad \quad \textrm{(Schwarzschild-AdS)}.
\end{equation}
By \cref{thm:horizontheorem}, $\vNext(\mc{H}^{-}_{\textrm{R}},\s_{0})$ is a Type II algebra. More precisely, since the spectrum of $\delta^{2}\mathcal{Q}_{-}/4G_{\textrm{N}}\beta$ is $\bb{R}$, the algebra is Type II$_{\infty}$. A ``classical-quantum'' state $\ket{\hat{\s}}$ corresponds to choosing a state $\ket{\s}$ of the Klein-Gordon field together with a wave function $f(\delta^{2}M_{\textrm{R}})$. If $\rho_{\hat{\s}}$ is the density matrix of $\ket{\hat{\s}}$ on $\vNext(\mc{R},\s_{0})$ then, since the algebra is Type II, its von Neumann entropy is now well-defined. For ``semiclassical'' states where $f(\delta^{2}M_{\textrm{R}})$ is (approximately) slowly varying (and so the conjugate time $t_{\textrm{R}}$ is sharply peaked), CLPW showed that the von Neumann entropy (up to a state-independent constant $C$) is\footnote{The von Neumann entropy of a density matrix on a Type II algebra is well-defined. In order to match $S_{\textrm{vN.}}(\rho_{\hat{\s}})$ to the generalized entropy one must split up this well-defined expression into the sum of two mutually divergent terms. Therefore, the right-hand side of \cref{eq:introvNSAdS} is necessarily a formal expression whereas the left-hand side of \cref{eq:introvNSAdS} is well-defined.} 
\begin{equation}
\label{eq:introvNSAdS}
S_{\textrm{vN.}}(\rho_{\hat{\s}}) \simeq S_{\textrm{gen.}} + S(\rho_{f}) + C \quad \quad \textrm{(Schwarzschild-AdS)},
\end{equation}
where $S(\rho_{f})$ is the von Neumann entropy of the ``classical'' density matrix $\rho_{f}=|f(\delta^{2}M_{\textrm{R}})|^{2}$ {which corresponds to a ``logarithmic'' correction to black hole entropy \cite{2022JHEP...10..008W,2002CQGra..19.2355D}}. We also show in section~\ref{sec_AdS} that a similar analysis can also be applied to Kerr black holes in AdS.

In \cref{subsec:deSitter}, we revisit the construction of the algebra of isometry invariant observables in the ``static patch'' $\mc{R}$ of de Sitter (see figure~\ref{fig:static_patch}) considered in \cite{Chandrasekaran:2022cip}. The algebra of field observables $\mf{A}(\mc{R},\s_{0})$ satisfies \cref{eq:introbulkbndry} where now $\mc{H}^{-}$ is the past cosmological horizon and $\ket{\s_{0}}$ is the Bunch-Davies vacuum. However, since de Sitter spacetime is a closed universe, the gravitational constraints vanish and one must introduce another gravitating body in order to obtain a non-trivial algebra of fields that satisfy these constraints. CLPW introduced an inertial ``observer'' in $\mc{R}$ carrying a ``clock'' with (bounded) energy $\varepsilon$ and time $\tau$. States of the clock are then wave functions in $\ms{H}_{\varepsilon}=L^{2}(\bb{R}_{+})$. The gravitational constraints then take the precise form of \cref{eq:introTUUA+A-} where now the observer's incoming energy reduces the area of the cosmological horizon by 
\begin{equation}
\frac{\delta^{2}\mathcal{Q}_{-}}{4G_{\textrm{N}}\beta} = -\varepsilon. 
\end{equation}
Since now the spectrum of $\delta^{2}\mathcal{Q}_{-}/4G_{\textrm{N}}\beta$ is bounded from above, the algebra of physical observables in $\mc{R}$ is Type II$_{1}$. Furthermore, the entropy of classical-quantum states $\ket{\hat{\s}}$ with slowly varying wave function $f(\varepsilon)$ is 
\begin{equation}
S_{\textrm{vN.}}(\rho_{\hat{\s}}) \simeq S_{\textrm{gen.}} + S(\rho_{f}) + C \quad \quad \textrm{(de Sitter)}.
\end{equation}

In \cref{sec_flat}, we consider the algebra of observables in the exterior region $\mc{R}$ of an asymptotically flat (one-sided) Kerr black hole (see figure~\ref{fig:Kerr_tikz}). As previously mentioned, the vacuum state $\ket{\s_{0}}$ is the ``Unruh vacuum'' which is stationary in $\mc{M}_{\textrm{R}}$ but not KMS. The modular flow for the Unruh state does not correspond to a local geometric flow in $\mc{R}$. Nevertheless, the flow is a geometric flow on $\mc{H}^{-}$ corresponding to translations along the orbits of the horizon Killing field. Furthermore, for a massless field, the algebra $\mf{A}(\mc{R},\s_{0})$ of local fields is equivalent \cite{Dappiaggi:2017kka} to the product of algebras of independent initial data specified on the past horizon and past null infinity 
\begin{equation}
\label{eq:introKerrdecomp}
\mf{A}(\mc{R},\s_{0})\cong \mf{A}(\mc{H}^{-}_{\textrm{R}},\s_{0}) \otimes \mf{A}(\scri^{-},\s_{0})\quad \quad \textrm{(Kerr)},
\end{equation}
where $\mf{A}(\scri^{-},\s_{0})$ is the Type I$_{\infty}$ algebra of initial data at $\scri^{-}$. The isometry group of $(\mc{R},g)$ is $\bb{R}\times \textrm{U}(1)$ and the horizon Killing field $\xi^{a}=t^{a}+\Omega_{\textrm{H}}\psi^{a}$ is a linear combination of an asymptotic time translation and an azimuthal rotation. $\Omega_{\textrm{H}}$ is the angular velocity of the black hole. Integrating \cref{eq:introFxi} over a Cauchy surface $\Sigma$ for $\mc{M}_{\textrm{R}}$ yields the global constraint 
\begin{equation}
\label{eq:introKillingconstraint1}
\int_{\Sigma}\sqrt{h}d^{3}x~\delta^{2}T_{ab}n^{a}\xi^{b}=\delta^{2}M_{i^{0}}-\Omega_{\textrm{H}}\delta^{2}J_{i^{0}}-\frac{\delta^{2}\mathcal{Q}_{+}}{4G_{\textrm{N}}\beta} \quad \quad \textrm{(Kerr)}
,\end{equation}
where $\delta^{2}M_{i^{0}}$ is the perturbed ADM mass and $\delta^{2}J_{i^{0}}$ is the perturbed angular momentum at spatial infinity. Evaluating \cref{eq:introFxi} on the boundary yields a separate constraint on $\mc{H}^{-}$ and $\scri^{-}$. The constraint on $\mc{H}^{-}$ is precisely of the form \cref{eq:introTUUA+A-}. Evaluating the integral \cref{eq:introFxi} at $\scri^{-}$ yields 
\begin{equation}
\label{eq:introKillingconstraint2}
\int_{\scri^{-}}dvd\Omega_{2}~\delta^{2}T_{vv}=\delta^{2}M_{i^{0}}-\Omega_{\textrm{H}}\delta^{2}J_{i^{0}} - \frac{\delta^{2}\mathcal{Q}_{-}}{4G_{\textrm{N}}\beta} \quad \quad \textrm{(Kerr)}
\end{equation}
where $v$ is the advanced time at past null infinity and $\delta^{2}T_{vv}$ is the stress-energy flux through $\scri^{-}$. {The sum of \cref{eq:introTUUA+A-} and \cref{eq:introKillingconstraint2} yields \cref{eq:introKillingconstraint1}.} In terms of initial data, the quantization of the intrinsic (second-order) fluctuations of the black hole are naturally specified at asymptotically early times. In \cref{subsec:algdressKerr} we quantize both the ``incoming'' perturbed area and angular momentum of the black hole, however, as in the previous cases considered in this section, we will restrict attention to the implications for the quantization of the area since it is the black hole charge associated to the horizon Killing field. As before, the perturbed area $\delta^{2}\mathcal{Q}_{-}/4G_{\textrm{N}}\beta$ can be quantized in the Hilbert space $\Hilb_{\mathcal{Q}}=L^{2}(\bb{R})$. The algebra of ``bulk'' fields dressed to the perturbed area which commute with \cref{eq:introKillingconstraint1,eq:introKillingconstraint2} is the algebra $\mf{A}_{\textrm{ext.}}(\mc{R},\s_{0})$ which we show can be decomposed as 
\begin{equation}
\label{eq:introKerrextdecomp}
\vNext(\mc{R},\s_{0})\cong \vNext(\mc{H}_{\textrm{R}}^-,\s_{0}) \otimes \mf{A}(\scri,\s_{0})\quad \quad \textrm{(Kerr)}
 \end{equation}
where, by \cref{thm:horizontheorem}, $\vNext(\mc{H}_{\textrm{R}}^-,\s_{0})$ is a Type II$_{\infty}$ algebra of dressed observables and so the full algebra $\vNext(\mc{R},\s_{0})$ is also Type II$_{\infty}$. The von Neumann entropy of a ``classical-quantum'' state $\ket{\hat{\s}}$ where the wavefunction $f(\delta^2 \mathcal{Q}_-/4G_{\textrm{N}}\beta)$ is ``slowly varying'' is 
\begin{equation}
\label{introeq:SvnKerr}
S_{\textrm{vN.}}(\rho_{\hat{\s}}) \simeq S_{\textrm{gen.}} +S(\rho_{f}) + C \quad \quad \textrm{(Kerr)},
\end{equation}
where now $\rho_{f}=|f(\delta^2 \mathcal{Q}_-/4G_{\textrm{N}}\beta)|^{2}$. We additionally demonstrate that dressing, instead, to the charges at $i^0$ yields a unitarily equivalent algebra.

Finally, in \cref{sec_dS}, we consider the case of Schwarzschild-de Sitter spacetime, which has two bifurcate Killing horizons that we label as $\mc{H}_{1}$ for the black hole horizon and $\mc{H}_{2}$ for the cosmological horizon (see figure~\ref{fig:SdS_tikz}). The isometry group of the region $\mc{R}$ bounded by the black hole and the cosmological horizon is $\bb{R}\times \textrm{SO}(3)$. We will again focus on the group of time translations because the inclusion of the rotations leaves the following results qualitatively unchanged. The von Neumann algebra in $\mc{R}$ can again be decomposed as 
\begin{equation}
\mf{A}(\mc{R},\s_{0})\cong \mf{A}(\mc{H}^{-}_{1,\textrm{R}},\s_{0})\otimes  \mf{A}(\mc{H}^{+}_{2,\textrm{L}},\s_{0})\quad \quad \textrm{(Schwarzschild-de Sitter)},
\end{equation}
where, by our conventions presented in \cref{subsec:KillHor}, $\mc{H}^{+}_{2,\textrm{L}}$ is the past cosmological horizon of $\mc{R}$ and we note that both $\mf{A}(\mc{H}^{-}_{1,\textrm{R}},\s_{0})$ and $\mf{A}(\mc{H}^{+}_{2,\textrm{L}},\s_{0})$ are Type III$_{1}$ algebras. For a black hole formed from collapse in de Sitter, the spacetime is closed and so one must again include another degree of freedom that we model as an ``observer'' with (bounded) energy $\varepsilon$. Following the analysis in de Sitter spacetime, one could attempt to ``solve'' the global constraints by dressing the quantum field to the observer. However, there are two physical, gravitational charges in $\mc{R}$ corresponding to the perturbed area of both the black hole and cosmological horizons which actually impose stronger constraints on the algebra of observables. The observer perturbs the black hole and cosmological horizon by 
\begin{equation}
\label{eq:introA1A2e}
\frac{\delta^{2}\mathcal{Q}_{1,-}}{4G_{\textrm{N}}\beta_{1}}+\frac{\delta^{2}\mathcal{Q}_{2,-}}{4G_{\textrm{N}}\beta_{2}} = -\varepsilon. 
\end{equation}
where $\beta_{1}$ is the inverse temperature of the black hole and $\beta_{2}$ is the inverse temperature of the cosmological horizon. In contrast to the case of de Sitter spacetime, there are now two horizons that yield two independent gravitational charges in $\mc{R}$ given by $\delta^{2}\mathcal{Q}_{1,-}/4G_{\textrm{N}}\beta_{1}$ and $\delta^{2}\mathcal{Q}_{2,-}/4G_{\textrm{N}}\beta_{2}$ whose spectrum is constrained by the bounded energy of the observer. Letting $\Hilb_{\mathcal{Q}_{1}}=L^{2}(\bb{R})$ and $\Hilb_{\mathcal{Q}_{2}}=L^{2}(\bb{R})$, we quantize these degrees of freedom on the Hilbert space 
\begin{equation}
\Hilb_{\mathcal{Q}_{1},\mathcal{Q}_{2}} = P_{\varepsilon>0}[\Hilb_{\mathcal{Q}_{1}}\otimes \Hilb_{\mathcal{Q}_{2}}] 
\end{equation}
where $P_{\varepsilon>0}$ is a projection operator that projects states onto $\frac{\delta^{2}\mathcal{Q}_{2,-}}{4G_{\textrm{N}}\beta_{2}}<\frac{\delta^{2}\mathcal{Q}_{1,-}}{4G_{\textrm{N}}\beta_{1}}$ (i.e., $\varepsilon>0$ in \cref{eq:introA1A2e}).
The ``local constraints'' on each Killing horizon are of the form 
\begin{align}
-\int_{-\infty}^{\infty}dU\int_{\bb{S}^{2}}d\Omega_{2}~U\delta^{2}T_{UU} &= \frac{\delta^{2}\mathcal{Q}_{1,-}}{4G_{\textrm{N}}\beta_{1}} - \frac{\delta^{2}\mathcal{Q}_{1,+}}{4G_{\textrm{N}}\beta_{1}}\quad \textrm{ (on $\mc{H}_{1}^{-}$)} \label{eq:SdSflux1} \\
\int_{-\infty}^{\infty}dV\int_{\bb{S}^{2}}d\Omega~V\delta^{2}T_{VV} &= \frac{\delta^{2}\mathcal{Q}_{2,-}}{4G_{\textrm{N}}\beta_{2}} - \frac{\delta^{2}\mathcal{Q}_{2,+}}{4G_{\textrm{N}}\beta_{2}}\quad \textrm{ (on $\mc{H}_{2}^{+}$)} \label{eq:SdSflux2}
\end{align}
where $V$ is the affine parameter of the past cosmological horizon of $\mc{R}$. The extended algebra of observables $\vNext(\mc{R},\s_{0})$ in $\mc{R}$ that simultaneously satisfy both constraints is 
\begin{equation}
\vNext(\mc{R},\s_{0}) \cong P_{\varepsilon>0}\vNext(\mc{H}^{-}_{1,\textrm{R}},\s_{0}){\otimes}\vNext(\mc{H}^{+}_{2,\textrm{L}},\s_{0})P_{\varepsilon>0}.
\end{equation}
where $\vNext(\mc{H}^{-}_{1,\textrm{R}},\s_{0})$ and $\vNext(\mc{H}^{+}_{2,\textrm{L}},\s_{0})$ are the crossed product with respect to the modular automorphism of $\mf{A}(\mc{H}^{-}_{1,\textrm{R}},\s_{0})$ and $\mf{A}(\mc{H}^{+}_{2,\textrm{L}},\s_{0})$. By \cref{thm:horizontheorem}, $\vNext(\mc{H}^{-}_{1,\textrm{R}},\s_{0})$ and $\vNext(\mc{H}^{+}_{2,\textrm{L}},\s_{0})$ are both Type II$_{\infty}$ algebras and so $\vNext(\mc{R},\s_{0})$ is Type II$_{\infty}$. The von Neumann entropy of a ``classical-quantum'' state $\ket{\hat{\s}}$ with slowly varying wave function $f(\delta^{2}\mathcal{Q}_{1,-},\delta^{2}\mathcal{Q}_{2,-})$ is given by 
\begin{equation}
\label{eq:introSdS1}
S_{\textrm{vN.}}(\rho_{\hat{\s}})\simeq S_{\textrm{gen.}}+ S(\rho_{f})+ C \quad \quad \textrm{(Schwarzschild-de Sitter)} ,
\end{equation} 
where in Schwarzschild-de Sitter spacetime the generalized entropy 
\begin{equation}
\label{eq:introSdS2}
S_{\textrm{gen.}} = \frac{A_{1}}{4G_{\textrm{N}}}+\frac{A_{2}}{4G_{\textrm{N}}}+S_{\textrm{vN.}}(\rho_{\s})
\end{equation}
is the sum of the area of the black hole horizon and the cosmological horizon together with the von Neumann entropy of the state $\rho_{\s}$ in $\mc{R}$. 

\subsection{Summary of contents}
While the reader may be familiar with some (or all) of the technical background reviewed in \cref{sec_classical,sec_quantization} necessary for our main results, we have taken care to provide a closed-form discussion as it pertains to this work since these concepts are not necessarily common knowledge. This paper is organized as follows.
\begin{itemize}
 \item In \cref{subsec:KillHor}, we review the local and global structure of spacetimes with bifurcate Killing horizons. We define the ``physically relevant'' region $\mc{M}_{\textrm{R}}$ and state the assumptions \ref{assump1}--\ref{assump3} that we will make regarding the global structure of the spacetime $(\mc{M},g)$ with regards to the initial value problem on $\mc{M}_{\textrm{R}}$. In \cref{subsec:scfieldth}, we review the covariant phase space structure of a Klein-Gordon field on an arbitrarily curved spacetime as well as on a Killing horizon.
 \item In \cref{sec_quantization}, we review quantum field theory on curved spacetimes as well as the further necessary structures to study entropy on a general, curved spacetime. In \cref{subsec:AlgQFTCS} we review the algebraic viewpoint and the $\ast$-algebra quantization of a free Klein Gordon field in this framework. We review the notion of Hadamard states and KMS states. Additionally, we introduce assumption \ref{assump4} regarding the existence of a stationary state on $\mc{M}_{\textrm{R}}$. In \cref{subsec:quantKillingscri} we construct the $\ast$-algebra of the Klein-Gordon fields on a Killing horizon and null infinity. In \cref{subsec:vNalgebra}, we introduce the structure of von Neumann algebras and Tomita-Takesaki theory. In \cref{subsec:crossprod}, we review the crossed product of a von Neumann algebra with its modular automorphism group.
 
 \item In \cref{sec:dressed} we discuss the algebras of observables and their properties both in the context of quantum fields in curved spacetime and including perturbative gravitational constraints. In \cref{subsec:VNmodKilling} we describe the modular properties of algebras of quantum fields on Killing horizons. In \cref{subsec:charge} we consider perturbative quantum gravitational effects by treating the Klein-Gordon field as a first-order perturbation of the background geometry. We obtain second-order charge-flux relations associated to any diffeomorphism and evaluate these relations on a Killing horizon. In \cref{subsec:dressed}, we present a structure theorem on Killing horizons which illustrates the origin of Type II algebras when one imposes the gravitational constraints of the Killing isometry. In \cref{subsec:dens_ent} we analyze density matrices and von Neumann entropies for the Type II algebra of dressed observables on $\mc{H}^{-}$. We compute the von Neuman entropy of classical-quantum states and show that the von Neumann entropy can be interpreted as the generalized entropy up to a state-independent constant.
 \item In \cref{sec_stationary} we revisit the algebra of observables in the exterior of a Schwarzschild black hole in \cref{subsec:sads_kms} and the interior of the de Sitter cosmological horizon in \cref{subsec:deSitter} which were originally analyzed in \cite{Chandrasekaran:2022cip,2022arXiv220910454C}. We illustrate how Type II algebras arise within the context of \cref{thm:horizontheorem}. We show in \cref{sec:obs_open} that while an observer was necessary in de Sitter space to obtain a non-trivial algebra, we illustrate that the addition of an observer in Schwarzschild-AdS does not change the type of the algebra. 

 \item In \cref{sec_flat} we construct the algebra of observables in the exterior of an asymptotically flat Kerr black hole by dressing the fields to the (perturbed) area and angular momentum of the black hole. While there are no globally KMS states on this spacetime we show in \cref{subsec:algdressKerr}, using \cref{thm:horizontheorem}, that the algebra is the tensor product of a Type II$_{\infty}$ algebra on the past horizon and a Type I algebra on past null infinity. The von Neumann entropy of ``classical-quantum'' states on this algebra is the generalized entropy. 

 \item In \cref{sec_dS}, we consider Schwarzschild-de Sitter spacetime and construct the algebra of physical observables in the exterior of the black hole but the interior of the cosmological horizon. Since the spacetime is (spatially) closed we again introduce an observer. However, in contrast to the case of pure de Sitter spacetime, there are now two physical gravitational charges corresponding to the (perturbed) area of the black hole and cosmological horizon. These charges impose ``local constraints'' on the algebra and we show that the algebra of physical observables is Type II$_{\infty}$. Furthermore, the entropy of ``classical quantum'' states is equivalent to the generalized entropy (see \cref{eq:introSdS1,eq:introSdS2}). 
 
 \item In \cref{sec_AdS}, we apply our general formalism to the entropy of Kerr black holes in asymptotically AdS spacetimes. While such black holes do not admit KMS states above the so-called ``Hawking-Reall bound,'' our constructions straightforwardly apply to black holes in the full sub-extremal range. We again find that the algebra of observables in the exterior is Type II$_{\infty}$ and the von Neumann entropy of classical-quantum states is equivalent to the generalized entropy. {We revisit the microcanonical thermofield double state in AdS/CFT in \cref{subsec:adscft}, making contact with the bulk perspective taken throughout the rest of the paper.}
 
 \item {In \cref{sec_discussion}, we outline the direct generalization of our results to higher spacetime dimensions and higher derivative gravity theories. We also discuss various subjects that we believe are important, but for which our current understanding is not sufficiently rigorous.} 

 \item In appendix \ref{app:Unruh}, we explicitly construct the modular flow for the Unruh state in the exterior of a black hole in an asymptotically flat spacetime. We show that the modular flow is non-local everywhere in $\mc{R}$ except on the past horizon $\mc{H}_{\textrm{R}}^{-}$. In appendix \ref{app1:timequant}, we provide the asymptotic quantization of massive fields at timelike infinity. In \cref{sec:grav_EM_class} we present the classical phase space and quantization of linearized gravitational perturbations on an arbitrary curved spacetime. In \cref{app:cov_phase} we review the covariant phase space formalism and derive general ``charge-flux'' relations relating the ``energy'' of (first-order) fields to (second-order) boundary charges. 
\end{itemize}

\subsection*{\centering{Notation and conventions}}
\label{subsec:notation}

We work in natural units, $c=\hbar=1$, however, we will not set $G_{\textrm{N}}$ to unity. We will use the notation and sign conventions of \cite{Wald:1984rg}. In particular, our metric signature is ``mostly positive'' and our sign convention for curvature is such that the scalar curvature of a round sphere is positive. Lowercase Latin indices $(a,b,c,\dots)$ will be used to denote tensors in the ``bulk'' spacetime which will be generally denoted as $\mc{M}$ and $y$ will denote a choice of arbitrary coordinates on $\mc{M}$. We will append subscripts to $\mc{M}$ to denote ``preferred'' submanifolds of $\mc{M}$ (e.g., for an asymptotically flat black hole spacetime $\mc{M}_{\textrm{R}}\subset \mc{M}$ denotes the union of interior and ``right'' exterior regions of the black hole). We will denote $\mc{H}$ as a general Killing horizon, $\mc{H}^{\pm}$ to denote the future/past horizons of a bifurcate Killing horizon with bifurcation surface $\mc{B}$. We will denote $\mc{H}^{+}_{\textrm{R/L}}$ to denote the portion of $\mc{H}^{+}$ to the future/past of $\mc{B}$. Similarly, $\mc{H}^{-}_{\textrm{L/R}}$ denotes the portion of $\mc{H}^{-}$ to the future/past of $\mc{B}$. The bifurcate Killing horizon divides the spacetime into ``wedges'' which we denote as $\mc{R}, \mc{L}, \mc{F}$ and $\mc{P}$. The null normal to $\mc{H}^{+}$ will be denoted as $\ell^{a}$ and the null normal to $\mc{H}^{-}$ as $n^{a}$. 

We will use the symbol $\mc{P}$ to denote a classical phase space and $\Alg$ to denote a $\ast$-algebra of quantum field observables. The symbol $\s$ will denote a state on this algebra, $\Hilb$ will denote a general Hilbert space representation and $\Fock$ a Fock space representation. 
A general von Neumann algebra will be denoted as $\mf{A}$. A von Neumann algebra on a spacetime $\mc{M}$ associated to a Hilbert space with ``vacuum state'' $\s_{0}$ will be denoted as $\mf{A}(\mc{M},\s_{0})$. We will append the subscript ``ext.'' to distinguish algebras on which the gravitational constraints have been imposed. Thus, for example, $\vNext(\mc{R},\s_{0})$ denotes the von Neumann algebra of observables in $\mc{R}$ with vacuum $\s_{0}$ that commute with the gravitational constraints. 

A general element of $\Alg$ or $\mf{A}$ will be denoted as $a$. When referring to a specific quantum observable, we will refer to such objects by the boldfaced version of the symbol for the corresponding classical observable; for example, the quantum observable corresponding to a classical scalar field \(\phi\) is denoted by \(\op\phi\). 
In our constructions, there will be a preferred ``gravitational charge'' in $\mc{R}$ which generates the horizon Killing field isometry. We will generally label this particular charge as $\op{X}$. The corresponding charge associated to this isometry in the causal complement of $\mc{R}$ will be labeled $\op{C}$. For example, in \cref{eq:introTUUA+A-}, $\op{X}=\delta^{2}\op{\mathcal{Q}}_{-}/4G_{\textrm{N}}\beta$ and $\op{C}=\delta^{2}\op{\mathcal{Q}}_{+}/4G_{\textrm{N}}\beta$. 

In an asymptotically flat spacetime, in order to analyze the behavior of massive/massless fields at timelike/null infinity, we will work in the Penrose conformal completion (see, e.g., \cite{Wald:1984rg,Geroch-asymp}). The conformal boundaries will be denoted at $\scri^{\pm}$ for future/past null infinity, $i^{\pm}$ for future/past timelike infinity and $i^{0}$ for spatial infinity. We will denote the conformal factor as $\Omega$ and impose the Bondi condition $\nabla_{a}\nabla_{b}\Omega=0$ at $\scri^{\pm}$. The null normal to $\scri^{\pm}$ will be denoted as $\tilde{n}^{a}=\nabla^{a}\Omega$.

We will often consider ``down'' index tensors on $\mc{H}^{-}$ which are orthogonal to $n^{a}$. We will denote such tensors with capital Latin letters $(A,B,C,\dots)$. For example, the degenerate metric on $\mc{H}^{-}$ satisfies $q_{ab}n^{b}=0$ and will be denoted as $q_{AB}$. We will also use such indices to denote the equivalence class of ``up'' index tensors on $\mc{H}^{-}$ where two tensors are equivalent if they differ by a multiple of $n^{a}$. On this equivalence class, the inverse metric $q_{AB}$ is non-degenerate with unique inverse $q^{AB}$. Similar conventions will apply to tensors on $\scri^{-}$ orthogonal to the null normal $\tilde{n}^{a}$. We will use coordinates $(U,x^{A})$ on $\mc{H}^{-}$ where $U$ is the affine parameter of $n^{a}$ and $x^{A}$ are arbitrary coordinates on the $2$-sphere cross-sections of $\mc{H}^{-}$. On $\scri^{-}$ we will use coordinates $(v,x^{A})$ where $v$ is an advanced time coordinate on $\scri^{-}$ and $x^{A}$ now denote coordinates on the $2$-sphere cross-sections of $\scri^{-}$.

\section{Classical Phase Space: Local Observables and Description on Killing Horizons}
\label{sec_classical}
The quantum theory of linear fields on curved spacetime is based upon the phase space structure of the classical theory. We shall be primarily interested in spacetimes with isometries that generate ``bifurcate Killing horizons.'' We will therefore first review the general structure of such spacetimes in \cref{subsec:KillHor}. In \cref{subsec:scfieldth}, we review the basic structure of the classical phase space, with emphasis on the precise description of local observables, and their corresponding representation in terms of initial data on Cauchy surfaces and Killing horizons. 

\subsection{Killing Horizons: Local and Global Structure}
\label{subsec:KillHor}
Our main interest in this paper will be spacetimes $(\mc{M},g)$ with a bifurcate Killing horizon. We now review the general structure of a bifurcate Killing horizon as well as the global properties of spacetimes with a Killing horizon.
We refer the reader to \cite{Kay_1988} for more details and proofs of certain lemmas reviewed in this subsection. 

Let $(\mc{M},g)$ be a $4$-dimensional, time-orientable spacetime and let $\mc{H}$ be a null hypersurface (i.e., a smooth codimension-one submanifold of $\mc{M}$ whose null normal is a smooth, future-directed null vector field $\xi^{a}$). The null generators of $\mc{H}$ are the null geodesic integral curves of $\xi^{a}$. We note that, for a general null surface, we will not assume that the null geodesics are affinely parameterized or that these null geodesics are complete. We further assume that the null surface $\mc{H}$ is locally diffeomorphic to the product $\mc{H}\simeq \mc{S}\times \bb{R}$ where $\mc{S}$ is the (spacelike) manifold of integral curves of $\xi^{a}$. The surface gravity $\kappa$ of $\xi^{a}$ measures the failure of $\xi^{a}$ to be affinely parameterized 
\begin{equation}
\label{eq:xinonaff}
\xi^{a}\nabla_{a}\xi^{b}\defn \kappa \xi^{b}.
\end{equation}
The submanifold $\mc{H}$ inherits a (degenerate) metric $q_{ab}$ which satisfies $q_{ab}\xi^{b}=0$. If $\ell^{a}$ corresponds to the affinely parameterized null vector tangent to the null geodesic integral curves of $\mc{H}$ (i.e., $\ell^{a}$ satisfies $\ell^{a}\nabla_{a}\ell^{b}=0$), then we define the second fundamental form $K_{ab}$ of the null surface $\mc{H}$ relative to $\ell^{a}$ as
\begin{equation}
K_{ab}\defn \frac{1}{2}\pounds_{\ell}q_{ab}.
\end{equation}
Finally it will be useful to define the ``rotational one-form'' on spacetime\footnote{By Frobenius' theorem, $\s_{a}$ ``measures'' the failure of $\xi^{a}$ to be hypersurface orthogonal.}
\begin{equation}
\label{eq:twist}
\s_{a}\defn \epsilon_{abcd}\xi^{b}\nabla^{c}\xi^{d}.
\end{equation} 

Since the tensor fields $\s_{a}$, $K_{ab}$ and $q_{ab}$ are orthogonal to $\xi^{a}$ we can view these quantities as tensors on cross-sections $\mc{S}$ of $\mc{H}$.\footnote{\label{foot:AB} For any $p\in \mc{H}$, let $\mc{V}_{p}$ be the space of vectors at $p$ which differ by multiples of $\xi^{a}$ and let $\mc{V}_{p}^{\ast}$ be the space of dual vectors $\mu_{a}$ at $p$ which satisfy $\mu_{a}\xi^{a}=0$. Then one can straightforwardly construct the space $\mathsf{W}_{p}$ of such tensors at $p$ and, thereby, the space of such tensor fields on $\mc{H}$.} Therefore, we may identify such tensors with tensor fields on any cross-section $\mc{S}$ whose coordinates we label with capital Latin indices, so we will denote the tensors $q_{ab}$, $K_{ab}$ and $\s_{a}$ on $\mc{H}$ as $q_{AB}$, $K_{AB}$ and $\s_{A}$ respectively. The shear $\sigma_{AB}$ and the expansion $\theta$ of the null surface $\mc{H}$ are defined through the decomposition of the second fundamental form 
\begin{equation}
\label{eq:expshear}
K_{AB}\defn \frac{1}{2}\theta q_{AB}+\sigma_{AB}
\end{equation}
where $q^{AB}\sigma_{AB}=0$. 
$\s_{a}$ is clearly orthogonal to $\xi^{a}$ and so its pullback may be expressed as $\s_{A}$ in our conventions. A null hypersurface $\mc{H}$ is a {\em Killing horizon} if the null normal $\xi^{a}$ corresponds to the pull-back of a Killing vector field on spacetime (i.e., on the spacetime, $\xi^{a}$ satisfies $\pounds_{\xi}g_{ab}=0$). Taking the pull-back of Killing's equation immediately yields that $K_{ab}$ must vanish and therefore, 
\begin{equation}
\label{eq:expshearbckgrnd}
\theta=0 \textrm{ and }\sigma_{AB} =0\quad \quad \quad \textrm{ (Killing horizon)}
\end{equation} 
on any Killing horizon. However the ``rotational one-form'' $\s_{A}$ is generally nonvanishing (e.g., $\s_{A}$ is non-vanishing on the horizon of a Kerr black hole). Finally, due to a theorem by Bardeen, Carter, and Hawking, it can be shown that the surface gravity $\kappa$ is constant on $\mc{H}$ in general relativity with any matter satisfying the dominant energy condition \cite{bardeen1973four}. 

\begin{figure}
 \centering
 \includegraphics[]{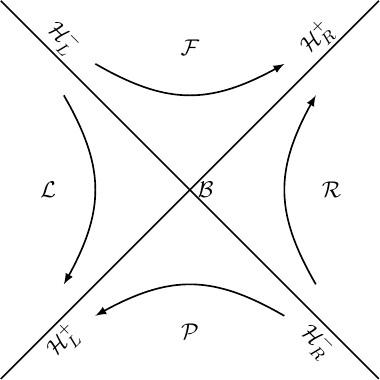}
 \caption{A spacetime diagram depicting a bifurcate Killing horizon given by the union of null surfaces $\mc{H}^{-}\cup \mc{H}^{+}$ where $\mc{H}^{-}=\mc{H}^{-}_{\textrm{R}}\cup \mc{H}^{-}_{\textrm{L}}$ and $\mc{H}^{+}=\mc{H}^{+}_{\textrm{R}}\cup \mc{H}^{+}_{\textrm{L}}$. The horizons intersect at the bifurcation surface $\mc{B}$. The orbits of the Killing field locally around the bifurcation surface are shown. The Killing horizon divides the spacetime into four regions $\mc{L}, \mc{R},\mc{P}$ and $\mc{F}$ as shown.}. 
 \label{fig:bifurcate_tikz}
\end{figure}

In this paper, we will be interested in spacetimes with (at least one) ``bifurcate Killing horizon''. Any neighborhood of a (connected) Killing horizon $\mc{H}$ with constant surface gravity $\kappa>0$ can be extended to a spacetime with a bifurcate Killing horizon \cite{Racz:1992bp}. So there is no loss of generality in considering such spacetimes. We assume that the spacetime $(\mc{M},g)$ posses a (nontrivial) one-parameter group of isometries $\chi_{t}$ generated by a spacetime Killing vector field $\xi^{a}$. We shall {\em not} assume $\xi^{a}$ to be hypersurface orthogonal on $\mc{M}$. Furthermore, we assume that these isometries leave fixed every point of a smooth, spacelike, codimension-two surface $\mc{B}$,\footnote{{In this section, we are considering $\mc{B}$ to be a connected Riemannian manifold. Later, we will consider spacetimes with multiple disconnected bifurcation surfaces, namely in Schwarzschild-de Sitter.}} which we will refer to as the ``bifurcation surface.'' Let $n^{a}$ and $\ell^{a}$ be a continuous choice of future-directed null vector fields on $\mc{B}$ which are orthogonal to $\mc{B}$. These vector fields generate a ``pair'' of Killing horizons in the following way. We extend $\ell^{a}$ off of $\mc{B}$ by choosing $\ell^{a}$ to be tangent to the congruence of affinely parameterized null geodesics with initial values given on $\mc{B}$. Since the isometries map each null curve into itself,\footnote{This follows from the fact that, under the isometry, any point $p\in\mc{B}$ is left fixed and $\ell^{a}$ remains null.} it follows that the Killing vector $\xi^{a}$ must be proportional to $\ell^{a}$. The resulting null surface $\mc{H}^{+}$ is a Killing horizon with $\mc{H}^{+} \simeq \mc{B}\times \bb{R}$ and null normal $\xi^{a}$ which satisfies
 \begin{equation}
 \label{eq:xifl}
 \xi^{a}=f \ell^{a} \quad \quad \quad \textrm{ (on $\mc{H}^{+}$)} \ ,
 \end{equation}
for some real-valued function $f$. We will denote the portion of $\mc{H}^{+}$ to the future of $\mc{B}$ as $\mc{H}_{\textrm{R}}^{+}$ and the portion of $\mc{H}^{+}$ to the past of $\mc{B}$ as $\mc{H}_{\textrm{L}}^+$. Repeating the same construction with $n^{a}$ yields another Killing horizon $\mc{H}^{-}$ and we denote the portions of $\mc{H}^{-}$ to the casual future and past of $\mc{B}$ as $\mc{H}_{\textrm{L}}^{-}$ and $\mc{H}_{\textrm{R}}^{-}$ respectively. The union $\mc{H}^{+}\cup \mc{H}^{-}$ will be referred to as a {\em bifurcate Killing horizon}. Since $\xi^{a}$ vanishes on $\mc{B}$ but $\nabla_{[a}\xi_{b]}$ is non-vanishing, it directly follows that, at least in a local neighborhood of $\mc{B}$, the orbits of the isometry $\chi_{t}$ will have the form of a ``boost around the surface $\mc{B}$'' as depicted in Figure \ref{fig:bifurcate_tikz}. 

It will also be useful to introduce coordinates on the horizons $\mc{H}^{+}$ and $\mc{H}^{-}$. We will initially define these coordinates near the bifurcation surface $\mc{B}$ however, as we will show, these coordinates will ultimately be defined on the full bifurcate Killing horizon. The vector field $\ell^{a}$ is an affinely parameterized vector field on $\mc{H}^{+}$ and let $V$ denote the affine parameter on the null geodesics (i.e., $V$ satisfies $\ell^{a}\nabla_{a}V=1$). We choose the parameter such that $V=0$ on $\mc{B}$. Furthermore, let $x^{A}$ be an arbitrary choice of coordinates on $\mc{B}$. We can extend these coordinates to all of $\mc{H}^{+}$ by parallel transport along the null generators, $\ell^{a}\nabla_{a}x^{A}=0$. This defines a local coordinate system $(V,x^{A})$ on $\mc{H}^{+}$ near $\mc{B}$. A similar construction can be done for the vector field $n^{a}$ with affine parameter $U$ 
 and, using $n^{a}$ to carry the coordinates $x^{A}$ on $\mc{B}$ to $\mc{H}^{-}$ thereby yields the coordinates $(U,x^{A})$ on $\mc{H}^{-}$ near $\mc{B}$. 

 The discussion thus far has focused on the local properties of a bifurcate Killing horizon in the neighborhood of the bifurcation surface $\mc{B}$. We now review some global properties of spacetimes with bifurcate Killing horizons. First, the local coordinates $(V,x^{A})$ and $(U,x^{A})$ extend to global coordinates on $\mc{H}^{+}$ and $\mc{H}^{-}$. The Killing orbits, by assumption, are complete and since $\kappa$ is constant it follows from \cref{eq:xinonaff,eq:xifl} that $\ell^{a}\nabla_{a}f$ is also constant along each generator and $\ell^{a}\nabla_{a}f>0$ at $\mc{B}$. Thus $f$ strictly increases along each generator and so $\xi^{a}$ cannot vanish anywhere on $\mc{H}^{+}$ or $\mc{H}^{-}$ except at $\mc{B}$. Therefore the null geodesic generators of a bifurcate Killing horizon are complete. If $v$ denotes the ``Killing parameter'' on $\mc{H}^{+}_{\textrm{R}}$ (i.e., $v$ satisfies $\xi^{a}\nabla_{a}v=1$) chosen such that $v=0$ on the surface $V=1$ then it directly follows from \cref{eq:xifl,eq:xinonaff} that 
\begin{equation}
V = e^{\kappa v}\quad \quad \quad \textrm{ (on $\mc{H}_{\textrm{R}}^{+}$).}
\end{equation}
If we denote $v^{\prime}$ as the ``Killing parameter'' of the null geodesics of $\xi^{a}$ on $\mc{H}_{\textrm{L}}^{+}$ then, by a similar analysis, the Killing parameter is related to the affine parameter on $\mc{H}_{\textrm{L}}^{+}$ by 
\begin{equation}
V = - e^{\kappa v^{\prime}}\quad \quad \quad \textrm{ (on $\mc{H}_{\textrm{L}}^{+}$).}
\end{equation}
where again we have chosen $v^{\prime}$ to vanish on the surface $V=-1$. The range of the Killing parameters are $-\infty < u < \infty$ and $-\infty < u^{\prime}<\infty$ and therefore the null generators are future and past complete. A exactly similar construction can be given for $\mc{H}^{-}$ where we denote $U$ as the affine parameter on the horizon, $u$ to be Killing parameter on $\mc{H}_{\textrm{R}}^{-}$ and $u^{\prime}$ to be the Killing parameter on $\mc{H}_{\textrm{L}}^{-}$. We obtain 
\begin{equation}
\label{eq:H-affkill}
U = -e^{-\kappa u} \quad \quad \textrm{ (on $\mc{H}_{\textrm{R}}^{-}$)}, \quad \quad \textrm{ and }\quad \quad U = e^{-\kappa u^{\prime}}\quad \quad \textrm{(on $\mc{H}_{\textrm{L}}^{-}$)}.
\end{equation}
where, again, we have chosen $U=0$ at $\mc{B}$ and $u=0~(u^{\prime} = 0)$ at the surface $U=1~(U = -1)$. Therefore, the coordinates $(V,x^{A})$ and $(U,x^{A})$ are well-defined coordinates on all of $\mc{H}^{+}$ and $\mc{H}^{-}$ respectively.

The bifurcate Killing horizon globally divides the spacetime into the regions $\mc{F}$, $\mc{P}$, $\mc{L}$ and $\mc{R}$ as shown in Figure \ref{fig:bifurcate_tikz}. The regions $\mc{F}$ and $\mc{P}$ are globally defined by 
\begin{equation}
\mc{F} \defn D^{+}(\mc{H}_{\textrm{R}}^{+}\cup \mc{H}_{\textrm{L}}^{-})\quad \textrm{ and }\quad \mc{P} \defn D^{-}(\mc{H}_{\textrm{R}}^{-}\cup \mc{H}_{\textrm{L}}^{+})
\end{equation}
where $D^{\pm}(S)$ denotes the future/past domain of dependence of any achronal set $S$. Furthermore, the regions $\mc{R}$ and $\mc{L}$ are globally defined by 
\begin{equation}
\mc{R} \defn I^{-}(\mc{H}_{\textrm{R}}^{+})\cap I^{+}(\mc{H}_{\textrm{R}}^{-}) \quad \textrm{ and }\quad \mc{L}\defn I^{-}(\mc{H}_{\textrm{L}}^{-})\cap I^{+}(\mc{H}_{\textrm{L}}^{+})
\end{equation}
where $I^{\pm}(S)$ denotes the chronological future/past of $S$. The intersections of the regions are $\mc{F}\cap \mc{P}=\mc{B}$ and $\mc{R}\cap \mc{L}=\varnothing$. 

We will be interested in the global evolution of quantum fields given initial data that is (at least) specified on $\mc{H}^{-}$ or $\mc{H}^{+}$. Such an evolution is well-defined for any globally hyperbolic\footnote{\label{fn:globalhyp} A spacetime $(\mc{M},g)$ is called ``globally hyperbolic'' if it has a Cauchy surface $\Sigma$, i.e., a spacelike or null codimension-one surface such that every inextendible causal curve hits $\Sigma$ precisely once, see \cite{Wald:1984rg} for details.} spacetime. By lemma 2.1 of \cite{Kay_1988}, the regions $(\mc{L},g)$ and $(\mc{R},g)$ are, in fact, globally hyperbolic and therefore one can expect to have a well-defined evolution and classical phase space in these regions as will be described in \cref{subsec:scfieldth}. If $\Sigma_{\textrm{R}}$ is a Cauchy surface of $\mc{R}$ and $\Sigma_{\textrm{L}}$ is a Cauchy surface of $\mc{L}$, then we will also be interested in the regions 
\begin{equation}
\label{eq:MRML}
\mc{M}_{\textrm{R}} \defn D(\mc{H}^{-}_{\textrm{L}}\cup \Sigma_{\textrm{R}}) \quad \textrm{ and }\quad \mc{M}_{\textrm{L}} \defn D(\mc{H}^{+}_{\textrm{R}}\cup \Sigma_{\textrm{L}}).
\end{equation}
where $D(S)=D^{+}(S)\cup D^{-}(S)$. 

We will now state our assumptions regarding the spacetime $\mc{M}$ as well as the regions $\mc{M}_{\textrm{R}}$ and $\mc{M}_{\textrm{L}}$. To avoid repetition, we will explicitly state our assumptions for $\mc{M}_{\textrm{R}}$ but similar assumptions shall apply to $\mc{M}_{\textrm{L}}$ with $\mc{H}^{-}$ replaced with $\mc{H}^{+}$. The main conditions are:
\begin{enumerate}
\item $(\mc{M},g)$ contains a bifurcate Killing horizon $\mc{H}^{-}\cup \mc{H}^{+}$ with bifurcation surface $\mc{B}$. \label{assump1}

\item The spacetime region $\mc{M}_{\textrm{R}}$ is globally hyperbolic.
 \label{assump2}

\item For linear fields of compact spatial support in $\mc{M}_{\textrm{R}}$, the initial data can be specified either on $\mc{H}^{-}$ or on a surface $\Sigma =\mc{H}^{-}\cup \Sigma^{\prime}$ where, in this case, the initial data can be independently specified on $\mc{H}^{-}$ and $\Sigma^{\prime}$ respectively. \label{assump3}
\end{enumerate}

\begin{figure}
 \centering
 \includegraphics[]{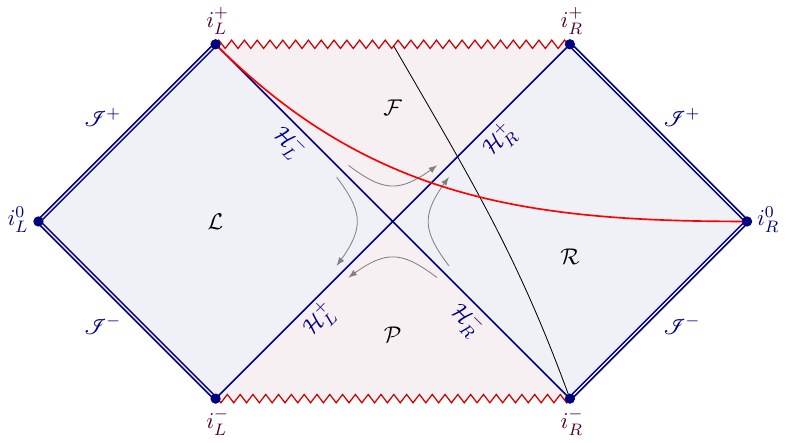}
 \caption{For an asymptotically flat spacetime, $\scri^{+}$ and $\scri^{-}$ denote the future and past null infinities of the spacetime, $i^{+}$ and $i^{-}$ denote the future and past timelike infinities and $i^{0}$ denotes spatial infinity. We also use the label $\textrm{L}$ and $\textrm{R}$ to denote whether these labels refer to the ``left'' or ``right'' asymptotically flat regions. The black surface represents the surface of a collapsing star and the red spacelike surface is a Cauchy surface for $\mc{M}_{\textrm{R}}=\mc{R}\cup \mc{F}$.}
 \label{fig:HH_tikz}
\end{figure}

In all cases of interest in this paper, we will consider spacetimes with bifurcate Killing horizons which contain a subregion $\mc{M}_{\textrm{R}}$ that is globally hyperbolic. However, for our general analysis and theorem, we state these assumptions \ref{assump1} and \ref{assump2} outright. We will sometimes refer to the region $\mc{M}_{\textrm{R}}$ as the ``physically relevant'' region. For example, for a black hole formed from gravitational collapse as depicted in \cref{fig:HH_tikz}, the region $\mc{M}_{\textrm{R}}$ will correspond to the black hole interior and exterior whereas the region $\mc{L}\cup \mc{P}$ is not physically relevant. Assumption \ref{assump3} is equivalent to assuming that the past evolution of initial data of compact spatial support in $\mc{M}_{\textrm{R}}$ decays sufficiently rapidly in the asymptotic past. In particular, in this paper, the surface $\Sigma^{\prime}$ will either correspond to another Killing horizon or to the past null or timelike infinity of an asymptotically flat spacetime. While this assumption has been rigorously proven for massive and/or massless fields in nearly all spacetimes considered in this paper, we shall explicitly assume this when considering general spacetimes. 

We now provide some relevant examples of spacetimes with bifurcate Killing horizons to which theorem \ref{thm:horizontheorem} of \cref{sec:dressed} will apply. 
The most familiar example is Minkowski spacetime, where the Killing field corresponds to Lorentz boosts and the bifurcate Killing horizon $\mc{H}^{+}\cup \mc{H}^{-}$ corresponds to a pair of intersecting null planes and the bifurcation surface $\mc{B}\simeq \bb{R}^{2}$ is a spacelike two-plane. The regions $\mc{L}$ and $\mc{R}$ correspond to the so-called left and right ``Rindler wedges,'' the spacetime is globally hyperbolic and the union of the four ``wedges'' is the entire spacetime. We also have that $\mc{M}_{\textrm{R}}=\mc{F}\cup \mc{R}$ and $\Sigma=\mc{H}^{-}$ is a Cauchy surface\footnote{The Rindler horizon $\mc{H}^{-}$ does not admit the strict requirements of a Cauchy surface for $\mc{M}_{\textrm{R}}$ since $\mc{H}^{-}$ lies to the past/future of any point at future/past null infinity except for a single generator. However, as emphasized in \cite{Unruh_1984}, it will essentially be a Cauchy surface for determining the evolution of solutions to the wave equation. The solution on the ``missing generator'' can be determined from initial data on $\mc{H}^{-}$ by continuity.} for this region. Similarly, $\mc{M}_{\textrm{L}}=\mc{L}\cup \mc{P}$ with Cauchy surface $\Sigma = \mc{H}^{+}$.

Another familiar spacetime that contains a bifurcate Killing horizon is the maximally extended Schwarzschild spacetime. The global structure is given in figure \ref{fig:HH_tikz}. The regions $\mc{L}$ and $\mc{R}$ correspond now to two disjoint asymptotically flat spacetimes. The region $\mc{F}$ is the black hole region with horizons $\mc{H}_{\textrm{R}}^{+}\cup \mc{H}_{\textrm{L}}^{-}$, $\mc{P}$ is the white hole region with horizons $\mc{H}_{\textrm{L}}^{+}\cup \mc{H}_{\textrm{R}}^{-}$ and they both contain spacelike singularities. The horizon Killing field normalized so that it is a unit time translation at $\scri^{\pm}$ will be denoted as $\xi^{a}=t^{a}$. The bifurcation surface $\mc{B}\simeq \bb{S}^{2}$ has the topology of a $2$-sphere\footnote{In $d=4$ spacetime dimension, the spatial topology of any stationary black hole is topologically $\bb{S}^{2}$ \cite{Hawking:1971vc,Chrusciel:1994tr} whereas in higher dimensions the topology of stationary black holes need not be spherical (see e.g., \cite{Galloway:2011np}). The structure theorem presented in \cref{sec:dressed} will apply to higher dimensional spacetimes (see \cref{sec_discussion}).}. Again, the region $\mc{M}_{\text{R}}$ corresponds to $\mc{F}\cup \mc{R}$ but now with Cauchy surface $\Sigma = \mc{H}^{-}_{\textrm{R}}\cup i_{{\textrm{R}}}^{-}\cup \scri^{-}_{\textrm{R}}$ and analogously $\mc{M}_{\text{L}}=\mc{L}\cup \mc{P}$ with Cauchy surface $\Sigma = \mc{H}^{+}_{\textrm{L}}\cup i_{\textrm{L}}^{-}\cup \scri_{\textrm{L}}^{-}$. For a black hole (with no charge or angular momentum) formed from gravitational collapse, the black surface in figure \ref{fig:HH_tikz} denotes the typical trajectory of the surface of a collapsing star. Therefore, the region to the right of the surface in figure \ref{fig:HH_tikz} is physically relevant for the gravitational collapse spacetime. However, the region to the left of the surface of the star is ``covered up'' by the collapsing body and is not physically relevant. Therefore, even though we shall work on the maximally extended spacetime, we will only consider the behavior of classical and quantum fields in the region $\mc{M}_{\textrm{R}}=\mc{R}\cup \mc{F}$ corresponding to the exterior and interior of the black hole. 

\begin{figure}
 \centering
 \includegraphics[height = 6cm]{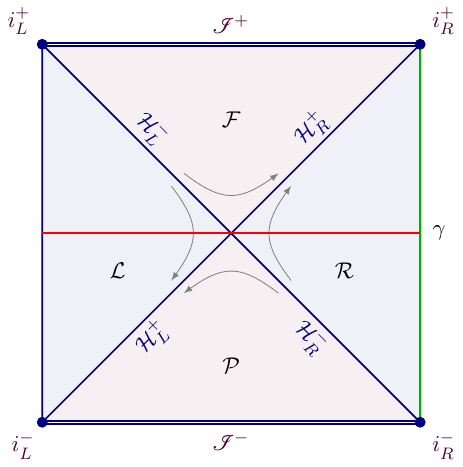} 
 \includegraphics[trim=-1cm 0 0 0,height = 5.5cm]{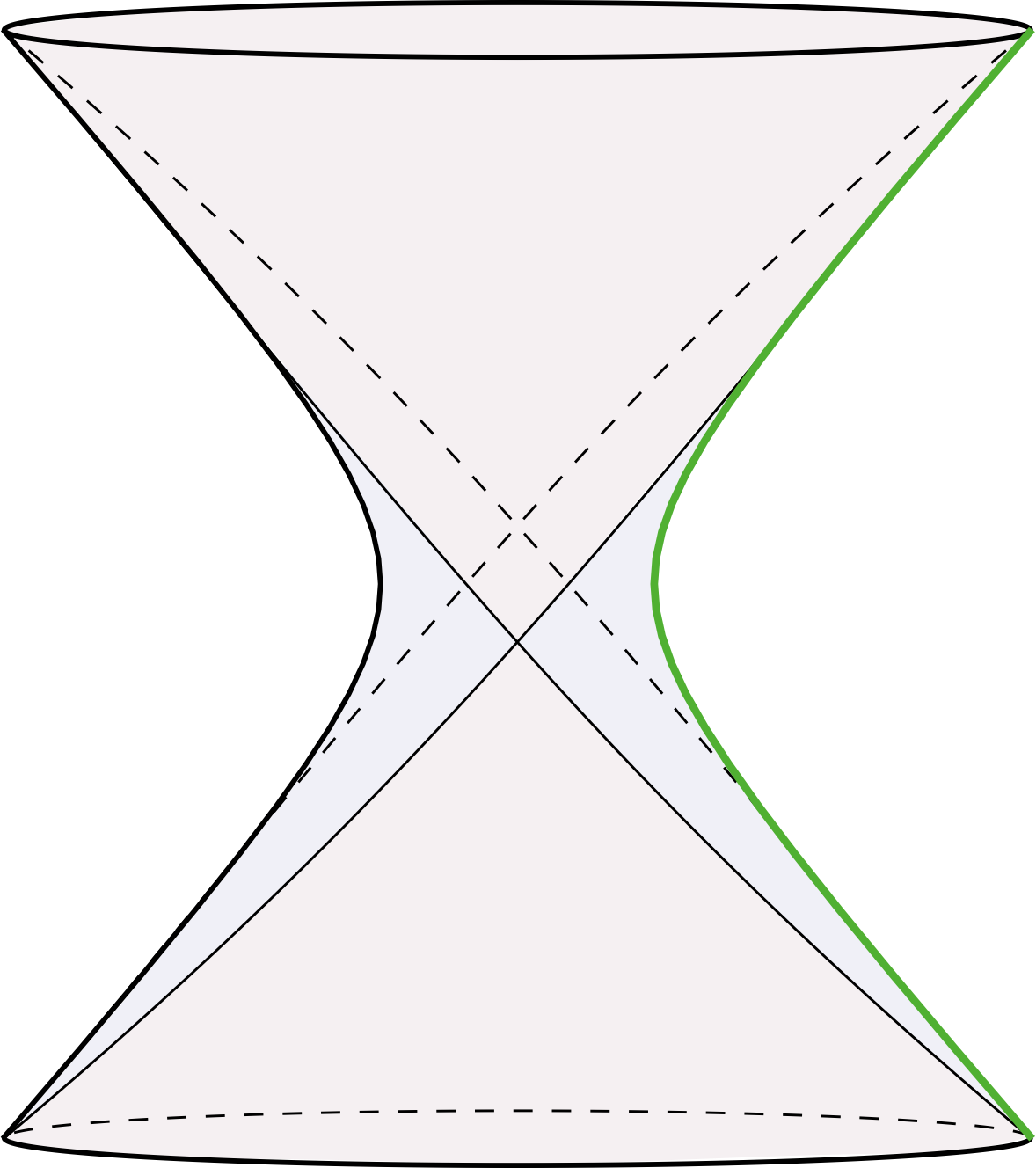}
 \caption{Left: The Penrose diagram for de Sitter space. The green line represents an inertial observer with $\mathcal{R}$ its associated static patch. The red spacelike surface is a Cauchy surface for the spacetime. Right: Embedding of de Sitter as a hyperboloid in flat space. Spatial slices are closed and increase in (proper) volume exponentially at late times.}
 \label{fig:static_patch}
\end{figure}

A third example of a spacetime with a bifurcate Killing horizon similar to the examples we have just given is de Sitter spacetime. The causal diagram of de Sitter spacetime is given in figure \ref{fig:static_patch}. Now the bifurcate Killing horizon is associated to a static world line $\gamma$, shown in green. To see the key differences of de Sitter spacetime compared to the previous examples, it is useful to embed the $4$-dimensional spacetime as a hyperboloid in $\bb{R}^{5}$ as depicted in figure \ref{fig:static_patch}. In contrast to the other cases, any spacelike Cauchy surface is closed with topology $\bb{S}^{3}$. The region $\mc{R}$ is known as the ``static patch'' and corresponds to $D^{+}(\gamma)\cap D^{-}(\gamma)$. The region $\mc{L}$ is the static patch of the antipodally reflected worldline. The full spacetime $\mc{M}$ is equivalent to the union of the four ``wedges'' and is globally hyperbolic. 

We now consider spacetimes whose global structure differs significantly from the above examples. One such example is the maximally extended {sub-extremal} Kerr spacetime (i.e., with $a<m$). While there are many similarities to the Penrose diagram of the maximally extended Schwarzschild spacetime, one key difference is that the singularities are now timelike and therefore the spacetime contains Cauchy horizons $\mc{H}_{\textrm{R}}^{\textrm{c}}$ and $\mc{H}_{\textrm{L}}^{\textrm{c}}$ respectively. The spacetime $\mc{M}$ is now not globally hyperbolic. In fact, any open neighborhood containing $\mc{H}_{\textrm{R}}^{\textrm{c}}$ or $\mc{H}_{\textrm{L}}^{\textrm{c}}$ is not globally hyperbolic. Consequently, our analysis will not apply to the full maximally extended spacetime. However, as in the case of the Schwarzschild black hole, the spacetime region $\mc{M}_{\textrm{R}}=\mc{F}\cup \mc{R}$ contains the ``physically relevant'' region for a Kerr black hole formed from gravitational collapse and is also globally hyperbolic. The surface $\mc{H}^{-}\cup i^{-}_{\textrm{R}}\cup \scri^{-}_{\textrm{R}}$ is a Cauchy surface for $\mc{M}_{\textrm{R}}$. The bifurcate Killing horizon $\mc{H}^{-}\cup \mc{H}^{+}$ is generated by the Killing vector $\xi^{a} = t^{a} + \Omega_{\textrm{H}}\psi^{a}$ where $\Omega_{\textrm{H}}$ is the angular velocity of the horizons $\mc{H}^{+}$ and $\mc{H}^{-}$ and $\psi^{a}$ is the spacelike rotational Killing field associated to azimuthal symmetries of the black hole. The Killing vector $\xi^{a}$ is not hypersurface orthogonal on $\mc{M}$ and therefore, the usual ``Euclidean methods'' for quantum field theory are inapplicable in this case. Additionally, in contrast to the case of Schwarzschild, there is no globally timelike Killing vector field in the regions $\mc{R}$ or $\mc{L}$. The regions in $\mc{L}$ and $\mc{R}$ where the vector field $t^{a}$ is spacelike are known as the ``ergoregion.''\footnote{The existence of an ergoregion directly implies that the classical Hamiltonian is not bounded from below, resulting in the well-known ``superradiant instabilities'' of the Kerr black hole.} In the quantum theory, the failure of $\xi^{a}$ to be globally timelike in $\mc{R}$ implies the non-existence of a globally stationary quantum state on the full, maximally extended spacetime \cite{Kay_1988}.

\begin{figure}
 \centering
 \includegraphics[height = 8cm]{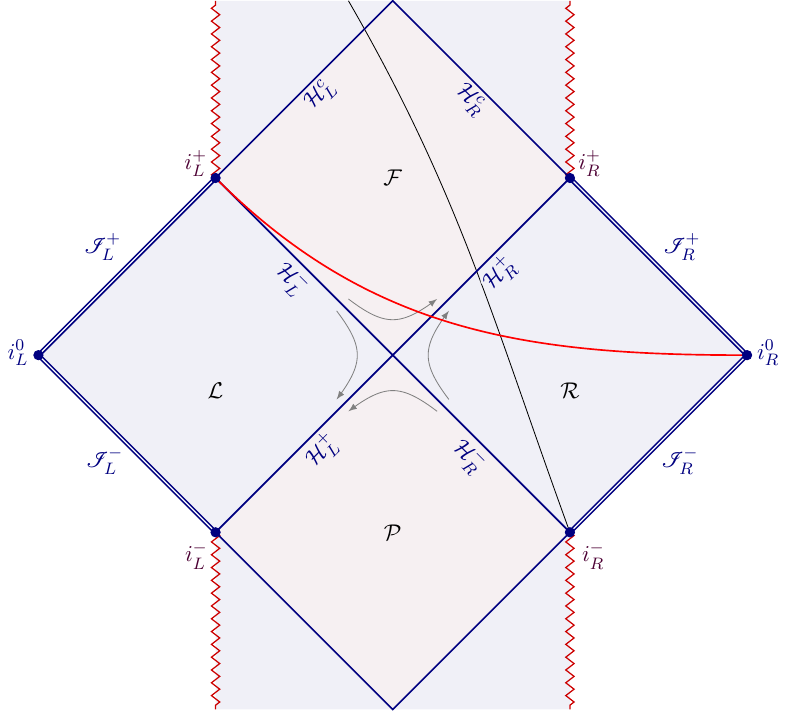}
 \caption{A portion of the maximally extended Kerr geometry. The black surface represents the surface of a collapsing star and the red spacelike surface is a Cauchy surface for the globally hyperbolic spacetime $\mc{M}_{\textrm{R}}=\mc{R}\cup \mc{F}$.}
 \label{fig:Kerr_tikz}
\end{figure}

\begin{figure}
 \centering
 \includegraphics[width = \textwidth]{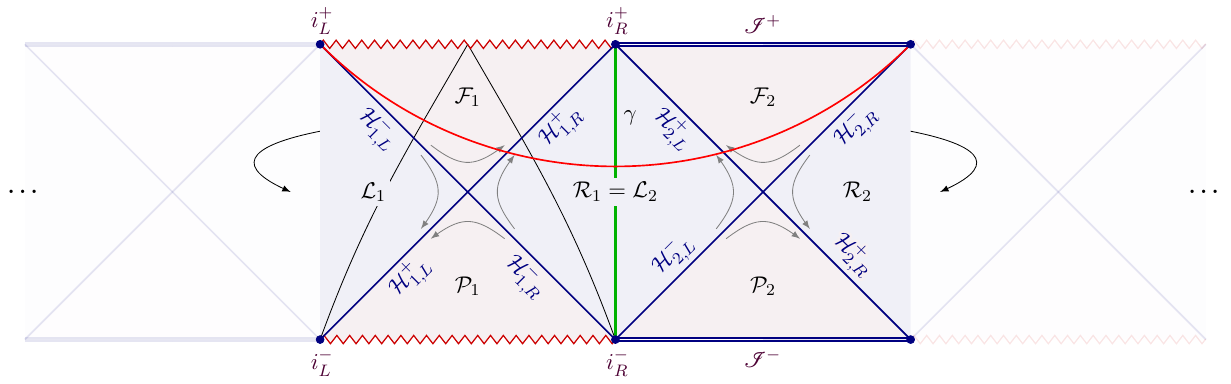}
 \caption{Penrose diagram of Schwarzschild-de Sitter spacetime. The maximally extended spacetime contains an infinite sequence of black holes and cosmological horizons depicted in the transparent regions. The spacetime of a single black hole can be obtained by identifying the timelike surfaces. The black surface denotes the surface of a (spherically symmetric) collapsing star. The green curve is the worldline $\gamma$ of an observer in $\mc{R}$ and the red line is a Cauchy slice for the physically relevant region $\mc{M}_{\textrm{R}}=\mc{F}_{1}\cup\mc{R}_{1}\cup \mc{F}_{2}$.}
 \label{fig:SdS_tikz}
\end{figure}

Finally, the last example we will review is Schwarzschild-de Sitter spacetime. The maximally extended spacetime contains an infinite number of bifurcate Killing horizons which alternate between a black hole horizon and a cosmological horizon. Due to this infinite chain of black holes and cosmological horizons, we must do more work to obtain the ``physical spacetime'' corresponding to a black hole formed from collapse in a (closed) de Sitter universe. Since the maximal extension clearly has a discrete $\bb{Z}$ isometry, one can straightforwardly obtain a spacetime of a single black hole by a quotient of $\mc{M}$ by this $\bb{Z}$ isometry \cite{Chrusciel:2020fql}. This is equivalent to identifying the (zero extrinsic curvature) timelike surfaces depicted in \cref{fig:SdS_tikz} \cite{2015CQGra..32b5004A}. We note that this spacetime is {\em closed} and contains compact spacelike hypersurfaces. We label the bifurcate Killing horizon of the black hole as $\mc{H}_{1}^{-}\cup \mc{H}_{1}^{+}$ and the ``neighboring'' bifurcate Killing horizon of the cosmological horizon as $\mc{H}_{2}^{-}\cup \mc{H}_{2}^{+}$. The horizons $\mc{H}_{1}^{-}\cup \mc{H}_{1}^{+}$ divide a portion of the spacetime into the regions $\mc{L}_{1},\mc{F}_{1},\mc{R}_{1}$ and $\mc{P}_{1}$ and $\mc{H}_{2}^{-}\cup \mc{H}_{2}^{+}$ divides a different portion of the spacetime into the regions $\mc{L}_{2},\mc{F}_{2},\mc{R}_{2}$ and $\mc{P}_{2}$ where now $\mc{R}_{1}=\mc{L}_{2}$ and $\mc{L}_{1}=\mc{R}_{2}$ as depicted in \cref{fig:SdS_tikz}. These neighboring Killing horizons have surface gravity $\kappa_{1}\geq \kappa_{2}$. The ``physical'' spacetime of a black hole formed from (spherical) collapse in de Sitter spacetime corresponds to the region\footnote{{Here we define $\mc{M}_{\rm R}$ with respect to the black hole bifurcate horizon.}} $\mc{M}_{\textrm{R}}=\mc{F}_{\textrm{1}}\cup \mc{R}_{1} \cup \mc{F}_2$ as depicted in \cref{fig:SdS_tikz}. It was proven in \cite{Kay_1988} that if the surface gravity of the two horizons are inequivalent (i.e., $\kappa_{1}\neq \kappa_{2}$) then there does not exist a globally stationary quantum state on Schwarzschild-de Sitter. However, in the identified spacetime, there does exist\footnote{
\label{foot:SdSpure}
A second proof given in \cite{Kay_1988} of the non-existence of a stationary state in the (unidentified) maximally extended Schwarzschild-de Sitter spacetime applies to the case of $\kappa_{1}=\kappa_{2}$. This proof uses the fact that such a state must be mixed when restricted to $\mc{L}_{1}$ or $\mc{L}_{2}$ and ``purified'' by its correlations with the right wedge $\mc{R}_{1}$ or $\mc{R}_{2},$ respectively. Since $\mc{R}_{1}=\mc{R}_{2},$ the state must be pure in both $\mc{L}_{1}\cup \mc{R}_{1}$ and $\mc{L}_{2}\cup \mc{R}_{1}.$ This is impossible if $\mc{L}_{1}\neq \mc{L}_{2}$. {This argument is evaded in the identified spacetime (since then $\mc{L}_{1} = \mc{L}_{2}$) where a stationary state does in fact exist.}} a globally stationary state in the case of $\kappa_{1}=\kappa_{2}$ (sometimes referred to as the Nariai solution) which can be obtained by ``Euclidean methods'' \cite{nariai1950some}.

\subsection{Klein Gordon Field}
\label{subsec:scfieldth}
In this paper, we will generally be concerned with linear fields on a globally hyperbolic (see \cref{fn:globalhyp}), curved spacetime $(\mc{M},g)$. For definiteness, we will consider a real scalar field $\phi$ with Lagrangian 
\begin{equation}
\label{eq:KG}
\mc{L}^{\textrm{KG}} = -\frac{1}{2}[\nabla^{a}\phi \nabla_{a}\phi + V\phi^2]
\end{equation}
where $V$ is any smooth function on $\mc{M}$.\footnote{{We leave $V$ as a general function for this section but, for concreteness, will later consider the cases $V = m^2$ and $V = 0.$}} Then $\phi$ satisfies 
\begin{equation}
\label{eq:EOMphi}
(\Box_{g} - V) \phi =0.
\end{equation}
where $\Box_{g}=g^{ab}\nabla_{a}\nabla_{b}$ and $\nabla_{a}$ is the covariant derivative compatible with $g_{ab}$. Since the spacetime is globally hyperbolic, the wave operator $\Box - V$ possesses a unique retarded and advanced propagator, which we denote as $A$ and $R$ respectively. For any test function $f:\mc{M}\to \bb{R}$, if $Af$ denotes the advanced solution to \cref{eq:EOMphi} with source $f$ and $Rf$ denotes the retarded solution with source $f$, then we may define the propagator 
\begin{equation}
 \label{eq:commutator}
 Ef = Af - Rf.
\end{equation}
which yields a source-free solution to \cref{eq:EOMphi}. The existence and uniqueness of $A$ and $R$ express the fact that the initial value problem of \cref{eq:EOMphi} is well-posed. The ``commutator function'' is often identified with a distributional kernel $E(y_{1},y_{2})$ such that for any test functions $f_{1}$ and $f_{2}$ 
\begin{equation}
 E(f_{1},f_{2}) = \int \sqrt{-g}d^{4}y~f_{1}(y)Ef_{2}(y).
\end{equation}
 The Lagrangian provides the theory with a phase space structure \cite{Lee:1990nz,Iyer:1994ys,Crnkovic:172498}. The symplectic product of two solutions $\phi_{1}$ and $\phi_{2}$ is given by\footnote{The symplectic form is a $2$-form on $\mc{P}$. In the case of a linear theory, $\mc{P}$ forms an affine space and so we may identify tangent vectors with points in $\mc{P}$.} 
\begin{equation}
\label{eq:symprod}
\symp^{\KG}_{\Sigma}(\phi_{1},\phi_{2})=\frac{1}{2}\int_{\Sigma}\sqrt{h}d^{3}x~[\phi_{2}n^{a}\nabla_{a}\phi_{1}-\phi_{1}n^{a}\nabla_{a}\phi_{2}]
\end{equation}
where $x$ denote a choice of arbitrary coordinates on the Cauchy surface $\Sigma$, $\sqrt{h}d^{3}x$ is the proper volume element and $n^{a}$ is the unit normal to $\Sigma$. The symplectic product is conserved (i.e., \cref{eq:symprod} is independent of the choice of Cauchy surface). The symplectic structure gives the space of initial data a phase space structure.\footnote{The phase space of a field theory is infinite-dimensional and has the structure of a Fr\'echet space \cite{Ashtekar:1981bq}. Since the suitable fall-off behavior of fields and topology of this space depends on the spacetime $(\mc{M},g)$ under consideration, we shall not attempt to give a more precise specification of $\mc{P}$.} The points of phase space $\mc{P}$ are given by specifying the pair $(\phi,n^{a}\nabla_{a}\phi)$ on any spacelike Cauchy surface $\Sigma$. Therefore, we can view $\symp^{\KG}$ as a bilinear map on $\mc{P}$. 

A classical observable is any smooth function $F$ on phase space (i.e., $F:\mc{P}\to \bb{R}$). If $\mc{P}$ were finite-dimensional, then $\symp_{\alpha \beta}$ is non-degenerate and there exists an inverse symplectic form $\symp^{\alpha \beta}$. The Poisson brackets of any two observables $F_{1}$ and $F_{2}$ can then be defined as 
\begin{equation}
\label{eq:PoissonF1F2}
\{F_{1},F_{2}\} = \symp^{\alpha\beta}\nabla_{\alpha}F_{1}\nabla_{\beta}F_{2}
\end{equation}
where Greek indices here refer to tangent vectors in phase space and, correspondingly, $\nabla_{\alpha}$ should be understood as a derivative on $\mc{P}$. However, on an infinite dimensional phase space $\symp$ is ``weakly non-degenerate'' (i.e. its inverse is not defined on all $\nabla_{\alpha}F$ for any $F$) and so one cannot straightforwardly obtain the Poisson bracket of any two observables $F_{1}$ and $F_{2}$ using \cref{eq:PoissonF1F2}. Nevertheless, there is a class of $F$'s which manifestly have a well-defined Poisson bracket. These correspond to observables that generate ``flows'' in phase space. Let $X^{\alpha}$ be a vector field on $\mc{P}$. Then the observable that generates the vector field $X^{a}$ is defined as 
\begin{equation}
\label{eq:FXobs}
\delta F \defn\symp(\delta z,X)
\end{equation}
for any smooth curves $z(\alpha)$ on $\mc{P}$ where $\delta z = dz/d\alpha\vert_{\alpha =0}$ and $\delta F = d F(z(\alpha))/d\alpha\vert_{\alpha =0 }$. The Poisson bracket of two observables $F_{1}$ and $F_{2}$ which generate the vector fields $X_{1}^{\alpha}$ and $X_{2}^{\alpha}$ is given by
\begin{equation}
\{F_{1},F_{2}\} = -\symp(X_{1},X_{2})
\end{equation}
and so the Poisson bracket can be defined for all such $F$. For the purposes of this paper, we shall only consider observables of this class since these are the only observables on $\mc{P}$ which will have a clear description in the corresponding quantum theory.

A particular class of observables that will be of primary importance for this paper is the local, smeared field observable 
\begin{equation}
\label{eq:foursmear}
\phi(f) \defn \int \sqrt{-g}d^{4}y~\phi(y)f(y)
\end{equation}
where $y$ denote a choice of arbitrary coordinates on $\mc{M}$ and $f\in C_{0}^{\infty}(\mc{M})$ is a test function. By lemma 3.2.1 of \cite{Wald_1995}, the smeared field observable can be manifestly expressed as a function on phase space
\begin{equation}
\label{eq:sympsmear}
\phi(f) = \symp_{\Sigma}^{\KG}(\phi,Ef)
\end{equation}
for any solution $\phi$. Thus, the ``smeared fields'' $\phi(f)$ are observables of the form \cref{eq:FXobs} with vector field $X$ which generates the infinitesimal map $\phi \to \phi + \epsilon Ef$ on phase space. It directly follows that the covariant Poisson brackets are given by 
\begin{equation}
\label{eq:Poissonbracket}
\{\phi(f_{1}),\phi(f_{2})\} = -\symp_{\Sigma}^{\KG}(Ef_{1},Ef_{2})= E(f_{1},f_{2})1
\end{equation}
where ``$1$'' is the identity on phase space (see, e.g., \cite{Dimock_1980,Wald_1995,Prabhu:2022zcr} for more details). 

By \cref{eq:sympsmear}, any local, four-dimensionally smeared field observable $\phi(f)$ can be equivalently expressed as a linear combination of ``three smeared'' observables in the following manner. For any test function $f$ of compact support, $F=Ef$ is a source-free solution to \cref{eq:EOMphi} with initial data on $\Sigma$, $[Ef]_{\Sigma}=s_{1}$ and $[n^{a}\nabla_{a}(Ef)]_{\Sigma}=s_{2}$. Conversely, by lemma 3.2.1 of \cite{Wald_1995}, given any solution $F$ to \cref{eq:EOMphi} there exists a test function $f$ --- unique up to the addition of $(\Box_{g}-V)g$ for any test function $g$ --- such that $F=Ef$. Therefore, if we define the linear functions $\varphi_{\Sigma}(s)$ on phase space corresponding to the ``three smeared'' observable  
\begin{equation}
\varphi_{\Sigma}(s) \defn \int_{\Sigma}\sqrt{h}d^{3}x~\phi(x)s(x)
\end{equation}
for any test function $s$ on $\Sigma$ then, by the above correspondence, it is equivalent to the ``four-smeared'' observable \cref{eq:sympsmear} by $\varphi_{\Sigma}(s) = -2\phi(f)$ where $f$ is a test function on $\mc{M}$ such that $[Ef]\vert_{\Sigma}=0$ and $[n^{a}\nabla_{a}(Ef)]_{\Sigma}=s$. We can similarly define the ``three-smeared'' observable $\pi_{\Sigma}(s)$ corresponding to $(n^{a}\nabla_{a}\phi)_{\Sigma}(x)$ smeared with the test function $s(x)$ on $\Sigma$ which can be related to the four-dimensionally smeared observable $\phi(f)$ where now $[Ef]_{\Sigma}=s$ and $[n^{a}\nabla_{a}(Ef)]_{\Sigma}=0$. Therefore, any field observable $\phi(f)$ can be equivalently expressed in terms of the ``three smeared'' observables on any spacelike Cauchy surface $\Sigma$ by
\begin{equation}
\label{eq:phivphipi}
\phi(f) = -\frac{1}{2}\varphi_{\Sigma}(s_{1}) +\frac{1}{2}\pi_{\Sigma}(s_{2}).
\end{equation}
where $s_{1}=[n^{a}\nabla_{a}Ef]\vert_{\Sigma}$ and $s_{2}=[Ef]\vert_{\Sigma}$. It's straightforward to show that the Poisson brackets \cref{eq:Poissonbracket} implies that, distributionally, $\{\pi_{\Sigma}(x_{1}),\varphi_{\Sigma}(x_{2})\}=\delta_{\Sigma}(x_{1},x_{2})1$ where $\delta_{\Sigma}$ is the delta-function on $\Sigma$ \cite{Wald_1995}. 

We now consider the case where the spacetime $(\mc{M},g)$ satisfies assumptions \ref{assump1}--\ref{assump3}. Since, by assumption, the spacetime $(\mc{M}_{\textrm{R}},g)$ is globally hyperbolic, the classical phase space $\mc{P}$ is well-defined on this submanifold. To describe the behavior of classical fields on the Killing horizon we may consider the symplectic form evaluated on a Cauchy surface $\Sigma$ of the form $\Sigma = \mc{H}^{-}\cup \Sigma^{\prime}$. By assumption \ref{assump3}, the initial data is independently specifiable on $\mc{H}^{-}$ and $\Sigma^{\prime}$ respectively. This implies that the symplectic form on $\Sigma$ can be expressed as the sum of the symplectic form on the space of initial data on $\mc{H}^{-}$ and the symplectic form on the space of initial data on $\Sigma^{\prime}$. Under these assumptions, the smeared field $\phi(f)$ for any $f\in C_{0}^{\infty}(\mc{M}_{\textrm{R}})$ can be expressed in terms of initial data on $\mc{H}^{-}$ and $\Sigma^{\prime}$
\begin{equation}
\label{eq:phifHS}
\phi(f) = \symp_{\Sigma}(\phi,Ef) = \symp_{\mc{H}^{-}}(\phi,Ef) + \symp_{\Sigma^{\prime}}(\phi,Ef).
\end{equation}
Where the first term on the right-hand side of \cref{eq:phifHS} corresponds to the ``three-smeared'' field observable smeared with initial data on $\mc{H}^{+}$ and the second term is the ``three-smeared'' field observable smeared with initial data on $\Sigma^{\prime}$. For any solution $\phi$, the initial data $\Phi(U,x^{A})$ on $\mc{H}^{-}$ 
\begin{equation}
\label{eq:phiH}
\Phi \defn \lim_{\mc{H}^{-}}\phi 
\end{equation}
is simply the limit of $\phi$ to the past horizon. The symplectic product of two solutions $\Phi_{1}$ and $\Phi_{2}$ on $\mc{H}^{-}$ is given by 
\begin{equation}
\label{eq:KGsympH}
\symp^{\KG}_{\mc{H}^{-}}(\Phi_{1},\Phi_{2}) = \frac{1}{2}\int_{\mc{H}^{-}}dUd\Omega_2~\bigg(\Phi_{2}\frac{\partial \Phi_{1}}{\partial U}-\Phi_{1}\frac{\partial \Phi_{2}}{\partial U}\bigg)
\end{equation}
where $d\Omega_2 = \sqrt{q}d^{2}x$ is the induced volume element on the constant $U$ cross-sections of $\mc{H}^{-}$. The convergence of the symplectic form implies that $\partial \Phi/\partial V$ decays as $O(1/|U|^{1+\epsilon})$ as $U\to \pm \infty$ for some $\epsilon>0$. It will also be convenient to define the normal derivative of $\Phi$ on $\mc{H}^{-}$
\begin{equation}
\label{eq:PiH}
\Pi \defn \partial_{U}\Phi.
\end{equation}
For any $s=Ef$ on $\mc{H}^{-}$, we have that 
\begin{equation} \label{eq:PiHorDefn}
\Pi(s) \defn \int_{\mc{H}^{-}}dUd\Omega_2~\frac{\partial \Phi}{\partial U}(U,x^{A})s(U,x^{A})=\symp^{\KG}_{\mc{H}^{-}}(\Phi,Ef)
\end{equation}
where, in the second equality, we have integrated by parts and assumed that the solution $s=Ef$ falls off sufficiently rapidly that we may neglect any boundary terms\footnote{The precise decay on the horizon of smooth solutions with initial data of compact spatial support depends upon the spacetime under consideration. For massless fields in Minkowski spacetime, $Ef$ will be of compact support on the Rindler horizon by Huygens' principle. Huygens' principle is violated in general curved spacetimes and so such solutions will have non-trivial decay on the horizon. Such decay has been analyzed in de Sitter \cite{Dappiaggi:2008dk}, Schwarzschild \cite{Dafermos:2010hd,Luk:2010jfs} as well as Reissner-Nordstrom de Sitter and Kerr-de Sitter \cite{Hintz:2015jkj}.}. Therefore, $\Pi(s)$ is an observable\footnote{We note that, by \cref{eq:phifHS}, $\Pi(s)$ is directly related to the local bulk observable $\phi(f)$. In contrast $\Phi(w)$ (i.e. $\Phi$ smeared with a test function $w$) will not correspond to a local bulk observable unless, of course, $w=\partial_{U}s$ for some test function $s$.} on $\mc{H}^{-}$ and, together with the observables $\symp^{\KG}_{\Sigma^{\prime}}(\phi,Ef)$, they are equivalent to the bulk field observables $\phi(f)$. The observables $\Pi(s)$ form a closed algebra with Poisson bracket 
\begin{align}
\label{eq:PBPi}
\{\Pi(s_{1}),\Pi(s_{2})\} =& -\symp^{\KG}_{\mc{H}^{-}}(Ef_{1},Ef_{2})1 \nonumber \\
=&\frac{1}{2}\int_{\mc{H}^{-}}dUd\Omega_2~\bigg(s_{1} \frac{\partial s_{2}}{\partial U}- s_{2}\frac{\partial s_{1}}{\partial U}\bigg)1
\end{align}
for any $s_{1}=Ef_{1}$ and $s_{2}=Ef_{2}$.

\section{Quantization of Free Fields, Quantization on Killing Horizons and Modular Theory}
\label{sec_quantization}

\subsection{Algebraic viewpoint: Quantum Field Theory on Curved Spacetimes}
\label{subsec:AlgQFTCS}
In this section, we review the necessary ingredients for a suitable mathematical and conceptual structure for a quantum field theory on a general curved spacetime. One often thinks of observables as represented as operators on a Hilbert space. However, it is problematic to fix a particular Hilbert space representation from the outset since one is often interested in considering different states that are not all expressible as vectors or density matrices on the same Hilbert space. An example of this that is relevant to the considerations of this paper is the Unruh, and Hartle-Hawking states in the exterior of an asymptotically flat Schwarzschild black hole. While these states are perfectly well-defined (i.e., the correlation functions of all field observables in these states are well-defined) neither state can be simultaneously represented in a single, irreducible Hilbert space representation. It is therefore essential to have, quite generally, a notion of the structure of the theory prior to a particular choice of background manifold or Hilbert space. The algebraic approach provides such a notion. In this subsection, we review the $\ast$-algebra structure of quantum fields on curved spacetime, the notion of ``Hadamard states'' which constitute the class of physical states as well as the quantization of the quantum field theory in terms of initial data. For further discussions of the algebraic viewpoint, we refer the reader to \cite{Wald_1995,Hollands:2014eia,Witten:2021jzq}.

In the algebraic approach, the basic structure of quantum field observables is that of a unital $\ast$-algebra $\Alg$.\footnote{A unital $\ast$-algebra $\ms{A}$ is a complex, associative algebra with a unit $\op{1}$ and an involution $a\to a^{\ast}$ for any $a\in \ms{A}$.} The quantum theory $\Alg(\mc{M},g)$ of a real scalar field $\phi$ associated to a spacetime $(\mc{M},g)$ is obtained by starting with the free algebra of ``smeared'' field observables $\op{\phi}(f)$, their formal adjoints $\op{\phi}(f)^{\ast}$, and the identity $\op{1}$ where $f$ is a real-valued, smooth function on $\mc{M}$ with compact support. The algebra $\Alg$ is obtained by factoring the free algebra by the following relations: 

\begin{enumerate}[label=(A.{\Roman*})]
\label{KGalg}
\item $\op{\phi}(c_{1}f_{1}+c_{2}f_{2})=c_{1}\op{\phi}(f_{1})+c_{2}\op{\phi}(f_{2})$ for any $f_{1},f_{2}$ and any $c_{1},c_{2}\in \mathbb{R}$ \label{A1}
\item $\op{\phi}((\Box - V)f)=0$ for all $f$  \label{A2}
\item $\op{\phi}(f)^{\ast}=\op{\phi}(f)$ for all $f$ \label{A3}
\item $[\op{\phi}(f_{1}), \op{\phi}(f_{2})]=iE(f_{1},f_{2})\op{1}$, for any $f_{1}$ and $f_{2}$\label{A4}
\end{enumerate}
\ref{A1} expresses the fact that $\phi(f)$ is linear in the test function and so is an algebra-valued distribution. In analogy to the conventions of distribution theory, we may formally write 
\begin{equation}
\label{eq:smearfield}
 \op{\phi}(f) = \int\sqrt{-g}d^{4}y~\op{\phi}(x)f(x).
\end{equation}
However, we note that the point-like object $\op{\phi}(x)$ is not an element of $\Alg(\mc{M},g)$.\footnote{In quantum field theory, the algebra cannot be extended to include elements of the form $\op{\phi}(x)$ since the expected value of $\op{\phi}^{2}(x)$ cannot be defined in any ``physical'' state (see, e.g., \cref{eq:Hadamardstate})} \ref{A2} says that $\op{\phi}$ satisfies the field equation in the distributional sense and \ref{A3} that $\op{\phi}$ is Hermitian. \ref{A4} are the covariant canonical commutation relations for $\op{\phi}$. 

States on $\Alg$ are defined as positive linear functionals on the algebra, i.e., a state, $\s$, is a linear map $\s:\Alg \to \bb{C}$ such that $\s(a^{\ast}a)\geq 0$ for all $a\in \Alg$ together with the normalization condition $\s(\op{1})=1$. Since an arbitrary algebra element $a\in \Alg(\mc{M},g)$ consists of finite sums of products of local fields, a specification of $\s$ is equivalent to providing the complete list of its smeared ``$n$-point correlation functions'' $\s(\op{\phi}(f_{1})\dots \op{\phi}(f_{n}))$. 

The notion of an algebraic state is equivalent to the more familiar Hilbert space notion of a state. To see this we first note that given a Hilbert space $\Hilb$ on which the algebra elements $a\in \Alg$ are represented as operators satisfying \ref{A1}-\ref{A4} then any normalized vector $\ket{\Psi}\in \Hilb$ gives rise to an algebraic state by $\s(a)=\braket{\Psi|\pi(a)|\Psi}$ for all $a\in \Alg(\mc{M},g),$ with $\pi$ the representation of $\Alg$. Conversely, given an algebraic state $\s: \Alg \to \bb{C}$ one can obtain a representation, $\pi,$ of $\ms{A}$ on $\Hilb$ and a vector $\ket{\Psi}\in \Hilb$ such that $\s(a)=\braket{\Psi|\pi(a)|\Psi}$ for all $a\in \Alg$. This construction is known as the Gelfand-Naimark-Segal (GNS) construction and consists of using $\s$ to define an inner product $\braket{a|b}\defn \s(a^{\ast}b)$ for any $a,b\in \Alg.$ Degenerate elements form a left ideal, $\mc{J},$ and one then completes the equivalence classes of $\Alg / \mc{J}$ in this inner product to obtain a Hilbert space $\Hilb$ \cite{GNS1,GNS2}. By construction the vector $\ket{\Psi}$ corresponding to the algebraic state $\s$ is cyclic, i.e., the action $\pi(a)$ on $\ket{\Psi}$ for all $a\in \Alg$ generates a dense subspace of states. If $\om$ is faithful,\footnote{A state is faithful if $\s(a^{\ast}a)= 0$ if and only if $a = 0$.} then $\ket{\Psi}$ will be separating, i.e., $\pi(a)\ket{\Psi} = 0 \Leftrightarrow \pi(a) = 0$. Furthermore, $a$ will be represented as a densely defined, unbounded operator on $\ms{H}$. We note that this construction utilizes only the $\ast$-algebra structure of $\Alg$. 

An algebraic state is ``mixed'' if it can be written as the (positive, convex) sum of two other states
\begin{align}
 \s=\lambda \s_{1}+(1-\lambda)\s_{2}, \quad 
\end{align}
for some $\lambda \in (0,1)$. If no such decomposition exists, the state is called ``pure.''
The GNS representation of a mixed state will correspond to a reducible representation of $\ms{A}$. For example, if one has a mixed state $\s$ corresponding to a density matrix $\rho$ on Hilbert space $\tilde \Hilb$ e.g., $\s(a) = \Tr \left(\rho a\right) $, the GNS construction will enlarge the Hilbert space $\tilde \Hilb$ in a way that will represent $\s$ as a vector. An important example is when $\s$ is a ``thermal state'' (defined below), in which case, $\om$ is represented on the GNS Hilbert space as the so-called thermofield double state \cite{haag1967equilibrium}.

An important class of states for the considerations of this paper are known as ``Gaussian states'' (also referred to as ``quasi-free states'' or ``vacuum states''). For a Gaussian state, the $n$-point functions for $n>2$ are given by formulas in terms of products of the $1$- and $2$-point functions analogous to the $n^{th}$ moments of a Gaussian distribution. We shall, in particular, consider Gaussian states whose $1$-point functions vanish and so all $n$-point functions are given by products of $2$-point functions. For example, for such a state, the $4$-point function is given by 
\begin{align}
\begin{aligned}
 \s(\op{\phi}(y_{1})\op{\phi}(y_{2})\op{\phi}(y_{3})\op{\phi}(y_{4})) = &\s(\op{\phi}(y_{1}) \op{\phi}(y_{2}))\cdot\s(\op{\phi}(y_{3})\op{\phi}(y_{4})) +\s(\op{\phi}(y_{1})\op{\phi}(y_{3}))\cdot\s(\op{\phi}(y_{2})\op{\phi}(y_{4}))\\ 
 &+\s(\op{\phi}(y_{1})\op{\phi}(y_{4}))\cdot\s(\op{\phi}(y_{2})\op{\phi}(y_{3}))
\end{aligned}
\end{align}
where all unsmeared formulas here and below should be interpreted in the distributional sense. We note that the GNS Hilbert space of any Gaussian state $\s$ has a Fock space structure $\Fock(\Hilb_{1})$ where $\Hilb_{1}$ is the ``one-particle'' Hilbert space with inner product $\braket{f_{1}|f_{2}}\defn \s(\op{\phi}(f_{1})^{\ast}\op{\phi}(f_{2}))$ for any test functions $f_{1},f_{2}$. In a general curved spacetime the norms obtained from two Gaussian states $\s_{1}$ and $\s_{2}$ will generically be inequivalent\footnote{The norms obtained from $\s_{1}$ and $\s_{2}$ are equivalent if there exists a constant $c$ such that $c^{-1}\s_{1}(\op{\phi}(f)\op{\phi}(f))\leq \s_{2}(\op{\phi}(f)\op{\phi}(f))\leq c~\s_{1}(\op{\phi}(f)\op{\phi}(f))$ for any choice of test function $f$.} and so the corresponding GNS Fock space representations will be unitarily inequivalent.\footnote{Any two GNS representations built from ``physical states'' (i.e., Hadamard states --- see \cref{eq:Hadamardstate}) will be unitarily equivalent if $\mc{M}$ is compact \cite{Verch:1992eg}. If the manifold is non-compact, then the representations will not be unitarily equivalent if the states have sufficiently different long-range behavior.}

A particular class of Gaussian states which shall be relevant to the considerations of this paper are Gaussian\footnote{The notion of a KMS state can be straightforwardly generalized to arbitrary (non-Gaussian) states (see, e.g., Section 2.1 of \cite{Hollands:2014eia})} ``KMS states'' (also referred to as ``thermal states'') associated to the spacetime. If the spacetime has a complete timelike or null Killing vector field $\xi^{a}$ --- as would be the case if, for example, the spacetime is the right ``wedge'' $\mc{R}$ of a bifurcate Killing horizon and $\xi^{a}$ is timelike in $\mc{R}$ --- one can define a KMS state relative to translations along the orbits of $\xi^{a}$. If $\chi_{t}:\mc{M}\to \mc{M}$ is the $1$-parameter group of isometries generating by the Killing vector $\xi$ then, since the retarded and advanced propagators are invariant under $\chi_{t}$, it follows that $\chi_{t}$ also generates a one-parameter family of automorphisms $\alpha_{t}:\Alg \to \Alg$ which acts on the smeared fields by 
\begin{equation}
\label{eq:alplhat}
\alpha_{t}[\op{\phi}(f)] = \op{\phi}(f_{t})
\end{equation}
where $f_{t}(y)=f(\chi_{-t}(y))$. It directly follows that these automorphisms compose by $\alpha_{t}\circ \alpha_{s}=\alpha_{t+s}$. On such a spacetime, a Gaussian KMS state $\s$ with inverse temperature $\beta$ is defined in the following manner: Consider the expected value $\s(a_{1}\alpha_{t}[a_{2}])$ for any $a_{1},a_{2}\in \ms{A}$. Viewing $\s(a_{1}\alpha_{t}[a_{2}])$ as a function $t\rightarrow \s_{0}(a_{1}\alpha_{t}[a_{2}])$ on $\bb{R}$ then this function has an analytic continuation to the ``strip'' in $\bb{C}$ corresponding to $0<\textrm{Im}t <\beta$, that is bounded and continuous on the boundary of the strip and, on the boundary, satisfies 
\begin{equation}
\label{eq:KMS}
\s(a_{1}\alpha_{t+i\beta}[a_{2}])=\s(\alpha_{t}[a_{1}]a_{2})
\end{equation}
for any $a_{1},a_{2}\in\ms{A}$ and $t \in \mathbb{R}$.\footnote{Tomita and Takesaki's modular theory guarantees that such an automorphism of a von Neumann algebra always exists, as we review in \cref{subsec:vNalgebra}. However, this \textit{modular automorphism} need not be geometric.}

The definition of an algebraic state given above is far too general in that it allows for states with too singular ultraviolet behavior. This can prevent nonlinear field observables, e.g., the local stress tensor, from being defined. It is therefore necessary to impose additional restrictions on the short distance behavior of states. The appropriate restriction is the Hadamard condition, which is a requirement that the short distance behavior of the $2$-point function of any ``physical'' state $\s$ is of the form 
\begin{equation}
\label{eq:Hadamardstate}
 \s(\op{\phi}(y_{1})\op{\phi}(y_{2})) = \frac{1}{4\pi^{2}}\frac{P(y_{1},y_{2})}{\sigma + i0^{+}T} + Q(y_{1},y_{2})\log(\sigma + i0^{+}T) + W(y_{1},y_{2}).
\end{equation}
The function $\sigma$ is the squared geodesic distance between $y_{1}$ and $y_{2}$, $T=t(y_{1})-t(y_{2})$ with $t$ a global time function on spacetime, $P$ and $Q$ are smooth symmetric functions that are locally constructed via the Hadamard recursion relations \cite{DeWitt} and $W$ is a state-dependent function which is also smooth and symmetric. Roughly speaking, the Hadamard condition is a mathematically precise version of the condition that all physical states should ``look like'' the vacuum state in Minkowski spacetime at sufficiently short distances.\footnote{A more useful formulation of the Hadamard condition is a ``microlocal spectrum condition'' on states \cite{Radzikowski:1996pa}.} The condition \eqref{eq:Hadamardstate} on the $2$-point function together with positivity of the state $\s$ implies that all of the connected $n$-point functions for $n\neq 2$ are smooth and symmetric \cite{Sanders:2009sw}. For a Gaussian state, the KMS condition implies that the state is Hadamard \cite{Sahlmann:2000fh}. 

Non-Hadamard states typically have infinite fluctuations for products of fields such as $\op{\phi}^{2}(f)$ even after ``normal ordering''\footnote{In a general curved space, the local and covariant prescription is ``Hadamard normal ordering'' which corresponds to subtracting a locally constructed Hadamard distribution \cite{Hollands:2001nf}} even though the expected value $\s(\op{\phi}^{2}(f))$ may be finite. While the discussion up until now has focused on free quantum fields, we note that (perturbative) interacting quantum fields can only be defined if the states are Hadamard \cite{Brunetti:1999jn,Hollands:2001fb}. The {\it dynamics} of interacting quantum fields can also be studied, since if a state is a Hadamard ``initially'' (i.e., in the neighborhood of any Cauchy surface) in a globally hyperbolic spacetime then it will remain Hadamard in the domain of dependence of the Cauchy surface \cite{Radzikowski:1996ei}.

While the above notion of a Hadamard state gives a physically satisfactory criterion on the local behavior of states, a natural question one can ask is if (1) such states exist on globally hyperbolic spacetimes and (2) if such states can be constructed given appropriate ``initial data.'' The answer to the first question is affirmative \cite{Fulling:1981cf,Junker:2001gx,Gerard:2012wb,Dappiaggi:2017kka,Hollandsthesis}. Given any single Hadamard state $\s$ on a globally hyperbolic spacetime, then the corresponding GNS representation contains a dense subspace of states including the vector representative $\ket{\s}$ and any finite products $\op{\phi}(f_{1})\dots \op{\phi}(f_{n})$ applied to $\ket{\s},$ for arbitrary test functions $f_{1}\dots f_{n},$ all of which are Hadamard.

The answer to the second question goes back to the earliest investigations of quantum field theory in curved spacetime \cite{hawking1975particle,wald1975particle,Parker:1969au,Birrell:1982ix,Wald_1995}. On a Cauchy surface $\Sigma$, one can attempt to define a Hadamard state by specifying suitable ``positive frequency'' initial data and then propagating that state to the future domain of dependence of $\Sigma$. Of particular relevance to the considerations of this paper is the specification of suitable initial data for quantum fields in the presence of a Killing horizon. On any spacetime with a Killing horizon, a uniqueness theorem for a stationary, Hadamard state defined on the full, maximally extended spacetime was proven by Kay and Wald \cite{Kay_1988}. They proved that there exists, at most, one state defined on the full maximally extended spacetime which is Hadamard and stationary. Furthermore, they proved that such a state must be KMS when restricted to $\mc{R}$. This state corresponds to the Bunch-Davies vacuum in de Sitter \cite{Bunch:1978yq,Chernikov:1968zm,Schomblond_1976} or the Hartle-Hawking vacuum in Schwarzschild \cite{Hartle:1976tp,Israel:1976ur}.\footnote{The existence of a Hartle-Hawking state on any static black hole spacetime in arbitrary dimensions was proven in \cite{Sanders:2013vza}.} In Schwarzschild, the initial data of the Hartle-Hawking state is a thermal state with a temperature equal to $\kappa/2\pi$. Thus, the Hartle-Hawking state corresponds to a black hole in thermal equilibrium. 
\par However, while Kay and Wald proved uniqueness, they also proved that such a state does not 
 generally exist. In particular, they proved that such a state cannot exist in Kerr or Schwarzschild-de Sitter spacetimes.\footnote{Due to the KMS property of these stationary (with respect to the horizon Killing field $\xi^{a}$) states defined on the maximally extended spacetime, a necessary condition for such a state to exist is that $\xi^{a}$ is timelike everywhere in $\mc{R}$. Thus, the existence of a globally stationary state on $(\mc{M},g)$ is directly related to the existence of an ``ergoregion.'' For Reissner-Nordstrom the horizon Killing field is timelike in the exterior but, for wave propagation of charged fields, such black holes can have a ``generalized ergoregion'' \cite{Green:2015kur,1999PhRvD..61b4014H,Bekenstein:1973mi,Denardo:1973pyo}. Therefore, for black holes with large enough charge, a globally stationary state will not exist. This was explained to us by R. M. Wald.} In the asymptotically flat case, this result is not in conflict with Hawking's original analysis establishing the thermal nature of black holes since Hawking's analysis essentially considered the propagation of an incoming stationary state $\s_{0}$ with vacuum initial data at $\scri^{-}$. Furthermore, this state is only defined in the interior and exterior of the black hole. {Therefore while a globally thermal state generally does not exist, for any spacetime with a bifurcate Killing horizon, we expect to be able to define a ``vacuum'' state $\s_{0}$ in $(\mc{M}_{\textrm{R}},g)$ which is stationary with respect to the horizon Killing field $\xi$. In this paper we will assume the following: }
\begin{enumerate}
\setcounter{enumi}{3}
\item There exists a state $\s_{0}$ on $\Alg(\mc{M}_{\textrm{R}},g)$ which is Hadamard, stationary and has zero energy with respect to the horizon Killing field $\xi^{a}$. \label{assump4}
\end{enumerate} 
While the notion of ``energy'' as a quantum observable will be made precise in the next section, on phase space $\mc{P}$ the energy, roughly speaking, corresponds to the integral $\int_{\Sigma}T_{ab}\xi^{b}$ over a Cauchy surface $\Sigma$ where $T_{ab}$ is the stress-energy tensor of the field. Including this observable in the quantum theory --- as we do in \cref{subsec:VNmodKilling} --- provides a well-defined notion of the energy of a quantum state. This ``zero energy'' condition also ensures that (a dense set of) states in the GNS representation above $\s_{0}$ have finite energy.
 
We note that the state $\s_{0}$ is only defined in $\mc{M}_{\textrm{R}}$ and generically cannot be extended to a Hadamard state outside this region. In some cases (e.g., de Sitter spacetime or a Schwarzschild black hole in AdS) the state $\s_{0}$ will also be KMS in $\mc{R}$ and, by Kay and Wald's uniqueness theorem, it can be extended to a state on the maximally extended spacetime \cite{Kay_1988}. For a black hole in an asymptotically flat spacetime, the state $\s_{0}$ is known as the ``Unruh state'' which can also be defined by specifying vacuum initial data with respect to asymptotic time translations at $\scri^{-}$ and vacuum initial data with respect to affine time translations on the past horizon $\mc{H}^{-}$ \cite{unruh1976notes}. For a black hole formed from gravitational collapse, the quantum state rapidly ``settles down'' to a stationary state \cite{haag1967equilibrium} and so the Unruh state provides a physically sensible initial data for a stationary state on any black hole background. 

We note that assumption \ref{assump4} has been proven for many spacetimes with bifurcate Killing horizons. The Unruh state has been explicitly shown to be Hadamard on the region $(\mc{M}_{\textrm{R}},g)$ of Schwarzschild \cite{Dimock:1987hi,2009arXiv0907.1034D} and this property is believed to continue to hold for Reissner-Nordstrom and Kerr black holes.\footnote{The Hadamard property of the Unruh state has been proven for massless fermions in slowly rotating Kerr \cite{Gerard:2020tdo}.} Furthermore, such a stationary state has been proven to be Hadamard in Schwarzschild-de Sitter \cite{Brum:2014nea}, Reissner-Nordstrom-de Sitter \cite{Hollands:2019whz} and Kerr-de Sitter \cite{Klein:2022jtb}.

\subsubsection{Quantization on Killing Horizons and Null Infinity}
\label{subsec:quantKillingscri}
In this paper, we consider spacetimes $(\mc{M},g)$ satisfying assumptions \ref{assump1}--\ref{assump4}. By assumptions \ref{assump2} and \ref{assump3}, the horizon $\mc{H}^{-}$ is a (partial) Cauchy surface for the globally hyperbolic region $\mc{M}_{\textrm{R}}\subset \mc{M}$. In all cases of interest, the initial data is freely specified on $\mc{H}^-$ and can be supplemented by additional initial data on either another Killing horizon or, in the case of an asymptotically flat spacetime, at past timelike or null infinity. In the case where the spacetime is asymptotically flat we shall, for simplicity, restrict attention to massless fields, however, the results of this paper carry over directly to the case of massive fields quantized at past timelike infinity (see \cref{app1:timequant}). 

We now give the quantization of the quantum field $\op{\phi}(f)$ in terms of data on $\mc{H}^{-}$ and/or $\scri^{-}$. The local field observable $\op{\phi}(f)$ can be expressed in terms of initial data on any Cauchy surface, as explained in \cref{subsec:scfieldth}. The field observable on the horizon is $\op{\Pi}(s)$ (see~\eqref{eq:PiHorDefn}), where $s$ is a test function corresponding to initial data on the past horizon. The algebra $\Alg_{\mc{H}^{-}}$ is defined by starting with the free algebra generated by $\op{\Pi}(s)$, $\op{\Pi}(s)^{\ast}$ and $\op{1}$ and factoring by relations analogous to that of \ref{A1}-\ref{A4}. Conditions \ref{A1} and \ref{A3} carry over straightforwardly to conditions on $\ms{A}_{\mc{H}^{-}}$. Condition \ref{A2} has no analog on $\ms{A}_{\mc{H}^{-}}$ since the test functions correspond to freely specified initial data on $\mc{H}^{-}$. Denoting points on $\mc{H}^-$ by $x = (U, x^A),$ with $U$ the affine time coordinate, the commutation relation, \ref{A4}, now becomes 
\begin{equation}
\label{eq:commKill}
[\op{\Pi}(x_{1}), \op{\Pi}(x_{2})] = i\delta^{\prime}(U_{1},U_{2})\delta_{\bb{S}^{2}}(x_{1}^{A},x_{2}^{A})\op{1}
\end{equation}
(see \cref{eq:PBPi}) where this equation is understood to hold in the distributional sense. 

We now turn to the regularity of states on $\ms{A}_{\mc{H}^{-}}$. Recall that for states on the bulk algebra, $\Alg$, we imposed the Hadamard condition on the singular behavior of the $n$-point correlation functions in \cref{eq:Hadamardstate}. This implies a corresponding necessary\footnote{By Hormander's ``propagation of singularities theorem'' \cite{Hormander_72}, the local singular behavior of a Hadamard state \cref{eq:Hadamardstate} implies that the necessary singular behavior of states on $\mc{H}^{-}$ is of the form of \cref{eq:2ptPI} \cite{Hollands_thesis}.} condition on the singular structure of the $n$-point correlation functions on $\ms{A}_{\mc{H}^{-}}$. In particular, the $2$-point function on $\mc{H}^{-}$ must be of the form 
\begin{equation}
\label{eq:2ptPI}
\s^{\mc{H}}(\op{\Pi}(x_{1})\op{\Pi}(x_{2})) = -\frac{2}{\pi}\frac{\delta_{\bb{S}^{2}}(x_{1}^{A},x_{2}^{A})}{(U_{1}-U_{2}-i0^{+})^{2}} + S(x_{1},x_{2})
\end{equation}
where $S$ is a smooth, symmetric function on $\mc{H}^{-}\times \mc{H}^{-}$. We also will require that $S$ as well as all connected $n$-point functions for $n\neq 2$ decay for any set of $U_{i}$ as $O((\sum_{i}U^{2}_{i})^{-1/2-\epsilon})$ as $|U_{i}|\to \infty$ for some $\epsilon>0$.\footnote{This decay ensures that there are no ``zero modes'' (i.e. states invariant under $\chi_{t}$) except the vacuum $\s_{0}$.} 

The unique, Gaussian state $\s^{\mc{H}}_{0}$ on $\Alg_{\mc{H}^{-}}$ invariant under affine time translations has a vanishing $1$-point function and the $2$-point function is given by \cref{eq:2ptPI} with $S=0$ \cite{Dappiaggi:2017kka,Kay_1988}, i.e., 
\begin{equation}
\label{eq:vacH-}
\s^{\mc{H}}_{0}(\op{\Pi}(x_{1})\op{\Pi}(x_{2})) = -\frac{2}{\pi}\frac{\delta_{\bb{S}^{2}}(x_{1}^{A},x_{2}^{A})}{(U_{1}-U_{2}-i0^{+})^{2}}.
\end{equation}
The state $\s^{\mc{H}}_{0}$ is KMS with inverse temperature $\beta = 2\pi/\kappa$ with respect to Killing time translations on $\mc{H}_{\textrm{R}}^{-}$. On the subalgebra $\ms{A}_{\mc{H}_{\textrm{R}}^{-}}\subset \ms{A}_{\mc{H}^{-}}$ of observables on $\mc{H}^{-}_{\textrm{R}}$, the Killing time translations $\chi_{t}$ are automorphisms given by $\alpha_{t}:\ms{A}_{\mc{H}_{\textrm{R}}^{-}} \to \ms{A}_{\mc{H}_{\textrm{R}}^{-}}$ 
\begin{equation} \label{eq:alphAutoH-R}
\alpha_{t}[\op{\Pi}(s)] \defn \op{\Pi}(s_{t})
\end{equation}
where $s_{t}(U,x^{A}) = s(e^{\kappa t}U,x^{A})$ for any test function $s$ on $\mc{H}_{\textrm{R}}^{-}$. To make the KMS property of \cref{eq:vacH-} restricted to $\mc{H}_{\textrm{R}}^{-}$ more explicit, it is more convenient to change variables to Killing time $u^{\prime}$ using $U=-e^{-\kappa u^{\prime}}$ (see \cref{eq:H-affkill}). It is then straightforward to show that, the $2$-point function restricted to $\mc{H}_{\textrm{R}}^{-}$ is 
\begin{equation}
\label{eq:vacH-R}
\s^{\mc{H}}_{0}(\op{\Pi}(s_{1})\op{\Pi}(s_{2})) = -\frac{\kappa^{2}}{2\pi}\int du^{\prime}_{1}du^{\prime}_{2}d\Omega_2~\frac{s_{1}(u_{1}^{\prime},x^{A})s_{2}(u_{2}^{\prime},x^{A})}{(\sinh(\frac{\kappa}{2}(u_{1}^{\prime}-u_{2}^{\prime}))-i0^{+})^{2}}
\end{equation}
for any $s_{1},s_{2}$ supported on $\mc{H}^{-}_{\textrm{R}}$. Using this form of the $2$-point function one can easily verify that $\s^{\mc{H}}_{0}(\op{\Pi}(s_{1})\alpha_{t}[\op{\Pi}(s_{2})])$ can be analytically continued to complex $t$ in the strip $0< \textrm{Im}t < 2\pi/\kappa$. It is bounded and continuous on the boundary of the strip with 
\begin{equation}
\s^{\mc{H}}_{0}(\op{\Pi}(s_{1})\alpha_{t+i\beta}[\op{\Pi}(s_{2})]) = \s^{\mc{H}}_{0}(\alpha_{t}[\op{\Pi}(s_{2})]\op{\Pi}(s_{1})), \quad \beta = \frac{2\pi}{\kappa} \ .
\end{equation}
The state $\s^{\mc{H}}_{0}$ is faithful on $\Alg_{\mc{H}^-_{\rm R}}$ and thus gives rise to a cyclic and separating vector in its GNS representation.

We conclude this subsection by giving the analogous quantization of massless fields at null infinity for any asymptotically flat spacetime. The phase space at $\scri^{-}$ and its quantization proceeds in nearly an identical manner to the ``null quantization'' on a Killing horizon \cite{Ashtekar:1987tt,Prabhu:2022zcr}. For a massless scalar field, the initial data at $\scri^{-}$ corresponds to the specification of a conformally weighted scalar field 
\begin{equation}
\tilde{\Phi}\defn \lim_{\scri^{-}}\conf^{-1}\phi 
\end{equation}
where $\conf$ is a conformal factor which, in Bondi coordinates, is $\conf = 1/r$ where the coordinate vector field $(\partial/\partial r)^{a}$ can be chosen to be tangent to affinely parameterized, past-directed null geodesics (see, e.g., \cite{Hollands:2016oma} for a construction of such coordinates near $\scri$). We shall consider asymptotically flat spacetimes where initial data can be independently specified at $\scri^{-}$ (i.e., for the purposes of this paper, the spacetime satisfies assumptions \ref{assump1}---\ref{assump3} with $\Sigma^{\prime}=\scri^{-}$). The symplectic form of two solutions $\phi_{1},\phi_{2}$ with corresponding initial data $\tilde{\Phi}_{1},\tilde{\Phi}_{2}$ on $\scri^{-}$ is 
\begin{equation}
\symp^{\KG}_{\scri^{-}}(\phi_{1},\phi_{2}) = \frac{1}{2}\int_{\scri^{-}}dvd\Omega_2~\bigg[\tilde{\Phi}_{2}\frac{\partial \tilde{\Phi}_{1}}{\partial v} - \frac{\partial \tilde{\Phi}_{2}}{\partial v}\tilde{\Phi}_{1}\bigg]
\end{equation}
where the coordinates on $\scri^{-}$ are $(v,x^{A})$ where $v$ denotes\footnote{We note the unfortunate duplication of notation where $v$ here denotes advanced time at $\scri^{-}$ as well as Killing time on the future horizon $\mc{H}^{+}_{\textrm{R}}$. This notation is standard and since, in this paper, we are primarily considering initial data on the past boundary the meaning of the coordinate $v$ will be clear in context.} the advanced time on $\scri^{-}$ and $x^{A}$ are arbitrary angular coordinates on the $2$-sphere cross-sections of past null infinity. For the symplectic form to converge $\partial \tilde{\Phi}/\partial v$ must decay as $O(1/|v|^{1+\epsilon})$ for some $\epsilon>0$. As before, given any ``bulk'' field observable $\phi(f),$ the corresponding observable on $\scri^{-}$ is $\tilde{\Pi}(\tilde{s})$ where $\tilde{\Pi} \defn \partial_{v}\tilde{\Phi}$ and $\tilde{s}=\lim_{\scri^{-}}\conf^{-1} Ef$. The Poisson brackets of $\tilde{\Pi}(\tilde{s})$ are given by an analogous formula to \cref{eq:PBPi} where the test functions are now functions on $\scri^{-},$ the affine time, $U$, is replaced by the advanced time at past null infinity, $v$, and the integral in now over $\scri^{-}$. 

The algebra $\Alg_{\scri^{-}}$ is defined by starting with the free algebra generated by $\tilde{\op{\Pi}}(\tilde{s}),\tilde{\op{\Pi}}(\tilde{s})^{\ast}$ and $\op{1}$ factored by conditions analogous to conditions \ref{A1} and \ref{A3} together with the commutation relation 
\begin{equation}
[\tilde{\op{\Pi}}(x_{1}),\tilde{\op{\Pi}}(x_{2})] = i\delta^{\prime}(v_{1},v_{2})\delta_{\bb{S}^{2}}(x_{1}^{A},x_{2}^{A})\op{1}.
\end{equation}
As before, there is no analog of condition \ref{A2} since $\op{\Pi}$ is smeared with free initial data. The condition on the singular behavior of the $2$-point function of states on $\Alg_{\scri^{-}}$ is also given by 
\begin{equation}
\label{eq:statescri}
\s^{\scri}(\tilde{\op{\Pi}}(x_{1})\tilde{\op{\Pi}}(x_{2})) = -\frac{2}{\pi}\frac{\delta_{\bb{S}^{2}}(x_{1}^{A},x_{2}^{A})}{(v_{1}-v_{2}-i0^{+})^{2}} + \tilde{S}(x_{1},x_{2})
\end{equation}
where $\tilde{S}$ is a smooth, symmetric function on $\scri^{-}\times \scri^{-}$. In addition to the ultraviolet regularity condition on states, we also will require that $\tilde{S}$ and all connected $n$-point functions for $n\neq 2$ decay for any set of $|v_{i}|\to \infty$ as $O((\sum_{i}v^{2}_{i})^{-1/2-\epsilon})$ for some $\epsilon>0$. The unique, Gaussian state $\s^{\scri}_{0}$ on $\Alg_{\scri^{-}}$ invariant under advanced time translations has vanishing $1$-point function and $2$-point function given by \cref{eq:statescri} with $\tilde{S}=0$.\footnote{For a massless scalar field in Minkowski spacetime, the global state with initial data $\s^{\scri}_{0}$ on $\scri^{-}$ corresponds to the Poincaré invariant vacuum.} 

\subsection{Von Neumann Algebras and Modular Theory}

\label{subsec:vNalgebra}

In \cref{subsec:AlgQFTCS}, we gave the necessary conceptual and mathematical structure to formulate quantum field theory on a general, curved spacetime; however, additional structure is required in order to discuss quantum information theoretic notions such as ``von Neumann entropy.'' The setting in which such notions are well-defined is that of a von Neumann algebra. From the $\ast$-algebra structure introduced in \cref{subsec:AlgQFTCS} it is straightforward to obtain an appropriate von Neumann algebra. One first forms a $C^{\ast}$-algebra from bounded functions of the quantum fields and then includes appropriate limits of these bounded operators. One may obtain a von Neumann algebra by considering (the weak closure of) a preferred representation of a $C^{\ast}$-algebra.

A $C^{\ast}$-algebra $\mf{A}$ is a particular type of normed $\ast$-algebra. Specifically, it is an algebra equipped with a norm $||\cdot ||:\mf{A} \to \bb{R}$ which additionally satisfies 
\be
 \norm{a^* a} = \norm{a}^2 \ 
\ee
for any $a\in \mf{A}$. 
Any $C^{\ast}$-algebra can be represented by an algebra of bounded operators acting on a Hilbert space $\Hilb$, though it is an important fact that such representations may be unitarily inequivalent for different choices of $\Hilb.$

As a concrete example, we may consider the construction of a $C^{\ast}$-algebra from the free scalar field, the so-called \textit{Weyl algebra}. For 
\begin{equation}
\label{eq:Weyl}
\op{W}(f)\defn e^{i\op{\phi}(f)},
\end{equation}
the Weyl algebra encodes the canonical commutation relations
\begin{align}
 \op{W}(f_1) \op{W}(f_2) = e^{\frac{i}{2} E(f_{1},f_{2})} \op{W}(f_1+f_2) ,\quad \op{W}(f)^* = \op{W}(-f)= \op{W}(f)^{-1}. 
\end{align}
The exponentiation makes clear the boundedness of the operators. This becomes a $C^{\ast}$-algebra by introducing a unique norm and completing in the norm topology (see, e.g., \cite{Bratteli:1996xq,Binz:2004}).

For any $\ast$-algebra, $\ms{A},$ on Hilbert space, $\Hilb$, it is natural to define the commutant algebra
\be 
\label{eq:commutant}
 \mathscr{A}' \defn \{b \in \sB\le(\Hilb\ri) ~|~ ba = ab,~ \forall ~a \in \mathscr{A}\} \ .
\ee
The commutant algebra $\mathscr{A}'$ is weakly closed and is therefore a von Neumann algebra. The von Neumann algebra, which we shall also label $\mf{A}$, associated to $\Alg$ and a choice of representation, $\pi$, can be straightforwardly obtained by the ``double commutant operation'': 

\be 
 \mf{A} \defn \pi[\Alg]'' \ .
\ee
Such a construction always yields a von Neumann algebra by the von Neumann double commutant theorem \cite{neumann1929beweis}.

Von Neumann algebras are well-studied in mathematics and thus have been classified based on their various properties \cite{connes1973classification}.\footnote{See \cite{2023arXiv230201958S} for a recent review aimed at a physics audience.} The classification is made for so-called ``factors,'' which are von Neumann algebras $\mf{A}$ whose center $\mf{A}\cap \mf{A}'$ is trivial, i.e., isomorphic to the complex numbers. The decomposition theorem \cite{von1949rings} states that all von Neumann algebras on separable Hilbert spaces are direct sums and/or integrals of factors. Thus, in the study of von Neumann algebras, it is sufficient to study factors. Factors are broadly classified into three types. We will not describe the classification in detail but instead only mention the distinctions relevant for the considerations of this paper. 

Type I factors are the algebras familiar from quantum mechanics. These algebras have irreducible Hilbert space representations and thus are equivalent to the algebra of {\it all} bounded operators on some Hilbert space. Type I algebras admit a trace, i.e., a functional, $\Tr$, such that $\Tr (a b) = \Tr (b a) $ for $a, b \in \mf{A}$. For matrix algebras, this is just the familiar matrix trace. In finite dimensional quantum mechanics, the algebra is called Type I$_d$ where $d$ is the dimension of the Hilbert space. For such algebras, all operators have finite-trace and are said to be \textit{trace-class}. In infinite-dimensional quantum mechanics, say a single harmonic oscillator, the algebra is called Type I$_{\infty}$, and not all operators will be trace-class. For example, the identity operator has infinite trace. Type I$_{\infty}$ algebras also arise in quantum field theory when considering global regions, such as entire Cauchy slices. The algebra of all bounded operators on the quantum field theory Hilbert space is a Type I$_{\infty}$ algebra. 

A Type II factor does not have any irreducible Hilbert space representation. The representation of a Type II factor always has a nontrivial commutant, of the same type, acting on the same Hilbert space. Despite this distinction from the Type I algebra, Type II algebras have a well-defined trace. However, these are no longer the familiar traces of linear algebra, but renormalized traces. If all operators in the Type II algebra are trace-class, then the algebra is called Type II$_1$. It is standard convention to normalize the trace such that the identity has unit trace
\begin{align}
\label{eq:tracenorm}
 \Tr( \textbf{1} )= 1.
\end{align}
If the identity operator is not trace-class, the algebra is called II$_\infty$. A Type II$_\infty$ factor may be considered to be a tensor product of a Type II$_1$ factor and a Type I$_\infty$ factor. Both Type II$_1$ and II$_\infty$ algebras play an important role in this paper.

The final type of von Neumann algebra is Type III, which neither has any irreducible representations nor admits a trace functional. While there exists a subclassification of Type III algebras into Type III$_{\lambda}$ for $0\leq \lambda \leq 1$ \cite{connes1973classification}, we will not be concerned with defining them because only Type III$_1$ algebras are relevant for us as these describe the algebras associated to local regions in quantum field theory.

The existence of a trace allows for the definition of density operators. Namely, for any state $\om$ on a Type I or II algebra there exists a positive, self-adjoint operator $\rho_{\om} \in \mf{A}$ such that
\be 
 \om(a) = \Tr (a \rho_{\omega}) \ ,~\forall a \in \mf{A}.
\ee

Equipped with a density operator, we may associate a von Neumann entropy to the algebraic state
\begin{align}
 S_{\textrm{vN.}} (\omega) = -\Tr (\rho_{\omega} \log \rho_{\omega} ).
\end{align}
For Type I algebras, the von Neumann entropy is positive semi-definite. It is zero if and only if the state is pure. It may be infinite for Type I$_\infty$ algebras. In contrast, the von Neumann entropy is negative semi-definite (upon normalization of the trace \eqref{eq:tracenorm}) for Type II$_1$ algebras. This implies that there exists a maximum entropy state, a novel concept for states on infinite dimensional Hilbert spaces. It is straightforward to check that the maximum entropy state corresponds to the identity operator.\footnote{An interesting feature of this state is that it has a so-called \textit{flat spectrum}, meaning that all of its R\'enyi entropies, defined as $S_{\alpha}\defn (1-\alpha)^{-1} \log \Tr \rho_{\omega}^{\alpha}$ are equal.} The generally negative von Neumann entropies may thus be considered as differences between the entropy of the state and the maximal entropy state. For Type II$_{\infty}$ algebras, the von Neumann entropy is not sign-definite. It can be arbitrarily positive or negative, so no maximum or minimum entropy exists. Of course, Type III algebras do not have well-defined von Neumann entropies because they do not admit a trace, and correspondingly, no density matrix that describes correlation functions for observables in the algebra exists for any state.

Now that we have introduced the additional structure of a von Neumann algebra, we may discuss Tomita-Takesaki modular theory which can be thought of as uplifting many results of quantum information theory from finite-dimensional quantum systems to the more general setting of von Neumann algebras. For more detailed recent reviews, see~\cite{2017arXiv170204924H, 2018arXiv180304993W}. The starting point for modular theory is a von Neumann algebra $\mf{A}$ and a cyclic-separating state $\ket{\omega_{0}} \in \Hilb.$ For any choice of state $\ket{\om} \in \Hilb$, one may then densely define the relative Tomita operator $\op{S}_{\s|\omega_{0}}$ by\footnote{Unfortunately, the ordering of $\s$ and $\omega_{0}$ in the subscript is not standardized in the literature. We emphasize that our notation is the opposite of \cite{2018arXiv180304993W}.}
\be 
 \op{S}_{\s|\omega_{0}} a \ket{\omega_{0}} = a^* \ket{\s},~ \forall a \in \mf{A} \ .
\ee
We will only consider $\s$ that are cyclic-separating, in which case $\op{S}_{\s|\omega_{0}}$
is invertible.
It can be shown that (an extension of) this relative Tomita operator has a polar decomposition
\be 
 \op{S}_{\s|\omega_{0}} = \op{J}_{\s|\omega_{0}} \op{\Delta}_{\s|\omega_{0}}^{\half} \ ,
\ee
with the \textit{relative modular operator}, $\op{\Delta}_{\s|\s_{0}}$, being positive and self-adjoint and the \textit{relative modular conjugation}, $\op{J}_{\s|\s_{0}}$, being antiunitary. 

A special case of the relative modular operator occurs when the second state is chosen to be the same as the first, i.e., $\s = \omega_{0}.$ In this case $\op{\Delta}_{\omega_{0}|\omega_{0}} = \op{\Delta}_{\omega_{0}}$ is the modular operator for the (cyclic-separating) state $\ket{\omega_{0}}$. An important property of the modular is that it preserves its defining state, i.e., $\op{\Delta}_{\om_0} \ket{\om_0} = \ket{\om_0}$. Taking the logarithm, we arrive at what is called the modular Hamiltonian
\begin{align}
 \op{H}_{\omega_0 }\defn -\log \op{\Delta}_{\omega_{0}}.
\end{align}
The modular Hamiltonian generates a one-parameter group of automorphisms on $\mf{A}$ via conjugation
\begin{align}
 \op{\Delta}_{\s_{0}}^{-it} a \op{\Delta}_{\s_{0}}^{it} \in \mf{A} ,\quad a\in\mf{A},~ t \in \mathbb{R}.
\end{align}
For Type III algebras, this is an outer automorphism because the modular operator cannot be split into a piece affiliated with $\mf{A}$ and a piece affiliated with $\mf{A}'$. 
Any state on the algebra satisfies the KMS-condition \cref{eq:KMS} with respect to its modular flow. In particular, viewing $\s_{0}(a\op{\Delta}_{\om_0}^{-it}b)$ as function $t\to \s_{0}(a\op{\Delta}_{\om_0}^{-it}b)$ on $\bb{R}$, then this function 
has an analytic continuation to the strip $0\leq \Im t \leq 1 $ with continuous boundary value and, on the boundary, satisfies 
\begin{align}
\label{eq:KMS_cond}
 \s_{0}(a\op{\Delta}_{\om_0}^{-i(s+i)}b)=\s_{0}(b\op{\Delta}_{\om_0}^{is}a), \quad s \in \mathbb{R} \ .
\end{align}
We emphasize that while all states satisfy the KMS condition with respect to their modular flow, this modular flow need {\em not} be a {natural}, geometric flow of the operator algebra. We will use the terminology ``KMS state'' to label states whose modular flows coincide with the flow generated by the Hamiltonian of the theory. 

If we construct the GNS Hilbert space for the state $\omega_0$, there is a special (real linear) subspace of states called the ``natural cone'' $\ms{P}^{\#}$, defined by
\begin{align}
\label{eq:naturalcone}
 \ms{P}^{\#}\defn \overline{\{\op{\Delta}_{\omega_{0}}^{1/4}a\ket{\omega_{0}}|~a \in \mf{A}\}}
\end{align}
{where the overbar denotes closure. Any ``normal state'' (i.e. continuous in the weak $\ast$-topology) on $\mf{A}$} has a unique vector representative in the natural cone and 
each element is invariant under modular conjugation $\op{J}_{\omega}\ket{\s} = \ket{\s}$. For all such states, $\op{J}_{\s|\omega_0}=\op{J}_{\s}=\op{J}_{\omega_{0}}$. 

The final important quantity in our considerations is the so-called ``relative entropy'' of states on a general von Neumann algebra. The relative entropy between two states $\omega_{0}$ and $\s$ with vector representatives $\ket{\omega_{0}}$, $\ket{\s}$ vector representatives $\ket{\omega_{0}}$, $\ket{\s}$ is \cite{araki1976relative}
\be 
 S_{\textrm{rel}}({\s|\omega_{0}}) \defn - \bra{\s_0}\log \op{\Delta}_{\s|\s_{0}}\ket{\s_0}.
\ee
This definition is independent of the vector representatives and agrees with the familiar definition for density matrices, $\rho_{\s}$ and $\rho_{\omega_{0}}$, in Type I and II algebras
\begin{align}
 S_{\textrm{rel}}({\s|\s_{0}}) = \Tr \rho_\s \log \rho_\s - \Tr \rho_\s \log \rho_{\omega_{0}}.
\end{align}
For completely general von Neumann algebras, this is as far as we can go since there does not exist a well-defined notion of von Neumann entropy in the Type III case.

\subsection{The Crossed Product}
\label{subsec:crossprod}
The key ingredient from von Neumann algebra theory that is used in this paper is a structure theorem for Type III$_1$ algebras in terms of a Type II$_\infty$ algebra and its modular automorphism group \cite{takesaki1973duality}.

Consider a Type III$_1$ von Neumann algebra $\mf{A}$ on Hilbert space $\mathscr{H}$ and a state $\omega$ with vector representation $\ket{\s}$. The modular automorphism group provides an action of $\mathbb{R}$ on $\mf{A}$ that is unitarily implemented by the conjugation $a(t) = U(t) a U(-t),$ with $U(t) \defn \op{\Delta}_{\omega}^{-it},~ t \in \mathbb{R}$. The crossed product algebra, $\hat{\mf{A}}_{\textrm{ext.}}$, of $\mf{A}$ by its modular automorphism group is a von Neumann algebra on $\mathscr{H} \otimes L^2(\mathbb{R})$ generated by $a \otimes \textbf{1}$ for $a \in \mf{A}$ and $U(t) \otimes e^{i\op{X}t}$ where $e^{i\op{X}t}$ acts on $L^2(\mathbb{R})$ {by multiplication by $e^{iXt}$}. Namely, 
\begin{align}
\hatvNext \defn \{a\otimes \textbf{1} ,\op{X} - \log \op{\Delta}_{\omega}; a\in \mf{A}, t\in \mathbb{R}\}.''
\end{align}
The algebraic structure is unchanged by conjugation with the unitary $U(-\op{t}) \defn \op{\Delta}_{\om}^{i \op{t}},$ where $\op{t}$ is the canonical conjugate to $\op{X}$ on $L^2(\mathbb{R})$ satisfying $[\op{X}, \op{t}] = i,$ which puts the generators of the algebra in a more useful form for the considerations of this paper
\begin{align}
\label{eq:xproductalg}
 \vNext\defn U(-\op{t})\hat{\mf{A}}_{\rm ext.}U(\op{t}) = \{ \op{\Delta}_{\omega}^{i\op{t}}a \op{\Delta}_{\omega}^{-i\op{t}}, \op{X}; a\in \mf{A}, t\in \mathbb{R}\}.''
\end{align}
where we have ``extended'' the von Neumann algebra $\mf{A}$ to $\vNext$ by including $\op{X}$. $\hat{\mf{A}}_{\rm ext.}$ is a Type II$_\infty$ algebra~\cite{takesaki1973duality} and the unitary will not change the type, so the conjugated algebra $\mf{A}_{\rm ext.}$ is also Type II$_{\infty}$. The most explicit way to see that the algebra is no longer Type III is by constructing a trace functional on the algebra. For $\hat{a}\in \mf{A}_{\rm ext.}$, we have
\begin{align}
\label{eq:gentraceII}
 \Tr (\hat{a}) = \int_{\bb{R}} dX~ e^{X} \bra{\s,X} \hat{a} \ket{\s,X}.
\end{align}
{where $\ket{\s,X}\defn \ket{\s}\otimes \ket{X}$ and $\ket{X}$ are (improper) eigenstates of $\op{X}$}
\begin{equation}
\label{eq:Xop}
\op{X}\ket{X}=X\ket{X}.
\end{equation}

It was shown in \cite{2022JHEP...10..008W} that \cref{eq:gentraceII} satisfies the properties of a trace. Importantly, there is no canonical normalization for this trace, so we could have just as easily multiplied the trace by an arbitrary constant, retaining the cyclicity property. The density matrix, $\rho_{\hat{\om}}$ on $\mf{A}_{\rm ext.},$ inherits the (inverse) arbitrary multiplicative constant via its definition
\begin{align}
 \Tr \rho_{\hat{\omega}}\hat{a}\defn \hat{\omega} (\hat{a}) , \quad \forall~\hat{a} \in \mf{A}_{\rm ext.} \ ,
\end{align}
which leads to an arbitrary additive constant for the von Neumann entropy. We note that $\Tr(\mathbf{1}) = \infty$ so there is no way to fix a normalization by rescaling the functional $\Tr(\cdot)$ such that identity has unit trace.

It is clear from the definition of the trace that not all operators will be trace-class. This will occur when the operators do not have sufficient decay at large positive $X$. If instead all operators were projected to have support with $X$ bounded from above, then all operators in the algebra would have a finite trace. This form of projection thus takes the Type II$_\infty$ algebra to a Type II$_1$ algebra for which the trace functional can be canonically normalized. We will see this projection appear from physical considerations in de Sitter space.

\section{Von Neumann Algebras on Killing Horizons, Gravitational Charges and Dressed Observables}
\label{sec:dressed}

In this section, we shall establish our first main result applicable to the quantum theory of a Klein-Gordon field on a spacetime $(\mc{M},g)$ satisfying assumptions \ref{assump1}---\ref{assump4}. In \cref{subsec:VNmodKilling}, we will first construct the von Neumann algebra of observables on a Killing horizon. As in the previous section, this algebra of observables corresponds to the algebra of test fields propagating on a fixed background spacetime. In \cref{subsec:charge}, we extend the phase space so that we may study perturbative ``gravitational constraints''. In particular, we treat the field as a first-order perturbation of the background spacetime which gives rise to global charge-flux relations involving ``second-order'' gravitational charges. For spacetimes satisfying assumptions \ref{assump1}---\ref{assump3}, we also obtain ``local gravitational constraints'' on a Killing horizon relating the second-order perturbed area on the horizon to the generator of Killing translations. In \cref{subsec:dressed}, we construct ``dressed observables'' on the horizon by extending the algebra on the horizon to include these second-order gravitational charges and ``dressing'' the field to these charges.

\subsection{Von Neumann Algebras and Modular Flow on Killing Horizons}
\label{subsec:VNmodKilling}
In this subsection, we explicitly construct the relevant von Neumann algebras and their corresponding modular flows on $\mc{H}^{-}$ that we shall need for the construction of dressed observables. In general, explicit computation of the modular flow is prohibitively difficult; however, there are a few cases in which it is known. Of particular interest in this paper are the cases of the exterior of a {(stationary)} black hole in the Hartle-Hawking state or a de Sitter static patch in the Bunch-Davies state. In both cases, modular flow is the family of automorphisms associated to a ``time-translation'' Killing vector that acts geometrically throughout the entire spacetime. These cases are discussed in detail in sec.~\ref{sec_stationary}. However, this will not be the case for general black hole spacetimes and so we must abandon the idea of the modular flow acting geometrically everywhere in the spacetime. As we will see, the modular flow of a stationary state in a black hole spacetime will act geometrically on $\sH^-$ and this much weaker condition is still sufficient for us to obtain a Type II algebra and thus compute an entropy. 

Returning to the classical phase space, consider the observable on $\mc{P}$ associated to the linear map 
\begin{equation}
\label{eq:Lxiphi}
\phi \longrightarrow \phi + \epsilon \pounds_{\xi}\phi
\end{equation}
where $\xi^{a}$ is the Killing vector field on $(\mc{M},g)$ associated to the bifurcate Killing horizon. We note that if $\phi$ is a classical solution then the field $\pounds_{\xi}\phi$ is also a solution since $\xi^{a}$ is a Killing vector and therefore \cref{eq:Lxiphi} is a well-defined map on phase space. The observable associated to the vector field $X=\pounds_{\xi}\phi$ on phase space is 
\begin{equation}
\label{eq:Fxidef}
F_{\xi}\defn \symp^{\KG}_{\Sigma}(\phi,\pounds_{\xi}\phi) = \frac{1}{2} \int_{\Sigma}\sqrt{h}d^{3}x~n^{a}[(\pounds_{\xi}\phi)\nabla_{a}\phi-\phi\nabla_{a}\pounds_{\xi}\phi ].
\end{equation}
A straightforward but lengthy calculation using the equations of motion and the decay of fields at spatial infinity\footnote{The precise decay condition required is that $|\phi\nabla_{a}\phi|\sim O(1/r^{2+\epsilon})$ for some $\epsilon>0$.} shows that this equals the flux of the following stress-energy flux through $\Sigma$ 
\begin{equation} \label{eq:FxiFromConsCurr}
{F_{\xi} = \int_{\Sigma}\sqrt{h}d^{3}x~T_{ab}n^{a}\xi^{b},}
\end{equation}
where
\begin{equation}
T_{ab}=\nabla_{a}\phi \nabla_{b}\phi - \frac{1}{2}g_{ab}(\nabla^{c}\phi \nabla_{c}\phi + V\phi^{2}) \ .
\end{equation}
The Poisson bracket on $\mc{P}$ of the observable $F_{\xi}$ with $\phi(f)$ is given by 
\begin{equation}
\label{eq:PBFxiphi}
\{F_{\xi},\phi(f)\} = \phi(\pounds_{\xi}f), 
\end{equation}
We now consider a spacetime $(\mc{M},g)$ satisfying \ref{assump1}---\ref{assump3}. In this case, we may consider the observable $F^{\mc{H}}_{\xi}$ which generates Killing translations on the phase space of initial data on $\mc{H}^{-}$. Evaluating $F^{\mc{H}}_{\xi}$ on the horizon yields 
\begin{align}
F^{\mc{H}}_{\xi}
=-\int_{\mc{H}^{-}}dUd\Omega_{2}~UT_{UU}=&-\kappa \int_{\mc{H}^{-}}dUd\Omega_2~U\Pi^{2}
\end{align}
where we used the fact that $\xi^{a}=-\kappa U n^{a}$ on $\mc{H}^{-}$.

In the quantum theory, we may then include the observable $\op{F}^{\mc{H}}_{\xi}$ satisfying the commutation relations associated to the Poisson brackets given by \cref{eq:PBFxiphi}. This directly yields an action of $\op{F}^{\mc{H}}_{\xi}$ on the initial data $\Alg_{\mc{H}^{-}}$ on $\mc{H}^{-}$ 
\begin{equation}
\label{eq:commFxis}
[\op{F}^{\mc{H}}_{\xi},\op{\Pi}(s)] = i\op{\Pi}(\pounds_{\xi}s) \ .
\end{equation}
In the quantum theory, the flux $\op{F}_{\xi}$ can be expressed as the integral 
\begin{equation}
\op{F}^{\mc{H}}_{\xi} =- \kappa \int_{\mc{H}^{-}}dUd\Omega_2~U:\op{\Pi}^{2}(x):
\end{equation}
where $:\op{\Pi}^{2}(x):$ is the suitably ``normal ordered'' (i.e., vacuum subtracted with respect to $\s^{\mc{H}}_{0}$) product of $\op{\Pi}(x)$.

We now construct the von Neumann algebra of observables on the horizon. In the ``bulk'' of the spacetime, we shall be primarily interested in the algebra of observables in the ``right wedge'' $(\mc{R},g)$ corresponding to the exterior of a black hole and/or the ``interior'' of the cosmological horizon. By assumption \ref{assump4}, there exists a stationary state $\s_{0}$ in $(\mc{M}_{\textrm{R}},g)$ which therefore satisfies
\begin{equation}
\s_{0}(\alpha_{t}[a]) = \s_{0}(a)
\end{equation}
for any $a\in \Alg_{\mc{M}_{\textrm{R}}}$ in the algebra of observables in $(\mc{M}_{\textrm{R}},g)$ and $t \in \mathbb{R}$. By assumption \ref{assump3} the Cauchy data on $\mc{H}^{-}$ can be specified independently, so we may construct the von Neumann algebra on $\mc{H}^-_{\textrm{R}}$ which corresponds to part of the initial data necessary to determine the observables in $(\mc{R},g)$. 

The construction of the von Neumann algebra on $\mc{H}^-_{\textrm{R}}$ proceeds as generally outlined in \cref{subsec:vNalgebra}. The (unique) state which is stationary under the Killing flow is the Gaussian state $\s^{\mc{H}}_{0}$ on $\Alg_{\mc{H}^{-}}$ with vanishing one-point function and two-point function given by \cref{eq:vacH-}. We denote the corresponding GNS Fock representation associated to $\s^{\mc{H}}_{0}$ on $\mc{H}^{-}$ as $\ms{F}$. The Fock space $\ms{F}$ contains a dense subspace of Hadamard states given by the space of $\ket{\s_{0}}$ and any finite products of $\op{\Pi}(s_{i})$ applied to it, where the $s_{i}$ are arbitrary test functions. The Fock space $\ms{F}$ is also a (reducible) representation of the subalgebra $\Alg_{\mc{H}_{\textrm{R}}^{-}}\subset \Alg_{\mc{H}^{-}}$of local observables $\op{\Pi}(s)$ with support on $\mc{H}_{\textrm{R}}^{-}$. The corresponding Weyl algebra generated by the ``exponentiated'' observables $\exp(i\op{\Pi}(s)),$ with the support of $s$ on $\sH^-_{\rm R}$ forms a $C^{\ast}$-algebra. Taking the double commutant on the GNS Hilbert space yields the von Neumann algebra $\mf{A}(\mc{H}_{\textrm{R}}^{-},\s_{0})$. The commutant of the algebra $\mf{A}(\mc{H}_{\textrm{R}}^{-},\s_{0})$ corresponds to the similarly constructed von Neumann algebra $\mf{A}(\mc{H}_{\textrm{L}}^{-},\s_{0})$ associated to $\mc{H}_{\textrm{L}}^{-},$ i.e., 
\begin{equation}
[\mf{A}(\mc{H}_{\textrm{R}}^{-},\s_{0})]^{\prime} = \mf{A}(\mc{H}_{\textrm{L}}^{-},\s_{0}).
\end{equation}
By the commutation relations \cref{eq:commKill} the algebra $\mf{A}(\mc{H}_{\textrm{R}}^{-},\s_{0})$ has a trivial center and is therefore a factor. Furthermore, due to the high degree of entanglement in $\s_{0}$ across the bifurcation surface (see \cref{eq:vacH-}) the algebra $\mf{A}(\mc{H}_{\textrm{R}}^{-},\s_{0})$ is a Type $\textrm{III}_{1}$ von Neumann algebra. 

We recall that, quite generally, if the action of the modular operator $\op{\Delta}_{\s_{0}}$ associated to $\s_{0}$ is strongly continuous on the Hilbert space, it can be represented in terms of a ``modular Hamiltonian'' 
\begin{equation}
\label{eq:modHam}
 \op{H}_{\s_{0}} \defn -\log \op{\Delta}_{\s_{0}}.
\end{equation}
We emphasize that while \cref{eq:modHam} is the general expression for the modular Hamiltonian, if the state $\s_{0}$ is not KMS, then $\op{H}_{\s_{0}}$ will generally not generate any local diffeomorphism of the spacetime. However, as explained in \cref{subsec:quantKillingscri}, the state $\s_{0}$ is KMS on $\Alg_{\mc{H}^-_{\textrm{R}}}$ with respect to Killing time translations with inverse temperature $\beta = 2\pi / \kappa,$. Recalling that a state is always KMS of unit inverse temperature with respect to its modular flow, it follows that the modular flow associated to the (cyclic-separating) state $\s^{\mc{H}}_{0}$ on $\mf{A}(\mc{H}_{\textrm{R}}^{-},\s_{0})$ is 
\begin{equation}
\label{eq:modauto}
\op{\Delta}^{-i\eta}_{\s_{0}} a \op{\Delta}^{i\eta}_{\s_{0}} = \alpha_{t = \beta \eta}[a] \textrm{ }\quad a\in \mf{A}(\mc{H}_{\textrm{R}}^{-},\s_{0}), \textrm{ }\eta\in \bb{R}.
\end{equation}
Expanding \eqref{eq:modauto} {(and the analogous equation on $\sH^-_{\rm L}$)} 
to first-order around $\eta = 0,$ one finds the modular Hamiltonian for $\mf{A}(\mc{H}_{\textrm{R}}^{-},\s_{0})$ in the state $\om_0$ is given by 
\begin{equation}
\label{eq:HmodFxi}
\op{H}_{\s_{0}} = \beta \op{F}^{\mc{H}}_{\xi} \quad \quad \quad \textrm{ (on $\mc{H}^{-}$)} \ .
\end{equation}
This is the statement of the Bisognano-Wichmann theorem \cite{bisognano1975duality}, or more generally, its generalization to Killing horizons by Sewell \cite{sewell1982quantum}.
We note that the modular automorphism \cref{eq:modauto} is an ``outer'' automorphism on $\mf{A}(\mc{H}_{\textrm{R}}^{-},\s_{0})$ since (any bounded function of) $\op{F}^{\mc{H}}_{\xi}$ is not an element of $\mf{A}(\mc{H}_{\textrm{R}}^{-},\s_{0})$. 

Since the vector field $\xi^{a}$ vanishes at the bifurcation surface $\mc{B}$, it is tempting to define the quantities $\op{F}^{\textrm{L}}_{\xi}$ and $\op{F}^{\textrm{R}}_{\xi}$ which would naively generate separate Killing time translations on $\mc{H}_{\textrm{L}}^{-}$ and $\mc{H}_{\textrm{R}}^{-}.$ The total flux would then be represented as 
\begin{equation}
\label{eq:FxiLR}
\op{F}^{\mc{H}}_{\xi} = \op{F}_{\xi}^{\mc{H},\textrm{R}}+\op{F}_{\xi}^{\mc{H},\textrm{L}}
\end{equation}
where 
\begin{equation}
\op{F}_{\xi}^{\mc{H},\textrm{R}}\defn - \kappa \int_{\mc{H}^{-}_R}dUd\Omega_2~U:\op{\Pi}^{2}(x):\textrm{ and }\op{F}_{\xi}^{\mc{H},\textrm{L}}\defn -\kappa \int_{\mc{H}^{-}_L}dUd\Omega_2~U:\op{\Pi}^{2}(x): \ .
\end{equation}
However, this splitting is impossible. Due to the infinite entanglement of the quantum field across the bifurcation surface, the quantities $\op{F}_{\xi}^{\textrm{L}}$ and $\op{F}_{\xi}^{\textrm{R}}$ have infinite fluctuations for any Hadamard state, so are not well-defined observables. Nevertheless, the sum given by \cref{eq:FxiLR} is a well-defined observable. From the point of view of von Neumann algebras, the failure of this splitting is equivalent to the failure in splitting the modular Hamiltonian for any Type III von Neumann algebra. However, a splitting of the modular Hamiltonian is possible for a von Neumann algebra of Type I or II, and we will see that such an algebra can be constructed once the perturbative gravitational constraints are taken into account.

\subsection{Perturbation Theory and Gravitational Charges}
\label{subsec:charge}

The discussion has thus far focused on the description of classical and quantum fields on a curved background spacetime where the metric is treated as non-dynamical. However, this picture changes considerably when one takes into account effects due to perturbative (quantum) gravity. Formally, one can view the previous discussion as describing the quantum theory in the strict $G_{\textrm{N}}\to 0$ limit and, in this section, we shall consider the ``small $G_{\textrm{N}}$'' corrections to this picture. The new key ingredient when the spacetime is dynamical is that diffeomorphisms are now gauged and the (conserved) charges associated with these diffeomorphisms correspond to ``boundary charges''. 

In the perturbative regime considered in this paper, the background spacetime $(\mc{M},g)$ will be kept fixed and non-dynamical. The matter fields will be treated as first-order perturbations propagating on the background spacetime. In this setting, the only diffeomorphisms that preserve the background structure are the isometries of the background metric corresponding to the vector fields $X^{a}$ on $\mc{M}$ satisfying
\begin{equation}
\pounds_{X}g_{ab}=0. 
\end{equation}
{A generic diffeomorphism will change the background metric, so the perturbative theory formulated around a single fixed background metric cannot be generally diffeomorphism invariant. A fully covariant treatment in which general diffeomorphisms are considered is beyond the scope of this paper. 

To construct the gravitational observables associated to the isometries of $(\mc{M},g)$ we must first extend the phase space $\mc{P}$ introduced in \cref{subsec:scfieldth} to include gravitational perturbations as well. More precisely, we now consider a one-parameter family of Klein-Gordon fields $\phi(\lambda)$ and metrics $g_{ab}(\lambda)$ on a manifold $\mc{M}$ which solve the Einstein field equations $G_{ab}(\lambda)=8\pi G_{\textrm{N}}T_{ab}(\lambda)$ and are smooth in the parameter $\lambda\in \bb{R}$. In the case where the spacetime has a particular asymptotic structure {we will require that this asymptotic structure is the same for each value of $\lambda.$ For example, for an asymptotically flat spacetimes, then we will require that $g_{ab}(\lambda)$ suitably decays to the Minkowski metric $\eta_{ab}$ in spacelike/null directions and, similarly, that $\phi(\lambda)$ suitably decays as well (see e.g., \cite{Hollands:2003ie,Satishchandran:2019pyc} for a complete discussion of the asymptotic conditions at spatial and null infinity). At $\lambda=0$ we have that 
\begin{equation}
\label{eq:bckgrnd}
g_{ab}=g_{ab}(0) \textrm{ and }\phi(0)=0
\end{equation}
where $g_{ab}$ is the ``background metric'' which satisfies the source-free Einstein equations. We will frequently make use of one-parameter and, sometimes, two-parameter variations $g_{ab}(\lambda_{1},\lambda_{2})$. In these cases, we will use the notation 
\begin{equation}
\delta g_{ab}\defn \frac{d}{d\lambda}g_{ab}\bigg\vert_{\lambda=0},\quad \delta^{2}g_{ab}\defn \frac{d^{2}}{d\lambda^{2}}g_{ab}\bigg\vert_{\lambda=0}, \quad \dots 
\end{equation}

Since the first-order perturbations shall play a distinguished role in our analysis, we shall separately label them as
\begin{equation}
\gamma_{ab}\defn \delta g_{ab} \textrm{ and }\phi \defn \delta \phi.
\end{equation}
The linearized metric perturbation $\gamma_{ab}$ satisfies the source-free linearized Einstein equation $\delta G_{ab}=0$ and $\phi$ satisfies the source-free Klein-Gordon equation \cref{eq:EOMphi}. Since $\gamma_{ab}$ and $\phi$ satisfy independent equations of motion we will, for simplicity, set the first-order metric perturbation to vanish but the first-order Klein-Gordon field will be non-vanishing. Therefore, the points in this phase space\footnote{This symplectic subspace is an affine space since $\delta g_{ab}$ can be changed by a linearized gauge transformation $\delta g_{ab} \to \delta g_{ab} + \nabla_{(a}X_{b)}$ for any vector field $X^{a}$. The subspace of relevance for the main body of this paper is the space of linearized perturbations with vanishing linearized Riemann tensor. Any such solution can be chosen as an ``origin'' of this space.} that we shall consider are points $(\gamma_{ab}=0,\phi)\in \mc{P}$. We will continue to refer to the phase space as $\mc{P}$ since there is a natural identification with the original phase space\footnote{In the quantization of $\mathcal{P}$ we cannot set $\gamma_{ab}$ to vanish as an operator since it is fundamentally subject to vacuum fluctuations. Quantizing the scalar field and the linearized metric (as explained in \cref{sec:grav_EM_class}) yields a Hilbert space $\Hilb\otimes \Hilb_{\textrm{grav.}}$ which contains an additional Hilbert space of gravitons. Thus the quantum analog of restricting to the phase space $\mathcal{P}$ is to fix the quantum state of the gravitons to be in the vacuum.} constructed in \cref{subsec:scfieldth}. Note that we only introduce the first-order perturbations in the phase space. Second-order and higher perturbations are not treated in our analysis.\footnote{As we will see, important exceptions to this rule are the second-order perturbed ``gravitational charges'' (for example, perturbed area of the horizon or the perturbed ADM mass) which will play a key role in constructing dressed observables in the quantum theory.} The phase space for $\gamma$ nonvanishing is constructed in \cref{sec:grav_EM_class}. Furthermore, the symplectic product of any elements of $\mc{P}$ is 
\begin{equation}
\label{eq:sympgrsympKG}
\symp_{\Sigma}((0,\phi_{1}),(0,\phi_{2})) = \symp_{\Sigma}^{\KG}(\phi_{1},\phi_{2})
\end{equation}

While the above considerations appear to be a minor modification to the phase space structure given in \cref{subsec:scfieldth}, the key difference corresponds to the phase space observable on $\mc{P}$ associated to the linear map
\begin{equation}
\label{eq:m}
\phi \to \phi +\epsilon \pounds_{X} \phi
\end{equation}
where $\gamma_{ab}$ is fixed to zero and $X^{a}$ is a Killing vector. On the phase space, $\mc{P}$, the symplectic product of $\phi$ with $\pounds_{X}\phi$ was, quite generally, shown\footnote{The analog of \cref{eq:BDYCharge} was originally derived in \cite{Hollands:2012sf} in vacuum GR where it was shown that the symplectic form evaluated on a pair of metric perturbations $\gamma_{1}$ and $\gamma_{2}$ is equivalent to the integral of a boundary term. The analysis can be straightforwardly extended to include scalar field perturbations (see \cref{app:cov_phase}).} in \cite{Hollands:2012sf} to be a ``boundary term'' 
given by (see Appendix \ref{app:cov_phase})
\begin{equation}
\label{eq:BDYCharge} 
\symp_{\Sigma}((0,\phi),(0,\pounds_{X}\phi))=\int_{\Sigma}~d\mc{Q}_{X}(g,\phi,\delta^{2}g,\delta^{2}\phi )
\end{equation}
where, for compactness, we have expressed the integrand in terms of differential form notation where the ``charge'' $\mc{Q}_{X}$ is a $2$-form. The precise form of $\mc{Q}_{X}$ can be found in Appendix \ref{app:cov_phase}, however, we will not need the precise definition in the present discussion. Note that the charge $\mc{Q}_{X}$ itself is not a function on $\mc{P},$ only the integration of $d\mc{Q}_X$ over an entire Cauchy slice is a genuine function on phase space which can be taken as being defined by the LHS of~\eqref{eq:BDYCharge}. Nonetheless,~\cite{Hollands:2012sf} established that a {\em local} two-form charge $\mc{Q}_X$ can be defined in such a way that~\eqref{eq:BDYCharge} holds. We shall express this flux integrated over a Killing horizon and null infinity in terms of the difference of charges defined on these surfaces. In those cases, we will give explicit formulas for these charges. 

There are two key observations of \cref{eq:BDYCharge}. The first is that the ``charge'' depends upon {\em second-order} perturbations of the field. Thus, these charges are not functions on the phase space $\mc{P}$ nor {are they promoted to operators on} its corresponding quantization $\Hilb$. The inclusion of these charges yields a ``global gravitational constraint'' relating the boundary values of charges $\mc{Q}_{X}$ to the bulk quantum fields. The second observation is that while in an open universe the right-hand side of \cref{eq:BDYCharge} will be generically non-vanishing, in a closed universe the integral clearly evaluates to zero. Therefore, on the full phase space $\mc{P}$ for a closed universe, \cref{eq:BDYCharge} does not yield a phase space observable but instead the ``global constraint'' that total flux given by the left-hand side of \cref{eq:BDYCharge} vanishes.

While the general formula \cref{eq:BDYCharge} applies to any isometry of $(\mc{M},g)$, the precise group of isometries depends on the specific spacetime under consideration. Specializing to the case where the spacetime $(\mc{M},g)$ satisfies assumptions \ref{assump1}---\ref{assump4}, the isometries $\chi_{t}$ generated by the Killing vector $\xi^{a}$ associated to the bifurcate Killing horizon will always exist. Therefore, we shall restrict attention to the ``gravitational charges'' associated to the Killing vector field $\xi^{a}$. 

By assumptions \ref{assump2} and \ref{assump3}, we may consider the phase space of initial data on $\mc{H}^{-}$ independently from observables associated to the rest of the Cauchy slice $\mc{H}^{-}\cup \Sigma^{\prime}$. The flux observable associated to this isometry to $\sH^-$ is obtained by evaluating the left-hand side of \cref{eq:BDYCharge} for $\Sigma=\mc{H}^{-}$ with $X=\xi$ which we recall yields \begin{equation}
\label{eq:BDYChargeH-} 
\symp_{\mc{H}^{-}}((0,\phi),(0,\pounds_{\xi}\phi))=\symp^{\KG}_{\mc{H}^{-}}(\phi,\pounds_{\xi}\phi) = F^{\mc{H}^{-}}_{\xi}.
\end{equation}
We now show that the observable \cref{eq:BDYChargeH-} can also be related to second-order ``boundary charges'' defined on $\mc{H}^{-}$. Indeed, it was shown in \cite{Chandrasekaran:2019ewn,Chandrasekaran:2018aop} that on a Killing horizon, one can obtain the global charge-flux relation associated to $\xi^{a}$ from the following local expression
\begin{equation}
\label{eq:pVQ}
\partial_{U}\delta^{2}\mc{Q}_{U,\xi}= 8\pi G_{\textrm{N}} \int_{S_{U}}d\Omega_2~U\delta^{2}T_{UU},
\end{equation}
where $\delta^{2}T_{ab}$ is the second-order variation of the stress tensor $T_{ab}(\lambda)$, $S_{U}$ is a constant $U$ cross-section of $\mc{H}^{-}$ and $\delta^{2}\mc{Q}_{U,\xi}$ is the second-order variation of\footnote{The charge \cref{eq:QVxi} and its associated conservation law was obtained on arbitrary smooth null surfaces in vacuum general relativity \cite{Chandrasekaran:2019ewn,Chandrasekaran:2018aop}. This result can be straightforwardly generalized to include matter field perturbations on a Killing horizon using Raychaudhuri's equation.} 
\begin{equation}
\label{eq:QVxi}
\mc{Q}_{U,\xi}(\lambda)\defn \int_{S_{U}}d^{2}x~\sqrt{q(\lambda)}~[1-U\theta(\lambda)],
\end{equation}
with $q(\lambda)\defn \det[q_{AB}(\lambda)]$ and $d^{2}x\sqrt{q(\lambda)}$ is the measure on the cross-section $S_{U}$ with fixed, arbitrary coordinates $x^{A}$. {For brevity we will henceforth drop the subscript $X=\xi$ 
 and write the perturbed charge as $\delta^{2}\mathcal{Q}_{U}$.} 
 
 Integrating~\eqref{eq:pVQ} along all of $\sH^-$ and recalling~\eqref{eq:FxiFromConsCurr} with $\xi^a = -\kappa U n^a$ on $\sH^-$, the flux $F^{\mc{H}}_{\xi}$ can be expressed as a difference of ``second-order'' charges 
\begin{equation} \label{eq:fluxSecondOrderCharges}
-\frac{8\pi G_{\textrm{N}}}{\kappa} F^{\mc{H}}_{\xi} = \lim_{U\to +\infty}\delta^{2}\mc{Q}_{U} - \lim_{U\to -\infty}\delta^{2}\mc{Q}_{U}.
\end{equation}
Defining the area of a cross-section $S_{U}$ for general $g_{ab}(\lambda)$ as 
\begin{equation}
A_{U}(\lambda)\defn \int_{S_{U}}d^{2}x~\sqrt{q(\lambda)}
\end{equation}
then, recalling that $\theta(0)=0=\delta \theta$, we have that 
\begin{equation}
\label{eq:QU}
\delta^{2}\mc{Q}_{U} = \delta^{2}A_{U} - \int_{S_{U}}d\Omega_2~U\delta^{2}\theta 
\end{equation}
and we recall that $d\Omega_{2}=d^{2}x\sqrt{q(0)}$.} Note that $\delta^{2}\mc{Q}_{U}$ depends on the second-order perturbations of the metric. 
Defining the limiting perturbed charges as\footnote{{The horizon becomes stationary at asymptotically late times and so the perturbed expansion vanishes and $\delta^{2}\mathcal{Q}_{U}$ approaches $\delta^{2}A_{U}$ at late times.}}
\begin{equation}
\delta^{2}\mathcal{Q}_{+} \defn \lim_{U\to +\infty}\delta^{2}\mathcal{Q}_{U},\quad \delta^{2}\mathcal{Q}_{-} \defn \lim_{U\to -\infty}\delta^{2}\mathcal{Q}_{U} \ ,
\end{equation}
and \eqref{eq:fluxSecondOrderCharges} becomes
\begin{equation}
\label{A+-F}
\delta^{2} \mathcal{Q}_{+} -\delta^{2}\mathcal{Q}_{-}= -4G_{\textrm{N}}\beta F^{\mc{H}}_{\xi}.
\end{equation}

Unless the global constraints \eqref{eq:BDYCharge} imply that $F^{\mc{H}}_{\xi}$ must vanish, the difference of charges in~\eqref{A+-F}, by virtue of the RHS, is a well-defined observable on the phase space of initial data on $\mc{H}^{-}$. Correspondingly, in the quantum theory, we may define the difference of ``asymptotic perturbed charge operators'' by
\begin{equation}
\label{eq:+-F}
\delta^{2} \op{\mathcal{Q}}_{+}-\delta^{2}\op{\mathcal{Q}}_{-} \defn -4 G_{\textrm{N}}\beta\op{F}^{\mc{H}}_{\xi} \ .
\end{equation}
This is a well-defined element of the algebra $\Alg$. Similarly, by \cref{eq:HmodFxi}, we also have that the perturbed charge difference is equivalent to the action of modular Hamiltonian 
\begin{equation}
 \delta^{2}\op{\mathcal{Q}}_{+}-\delta^{2}\op{\mathcal{Q}}_{-} = -4 G_{\textrm{N}}\op{H}_{\s_{0}}\quad \quad \quad \textrm{ (on $\mc{H}^{-}$)}.
\end{equation}
However, we note that the individual perturbed charges $\delta^{2}\mathcal{Q}_{+}$ and $\delta^{2}\mathcal{Q}_{-}$ are not elements of phase space $\mc{P}$ and similarly, $\delta^{2}\op{\mathcal{Q}}_{+}$ and $\delta^{2}\op{\mathcal{Q}}_{-}$ are not elements of $\Alg$. In the next section, we will extend the algebra $\Alg$ to explicitly include $\delta^{2}\op{\mathcal{Q}}_{+}$ and $\delta^{2}\op{\mathcal{Q}}_{-}$ subject to the constraint \cref{eq:+-F}.

\subsection{Algebra of Dressed Observables on Killing Horizons}
\label{subsec:dressed}
The observables in quantum gravity are diffeomorphism invariant. In this perturbative regime with a fixed background metric, general diffeomorphism invariance is reduced to invariance under the isometries of the background geometry. As is well-known, the local observables $\phi(f)$ are not diffeomorphism invariant, and this remains the case even if one restricts to isometries. However, it has long been suggested that one can construct ``gravitationally dressed observables'' \cite{DeWitt:1962cg,DeWitt:1967yk,2016PhRvD..93b4030D, 2016JHEP...09..102D,Brown:1994py,Ciambelli:2023mir} or ``relational observables''\footnote{The equivalence of ``gravitationally dressed observables''  and ``relational observables'' was shown in \cite{Goeller:2022rsx}} \cite{Rovelli:1995fv,Dittrich:2005kc,2012SIGMA...8..017T,Goeller:2022rsx}
using either gravitational degrees of freedom or other gravitating degrees of freedom to construct ``joint'' observables invariant under the action of diffeomorphisms.\footnote{ In the non-relativistic case, there has been significant progress in the construction of ``quantum reference frames'' (see, e.g., \cite{Giacomini:2017zju,delaHamette:2021oex,hoehn:2023ehz}). For a formulation of relational observables within the algebraic framework outlined in \cref{subsec:AlgQFTCS} see, e.g.,  \cite{Brunetti:2022itx,Brunetti:2016hgw}.} 

Additionally, we will be interested in observables restricted to a particular subregion $\mc{R}$ of the full spacetime. One can only hope to describe such `$\mc{R}-$localized' observables in a diffeomorphism invariant way if the region $\mc{R}$ itself is invariantly defined and so we will demand that $\mc{R}$ be invariantly defined. For example, we will consider the (right) exterior of an asymptotically flat black hole which can be defined invariantly as $\mc{R} = I^-(\scri^+_{\rm R}) \cap I^+(\scri^-_{\rm R})$ or in cases in which an observer traveling along a worldline $\gamma$ is present, we consider their domain of communication $\mc{R} = I^+(\gamma) \cap I^-(\gamma).$ Diffeomorphism invariance is then replaced by invariance under the isometries of $(\mc{R},g)$. In this subsection, we shall focus on the construction of observables invariant under the isometries generated by the horizon Killing field $\xi^{a}$. Since by assumptions \ref{assump1}---\ref{assump3}, the data for region $(\mc{R},g)$ can be independently specified on $\mc{H}_{\textrm{R}}^{-}$ we will construct the von Neumann algebra of observables on $\mc{H}_{\textrm{R}}^{-}$. With this ingredient, the construction of the global, invariant algebra in $\mc{R}$ can be straightforwardly constructed in each spacetime of interest. The construction of observables invariant under the full group of isometries will depend upon the specific spacetime under consideration and can be constructed on a case-by-case basis using the methods developed in this section. 

A key ingredient in our general analysis will be a ``structure theorem'' which illustrates the origin of Type II algebras in perturbative quantum gravity for spacetimes with bifurcate Killing horizons. As we will now show, this structure theorem arises naturally when one constructs observables on $\mc{R}$ invariant under the action of the Killing flow $\chi_{t}$. To construct such observables in $\mc{R}$ we will need to ``dress'' the local field $\phi(f)$ to a (second-order) ``gravitational charge'' in $\mc{R}$ associated to the Killing isometry. Let $X$ be the gravitational charge associated to $\xi$ (e.g., For a Schwarzschild black hole, $X$ would be the (second-order) perturbed ADM mass). Since these charges depend upon {\em second-order} metric perturbations and are therefore not elements of $\mc{P}$. Consequently, prior to imposing the constraints, we must first extend the phase space $\mc{P}$ to include the one-dimensional degree of freedom $X\in \mc{P}_{\textrm{X}}$ such that the full phase space is 
\begin{equation}
\label{eq:phaseX}
\mc{P}\oplus \mc{P}_{\textrm{X}} \ ,
\end{equation}
where $X$ Poisson commutes with all observables of $\mc{P}$ and, of course, commutes with itself. We denote the corresponding $\ast$-algebra of observables generated by $\op{X}$ as $\Alg_{\textrm{X}}$ so that the full algebra is simply the tensor product
\begin{equation}
\label{eq:AlgxAlgX}
\Alg \otimes \Alg_{\textrm{X}}. 
\end{equation}
States on this larger algebra now also require the specification of a wave function of the one-dimensional degree of freedom $X$. The group of Killing translations is $\bb{R}$ and so we quantize $\op{X}$ on the Hilbert space $\Hilb_{\textrm{X}}\defn L^{2}(\bb{R})$ by of square-integrable wave functions $\bb{R}$, where $\op{X}$ acts as in \eqref{eq:Xop}.  The full, extended Hilbert space is 
\begin{equation}
\label{eq:HilbextX}
\Hilb_{\textrm{ext.}}\defn \Fock \otimes \Hilb_{\textrm{X}}.
\end{equation}
The Hilbert space $\Hilb_{\textrm{X}}$ has a (densely defined) ``conjugate'' operator $\op{t}$ which satisfies 
\begin{equation}
[\op{X},\op{t}]=i
\end{equation}
where we have labeled the conjugate variable as $\op{t}$ since $\op{X}$ will generally have dimensions of energy and so $\op{t}$ will have the interpretation of a ``time'' associated to the Killing flow $\chi_{t}$. 

Using Stokes' theorem to evaluate the RHS of \cref{eq:BDYCharge} one obtains the difference between the values of the charge at different boundaries of the Cauchy surface for $\mc{M}_{\textrm{R}}$. Denoting the boundary values of the charge by $C$ and $X$ where $C$ is a gravitational charge that lies in the causal complement of $\mc{R}$ 
\begin{equation}
F_{\xi}= X - C
\end{equation}
In the quantum theory, since we have already included $\op{X}$ into the algebra, the inclusion of $\op{C}$ defined as 
\begin{equation}
\label{eq:CXF}
\op{C}\defn \op{X} - \op{F}_{\xi}
\end{equation}
yields a non-trivial constraint on the algebra since $\op{C}$ commutes with all ``physical'' observables in $\mc{R}$. We will refer to such observables that commute with the constraint \cref{eq:CXF} as ``dressed observables'' in $\mc{R}$. 

\Cref{eq:CXF} yields a non-trivial constraint on the observables $\op{\Pi}(s)$ on $\mc{H}_{\textrm{R}}^{-}$ since the action of $\op{F}_{\xi}$ on such observables is equivalent to the action of $\op{F}_{\xi}^{\mc{H}}$. The observables on the subalgebra $\Alg_{\mc{H}_{\textrm{R}}^{-}}\otimes \Alg_{\textrm{X}}$ which commute with the constraints are the identity $\op{1}$, functions of $\op{X}$ and the ``dressed observables''\footnote{The operator $e^{i \op{F}^{\mc{H}}_{\xi}\op{t}}$ is a well-defined operator on $\Fock^{\mc{H}}\otimes \Hilb_{\textrm{X}}$. In particular, diagonalizing $\op{t}$ on $\Hilb_{\textrm{X}}$ yields a well-defined operator on $\Fock^{\mc{H}}$.}
\begin{equation}
\op{\Pi}(s,\op{t})\defn e^{-i \op{F}^{\mc{H}}_{\xi}\op{t}}\op{\Pi}(s)e^{i\op{F}^{\mc{H}}_{\xi}\op{t}}.
\end{equation}
where the support of $s$ is on $\mc{H}_{\textrm{R}}^{-}$. The unital $\ast$-algebra $\Alg^{\textrm{dress.}}_{\mc{H}^{-}_{\textrm{R}}}$ of dressed observables $\op{\Pi}(s;\op{t})$ satisfy 
\begin{equation}
[\op{\Pi}(s_{1};\op{t}),\op{\Pi}(s_{2};\op{t})]=-i\symp^{\KG}_{\mc{H}^{-}}(s_{1},s_{2})\op{1}.
\end{equation}
Therefore, the algebra of dressed observables is isomorphic to the algebra of undressed observables
\begin{equation}
\Alg^{\textrm{dress.}}_{\mc{H}^{-}_{\textrm{R}}} \cong \Alg_{\mc{H}^{-}_{\textrm{R}}}.
\end{equation}
However, the extended algebra of dressed observables together with the charge $\op{X}$ forms a ``crossed product'' algebra 
\begin{equation}
\Alg_{\mc{H}^{-}_{\textrm{R}}}^{\textrm{ext.}}\defn \Alg^{\textrm{dress.}}_{\mc{H}^{-}_{\textrm{R}}}\rtimes \Alg_{\textrm{X}}
\end{equation}
where the observables $\op{X}$ and $\op{\Pi}(s,\op{t})$ satisfy the additional commutation
\begin{equation}
[\op{X},\op{\Pi}(s;\op{t})] 
=i\op{\Pi}(\pounds_{\xi}s,\op{t}) \ .
\end{equation}

We now consider the algebra of ``gravitationally dressed'' observables on $\mc{H}^{-}_{\textrm{R}}$. The observables in the region $\mc{H}_{\textrm{R}}^{-}$ which commute with $\op{C}$ are the identity, functions of $\op{X}$ itself, and the dressed quantum field theory observables

The corresponding von Neumann algebra of observables localized to the subregion $\sR$ is
\begin{equation}
\label{eq:VNdressobs}
\vNext(\mc{H}^{-}_{\textrm{R}},\s_{0}) \defn \{\op{\Pi}(s;\op{t}),\op{X}\}^{\prime \prime} \quad \quad \quad \textrm{supp}(s)\subset \mc{H}^{-}_{\textrm{R}} \ ,
\end{equation}
where the commutant is taken in $\Fock^{\mc{H}}\otimes \Hilb_{\textrm{X}}$.

Recall that $\beta \op{F}^{\mc{H}}_{\xi} = \op{H}_{\s_{0}}$ on $\mc{H}^{-}$, where $\op{H}_{\s_{0}}$ is the modular Hamiltonian for the algebra $\mf{A}(\sH^-_{\rm R}, \om_0)$ in the state $\om^{\mc{H}}_0$. Conjugating the operators $\{\op{\Pi}(s,\op{t}), \op{X}\}$ by $e^{i \op{F}^{\mc{H}}_{\xi} \op{t}}$ yields a unitarily equivalent algebra $\hat{\mf{A}}_{\textrm{ext.}}(\sH^-_{\rm R}, \om_0) \defn \{\op{\Pi}(s),\op{X} + \beta^{-1} \op{H}_{\om_0}\}''$. In this presentation, we see that the second factor generates the modular automorphism group of $\mf{A}(\sH^-_{\rm R}, \om_0)$ in the state $\om^{\mc{H}}_0$, so this algebra (which is unitarily equivalent to $\vNext(\mc{H}^{-}_{\textrm{R}}, \om_0)$) is the crossed product of $\mf{A}(\sH^-_{\rm R}, \om_0)$ with its modular automorphism group. Since the algebra $\mf{A}(\sH^-_{\rm R}, \om_0)$ is of Type III$_1$, the crossed product with its modular group (for our faithful normal state $\om_0$) is a von Neumann algebra of Type II$_{\infty}$~\cite{takesaki1973duality}. The unitary conjugation does not affect the type, so we conclude that $\vNext(\mc{H}^{-}_{\textrm{R}},\s_{0})$ is also of Type II$_{\infty}$.

We associate the commutant of $\vNext
(\mc{H}^{-}_{\textrm{R}},\s_{0})$ with $\mathcal{H}_L^-$
\begin{equation}
\vNext(\mc{H}^{-}_{\textrm{L}},\s_{0})\defn \vNext(\mc{H}^{-}_{\textrm{R}},\s_{0})^{\prime} = \{\op{\Pi}(s),\op{X} - \op{F}^{\mc{H}}_{\xi}\}^{\prime \prime} \quad \supp(s)\subset \mc{H}_{L}^{-} \ .
\end{equation}
This algebra is also Type II$_{\infty}$. Since ${\mf{A}}_{\textrm{ext.}}(\mc{H}^{-}_{\textrm{R}},\s_{0})$ is of Type II, a trace functional can be defined on it. The explicit formula is
\begin{align} \label{eq:traceUnconj}
 \Tr(\hat{a}) = \beta\int_{\bb{R}}dX~e^{\beta X}\bra{\s_0,X}\hat{a}\ket{\s_0,X}, \quad \hat{a}\in {\mf{A}}_{\textrm{ext.}}(\mc{H}^{-}_{\textrm{R}},\s_{0}) 
\end{align}
where we recall that $\ket{\s_{0},X}\defn \ket{\s_{0}}\otimes \ket{X}$. We note that, due to the integral over the entire real line, many elements of $\vNext(\mc{H}^{-}_{\textrm{R}},\s_{0})$ --- including all operators in the Type III$_1$ subfactor $\le(\Alg^{\textrm{dress.}}_{\mc{H}^{-}_{\textrm{R}}}\ri)''$ --- have infinite trace. 

We remark that in the crossed product construction thus far, it was crucial for the auxiliary variable, $\op{X}$, to take values on the entire real line. We will encounter scenarios where $\op{X}$ naturally takes values only on a portion of the real line. If the spectrum is upper bounded, then the trace will be finite. In particular, the trace of the identity operator is
\begin{align}
 \Tr(\op{1}) = \beta \int_{-\infty}^0 dX e^{\beta X} = 1,
\end{align}
where we have taken the upper bound to be zero for simplicity. In this case, the algebra is of Type II$_1$ \cite{Chandrasekaran:2022cip}.

We summarize our above results on the von Neumann algebra of dressed observables on $\mc{H}^{-}_{\textrm{R}}$ in the following theorem: 
\begin{theorem}
\label{thm:horizontheorem}
 Let $(\mc{M},g)$ be a spacetime satisfying assumptions \ref{assump1} -- \ref{assump4}. The von Neumann algebra $\vNext(\mc{H}^{-}_{\textrm{R}},\s_{0})$ 
 \cref{eq:VNdressobs} is Type II$_{\infty}$ if the spectrum of $\op{X}$ is unbounded and if the spectrum of $\op{X}$ is bounded from above, then $\vNext(\mc{H}^{-}_{\textrm{R}},\s_{0})$ is Type II$_{1}$.
\end{theorem}

Note that, mathematically, this theorem is a direct application of Takesaki's theorem on crossed products~\cite{takesaki1973duality} applied to Killing horizons; however, we will see such algebras arise from physical considerations in the following sections.

\subsection{Density matrices and entropies on Killing Horizons}
\label{subsec:dens_ent}
By \cref{thm:horizontheorem}, the algebra $\vNext(\mc{H}^{-}_{\textrm{R}},\s_{0})$ of ``dressed observables'' is Type II and has a well-defined trace. A special class of states on $\Fock^{\mc{H}} \otimes L^{2}(\bb{R}),$ so-called classical-quantum states, take the form 
\begin{equation}
\label{eq:hats}
\ket{\hat{\omega}^{\mc{H}}} =\int_{\bb{R}}dX~f(X)\ket{\s^{\mc{H}}}\otimes \ket{X}
\end{equation}
where $\ket{\s^{\mc{H}}}\in \Fock^{\mc{H}}$ is a (normalized) Hadamard state. For simplicity,\footnote{This has no effect on the entropies. All vector representations of $\s^{\mc{H}}$ will be related to $\ket{\s^{\mc{H}}}$ by local unitary operators in the commutant algebra.} we will assume $\ket{\om^{\mc{H}}}$ lies in the ``natural cone'' $\ms{P}^{\#}$ of $\s^{\mc{H}}_{0}$. We also take $f(X)\in L^2(\mathbb{R})$ to be Schwartz so that $\ket{\hat{\s}^{\mc{H}}}$ lies in the (dense) domain on which (the unbounded operator) $\op{X}$ is defined. We can use the trace \eqref{eq:traceUnconj} to define a density matrix $\rho_{\hat{\s}}$ on $\vNext(\mc{H}^{-}_{\textrm{R}},\s_{0})$ associated to $\hat{\s}^{\mc{H}}$ by the relation 
\begin{equation}
\hat{\s}^{\mc{H}}(a) \defn \textrm{Tr}(\rho_{\hat{\s}}a), \quad \forall~a\in \vNext(\mc{H}^{-}_{\textrm{R}},\s_{0}).
\end{equation}
It is clear from this definition that the density matrix inherits the scaling ambiguity of the trace.
To obtain the density matrix $\rho_{\hat{\s}}$, it is useful to first obtain the modular operator $\op{\Delta}_{\hat{\s}}$. For $\omega^{\mc{H}}=\omega^{\mc{H}}_{0}$ in \cref{eq:hats}, the modular operator was derived in \cite{2022JHEP...10..008W}. In \cite{Chandrasekaran:2022cip}, it was derived for general $\s$ but assuming $f(X)$ ``slowly varies'' in $X$. This was improved in \cite{Jensen:2023yxy}, where it was evaluated for general $f(X)$. Since $\vNext(\mc{H}^{-}_{\textrm{R}},\s_{0})$ is a Type II von Neumann algebra, there is a splitting of the modular operator into density matrices
\begin{align}
 \op{\Delta}_{\hat{\s}} \defn \rho_{\hat{\s}} (\rho^{\prime}_{\hat{\s}})^{-1}
\end{align} 
where $\rho_{\hat{\s}}$ is affiliated with ${\mf{A}}_{\textrm{ext.}}(\mc{H}^{-}_{\textrm{R}},\s_{0})$ and $\rho_{\hat{\s}}'$ is affiliated with ${\mf{A}}_{\textrm{ext.}}(\mc{H}^{-}_{\textrm{L}},\s_{0})$. 
The density matrix, obtained by \cite{Jensen:2023yxy}, for any state $\hat{\s}$ of the form \cref{eq:hats} (with the vector representative of $\omega$ taken in the natural cone of $\omega_0$) is
\begin{align}
\label{eq:rhoR}
 \rho_{\hat{\s}} =\beta^{-1}e^{-\frac{i\op{t}\op{H}_{\s_{0}}}{\beta}}\ {{f}(\op{X}+\beta^{-1}\op{H}_{\omega_{0}})e^{-\frac{ \beta\op{X}}{2}}\op{\Delta}_{\s|\s_{0}}e^{-\frac{ \beta\op{X}}{2}}\bar{f}(\op{X}+\beta^{-1}\op{H}_{\omega_{0}})}e^{\frac{i\op{t}\op{H}_{\s_{0}}}{\beta}},
\end{align} 
where $\op{H}_{\s}$ is the modular Hamiltonian associated to the state $\s,$ $\op{H}_{\s|\s_{0}} \defn -\log \op{\Delta}_{\s|\s_{0}}$ is the relative modular Hamiltonian, and $\op{t}$ is the conjugate to $\op{X}$. One can check that $\Tr \rho_{\hat{\s}} = 1$. 

The von Neumann entropy of \cref{eq:rhoR} is manifestly well-defined but generally cumbersome to explicitly calculate. However, as pointed out in \cite{2022arXiv220910454C}, one can obtain a simple expression for the von Neumann entropy in the regime where $f(X)$ is slowly varying in $X$, which implies that it is sharply peaked in the conjugate ``time'' variable $t$. In this regime, the von Neumann entropy of \cref{eq:rhoR} 
agrees with the ill-defined, but more familiar, generalized entropy of \cite{bekenstein1974generalized}.
In the slowly varying regime, $\op{\Delta}_{\omega|\omega_0}$ and $f(\op{X} + \beta^{-1}\op{H}_{\omega_{0}})$ will approximately commute and the logarithm may be taken \cite{Chandrasekaran:2022cip,Jensen:2023yxy}\footnote{Proving eq.~\ref{eq:logrho} is highly non-trivial due to the unboundedness of the operator $\rho_{\hat{\omega}}$. A rigorous derivation of this logarithm and the control of the associated errors can be found in \cite{2023arXiv231207646K}}
\begin{equation}
\label{eq:logrho}
-\log \rho_{\hat{\s}}\simeq e^{-\frac{i\op{t}\op{H}_{\s_{0}|\s}}{\beta}}\op{H}_{\s}e^{\frac{i\op{t}\op{H}_{\s_{0}|\s}}{\beta}}-\op{H}_{\s_{0}|\s}+ \beta\op{X}-\log(|f(\op{X})|^{2}) + \log \beta.
\end{equation} 
Here, the identity $\op{H}_{\s_0}-\op{H}_{\s|\s_0}=\op{H}_{\s_0|\s}-\op{H}_{\s} $ has been used to put the equation in a form where Araki's relative entropy is manifest. Thus, the von Neumann entropy, at leading order, is given by 
\begin{align}
\label{eq:Svn_leading}
S_{\textrm{vN.}}(\rho_{\hat{\s}})&\defn -\textrm{Tr}(\rho_{\hat{\s}}\log\rho_{\hat{\s}})\nonumber \\
&\simeq \bra{\hat{\s}^{\mc{H}}}e^{-\frac{i\op{t}\op{H}_{\s_{0}|\s}}{\beta}}\op{H}_{\s}e^{\frac{i\op{t}\op{H}_{\s_{0}|\s}}{\beta}}\ket{\hat{\s}^{\mc{H}}}+\hat{\s}^{\mc{H}}(\beta\op{X})-S_{\textrm{rel.}}(\s^{\mc{H}}|\s^{\mc{H}}_{0})+S_{\textrm{vN.}}(\rho_{f}) + \log \beta\nonumber \\
&\simeq \hat{\s}^{\mc{H}}(\beta\op{X})-S_{\textrm{rel.}}(\s^{\mc{H}}|\s^{\mc{H}}_{0})+S(\rho_{f}) + \log \beta
\end{align}
where, using the slowly varying property once more, the first term in the second line is approximately zero \cite{Jensen:2023yxy}.
Here, $\rho_{f}$ is the ``classical'' density matrix on $L^{2}(\bb{R})$ defined by 
\begin{equation}
\rho_{f}\defn |f(\op{X})|^{2}
\end{equation}
and $S(\rho_{f})$ is the entropy of this density matrix on $L^{2}(\bb{R})$
\begin{equation}
S(\rho_{f})\defn -\textrm{Tr}(\rho_{f}\log \rho_{f})=-\int_{-\infty}^{\infty}dX~|f(X)|^{2}\log(|f(X)|^{2}).
\end{equation}

We now adapt an argument of Wall \cite{2012PhRvD..85j4049W} which relates the relative entropy in \cref{eq:Svn_leading} to the more familiar generalized entropy \cite{bekenstein1974generalized}. We caution the reader that any such decomposition is necessarily heuristic because it involves decomposing the well-defined expression $S_{\textrm{vN.}}(\rho_{\hat{\s}})$ into the sum of divergent quantities. 

To obtain the generalized entropy, we must be more explicit as to which gravitational charge $\op{X}$ in $\mc{R}$ that the fields are dressed to. A gravitational charge associated to the Killing field $\xi$ which, by definition, lies in $\mc{R}$ is 
\begin{equation}
\op{X} \defn \frac{\delta^{2}\op{\mathcal{Q}}_{-}}{4G_{\textrm{N}}\beta}.
\end{equation}
In particular, in all spacetimes considered in this paper, we will either dress to the perturbed charge $\delta^{2}\op{\mathcal{Q}}_{-}/4G_{\textrm{N}}\beta$ explicitly or dress to another charge in the spacetime which will be related to the perturbed charge by the global constraints. In this sense, there is no loss of generality in dressing to $\delta^{2}\op{\mathcal{Q}}_{-}/4G_{\textrm{N}}\beta$. 

Including $\op{X}$ in the algebra $\vNext(\mc{H}_{\textrm{R}}^{-}, \s_{0})$ implies that we can now integrate \cref{eq:pVQ} and define the charge $\delta^{2}\op{\mc{Q}}_{U}$ on a finite $U$ cut of $\mc{H}^{-}$ by
\begin{equation}
\label{eq:QVT}
\delta^{2}\op{\mc{Q}}_{U}\defn 4G_{\textrm{N}} \beta \op{X} +8\pi G_{\textrm{N}}\int_{-\infty}^{U}dU\int_{\bb{S}^{2}}d\Omega_{2}~U\op{T}_{UU} \ ,
\end{equation}
which we note is only well-defined as a quadratic form on $\Fock^{\mc{H}}\otimes L^{2}(\bb{R})$ (i.e., its expectation values in Hadamard states are well-defined). While the first term on the right-hand side of \cref{eq:QVT} is a well-defined observable affiliated to $\vNext(\mc{H}_{\textrm{R}}^{-},\s_{0}),$ the second term has divergent fluctuations (and is therefore not a well-defined operator on $\Hilb_{\rm ext.}$) unless $U\to +\infty$. We (heuristically) define the generalized entropy on a constant $U$ cut of the horizon to be 
\begin{equation}
\label{eq:Sgen}
S_{\textrm{gen.},U}(\hat{\s}^{\mc{H}})\defn \frac{A}{4G_{\textrm{N}}}+\hat{\s}^{\mc{H}}\le(\frac{\delta^{2}\op{\mc{Q}}_{U}}{4G_{\textrm{N}}}\ri)+S_{\textrm{vN.}}(\s\vert_{\mc{H}^{-}_{U}})
\end{equation}
where $A$ is the ``background area'' of the bifurcation surface, the second term is the expected value of $\delta^2\op{\mc{Q}}_{U}/4G_{\textrm{N}}$ and $S_{\textrm{vN.}}(\s\vert_{\mc{H}^{-}_{U}})$ is the von Neumann entropy of $\s^{\mc{H}}$ restricted to the past of a constant $U$ cut of $\mc{H}^{-}$ which we denote as $\s\vert_{\mc{H}^{-}_{U}}$. The second term in \cref{eq:Sgen} is well-defined since $\delta^{2}\op{\mc{Q}}_{U,\xi}$ is well-defined as a quadratic form and $\hat{\s}^{\mc{H}}(\delta^{2}\op{\mc{Q}}_{U,\xi})$ is $O(G_{\textrm{N}})$ so this term is well-defined in the $G_{\textrm{N}} \to 0$ limit. Famously, the first and third terms are individually undefined, however, it has been proposed that their sum is possibly better defined \cite{1994PhRvD..50.2700S}. The first term diverges in the $G_{\textrm{N}}\to 0$ limit and the third term diverges due to the infinite vacuum entanglement of the quantum field across the cut of the horizon. We will be particularly interested in the generalized entropy at the bifurcation surface $\mc{B}$ (i.e. $U=0$) and so it will be useful to define $S_{\textrm{gen.},\mc{B}}(\hat{\s}^{\mc{H}})\defn S_{\textrm{gen.},0}(\hat{\s}^{\mc{H}})$ which can be compactly expressed in terms of the expectation value of the $\op{A}_{\mc{B}}$ area of the bifurcation surface (to second-order) and the von Neumann entropy 
\begin{equation}
\label{eq:Sgendef}
S_{\textrm{gen.},\mc{B}}(\hat{\s}^{\mc{H}})= \hat{\s}\bigg(\frac{\op{A}_{\mc{B}}}{4G_{\textrm{N}}}\bigg)+S_{\textrm{vN.}}(\s\vert_{\mc{H}^{-}_{\textrm{R}}})
\end{equation}
and 
\begin{equation}
\label{eq:ABdef}
\op{A}_{\mc{B}}\defn A\op{1}+\delta^{2}\op{A}_{\mc{B}}
\end{equation}
where $\delta^{2}\op{A}_{\mc{B}}\defn \delta^{2}\op{\mc{Q}}_{0}$ is the perturbed area of the bifurcation surface.

We now present a heuristic argument, following \cite{2012PhRvD..85j4049W}, that the sum of the first and second terms in \cref{eq:Svn_leading} can be interpreted as an expression of the generalized entropy of the bifurcation surface
\begin{equation}
\label{eq:SrelGS}
\hat{\s}^{\mc{H}}(\beta\op{X})-S_{\textrm{rel.}}(\s^{\mc{H}}|\s^{\mc{H}}_{0}) = S_{\textrm{gen.},\mc{B}}(\hat{\s}^{\mc{H}})+C
\end{equation}
where $C$. We first note that, evaluating \cref{eq:QVT} at the bifurcation surface $U=0$ implies that 
\begin{equation}
\label{eq:ABXTVV}
\hat{\s}^{\mc{H}}(\delta^{2}\op{A}_{\mc{B}}) = 4G_{\textrm{N}} ~\hat{\s}^{\mc{H}}(\beta\op{X}) +8\pi G_{\textrm{N}}\int_{\mc{H}_{\textrm{R}}^{-}}dUd\Omega_{2}~U\s^{\mc{H}}(\op{T}_{UU})
\end{equation}
where $\delta^{2}\op{A}_{\mc{B}}\defn \delta^{2}\op{\mc{Q}}_{0}$ which, evaluating \cref{eq:QVxi} at $V=0$, is the perturbed area at the bifurcation surface. In the last term of \cref{eq:ABXTVV} we used the fact that $\hat{\s}^{\mc{H}}(\op{T}_{UU})=\s^{\mc{H}}(\op{T}_{UU})$.

In order to express the well-defined left-hand side of \cref{eq:SrelGS} in terms of the ill-defined right-hand side we must forgo rigor and express \cref{eq:SrelGS} and \cref{eq:ABXTVV} in terms of formal density matrices for the quantum fields. With this caveat in mind, the stress-energy integral over $\mc{H}_{\textrm{R}}^{-}$ is (formally) the expected value of the ``right modular Hamiltonian'' for the vacuum state $\om_0$ which can be expressed as
\begin{equation}
\label{eq:TVVrhos0}
2\pi \int_{\mc{H}_{\textrm{R}}^{-}}dUd\Omega_{2}~U\op{T}_{UU} = \log(\rho_{\s_{0}}) + C^{\prime}
\end{equation}
where $\rho_{\s_{0}}$ is the density matrix associated to $\s_{0}\vert_{\mc{H}^{-}_{\textrm{R}}}$ and the constant $C^{\prime}=-\s^{\mc{H}}_{0}(\log \rho_{\s_{0}})=S_{\textrm{vN.}}(\s_{0}\vert_{\mc{H}^{-}_{\textrm{R}}})$ is fixed by taking the expected value of \cref{eq:TVVrhos0} in $\s^{\mc{H}}_{0}$ using the fact that $\s^{\mc{H}}_{0}(\op{T}_{UU})=0$.
Additionally, we recall that the relative entropy can be expressed as 
\begin{equation}
S_{\textrm{rel.}}(\s^{\mc{H}}|\s^{\mc{H}}_{0}) = \s^{\mc{H}}(\log \rho_{\s}) - \s^{\mc{H}}(\log\rho_{\s_{0}}). 
\end{equation}
where $\rho_{\s}$ is the (formal) density matrix associated to $\s\vert_{\mc{H}^{-}_{\textrm{R}}}$. Using these two formal relations we can re-express the relative entropy as 
\begin{align}
\label{eq:Srel}
-S_{\textrm{rel.}}(\s^{\mc{H}}|\s^{\mc{H}}_{0})=&2\pi \int_{\mc{H}_{\textrm{R}}^{-}}dUd\Omega_{2}~U\s^{\mc{H}}(\op{T}_{UU})-\s^{\mc{H}}(\log\rho_{\s})-S_{\textrm{vN.}}(\s_{0}\vert_{\mc{H}_{\textrm{R}}^{-}}) \nonumber \\
=&\frac{\hat{\s}^{\mc{H}}(\delta^{2}\op{A}_{\mc{B}})}{4G_{\textrm{N}}}-\frac{\hat{\s}^{\mc{H}}(\beta \op{X})}{4G_{\textrm{N}}}-\s^{\mc{H}}(\log\rho_{\s})-S_{\textrm{vN.}}(\s_{0}\vert_{\mc{H}_{\textrm{R}}^{-}}) \nonumber \\
=&\frac{\hat{\s}^{\mc{H}}(\delta^{2}\op{A}_{\mc{B}})}{4G_{\textrm{N}}}-\frac{\hat{\s}^{\mc{H}}(\beta \op{X})}{4G_{\textrm{N}}}+S_{\textrm{vN.}}(\s\vert_{\mc{H}_{\textrm{R}}^{-}})-S_{\textrm{vN.}}(\s_{0}\vert_{\mc{H}^{-}_{\textrm{R}}})
\end{align}
where in the second line we used \cref{eq:ABXTVV} in the third line we used the fact that $S_{\textrm{vN.}}(\s\vert_{\mc{H}_{\textrm{R}}^{-}})=-\s^{\mc{H}}(\log \rho_{\s})$. Adding and subtracting the ``background area'' $A/4G_{\textrm{N}}$ we obtain 
\begin{equation}
\label{eq:Srel2}
-S_{\textrm{rel.}}(\s^{\mc{H}}|\s^{\mc{H}}_{0})=S_{\textrm{gen.},\mc{B}}(\hat{\s}^{\mc{H}}) - \frac{A}{4G_{\textrm{N}}}-\frac{\hat{\s}^{\mc{H}}(\beta \op{X})}{4G_{\textrm{N}}}-S_{\textrm{vN.}}(\s_{0}\vert_{\mc{H}^{-}_{\textrm{R}}}).
\end{equation}
We note that since $\s^{\mc{H}}_{0}$ is stationary the last term in \cref{eq:Srel2} is equal to the entropy of $\om^{\mc{H}}_0$ restricted to observables at early affine times. By the decay of correlation functions given by \cref{eq:2ptPI}, any state, $\om^{\mc{H}}$, approaches the vacuum, $\om^{\mc{H}}_0$, at asymptotically early times, thus the last term in \cref{eq:Srel2} is equivalent to the von Neumann entropy of $\s^{\mc{H}}$ at asymptotically early times. Therefore, the last three terms in \cref{eq:Srel2} can be interpreted as $S_{\textrm{gen.},U}(\hat{\s}^{\mc{H}})$ at asymptotically early affine times. 

\Cref{eq:Srel2} yields the desired relation~\eqref{eq:SrelGS} where the (divergent) state-independent constant is $C=-A/4G_{\textrm{N}}-S_{\textrm{vN.}}(\s_{0}\vert_{\mc{H}^{-}_{\textrm{R}}})$. We can therefore express the entropy of $\rho_{\hat{\s}}$ as 
\begin{equation}
S_{\textrm{vN.}}(\rho_{\hat{\s}})\simeq S_{\textrm{gen.},\mc{B}}(\hat{\om}^{\mc{H}}) +S(\rho_{f})+C.
\end{equation}
We stress again that the von Neumann entropy $S_{\textrm{vN}}(\rho_{\hat{\s}})$ given by \cref{eq:Svn_leading} is manifestly well-defined. The above manipulations illustrate that $S_{\textrm{vN}}(\rho_{\hat{\s}})$ can be interpreted as the generalized entropy $S_{\textrm{gen.},\mc{B}}$ including a ``fluctuation entropy'' of the horizon area given by $S(\rho_{f})$. For AdS Schwarzschild black holes in the canonical ensemble, this fluctuation entropy gives the universal logarithmic correction to black hole entropy \cite{2022JHEP...10..008W,2002CQGra..19.2355D}. {We believe it may be reasonable to \textit{define} the generalized entropy as the von Neumann entropy of $\rho_{\hat{\omega}}$.

\section{Globally KMS States}
\label{sec_stationary}
In this section, we apply the general analysis of the previous section to construct the algebra of ``dressed observables'' for spacetimes that admit stationary states that are globally KMS. In this case, there exists a state for which the modular Hamiltonian of the algebra of quantum fields in $(\mc{R},g)$ is globally equivalent to the generator of the Killing flow $\xi^{a}$. Namely, in these cases, there exists a stationary state $\s_{0}$ on $(\mc{M},g)$ such that the relation
\begin{equation}
\label{eq:modF}
\op{H}_{\s_{0}}=\beta \op{F}_{\xi}
\end{equation}
is valid {\em globally}. Examples of such globally KMS states are the Hartle-Hawking state in AdS-Schwarzschild and the Bunch-Davies state in de Sitter spacetime. In such cases, dressed observables invariant under the Killing flow were constructed in \cite{2022arXiv220910454C,Chandrasekaran:2022cip}. The main goal of this section is to re-derive the main results of these references in a manner that will ultimately generalize to arbitrary spacetimes with Killing horizons. In both cases, the construction of the algebra of dressed observables will simply amount to a direct application of \cref{thm:horizontheorem}. The case of de Sitter spacetime is somewhat subtle, as one must introduce an observer in order to obtain a non-trivial, isometry invariant algebra.

\subsection{Schwarzschild Black Hole in AdS}
\label{subsec:sads_kms}

\begin{figure}
 \centering
 \includegraphics[width = .7\textwidth]{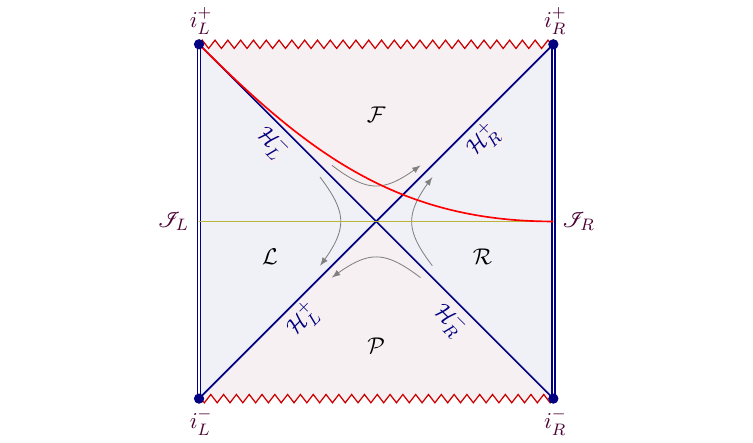}
 \caption{The Penrose diagram of maximally extended Schwarzschild AdS. The yellow surface is a Cauchy surface for the entire spacetime, while the red surface is a Cauchy surface for $\mc{R} \cup \mc{F}$.}
 \label{fig:AdS_tikz}
\end{figure}

In this subsection, we consider the maximally extended, Schwarzschild black hole in Anti-de Sitter spacetime. In this case, the bifurcate Killing horizon $\mc{H}^{-}\cup \mc{H}^{+}$ of the black hole globally divides the spacetime into four regions as depicted in \cref{fig:bifurcate_tikz} and we shall construct the algebra of dressed observables in the ``right wedge'' $(\mc{R},g)$. The region $\mc{R}$ is the exterior of the black hole and is therefore invariantly defined $I^{+}(\scri_{\textrm{R}})\cap I^{-}(\scri_{\textrm{R}})$ (see figure~\ref{fig:AdS_tikz}). The isometries of $(\mc{R},g)$ are $\bb{R}\times \textrm{SO}(3)$ corresponding to the isometry $\chi_{t}$ and rotations. We will construct an algebra of ``dressed observables'' invariant under $\chi_t$, noting, as in CLPW, that including $\textrm{SO}(3)$ will not affect the nature of the algebra \cite{2022arXiv220910454C}. 

The von Neumann algebra of ``dressed observables'' in $\mc{R}$ was previously constructed in \cite{2022arXiv220910454C} and the main goal of this section is to repeat this construction using the framework developed in the previous section. Anti-de Sitter spacetime is not globally hyperbolic, and therefore the construction of quantum field theory in curved spacetime presented in \cref{subsec:AlgQFTCS} cannot be directly applied. However, one can obtain a well-posed initial value problem by imposing appropriate boundary conditions\footnote{For a scalar field Dirichlet boundary conditions, Neumann boundary conditions or a linear combination of Dirichlet and Neumann boundary conditions (known as ``Robin'' boundary conditions) yield a well-posed initial value problem \cite{Ishibashi:2004wx}.} (e.g., Dirichlet boundary conditions) at the spatial boundary \cite{Ishibashi:2003jd}. After imposing such boundary conditions, there exist unique advanced and retarded Greens functions and so the construction of the phase space $\mc{P}$ given in \cref{subsec:scfieldth} and its corresponding quantization in \cref{sec_quantization} yields an algebra $\Alg$ of local observables $\op{\phi}(f)$ with commutation relations determined by the (unique) advanced minus retarded Greens function $E$ (see \ref{A4}). 

On this spacetime, there exists a globally defined, Gaussian, stationary state $\s_{0}$ for Klein-Gordon fields with Dirichlet boundary conditions which is Hadamard on the maximally extended spacetime $(\mc{M},g)$. This state is the Hartle-Hawking state which can be straightforwardly obtained by Euclidean methods and is a {\em globally} KMS state with respect to the Killing field $\xi$ \cite{Hartle:1976tp,Israel:1976ur,Kay_1988,Sanders:2013vza}. Physically, this state corresponds to a black hole in ``thermal equilibrium.'' Therefore, the state $\s_{0}$ globally satisfies \cref{eq:KMS}, and the corresponding GNS representation is a Fock space $\Fock$. The von Neumann algebra $\mf{A}(\mc{R},\s_{0})$ of (bounded functions of) the field observables $\op{\phi}(f)$ supported in the right wedge is a Type III$_1$ algebra. The modular Hamiltonian $\op{H}_{\s_{0}}$ is given by \cref{eq:modF} and 
is precisely equivalent to the observable which generates the Killing flow on $\mf{A}(\mc{R},\s_{0})$.

Assumption \ref{assump2} asserts that any smooth solution with initial data of compact spatial support in $\mc{R}$ decays sufficiently rapidly such that the past horizon $\mc{H}^{-}$ is an initial data surface for such solutions.\footnote{If the symplectic flux of solutions was non-vanishing at $i_{\textrm{R}}^{-}$ then one would have to supplement the data on $\mc{H}^{-}$ with data at $i_{\textrm{R}}^{-}$. The vanishing of the symplectic flux through $i^{-}$ was proven for massless fields on an asymptotically flat Schwarzschild black hole background \cite{2009arXiv0907.1034D}. } As explained in \cref{subsec:quantKillingscri}, bulk quantum fields $\op{\phi}(f)$ are related to the ``initial data'' $\op{\Pi}(s)$ on $\mc{H}^{-}$ by 
\begin{equation}
\label{eq:AdSphiPi}
\op{\phi}(f) = \op{\Pi}(s) \ ,
\end{equation}
where we recall that $s\defn [Ef]_{\mc{H}^{-}}$. Therefore, the algebra of observables on $(\mc{M},g)$ is isomorphic to the algebra of observables on $\mc{H}^{-}$, i.e., 
\begin{equation}
\Alg\cong \Alg_{\mc{H}^{-}},
\end{equation}
and the initial data for the Hartle-Hawking state on $\Alg_{\mc{H}^{-}}$ is given by \cref{eq:vacH-}. The Type III$_1$ von Neumann algebra associated with the globally hyperbolic, exterior region $(\mc{R},g)$ is isomorphic to the von Neumann algebra on part of the past horizon $\mc{H}^{-}_{\rm R}$ so that 
\begin{equation}
\label{eq:Righthorizon}
\mf{A}(\mc{R},\s_{0})\cong \mf{A}(\mc{H}_{\textrm{R}}^{-},\s_{0}).
\end{equation}

We now construct ``dressed observables'' in the region $(\mc{R},g)$ by ``including'' the perturbed ADM mass $\delta^{2}M_{\textrm{R}}$ in the algebra. The ADM mass generates asymptotic time translations in the right wedge. Since $(\mc{R},g)$ has a timelike isometry $\xi^{a}$ we can naturally extend its action onto bulk fields in $\mc{R}$ by integrating the constraints \cref{eq:BDYCharge} on a spatial slice which yields (with $X^a = \xi^a$)
\begin{equation}
F_{\xi}=\delta^{2}M_{\textrm{R}}-\delta^{2}M_{\textrm{L}} ,
\end{equation}
where $M_{R}$ and $M_{L}$ are the ADM mass of the regions $\mc{R}$ and $\mc{L}$ respectively. 

To satisfy the gravitational constraint in the quantum theory,
\begin{equation}
\label{eq:FxiMRML}
\op{F}_{\xi}=\delta^{2}\op{M}_{\textrm{R}}-\delta^{2}\op{M}_{\textrm{L}} ,
\end{equation}
the second-order perturbed charge $\delta^{2}\op{M}_{\textrm{R}}$ must be promoted to a quantum observable used to dress the field operators in $\Alg$. Therefore, the global constraint is of the form \cref{eq:CXF} with 
\begin{equation}
\op{X} = \delta^{2}\op{M}_{\textrm{R}} \quad \textrm{and} \quad \op{C}= \delta^{2}\op{M}_{\textrm{L}}
\end{equation}

The Hilbert space representation of $\op{X}$ is $\Hilb_{\textrm{X}}=L^{2}(\bb{R})$. We have a ``conjugate operator'' on this representation $\op{t}$ which satisfies
\begin{equation}
[\op{X},\op{t}]=i.
\end{equation}
One can then construct a ``dressed'' scalar field observable which commutes with $\delta^{2}\op{M}_{\textrm{L}}$ and obeys the constraint as
\begin{equation}
\label{eq:dressMR}
\op{\phi}(f;\op{t})\defn e^{-i\op{F}_{\xi}\op{t}}\op{\phi}(f)e^{i\op{F}_{\xi}\op{t}}
\end{equation}
which acts on $\Hilb_{\textrm{ext.}}\defn \Fock\otimes L^{2}(\bb{R})$. {We note that, indeed, }
\begin{equation}
{[\op{X},\op{\phi}(f;\op{t})]=i\op{\phi}(\pounds_{\xi}f;\op{t})}
\end{equation}
{so the perturbed ADM mass generates time translations of the dressed observables.}
The von Neumann algebra of dressed observables on $(\mc{R},g)$ is 
\begin{equation}
\vNext(\mc{R},\s_{0})\defn \{\op{\phi}(f;\op{t}),\op{X}\}^{\prime\prime} , \quad \supp{f} \subset \mc{R} \ .
\end{equation}
Using the fact that \cref{eq:modF} holds globally, it directly follows from the considerations of \cref{subsec:crossprod} that $\vNext(\mc{R},\s_{0})$ is Type II$_{\infty}$. 

To obtain this result in a manner that is generalizable to generic spacetimes with bifurcate Killing horizons, we use the fact that any local quantum field $\op{\phi}(f)$ can be expressed in terms of its initial data $\op{\Pi}(s)$ given by \cref{eq:AdSphiPi}.

\begin{equation}
\op{\phi}(f;\op{t}) = e^{-i\op{F}_{\xi}\op{t}}\op{\phi}(f)e^{i\op{F}_{\xi}\op{t}} = e^{-i\op{F}^{\mc{H}}_{\xi}\op{t}}\op{\Pi}(s)e^{i\op{F}^{\mc{H}}_{\xi}\op{t}}\defn \op{\Pi}(s;\op{t}).
\end{equation}
where we used the fact that the flux is conserved so that $\op{F}_{\xi}=\op{F}^{\mc{H}}_{\xi}$. It follows that the von Neumann algebra of dressed observables on the right part of the past horizon
\begin{equation}
\vNext(\mc{H}_{\textrm{R}}^-,\s_{0})\defn \{\op{\Pi}(s;\op{t}),\op{X}\}^{\prime\prime} , \quad \supp{s} \subset \mc{H}_{\textrm{R}}^- \ ,
\end{equation}
is equivalent to the algebra of dressed observables in $(\mc{R},g)$ 
\begin{equation}
\vNext(\mc{R},\s_{0})\cong \vNext(\mc{H}_{\textrm{R}}^{-},\s_{0}).
\end{equation}
Therefore, \cref{thm:horizontheorem} directly implies that the von Neumann algebra of dressed bulk observables in $\mc{R}$ is Type II$_{\infty}$. For a ``classical-quantum'' state, $\hat{\s},$ of the form of \cref{eq:hats}, the entropy is 
\begin{equation}
\label{eq:Srho}
S_{\textrm{vN.}}(\rho_{\hat{\s}}) \simeq \hat{\s}(\beta \op{X})-S_{\textrm{rel.}}(\s|\s_{0})+S_{\textrm{vN.}}(\rho_{f}) + \log \beta.
\end{equation}
Finally, in order to interpret $S_{\textrm{vN.}}(\rho_{\hat{\s}})$ as the generalized entropy as in \cref{eq:Srel2}, we note the perturbed ADM masses $\delta^{2}{M}_{\textrm{R}}$ and $\delta^{2}M_{\textrm{L}}$ are conserved and can be related to the perturbed charges on the past horizon by the ``matching condition''
\begin{equation}
\delta^{2}M_{\textrm{R}}=\frac{\delta^{2}\mathcal{Q}_{-}}{4G_{\textrm{N}}} \quad \textrm{and} \quad \delta^{2}M_{\textrm{R}}=\frac{\delta^{2}\mathcal{Q}_{+}}{4G_{\textrm{N}}} \ .
\end{equation}
Therefore, the ``global constraint'' \cref{eq:dressMR} is precisely equivalent to the ``local constraint'' \cref{eq:+-F}. Consequently, dressing to the perturbed mass is equivalent to dressing to the perturbed charge $\op{X}=\delta^{2}\op{\mathcal{Q}}_{-}/4G_{\textrm{N}}\beta$ and so, by \cref{eq:Srel2}, the von Neumann entropy of ``classical-quantum'' states can be (heuristically) expressed as
\begin{equation}
 S_{\textrm{vN.}}(\rho_{\hat{\s}})\simeq S_{\textrm{gen.},\mc{B}}(\hat{\om})+S_{\textrm{vN.}}(\rho_{f})+C
\end{equation}
and has the interpretation of a generalized black hole entropy.

\subsection{de Sitter Spacetime}
\label{subsec:deSitter}

In this subsection, we consider the case where $(\mc{M},g)$ is de Sitter spacetime. On such a spacetime one can invariantly define, relative to a timelike, inextendable, geodesic worldline $\gamma$, a causal region $(\mc{R},g)$ corresponding to the domain of communication of $\gamma$. Associated with this causal region is a bifurcate Killing horizon $\mc{H}^{-}\cup \mc{H}^{+},$ called the future and past cosmological horizon of $\gamma,$ which globally divides the spacetime as depicted in \cref{fig:static_patch}. However, as opposed to the previous section, the spacetime is a (spatially) closed universe. The isometries of $(\mc{R},g)$ are $\bb{R}\times \textrm{SO}(3)$ corresponding to translations along the observer's worldline and rotations about the worldline. We will primarily focus on the subgroup of translations which are also the orbits of $\xi^{a},$ the horizon Killing field. 

The spacetime is globally hyperbolic and, consequently, one can straightforwardly construct the $\ast$-algebra $\Alg$ of local observables, $\op{\phi}(f)$. For a massive scalar field (i.e., $V=m^{2}$ in \eqref{eq:KG}), there exists a unique de Sitter invariant state $\s_{0}$ known as the ``Bunch-Davies'' vacuum. However, this state does not exist for a massless scalar field \cite{Allen_1985_scalar}.\footnote{A unique de Sitter invariant state does exist on the algebra of electromagnetic fields \cite{Allen:1985_vector} and linearized gravitational fields \cite{Allen:1986_graviton}.} {While the Bunch-Davies state is a stationary state in $\mc{M}_{\textrm{R}}=\mc{R}\cup \mc{F}$, in contrast to most of the other examples considered in this paper, this state can be extended as a global state on the full de Sitter spacetime $(\mc{M},g)$.} Therefore, in this subsection, we will consider the quantization of massive scalar fields on $(\mc{M},g)$. Using the state $\om_0$ one can directly obtain a von Neumann algebra $\mf{A}(\mc{R},\om_0)$ of (bounded functions of) operators $\op{\phi}(f)$ supported in $(\mc{R},g)$ with GNS Fock space $\Fock$ associated to $\s_{0}$. It was proven in Theorem 3.3.2 of \cite{Dappiaggi:2017kka} that the past horizon is an initial data surface for massive fields in de Sitter. Therefore, the bulk quantum field is again related to the corresponding initial data by $\op{\phi}(f) = \op{\Pi}(s),$ where $s$ is the source-free solution $Ef$ restricted to $\mc{H}^{-}$ and consequently, the algebras $\Alg$ and $\Alg_{\mc{H}^{-}}$ are isomorphic. Thus, the von Neumann algebras $\mf{A}(\mc{R},\s_{0})$ and $\mf{A}(\mc{H}^{-}_{\textrm{R}},\s_{0})$ are also isomorphic. 

We now consider the construction of observables invariant under the isometries of $(\mc{R},g)$. The quantum field observables $\op{\phi}(f)$ are not invariant under the action of these isometries and one must appropriately ``gravitationally dress'' $\op{\phi}(f)$ using the gravitational constraints to obtain a diffeomorphism invariant observable. However, as explained in \cite{Chandrasekaran:2022cip}, one key difference between this construction and the construction presented in \cref{subsec:sads_kms} is that the spacetime is closed and so the Killing isometries are degeneracies of the symplectic form. Indeed, in the absence of any other degrees of freedom, \cref{eq:BDYCharge} implies the global constraint
\begin{equation}
{\symp((0,\phi),(0,\pounds_{X}\phi)=0}
\end{equation}
{where $X$ is any isometry of $(\mc{R},g)$. In particular, we have that,}
\begin{equation}
{\symp((0,\phi),(0,\pounds_{\xi}\phi)=F_{\xi}=0}.
\end{equation}
{Since the symplectic form is degenerate on the (pre-)phase space, we must factor out by the degenerate elements to obtain the phase space \cite{Lee:1990nz}. After removing the degenerate elements, the isometries have a trivial action on the phase space $\mc{P}$.}

Following CLPW, in order to nontrivially impose the gravitational constraint, it seems necessary to introduce an extra degree of freedom which interacts with the quantum field $\op{\phi}(f)$ through the gravitational constraints \cite{Chandrasekaran:2022cip}. For the constraints arising from the Killing field $\xi$, we need only consider the following simple model of a ``clock'' carried by the observer $\gamma$ following the time-like Killing flow in the static patch (as well as a complementary observer with their own ``clock'' in the other static patch).

Within our perturbative scheme explained in \cref{subsec:charge}, the clock is an externally specified body with a one-parameter family of stress-energy tensors $T_{ab}^{\textrm{obs.}}(\lambda)$ which satisfies $T_{ab}^{\textrm{obs.}}(0)=0=\delta T_{ab}^{\textrm{obs.}}$ however $\delta^{2} T_{ab}^{\textrm{obs.}}$ is non-vanishing. In particular, the energy of the clock at second-order is}
\begin{equation}
\varepsilon \defn \int_{\Sigma}\sqrt{h}d^{3}x~\delta^{2}T_{ab}^{\textrm{obs.}}n^{a}\xi^{b},
\end{equation}
where $\Sigma$ is a spacelike Cauchy surface for $(\mc{M},g)$ and we denote the time of the clock along $\gamma$ as $\tau$. We will assume that the stress-energy satisfies, at least, the weak energy condition so that $\varepsilon>0$. In this minimal model, we will assume that these are the only degrees of freedom of the observer. Therefore, the points in phase space are $(\varepsilon,\tau)\in \mc{P}_{\textrm{obs.}}$ which we endow with the following symplectic structure
\begin{equation}
\symp_{\textrm{obs.}}((\varepsilon_{1},\tau_{1}),(\varepsilon_{2},\tau_{2}))\defn \varepsilon_{2}\tau_{1}-\varepsilon_{1}\tau_{2},
\end{equation}
which clearly yields the following Poisson brackets on $\mc{P}_{\textrm{obs.}}$
\begin{equation}
\{\varepsilon,\tau\}=1.
\end{equation}
{Since $\tau$ measures time along the observer's inertial worldline and $\varepsilon$ is conserved along the worldline, we extend the infinitesimal action of the isometry $\xi$ to the phase space $\mc{P}_{\textrm{obs.}}$ by }
\begin{equation}
\delta_{\xi}\varepsilon = 0 \quad \quad \delta_{\xi}\tau =1.
\end{equation}
Therefore, the observable on $\mc{P}$ associated to this infinitesimal flow is 
\begin{equation}
\symp_{\textrm{obs.}}((\varepsilon,\tau),\delta_{\xi}(\varepsilon,\tau))=\varepsilon.
\end{equation}
The quantization of $\mc{P}_{\textrm{obs.}}$ yields the $\ast$-algebra $\Alg_{\textrm{obs.}}$ generated by $\op{\varepsilon}$ and $\op{\tau}$ factored by the commutation relation 
\begin{equation}
[\op{\varepsilon},\op{\tau}]=i.
\end{equation}
Furthermore, the condition $\varepsilon>0$ is a condition on the wave function of the energy to lie in 
\begin{equation}
\ms{H}_{\textrm{obs.}}\defn L^{2}(\bb{R}_{+}),
\end{equation}
corresponding to square-integrable functions on the positive real line. 
To obtain the gravitational constraints, we note that the observer degrees of freedom are independent of the first-order fields and so the (pre-)phase space of the full theory is $\mc{P}\oplus \mc{P}_{\textrm{obs.}}$. The symplectic form on the (pre-)phase space is simply the sum of the symplectic form of the first-order fields and the observer 
\begin{equation}
\symp_{\Sigma}((0,\phi_{1},\varepsilon_{1},\tau_{1}),(0,\phi_{2},\varepsilon_{2},\tau_{2})) = \symp_{\Sigma}^{\KG}(\phi_{1},\phi_{2})+\symp_{\textrm{obs.}}((\varepsilon_{1},\tau_{1}),(\varepsilon_{2},\tau_{2}))
\end{equation}
where $\Sigma$ is a spatial slice of $\mc{M}$ and each term in the sum on the right-hand side is separately conserved. The observer's stress energy now enters into the gravitational constraints and, applying a similar analysis which led to \cref{eq:BDYCharge}, the Killing symmetry $\xi$ yields a gravitational constraint on the (pre-)phase space $\mc{P}\oplus \mc{P}_{\textrm{obs.}}$
\begin{equation}
\label{eq:gengravconst}
\symp_{\Sigma}((0,\phi,\varepsilon,\tau),\delta_{\xi}(0,\phi,\varepsilon,\tau)) = F_{\xi} +\varepsilon =0 \ .
\end{equation}
Removing the degeneracies of the symplectic form is equivalent to removing the observer itself. In order to obtain a sensible phase space that obeys the gravitational constraints, we also include an observer, $\varepsilon'$ , in the complementary region $\mc{L},$ such that 
\begin{equation}
\label{eq:gengravconst2}
\symp_{\Sigma}((0,\phi,\varepsilon,\tau),\delta_{\xi}(0,\phi,\varepsilon,\tau)) = F_{\xi} +\varepsilon =\varepsilon' \ .
\end{equation}
Clearly, the symplectic form is no longer degenerate on $\mc{P}\oplus \mc{P}_{\textrm{obs.}}$.

{In this case, the quantum constraint is then of the form}
\begin{equation}
{\op{C}=\op{X} - \op{F}_{\xi}}
\end{equation}
where 
\begin{equation}
\op{X} = -\op{\varepsilon}
\end{equation}
{and $\op{C}=-\op{\varepsilon}^{\prime}$ or the energy of any other gravitating body in $\mc{L}$. The key point is that $\op{C}$ commutes with observables in $\mc{R}$ and yields a non-trivial constraint. The algebra of observables that obey the constraint and commute with $\op{C}$ is generated by operators $\op{X}$ and the dressed quantum field}
\begin{equation}
\label{eq:phiftau}
\op{\phi}(f;\op{\tau})\defn \op{P}_+ e^{i\op{F}_{\xi}\op{\tau}}\op{\phi}(f)e^{-i\op{F}_{\xi}\op{\tau}} \op{P}_+ ,
\end{equation}
where $\op{P}_+ $ is the projection onto the negative support of $\op{X}$ due to the energy condition on $\op{\varepsilon}$. Since the observer is within the static patch, the full von Neumann algebra of dressed bulk observables in the static patch $(\mc{R},g)$ is given by 
\begin{equation}
\vNext(\mc{R},\s_{0})\defn \{\op{\phi}(f;\op{\tau}),\op{X}\}^{\prime \prime}.
\end{equation}
Here, the commutant operation is with respect to $\Fock \otimes L^2(\mathbb{R})$.
As in the previous subsection, since $\om_0$ is globally KMS, $\op{F}_{\xi}$ is proportional to the modular Hamiltonian globally by \eqref{eq:modF} and this implies that $\vNext(\mc{R},\s_{0})$ is a Type II algebra. We also again note that the bulk dressed observable is equivalent to the corresponding ``dressed initial data'' 
\begin{equation}
\op{\phi}(f;\op{\tau})=\op{\Pi}(s;\op{\tau})
\end{equation}
and therefore 
\begin{equation}
\vNext(\mc{R},\s_{0}) \cong \vNext(\mc{H}^{-}_R,\s_{0})\defn \{\op{\Pi}(s;\op{\tau}),\op{X}\}^{\prime\prime}.
\end{equation}

It directly follows from \cref{thm:horizontheorem} that, since the spectrum of $\op{X}$ is bounded from above (recall $\op{X} = -\op{\varepsilon}$), $\vNext(\mc{R},\s_{0})$ is a Type II$_{1}$ algebra. 
The entropy of any classical-quantum state $\hat{\s}$ of the form of \cref{eq:hats} is given by \eqref{eq:Svn_leading}. As in the case of Schwarzschild-AdS, the type of the algebra is unchanged by imposing the gravitational constraints from the $\textrm{SO}(3)$ isometry.\footnote{As noted in \cite{Chandrasekaran:2022cip}, one would have to endow the observer with additional degrees of freedom such as the ``frame'' in order to impose the constraints from the $\textrm{SO}(3)$ isometry.} In order to interpret this formula
for $S(\rho_{\hat{\s}})$ as the generalized entropy, we note that the presence of the energy of the observer in $\mc{R}$ perturbs {the ``incoming'' charge} of the cosmological horizon by 
\begin{equation}
\label{eq:d2A=eps}
 \frac{\delta^{2}\mathcal{Q}_{-}}{4G_{\textrm{N}}\beta}= -\op{\varepsilon}.
\end{equation}
The von Neumann entropy of any classical-quantum state can be (heuristically) expressed as the generalized entropy of the cosmological horizon 
\begin{equation}
 S_{\textrm{vN.}}(\rho_{\hat{\s}})\simeq S_{\textrm{gen.},\mc{B}}(\hat{\om})+S_{\textrm{vN.}}(\rho_{f})+C.
\end{equation}

A particularly interesting aspect of this algebra is that there exists a maximally entropic state, namely the state with density matrix equal to the identity operator. Under our normalization of the trace, this state has zero entropy. Thus, all other states on the algebra will have negative entropy. The notion that the Bunch-Davies state is of maximal entropy has been considered from multiple perspectives \cite{2000JHEP...11..038B,2001JHEP...04..035B,1998PhRvD..57.3503M,2018JHEP...07..050D,2022arXiv220601083L}. 

\subsection{Adding an unnecessary observer in Schwarzschild AdS}
\label{sec:obs_open}
The constructions of the crossed product algebras for the previous two sections had very similar flavors. However, there was a crucial difference in that we were forced to explicitly include an observer in de Sitter in order to obtain a nontrivial algebra of observables. For spacetimes with asymptotic structure for which the flux $\op{F}_{\xi}$ is nonvanishing one can dress the fields to the boundary or some other feature of the spacetime. However, we are certainly free to add an observer in these spacetimes as well, and it would be worrisome if this changed the structure of the algebra or entropies significantly. Physically, the entropies should only change by the added entropy of the observer and its potential backreaction that changes horizon areas. In this subsection, we demonstrate the effect of adding such an observer in AdS-Schwarzschild spacetime. 

We augment the (nontrivial) Hamiltonian to include the energy of an observer. The constraint equation, \eqref{eq:FxiMRML}, is augmented as
\begin{align}
\op{F}_{\xi} + \op{\varepsilon}= \delta^{2}\op{M}_{\textrm{R}} - \delta^{2}\op{M}_{\textrm{L}}. 
\end{align}
We can dress to either $\op{\varepsilon}$ or $\op{X} \defn \delta^2 \op{M}_R$ to satisfy the constraints. We are motivated to dress to $\op{X}$ such that the perturbed mass generates time translations 
\begin{align}
 \op{\phi}(f,\op{t}) = e^{-i\op{F}_{\xi}\op{t}}\op{\phi}(f) e^{i\op{F}_{\xi}\op{t}}.
\end{align}
The von Neumann algebra is
\begin{equation}
\vNext(\mc{R},\s_{0})\defn \{\op{\phi}(f;\op{t}),\op{X},\op{\varepsilon}\}^{\prime \prime}
\end{equation}
acting on the Hilbert space $\Fock\otimes \Hilb_{\textrm{M}}\otimes \Hilb_{\textrm{obs.}}$.
The algebra has a trace
\begin{align}
 \Tr (a) = \beta \int_{\mathbb{R}}dX \int_{\mathbb{R}_+} d\varepsilon ~e^{\beta X }\bra{\s_{0},X,\varepsilon} a \ket{\s_{0},X,\varepsilon},
\end{align}
where $\ket{\s_{0},X,\varepsilon}=\ket{\s_{0}}\otimes \ket{X}\otimes \ket{\varepsilon}$ where $\ket{\varepsilon}$ are eigenstates of $\op{\varepsilon}$ with eigenvalue $\varepsilon$ and we have additionally imposed positive energy $(\varepsilon>0)$ for the observer. It is clear that the trace of the identity is infinite, so the algebra remains Type II$_{\infty}$ even when including the observer. There will be an additional contribution to the entropy from the observer if taken to be in a mixed state, though there will be no fluctuation term of the observer energy.

\section{Black Holes in Asymptotically Flat Spacetime}
\label{sec_flat}

In the previous section, we revisited the construction of dressed observables in Schwarzschild Anti-de Sitter spacetime as well as in empty de Sitter spacetime. Both spacetimes admit states which are globally KMS and, consequently, a direct application of the gravitational constraints yields a Type II algebra. In this section, we consider the construction of dressed observables in spacetimes that do not admit globally KMS states. Even in cases where such a globally KMS state exists --- such as the Hartle-Hawking state in asymptotically flat Schwarzschild spacetime --- such a state has infinite energy due to the incoming thermal radiation from $\scri^{-}$ and assumption \ref{assump4} will fail to hold. Therefore, one cannot construct dressed observables in the exterior of an asymptotically flat Schwarzschild black hole in this manner. Nevertheless, in all such cases, the ``Unruh state'' can be defined in the ``physically relevant'' region $\mc{M}_{\textrm{R}}$ and, for massless fields, is equivalent to the ``vacuum'' at $\scri^{-}$ and $\mc{H}^{-}$. 

In this section, we shall construct the algebra of dressed observables in the exterior of a Kerr black hole in an asymptotically flat spacetime. As we shall show, despite the fact that $\s_{0}$ is not globally KMS, the algebra of dressed observables in these regions is Type II$_{\infty}$, and we will obtain explicit formulas for the trace of any observable and the von Neumann entropy of density matrices on the algebra. 

\subsection{Algebra of Dressed Observables in the Exterior of a Kerr Black Hole }
\label{subsec:algdressKerr}

Let $(\mc{M}_{\textrm{R}},g)$ be the spacetime of an asymptotically flat Kerr black hole. More precisely, as explained in \cref{subsec:KillHor}, the spacetime $\mc{M}_{\textrm{R}}$ is the ``physically relevant region'' corresponding to the future interior and right exterior of a (one-sided) Kerr black hole. This region is globally hyperbolic, and so the construction of the phase space $\mc{P}$ as well as the algebra $\Alg$ of the local fields $\op{\phi}(f)$ proceeds as explained in \cref{subsec:scfieldth,subsec:AlgQFTCS}. In this subsection, we will restrict to massless fields (i.e., $V=0$ in \cref{eq:EOMphi}) for simplicity. However, massive fields can be similarly treated as explained in \cref{app1:timequant}. The region $\mc{R}$ is invariantly defined as $I^{-}(\scri^{+})\cap I^{+}(\scri^{-})$ and the isometry group of $(\mc{R},g)$ is $\bb{R}\times \textrm{U}(1)$ where the orbits of $\bb{R}$ are the orbits of the asymptotically timelike Killing vector $t^{a}$ and the orbits of $\textrm{U}(1)$ correspond to the orbits of the rotational Killing vector $\psi^{a}$ associated to the azimuthal symmetries of $(\mc{R},g)$. While $t^{a}$ and $\psi^{a}$ are both spacelike on the horizon, the horizon Killing field $\xi^{a}=t^{a}+\Omega_{\textrm{H}}\psi^{a}$ is null on the past and future horizon. 

As before, a key ingredient in our arguments will be the relation between $\op{\phi}(f)$ and its ``initial data'' in the past. For a massless field, this data is most naturally specified on the horizon and past null infinity. However, in the exterior of a black hole, there exist past-directed null geodesics that orbit the black hole forever and never approach these asymptotic boundaries. The orbits of these geodesics form an (unstable) ``photon sphere.'' For wave propagation, this leads to a ``trapping'' of solutions, which leads to a slower rate of decay. For our purposes, a slower decay may lead to a ``leakage of symplectic flux'' through $i^{-}$. For massless fields on a Schwarzschild spacetime, the lack of such leakage was proven in Theorem 2.2 of \cite{2009arXiv0907.1034D} which uses nontrivial point-wise decay estimates for solutions obtained in \cite{Dafermos:2010hd}.\footnote{Point-wise decay estimates obtained in \cite{Luk:2010jfs} imply that Theorem 2.2 of \cite{2009arXiv0907.1034D} also holds in higher dimensions.} Therefore, in Schwarzschild, initial data of compact spatial support decays sufficiently rapidly such that one can decompose the algebra $\Alg$ into the tensor product 
\begin{equation}
\label{eq:algHscri}
\Alg \cong \Alg_{\mc{H}^{-}}\otimes \Alg_{\scri^{-}}
\end{equation}
and the observable $\op{\phi}(f)$ is determined by the ``initial data'' specified on the past horizon $\op{\Pi}(s)\otimes \op{1}$ and past null infinity $\op{1}\otimes\tilde{\op{\Pi}}(\tilde{s})$ (see \cref{subsec:quantKillingscri}). For the sake of brevity, we will drop the identity element on each algebra and write 
\begin{equation}
\label{eq:phiPiPi}
\op{\phi}(f) = \op{\Pi}(s) + \tilde{\op{\Pi}}(\tilde{s}),
\end{equation}
where $s\defn [Ef]\vert_{\mc{H}^{-}}$, $\tilde{s}\defn [Ef]\vert_{\scri^{-}}$ and it is understood that we have extended $\op{\Pi}(s)$ and $\tilde{\op{\Pi}}(\tilde{s})$ to $\Alg$ in the obvious way. While such point-wise decay has not been rigorously proven for massless fields propagating on a Kerr background, we shall simply assume that the necessary decay conditions are satisfied such that the decomposition \cref{eq:phiPiPi,eq:algHscri} is valid\footnote{We note that this assumption is stronger than assumption \ref{assump3} since we assume that the massless Klein Gordon field vanishes at $i^{-}$. If there is a non-trivial symplectic flux through $i^{-}$ then the following analysis can be straightforwardly amended by including data at past timelike infinity.}. 

As previously mentioned, the key difference between the Kerr spacetime and the previous cases considered in this paper is that there is no ``Hartle-Hawking'' state (i.e., no globally KMS state) on $\mc{M}_{\textrm{R}}$ with respect to the horizon Killing vector $\xi^{a}=t^{a} + \Omega_{\textrm{H}}\psi^{a}$. The Killing vector field $\xi$ becomes spacelike at large distances from the black hole and therefore one cannot have a state that is KMS in $\mc{R}$ with respect to $\xi$. Nevertheless, one can define a stationary --- but not globally KMS --- state $\s_{0}$ in the region $\mc{M}_{\textrm{R}}$ of any asymptotically flat black hole spacetime. Such a state has the following ``initial conditions'' on $\Alg_{\mc{H}^{-}}\otimes \Alg_{\scri^{-}}$
\begin{equation}
\s_{0}\defn \s_{0}^{\mc{H}}\otimes \s_{0}^{\scri},
\end{equation}
where $\s_{0}^{\mc{H}}$ is the Gaussian state on $\Alg_{\mc{H}^{-}}$ with $2$-point function given by \cref{eq:vacH-} and $\s_{0}^{\scri}$ is the Gaussian state on $\Alg_{\scri^{-}}$ with $2$-point function given by \cref{eq:statescri} with $\tilde{S}=0$. The state $\s_{0}$ is manifestly globally stationary since $\s_{0}^{\mc{H}}$ and $\s_{0}^{\scri}$ are both invariant under the action of $\chi_{t}$. Physically, $\s_{0}$ is the stationary state achieved after the gravitational collapse of a black hole; all Hadamard states that approach the vacuum state near spatial infinity also approach the Unruh state at late retarded times \cite{Fredenhagen:1989kr}. The requirement that the state approach the vacuum state near spatial infinity is both physically motivated and necessary for including the backreaction of quantum fields. 

The corresponding GNS Fock representation of the algebra $\Alg$ with the Unruh state is $\Fock$. The Type III$_{1}$ von Neumann algebra of field observables in $(\mc{R},g)$ is $\mf{A}(\mc{R},\s_{0})$. To see the origin of the Type III$_{1}$ quality, it will be useful to also consider the von Neumann algebras associated to the past horizon and past null infinity. Let $\ms{F}_{\mc{H}}$ be the GNS Fock representations of $\Alg_{\mc{H}^{-}}$ with respect to $\s_{0}^{\mc{H}}$ and let $\ms{F}_{\scri}$ be the GNS Fock representation of $\Alg_{\scri^{-}}$ with respect to $\s_{0}^{\scri}$. The von Neumann algebra $\mf{A}(\scri^{-},\s_{0})$ of field observables on $\scri^{-}$ is Type I$_{\infty}$ and the von Neumann algebra $\mf{A}(\mc{H}_{\textrm{R}}^{-},\s_{0})$ of field observables on $\mc{H}_{\textrm{R}}^{-}$ is Type III$_{1}$. The decomposition of the bulk algebra into the boundary algebra \cref{eq:algHscri} implies that 
\begin{equation}
\label{eq:RVHscri}
\mf{A}(\mc{R},\s_{0}) = \mf{A}(\mc{H}_{\textrm{R}}^{-},\s_{0})\otimes \mf{A}(\scri^{-},\s_{0}).
\end{equation}
The commutant is a Type III$_{1}$ von Neumann on $\mc{H}_{\textrm{L}}^{-}$ defined as
\begin{equation}
\label{eq:commHL}
\mf{A}(\mc{R},\s_{0})^{\prime}= \mf{A}(\mc{H}_{\textrm{L}}^{-},\s_{0}).
\end{equation}
\Cref{eq:RVHscri,eq:commHL} express the fact that the Type III$_{1}$ nature of the von Neumann algebra of observables in $\mc{R}$ arises from correlations across the future horizon which, in terms of initial data, arise from correlations across the bifurcation surface $\mc{B}$. 

In contrast to the spacetimes that we have previously considered, the Killing vector $\xi^{a}=t^{a}+\Omega_{\textrm{H}}\psi^{a}$ is a linear combination of two isometries. While we could obtain these constraints by directly integrating \cref{eq:BDYCharge} for $X=\psi,t$ respectively, it will be useful to first construct the ``local constraints'' on past null infinity and the past horizon. On $\scri^{-}$ we have that 
\begin{equation}
F_{t}^{\scri}\defn \symp^{\KG}_{\scri^{-}}(\phi,\pounds_{t}\phi)=\int_{\scri^{-}}dvd\Omega_{2}~\Omega^{-2}\delta^{2}T_{vv}
\end{equation}
and 
\begin{equation}
F_{\psi}\defn \symp^{\KG}_{\scri^{-}}(\phi,\pounds_{t}\phi)=\int_{\scri^{-}}dvd\Omega_{2}~\Omega^{-2}\psi^{A}\delta^{2}T_{Av}
\end{equation}
where $\psi_{A}$ is the pull-back of $\psi_{a}$ to $\scri^{-}$ {and we recall that $\Omega \sim 1/r$} {near null infinity}. These fluxes through $\scri^{-}$ can be related to the difference of perturbed charges in the limit as $v\to \pm \infty$. For a general asymptotic symmetry $X^{a}$, the charge on a constant $v$ cut of $\scri^{-}$ is \cite{2000PhRvD..61h4027W} 
\begin{equation}
\delta^{2}\mc{Q}_{v,X}\defn \frac{1}{8\pi}\int_{S_{v}}d\Omega_{2}~\Omega^{-1}\delta^{2}C_{abcd}X^{a}\tilde{\ell}^{b}\tilde{n}^{c}\tilde{\ell}^{d}
\end{equation}
where $\delta^{2}C_{abcd}$ is the second-order perturbed Weyl tensor of the conformally completed spacetime, $\tilde{n}^{a}=(\partial/\partial v)^{a}$ and, choosing a foliation of $\scri^{-}$ by $v=\textrm{constant}$ cross-sections $S(v)$, $\tilde{\ell}^{a}$ is the unique null vector on $\scri^{-}$ which is normal to cross-sections and satisfies $\tilde{\ell}^{a}\tilde{n}_{a}=0$. Choosing cross-sections tangent to the orbits of $\psi^{a}$ at $\scri^{-}$, the perturbed (Bondi) mass and angular momentum on a constant $v$ cut of $\scri^{-}$ are 
\begin{equation}
\label{eq:MvJv}
\delta^{2}M_{v} = \delta^{2}\mc{Q}_{v,t},\quad \delta^{2}J_{v}\defn -\delta^{2}\mc{Q}_{v,\psi} \ .
\end{equation}
The fluxes through $\scri^{-}$ can be expressed as 
\begin{equation}
\label{eq:FtnullFpsinull}
F_{t}^{\scri}=\delta^{2}M_{i^{0}}-\delta^{2}M_{i^{-}},\quad F_{\psi}^{\scri}=\delta^{2}J_{i^{-}}-\delta^{2}J_{i^{0}},
\end{equation}
where the subscripts $i^{0}$ and $i^{-}$ correspond to limits of \cref{eq:MvJv} as $v\to \pm \infty$. A similar decomposition of the flux $F_{t}^{\mc{H}}$ and $F_{\psi}^{\mc{H}}$ in terms of the difference of perturbed charges can be obtained on past horizon $\mc{H}^{-}$. Indeed, as we have already shown in \cref{A+-F}, the flux $F_{\xi}^{\mc{H}}$ can be expressed as the difference of perturbed charges. Choosing a foliation such that $\psi^{a}$ is tangent to the constant $U$ cross-sections $S_{U}$ on $\mc{H}^{-}$, then the perturbed charge on $\mc{H}^{-}$ associated to $\psi$ is determined by the ``rotational one-form'' $\omega_{A}$ of $\xi^{a}$ \cite{Ashtekar:2001is,Ashtekar:2001jb,Chandrasekaran:2018aop,Chandrasekaran:2019ewn}
\begin{equation}
\delta^{2}J_{U}\defn -\frac{1}{8\pi}\int_{S_{U}}d\Omega_{2}~\epsilon^{AB}\psi_{A}\delta^{2}\omega_{B}
\end{equation}
where $\s_{A}$ is defined in \cref{eq:twist}. Defining the limiting perturbed charges as 
\begin{equation}
\delta^{2}J_{+}\defn \lim_{U\to +\infty}\delta^{2}J_{U},\quad \textrm{ and }\quad \delta^{2}J_{-}\defn \lim_{U\to -\infty}\delta^{2}J_{U}
\end{equation}
and the total flux $F^{\mc{H}}_{\psi}$ is 
\begin{equation}
\label{eq:Fpsihor}
F^{\mc{H}}_{\psi}=\delta^{2}J_{-}-\delta^{2}J_{+}.
\end{equation}
Furthermore, since $t^{a}=\xi^{a}-\Omega_{\textrm{H}}\psi^{a}$ it follows that the flux $F_{t}^{\mc{H}}=F^{\mc{H}}_{\xi}-\Omega_{\textrm{H}}F^{\mc{H}}_{\psi}$ through the past horizon is 
\begin{equation}
\label{eq:Fthor}
F_{t}^{\mc{H}} = \bigg(\frac{\delta^{2}\mathcal{Q}_{-}}{4G_{\textrm{N}}\beta}+\Omega_{\textrm{H}}\delta^{2}J_{-}\bigg)-\bigg(\frac{\delta^{2}\mathcal{Q}_{+}}{4G_{\textrm{N}}\beta}+\Omega_{\textrm{H}}\delta^{2}J_{+}\bigg)
\end{equation}
\Cref{eq:FtnullFpsinull,eq:Fpsihor,eq:Fthor} yield ``local constraints'' on $\mc{H}^{-}$ and $\scri^{-}$ associated to the isometries of Kerr. These fluxes $\op{F}_{t}^{\mc{H}}$, $\op{F}_{\psi}^{\mc{H}}$, $\op{F}_{t}^{\scri}$ and $\op{F}_{\psi}^{\scri}$ are elements of the global quantum field algebra. Integrating \cref{eq:BDYCharge} over a spacelike Cauchy surface for $X=\psi,t$ yields the following global constraints
\begin{equation}
\label{eq:quantfluxscri}
\op{F}_{\psi} = -\delta^{2}\op{J}_{i^{0}}+\delta^{2}\op{J}_{+}, \quad \textrm{and }\op{F}_{t}=\delta^{2}\op{M}_{i^{0}} - \frac{\delta^{2}\op{\mathcal{Q}}_{+}} {4G_{\textrm{N}}\beta}-\Omega_{\textrm{H}}\delta^{2}\op{J}_{+} \ .
\end{equation}
Adding these equations together we get the identity 
\begin{equation}
\label{eq:FxitotKerr}
\op{F}_{\xi}=\delta^{2}\op{M}_{i^{0}}-\Omega_{\textrm{H}}\delta^{2}\op{J}_{i^{0}}-\frac{\delta^{2}\op{\mathcal{Q}}_{+}}{4G_{\textrm{N}}\beta} \ .
\end{equation}
Finally, we note that the above relations imply a ``matching condition'' for the perturbed charges at past timelike infinity\footnote{If we had incoming matter from time-like infinity these relations would be amended to include the fluxes $\op{F}_{\psi}^{i^{-}}$ and $\op{F}_{t}^{i^{-}}$ through $i^{-}$ (see \cref{app1:timequant}).}
\begin{equation}
\label{eq:i-match}
\frac{\delta^{2}\op{\mathcal{Q}}_{-}}{4G_{\textrm{N}}\beta}=\delta^{2}\op{M}_{i^{-}}-\Omega_{\textrm{H}}\delta^{2}\op{J}_{i^{-}} \quad \textrm{and} \quad \delta^{2}\op{J}_{-} = \delta^{2}\op{J}_{i^{-}}
\end{equation}
The fluxes discussed above are included in the global quantum field algebra. We now extend the algebra to include the gravitational charges in $\mc{R}$. There are, naively, six gravitational charges in $\mc{R}$ associated to its two independent isometries. However, the charge-flux relations~\eqref{eq:FtnullFpsinull} and ``matching conditions''~\eqref{eq:i-match} allow four of these to be determined, leaving only two independent gravitational charges in $\mc{R}$. As in section \cref{subsec:sads_kms}, the quantization of these charges corresponds to {\em intrinsic} fluctuations of the gravitational charges of the black hole specified in the initial data of the second-order gravitational field. Therefore, in terms of initial data, it is natural to include --- and thereby dress to --- the gravitational charges $\delta^{2}\op{\mathcal{Q}}_{-}$ and $\delta^{2}\op{J}_{-}$ of the black hole at asymptotically early times. The other charges in $\mc{R}$, $\delta^{2}\op{M}_{i^0},\delta^{2}\op{J}_{i^0},\delta^{2}\op{J}_{i^{-}}$ and $\delta^{2}\op{M}_{i^{-}}$ are then determined by \cref{eq:FtnullFpsinull,eq:i-match}. As we shall explain at the end of this subsection, dressing to the other charges yields a unitarily equivalent algebra.

Let $\Alg_{\textrm{M}}$ and $\Alg_{\textrm{J}}$ be the one-dimensional abelian algebras generated by the generators of time-translations and rotations on the black hole horizon respectively
\begin{equation}
{\op{X}_t := \frac{\delta^{2}\op{\mathcal{Q}}_{-}}{4G_{\textrm{N}}\beta}+ \Omega_\textrm{H} \delta^2 \op{J}_- \quad \textrm{ and }\quad \op{X}_{\psi}\defn -\delta^{2}\op{J}_{-}} \ .
\end{equation} 
We are interested in representations of these algebras in which these operators generate left-translation on the groups $\bb{R}$ and $\textrm{U}(1),$ respectively. $\op{X}_t$ is quantized in $\Hilb_{\textrm{{M}}}\cong L^{2}(\bb{R})$ on which both the operator and its ``conjugate'' variable $\op{t}$ are both densely defined and satisfy 
\begin{equation}
\label{eq:Mt}
[\op{X}_t,\op{t}]=i \ .
\end{equation}
It is tempting to seek analogous representations of $\textrm{U}(1)$ on which ${\op{X}_{\psi}}$ has a ``conjugate'' (periodic) variable $\op{\psi}$ which satisfies 
\begin{equation}
[\op{X}_{\psi},\op{\psi}]=i \ .
\end{equation}
However, as is well known, there do not exist any irreducible representations on which both ${\op{X}_{\psi}}$ and $\op{\psi}$ are well-defined operators. To see this, consider the Hilbert space representation $\ms{H}_{\textrm{J}}$ of $\Alg_{\textrm{J}}$ which consists of square-integrable functions $f(\psi)$ on $\bb{S}$ with inner product
\begin{equation}
\braket{f_{1}|f_{2}}_{\ms{H}_{\textrm{J}}}\defn \int_{\bb{S}}d\psi~\bar{f}_{1}(\psi)f_{2}(\psi)
\end{equation}
where the representation of the (densely defined) ${\op{X}_{\psi}}$ is ${\op{X}_{\psi}}=-i(\partial/\partial \psi)$. The functions $e^{ij\psi}$ where $j\in \bb{Z}$ form an orthonormal basis of $\ms{H}_{\textrm{J}}$ and are eigenstates of ${\op{X}_{\psi}}$ with eigenvalue $j$ and so $\ms{H}_{\textrm{J}}$ can be expressed as the direct sum of eigenstates of ${\op{X}_{\psi}}$
\begin{equation}
\Hilb_{\textrm{J}}= \bigoplus_{j}\ms{H}_{\textrm{j}}
\end{equation}
where $\ms{H}_{\textrm{j}}$ is the one-dimensional Hilbert space of complex, periodic functions $e^{ij\psi}$. There exists a family of well-defined unitary operators $\op{T}_{m},~m \in \bb{Z},$ acting by $(\op{T}_{m} f)(\psi) = e^{im\psi} f(\psi),$ satisfying 
\begin{equation}
\label{eq:JT}
[{\op{X}_{\psi}},\op{T}_{m}]=m\op{T}_{m}.
\end{equation}
However, these operators do not admit a strongly continuous action on $\ms{H}_{\textrm{J}}$ and so, strictly speaking, there does not exist an operator $\op{\psi}$ such that $\op{T}_{m}=e^{im\op{\psi}}$. Nevertheless, we will continue to use the symbol $\op{\psi}$ with the caveat that only its ``exponential'' $\op{T}_{m}$ is actually well-defined. 

We now consider the von Neumann algebra of ``dressed observables'' in $(\mc{R},g)$ invariant under the action of the isometry group $\bb{R}\times \textrm{U}(1)$. We now have two constraints in that the physical observables in $\mc{R}$ must commute with {$\op{C}_{\psi} = \delta^2 \op{J}_+$ and $\op{C}_t = \delta^2 \op{\mathcal{Q}}_+/4G_{\rm N}\beta + \Omega_{\rm H} \delta^2 \op{J}_+$ whose action on the extended Hilbert space is defined by}
\begin{equation}
\label{eq:KerrCops}
\op{C}_{\psi}\defn {\op{X}_{\psi}} -\op{F}_{\psi}^{\mc{H}}\quad \textrm{ and }\quad \op{C}_{t}\defn  \op{X}_t-\op{F}_{t}^{\mc{H}} \ .
\end{equation}
The observables which obey the constraints \cref{eq:quantfluxscri} and commute with $\op{C}_{\psi}$ and $\op{C}_{t}$ are the charges $\op{X}_{\psi}$, $\op{X}_{t}$ as well as the dressed quantum field 
\begin{equation}
\label{eq:dressKerr}
\op{\phi}(f;\op{t},\op{\psi})\defn e^{i\op{F}^{\mc{H}}_{t} \op{t}}e^{i\op{F}^{\mc{H}}_{\psi}\op{\psi}}\op{\phi}(f)e^{-i\op{F}^{\mc{H}}_{\psi}\op{\psi}}e^{-i\op{F}^{\mc{H}}_{t}\op{t}}
\end{equation}
where the expression ``$e^{i\op{F}^{\mc{H}}_{\psi}\op{\psi}}$'' should be interpreted as the unitary operator $\op{T}_{\op{F}^{\mc{H}}_{\psi}}$ on $\Fock\otimes \ms{H}_{\textrm{J}}$ which is well-defined since the spectrum of $\op{F}^{\mc{H}}_{\psi}$ on $\Fock$ is the integers. The action of $\op{F}_{t}^{\mc{H}}$ and $\op{F}_{\psi}^{\mc{H}}$ on ``bulk'' observables $\op{\phi}(f)$ is defined by their action on initial data (e.g., $\op{F}_{\psi}^{\mc{H}}$ generates infinitesimal rotations of $\op{\Pi}(s)$ and has a trivial action on $\tilde{\op{\Pi}}(\tilde{s})$). Using the decomposition \cref{eq:phiPiPi} we have that 
\begin{equation}
\label{eq:phifdress}
\op{\phi}(f;\op{t},\op{\psi}) = \op{\Pi}(s;\op{t},\op{\psi}) + \tilde{\op{\Pi}}(\tilde{s}) \ ,
\end{equation}
where $\op{\Pi}(s;\op{t},\op{\psi})$ is the dressed initial data on $\mc{H}_{\textrm{R}}^{-}$ 
\begin{equation}
\label{eq:dressPiKerr}
\op{\Pi}(s,\op{t},\op{\psi})\defn e^{i\op{F}^{\mc{H}}_{t} \op{t}}e^{i\op{F}^{\mc{H}}_{\psi}\op{\psi}}\op{\Pi}(s)e^{-i\op{F}^{\mc{H}}_{\psi}\op{\psi}} e^{-i\op{F}^{\mc{H}}_{t}\op{t}} \ .
\end{equation}
We note that the perturbed ADM mass and angular momentum at $i^0$ are now {\em defined} by 
\begin{equation}
\label{eq:ADM}
\delta^{2}\op{M}_{i^{0}}\defn \op{F}_{t}^{\scri} + \op{X}_{t} \quad \textrm{ and }\quad -\delta^{2}\op{J}_{i^{0}}\defn \op{F}_{\psi}^{\scri} + \op{X}_{\psi} \ ,
\end{equation}
which, indeed, generate time translations and rotations of bulk observables, via
\begin{align}
[\delta^{2}\op{M}_{i^{0}},\op{\phi}(f,\op{t},\op{\psi})] =&[\op{X}_{t},\op{\Pi}(s;\op{t},\op{\psi})] + [\op{F}_{t}^{\scri} ,\tilde{\op{\Pi}}(\tilde{s})] \nonumber \\
=&~i\op{\Pi}(\pounds_{t}s;\op{t},\op{\psi})+i\tilde{\op{\Pi}}(\pounds_{t}\tilde{s}) \nonumber \\
=&~i\op{\phi}(\pounds_{t}f,\op{t},\op{\psi})
\end{align}
and, similarly, 
\begin{equation}
\label{eq:admconstraint}
[\delta^{2}\op{J}_{i^{0}},\op{\phi}(f,\op{t},\op{\psi})] = -i\op{\phi}(\pounds_{\psi}f,\op{t},\op{\psi}) \ .
\end{equation}
The ``extended'' $\ast$-algebra is the ``crossed product'' algebra {$\Alg_{\textrm{ext.}}\defn \Alg_{\textrm{dress.}}\rtimes (\Alg_{\textrm{J}}\otimes \Alg_{\textrm{X}})$ } where $\Alg_{\textrm{dress.}}$ is the $\ast$-algebra generated by $\op{\phi}(f;\op{t},\op{\psi})$ and $\op{1}$. The full von Neumann algebra of dressed observables in $\mc{R}$ is thus 
\begin{equation}
\label{eq:extKerrAlg}
\mf{A}_{\textrm{ext.}}(\mc{R},\s_{0})\defn \{\op{\phi}(f;\op{t},\op{\psi}),\op{X}_{t},\op{X}_{\psi}\}'' \ ,
\end{equation}
where the support of $f$ is in $\mc{R}$. As previously emphasized, the modular flow for $\s_{0}$ does not correspond to a geometric flow with respect to any vector field in $\mc{R}$ and therefore, we cannot immediately deduce the type of the algebra from \cref{eq:extKerrAlg}. However, we can straightforwardly determine its type by using the decomposition {\cref{eq:phifdress}} together with the analogous decomposition of $\op{F}_{t}$ and $\op{F}_{\psi}$ on $\Alg_{\mc{H}^{-}}\otimes \Alg_{\scri^{-}}$. {In particular the algebra $\mf{A}_{\textrm{ext.}}(\mc{R},\s_{0})$ is 
\begin{equation}
\label{eq:extKerrAlg2}
\mf{A}_{\textrm{ext.}}(\mc{R},\s_{0})= \mf{A}_{\textrm{ext.}}(\mc{H}_{\textrm{R}}^{-},\s_{0}) \otimes \mf{A}(\scri^{-},\s_{0})
\end{equation}
the tensor product of the algebra of dressed observables on the horizon 
\begin{equation}
\mf{A}_{\textrm{ext.}}(\mc{H}_{\textrm{R}}^{-},\s_{0})\defn \{\op{\Pi}(s;\op{t},\op{\psi}),\op{X}_{t},\op{X}_{\psi}\}'' 
\end{equation}
with the Type I$_{\infty}$ algebra $\mf{A}(\scri^{-},\s_{0})$ of quantum field theory observables at $\scri^{-}$. }
 
{The} isometries are outer automorphisms of $\mf{A}(\mc{H}_{\textrm{R}}^{-},\s_{0})$ and therefore, the ``crossed product'' $\mf{A}_{\textrm{ext.}}(\mc{H}_{\textrm{R}}^{-},\s_{0})$ can change the type of the algebra. Indeed, we now show that this algebra is Type II$_{\infty}$. To see this, it {is} useful to conjugate the algebra by the unitaries {$U(\op{t})\defn e^{-i\op{F}_{t}^{\mc{H}}\op{t}}$ and $U(\op{\psi})\defn e^{-i\op{F}_{\psi}^{\mc{H}}\op{\psi}}$} to obtain
\begin{equation}
\hat{\mf{A}}_{\textrm{ext.}}(\mc{H}_{\textrm{R}}^{-},\s_{0})\defn U(\op{t})U(\op{\psi}){\mf{A}}_{\textrm{ext.}}(\mc{H}_{\textrm{R}}^{-},\s_{0})U({-}\op{t})U({-}\op{\psi}) = \{\op{\Pi}(s),\op{X}_{\psi}-\op{F}_{\psi}^{\mc{H}},\op{X}_{t}-\op{F}_{t}^{\mc{H}}\}'' \ .
\end{equation}
This is the crossed product with respect to time translations and rotations, i.e., the group $\bb{R}\times \textrm{U}(1),$ on $\mc{H}^{-}$. One can equivalently express this algebra as that obtained by including the linearly independent generators 
\begin{equation}
\op{X}_{\xi}-\op{F}^{\mc{H}}_{\xi} \quad \textrm{and }\quad \op{X}_{\psi}-\op{F}^{\mc{H}}_{\psi}
\end{equation}
where $\op{X}_{\xi}\defn \op{X}_{t}+\Omega_{\textrm{H}}\op{X}_{\psi}$. Therefore, the (conjugated) algebra is isomorphic to 
\begin{equation}
\hat{\mf{A}}_{\textrm{ext.}}(\mc{H}_{\textrm{R}}^{-},\s_{0}) \cong \{\op{\Pi}(s),\op{X}_{\xi}-\op{F}^{\mc{H}}_{\xi},\op{X}_{\psi}-\op{F}^{\mc{H}}_{\psi}\}'' \ .
\end{equation}
This is precisely the crossed product of the type III$_1$ algebra $\mf{A}(\sH^-_{\rm R}, \om_0)$ with respect to its modular automorphism group (as seen by the addition of $\op{X}_{\xi}-\op{F}^{\mc{H}}_{\xi}$) together with the compact group $\textrm{U}(1)$. Since the unitary conjugation does not change the type of algebra, it directly follows {from \cref{thm:horizontheorem}} that $\mf{A}_{\textrm{ext.}}(\mc{H}_{\textrm{R}}^{-},\s_{0})$ is Type II$_{\infty}$. The {full} algebra is the {tensor} product of a Type II$_{\infty}$ algebra and a Type I$_{\infty}$ algebra, and so is Type II$_{\infty}$.

{In the above discussion, we have explicitly dressed the quantum fields to the black hole charges at $i^-$. However, we can equivalently explicitly dress the quantum fields to the charges at $i^0$ by conjugating $\mf{A}_{\textrm{ext.}}(\mc{R},\s_0)$ by a unitary
\begin{align}
\label{eq:i0algebra}
e^{i\op{F}_{t}^{\scri}\op{t}}e^{i\op{F}_{\psi}^{\scri}\op{\psi}}\mf{A}_{\textrm{ext.}}(\mc{R},\s_0)e^{-i\op{F}_{t}^{\scri}\op{t}}e^{-i\op{F}_{\psi}^{\scri}\op{\psi}}=  \{\op{\Pi}(s;\op{t},\op{\psi}),\tilde{\op{\Pi}}(\tilde{s};\op{t},\op{\psi}),\op{X}_{t}+\op{F}_{t}^{\scri},\op{X}_{\psi}+\op{F}_{\psi}^{\scri}\}''
\end{align}
where $\tilde{\op{\Pi}}(\tilde{s};\op{t},\op{\psi})$ is given by \cref{eq:dressPiKerr} with $\op{\Pi}(s)$ replaced with $\tilde{\op{\Pi}}(\tilde{s})$ and $\op{F}_{t/\psi}^{\mc{H}}$ replaced with $\op{F}^{\scri}_{t/\psi}$. From \eqref{eq:ADM}, we see that the ADM mass and angular momentum have been explicitly included in the algebra. The type is unaffected by unitary conjugation, so \eqref{eq:i0algebra} is Type II$_{\infty}$. 
}

{As we have just shown, the algebra $\mf{A}_{\textrm{ext.}}(\mc{R},g)$ is a Type II$_{\infty}$ algebra and therefore has a well-defined trace.  As in the previous sections, to straightforwardly obtain the generalized entropy we now consider the subalgebra of observables gauged under only the horizon Killing field $\xi^{a}=t^{a}+\Omega_{\textrm{\textrm{H}}}\psi^{a}$. Therefore, the relevant Hilbert space is obtained by identifying all eigenvectors of $\op{X}_{\xi}$ in $\Hilb_{\textrm{M}}\otimes \Hilb_{\textrm{J}}$ with the same eigenvalue. The resulting Hilbert space $\Hilb_{\textrm{X}_{\xi}}\cong L^{2}(\bb{R})$ admits a (densely defined) conjugate variable $\op{t}_{\xi}$. Therefore, the relevant von Neumann algebra is}
\begin{equation}
{\bar{\mf{A}}_{\textrm{ext.}}(\mc{R},\s_{0})\cong \bar{\mf{A}}_{\textrm{ext.}}(\mc{H}_{\textrm{R}}^{-},\s_{0})\otimes \mf{A}(\scri^{-},\s_{0})}
\end{equation}
 where 
\begin{equation}
\label{eq:Ubar}
{\bar{\mf{A}}_{\textrm{ext.}}(\mc{H}_{\textrm{R}}^{-},\s_{0})\defn \{\op{\Pi}(s;\op{t}_{\xi}),\op{X}_{\xi}\}^{\prime\prime}}
\end{equation}
on the the Hilbert space $\Fock\otimes \Hilb_{\mathcal{Q}}$ where $\op{\Pi}(s;\op{t}_{\xi})$ is $\op{\Pi}(s)$ conjugated with $e^{i\op{t}_{\xi}\op{F}_{\xi}^{\mc{H}}}$.

A general element {{$a \in  \bar{\mf{A}}_{\textrm{ext.}}(\mc{R},\s_{0})$}} can be expressed as the sum of elements of the form
\begin{equation}
{{a=\bar{a} \otimes \tilde{a}}}
\end{equation} where {$\bar{a}\in \bar{\mf{A}}_{\textrm{ext.}}(\mc{H}_{\textrm{R}}^{-},\s_{0})$} and {$\tilde{a}\in \mf{A}(\scri^{-},\s_{0})$}. Therefore, it suffices to define the trace on such elements. The trace is 
\begin{equation}
\label{eq:Tracekerr}
{\textrm{Tr}(a)=\beta \int_{\bb{R}}dX_{\xi}~e^{\beta X_{\xi}}\braket{\s_{0}^{\mc{H}},X_{\xi}| \textrm{Tr}_{\scri}(\tilde{a})\cdot \bar{a}|\s_{0}^{\mc{H}},X_{\xi}}}
\end{equation}
where $\textrm{Tr}_{\scri}(\cdot)$ is the standard Hilbert space trace over $\Fock^{\scri}$ and {$\ket{\s^{\mc{H}}_{0},X_{\xi}}=\ket{\s^{\mc{H}}_{0}}\otimes \ket{X_{\xi}} $ and $\ket{X_{\xi}}$ are eigenstates of $\op{X}_{\xi}$ with eigenvalue $X_{\xi}$.}
The trace is then extended to general elements of {$\bar{\mf{A}}_{\textrm{ext.}}(\mc{R},\s_{0})$} by linearity. 

Considering a ``classical-quantum'' state on $\Fock \otimes \Hilb_{\textrm{A}}$ 
\begin{equation}
\label{eq:classqstatekerr}
{\ket{\hat{\s}}\defn \int_{\bb{R}}dX_{\xi}~f(X_{\xi})\ket{\s,X_{\xi}}}
\end{equation}
where $f$ is a state on {$\Hilb_{\textrm{X}_{\xi}}$} with Schwartz-decay in {$X_{\xi}$} and $\ket{\s}\in \Fock$ a Hadamard state which, for simplicity, we take to be a product state
\begin{equation}
\ket{\s}=\ket{\s^{\mc{H}}}\otimes \ket{\s^{\scri}}.
\end{equation}
Following the same analysis as in \cref{subsec:dens_ent}, the corresponding density matrix $\rho_{\hat{\s}}$ on $\bar{\mf{A}}_{\textrm{ext.}}(\mc{R},\s_{0})$ is well-defined {and equal to}
{
\begin{align}
\begin{aligned}
\label{eq:rhoRkerr}
 \rho_{\hat{\s}} =&\beta^{-1}e^{-\frac{i\op{t}_{\xi}\op{H}_{\s_{0}^{\mc{H}}}}{\beta}}\ {{f}(\op{X}_{\xi}+\beta^{-1}\op{H}_{\omega_{0}^{\mc{H}}})e^{-\frac{ \beta\op{X}_{\xi}}{2}}\op{\Delta}_{\s^{\mc{H}}|\s_{0}^{\mc{H}}}e^{-\frac{ \beta\op{X}_{\xi}}{2}}}
 \bar{f}(\op{X}_{\xi}+\beta^{-1}\op{H}_{\omega_{0}^{\mc{H}}})e^{\frac{i\op{t}_{\xi}\op{H}_{\s_{0}^{\mc{H}}}}{\beta}}
 \\& 
  \otimes \ket{\s^{\scri}}\bra{\s^{\scri}}.
 \end{aligned}
\end{align} }
The von Neumann entropy of the state, assuming $f(X_{\xi})$ is slowly varying in $X_{\xi}$, is 
{\begin{align}
\label{eq:SvNKerr}
S_{\textrm{vN.}}(\rho_{\hat{\s}})\defn &-\textrm{Tr}(\rho_{\hat{\s}}\log(\rho_{\hat{\s}}))
\nonumber \\
\simeq&~\hat{\s}(\beta \op{X}_{\xi})-S_{\textrm{rel.}}(\s^{\mc{H}}|\s^{\mc{H}}_{0})
+S(\rho_{f})+\log \beta \ ,
\end{align}}
where $S(\rho_{f})$ is the von Neumann entropy of {$\rho_{f}\defn |f(\op{X}_{\xi})|^{2}$} on ${\Hilb_{\mathcal{Q}}}$. If instead of a pure state, we considered initial data where the density matrix of the quantum field is $\ket{\s^{\mc{H}}}\bra{\s^{\mc{H}}}\otimes \rho^{\scri}$  where $\rho^{\scri}$ is the density matrix associated a mixed state $\scri^{-}$, then \cref{eq:SvNKerr} would receive an additional contribution from the von Neumann entropy $S(\rho^{\scri})$ of $\rho^{\scri^{-}}$.

Following the manipulations of \cref{subsec:dens_ent} to formally decompose the von Neumann entropy \cref{eq:SvNKerr} in terms of the generalized entropy as in \cref{subsec:dens_ent} yields 
{\begin{equation}
S_{\textrm{vN.}}(\rho_{\hat{\s}}) \simeq S_{\textrm{gen.},\mc{B}}(\hat{\s}) + S(\rho_{f})+ C \ .
\end{equation}}
where 
\begin{equation}
S_{\textrm{gen.},\mc{B}}(\hat{\s}) = \hat{\s}\bigg(\frac{\op{A}_{\mc{B}}}{4G_{\textrm{N}}}\bigg)+S_{\textrm{vN.}}(\s\vert_{\mc{R}})
\end{equation}
and we recall that $\op{A}_{\mc{B}}$ is the area (\cref{eq:ABdef}) of the bifurcation surface up to second-order. 
Therefore, the von Neumann entropy of $\rho_{\hat{\s}}$ is the generalized entropy

\section{Black Holes in Asymptotically de Sitter Spacetime}
\label{sec_dS}
\subsection{Schwarzschild-de Sitter}

\label{subsec:sds}
In this section, we consider stationary black hole spacetimes $(\mc{M},g)$ with a positive cosmological constant. For simplicity, we will consider Schwarzschild-de Sitter spacetime, however, the construction of dressed observables carries over straightforwardly to the case of charged, rotating black holes in de Sitter spacetime. The Penrose diagram of a single black hole formed from (spherically symmetric) collapse in de Sitter spacetime is depicted in \cref{fig:SdS_tikz}. The spacetime is a closed universe where the topology of spatial slices is $\bb{S}^{1}\times\bb{S}^{2}$. The region $\mc{M}_{\textrm{R}}=\mc{F}_{1}\cup \mc{R}_{1}\cup \mc{F}_{2}$ corresponds to the future interior of the black hole, the future interior of the cosmological horizon, and the region $(\mc{R}_{1},g),$ which is the globally hyperbolic region ``between'' the two horizons. $(\mc{M}_{\textrm{R}},g)$ is globally hyperbolic and so, as before, the construction of the phase space $\mc{P}$ and algebra of observables $\Alg$ of field $\op{\phi}(f)$ is well-defined. In contrast to the case of de Sitter spacetime, we will consider massive or massless fields (i.e., $V=m^{2}$ or $V=0$ in \eqref{eq:EOMphi}). The region $\mc{R}_{1}$ is invariantly defined as $I^+(\mc{H}_1^-) \cap I^-(\mc{H}_1^+)$ (see figure \ref{fig:SdS_tikz}). The isometry group of $(\mc{R}_{1},g)$ is $\bb{R}\times \textrm{SO}(3)$ where the time translations correspond to the isometry $\chi_{t}$. 

We now consider the ``initial data'' corresponding to field observable $\op{\phi}(f)$. In Schwarzschild-de Sitter spacetime, it was proven in Theorem 3.2.2 of \cite{Brum:2014nea} that there is no ``leakage'' of symplectic flux of the fields through $i_{\textrm{R}}^{-}$ and so one can decompose\footnote{This decomposition was also rigorously proven for Klein Gordon fields in Reissner-Nordstrom de Sitter \cite{Hollands:2019whz} and (slowly rotating) Kerr de Sitter \cite{Klein:2022jtb} using point-wise decay estimates obtained by Hintz and Vasy \cite{Hintz:2015jkj}.} the field observable $\op{\phi}(f)$ into initial data specified on the past black hole horizon $\mc{H}_{1}^{-}$ and the ``future'' cosmological horizon $\mc{H}_{2}^{+}$ 
\begin{equation}
\op{\phi}(f) = \op{\Pi}_{1}(s_{1}) + \op{\Pi}_{2}(s_{2})
\end{equation}
where $\op{\Pi}_{1}$ is the initial data on $\mc{H}_{1}^{-}$, $\op{\Pi}_{2}$ is the initial data on $\mc{H}_{2}^{+}$, $s_{1}\defn [Ef]\vert_{\mc{H}_{1}^{-}}$ and $s_{2}\defn [Ef]\vert_{\mc{H}_{2}^{+}}$. 

Thus, there is an isomorphism between the $\ast$-algebra for $\mc{M}_{\rm R}$, $\Alg$, and the tensor product of the $\ast$-algebras for $\mc{H}_{1}^-$ and $\mc{H}^-_{2}$
\begin{align}
 \Alg \cong \Alg_{\mc{H}_{1}^-} \otimes \Alg_{\mc{H}_{2}^+} \ .
\end{align}
While there is no globally KMS state on $\Alg$, there is a stationary state $\s_{0}$ for quantum fields on Schwarzschild-de Sitter. This state is defined to be the vacuum for initial data on the past black hole and cosmological horizons
\begin{align}
 \omega_0 = {\omega}_{0}^{\mc{H}_{1}} \otimes {\omega}_{0}^{\mc{H}_{2}},
\end{align}
where ${\omega}_{0}^{\mc{H}_{1}}$ and $\omega_{0}^{\mc{H}_{2}}$ are the Gaussian states on $\Alg_{\mc{H}_{1}^-}$ and $\Alg_{\mc{H}_{2}^+}$ with $2$-point functions given by \cref{eq:vacH-}. We note that the state is invariant under the isometries of the spacetime and was rigorously proven to be Hadamard in $(\mc{M}_{\textrm{R}},g)$ in \cite{Brum:2014nea}. We note that $\s_{0}^{\mc{H}_{1}}$ is KMS on $\mc{H}_{1,\textrm{R}}^{-}$ with inverse temperature $\beta_{1}$ and $\s_{0}^{\mc{H}_{2}}$ is KMS on $\mc{H}_{2,\textrm{L}}^{+}$ with inverse temperature $\beta_{2}$. Therefore, the full 
state $\s_{0}$ cannot be globally KMS in $\mc{R}_{1}$ unless $\beta_{1}=\beta_{2}$. 

The GNS Fock representation $\Fock$ obtained from $\Alg$ with respect to $\omega_0$ is the tensor product of the Fock spaces on the two past horizons $\Fock_{\mc{H}_{1}}\otimes \Fock_{\mc{H}_{2}}$. The von Neumann algebra of quantum field observables in $\mc{R}_1$, $\mf{A}(\mc{R}_1, \omega_0)$ is Type III$_1$. This can be seen because it is generated by the two Type III$_1$ von Neumann algebras on the horizons
\begin{align}
 \mf{A}(\mc{R}_1, \omega_0) = \mf{A}(\mc{H}_{R,1}^-, \omega_0) \otimes \mf{A}(\mc{H}_{L,2}^+, \omega_0) \ .
\end{align}
Integrating \cref{eq:BDYCharge} over a Cauchy surface for the region $\mc{M}_{\textrm{R}}$ yields the global constraint\footnote{In reference \cite{Gibbons:1977mu}, Gibbons and Hawking obtain the relationship between the first-order perturbed areas of the cosmological horizon and black hole horizon due to the presence of a (first-order) stress tensor. However, their analysis can be straightforwardly extended to second-order perturbations following \cite{Hollands:2012sf}.}  \cite{Gibbons:1977mu}
\begin{equation}
\label{eq:globconst}
\op{F}_{\xi}=-\frac{\delta^{2}\op{\mathcal{Q}}^{\mc{H}_{2}}_{+}}{4G_{\textrm{N}}\beta_{2}} - \frac{\delta^{2}\op{\mathcal{Q}}^{\mc{H}_{1}}_{+}}{4G_{\textrm{N}}\beta_{1}} \ .
\end{equation}
This global constraint is of the form of \cref{eq:CXF} with $\op{X}=0$ --- there are no gravitational charges in \cref{eq:globconst} which lie in $\mc{R}_{1}$ --- and there are now two charges which lie in the causal complement of $\mc{R}_{1},$
\begin{equation}
\op{C}_{1}\defn \delta^{2}\op{\mathcal{Q}}_{+}^{\mc{H}_{1}}/4G_{\textrm{N}}\beta_{1} \quad \textrm{and}\quad \op{C}_{2}\defn \delta^{2}\op{\mathcal{Q}}_{+}^{\mc{H}_{2}}/4G_{\textrm{N}}\beta_{2} \ ,
\end{equation}
 which thereby commute with all observables in $\mc{R}_{1}$. As before, in order to nontrivially impose the gravitational constraint we introduce an observer {that follows the orbits of the timelike Killing field} with (bounded) energy $\op{\varepsilon}$, time $\op{\tau}$ and Hilbert space $\ms{H}_{\textrm{obs.}}=L^{2}(\bb{R}_{+})$ (see \cref{subsec:deSitter}). Introducing the observer changes the global constraint to\footnote{The identification of the spacetime depicted in figure~\ref{fig:SdS_tikz} additionally implies that $\op{C}_{1}=-\op{C}_{2}$ and therefore, by \cref{eq:FxieC1C2}, the observer is not an independent degree of freedom; their energy is completely determined by the quantum field. As in de Sitter spacetime, this can be remedied by introducing an additional gravitating body (e.g., an observer) with energy $\op{\varepsilon}^{\prime}$ that approaches $i^{+}_{\textrm{L}}$. The constraint at $i^+_L$ then becomes $\op{\varepsilon}^{\prime}=-\op{C}_{1}-\op{C}_{2}$.} 
\begin{equation}
\label{eq:FxieC1C2}
\op{F}_{\xi} + \op{\varepsilon} = -\op{C}_{1}-\op{C}_{2}.
\end{equation}

Since this global constraint is similar to the constraint obtained in de Sitter spacetime in \cref{subsec:deSitter}, it would appear that the construction of dressed observables would mirror that case. However, there are actually three independent ``local constraints'' that imply the global constraint. Satisfying these local constraints will impose stronger conditions on the algebra of the dressed observables than \cref{eq:FxieC1C2}. 

The local constraints will involve the two ``gravitational charges'' in $\mc{R}_{1}$ which we will suggestively label as 
\begin{equation}
\op{X}_{1} \defn \frac{\delta^{2}\op{\mathcal{Q}}_{-}^{
\mc{H}_{1}}}{4G_{\textrm{N}}\beta_{1}}\quad \textrm{and}\quad 
\op{X}_{2} \defn \frac{\delta^{2}\op{\mathcal{Q}}_{-}^{
\mc{H}_{2}}}{4G_{\textrm{N}}\beta_{2}}.
\end{equation}
The local fluxes $\op{F}_{\xi}^{\mc{H}_{1}}$ and $\op{F}_{\xi}^{\mc{H}_{2}}$ then satisfy local constraints given by
\begin{equation}
\label{eq:C1C2X1X2}
 \op{C}_{1}= \op{X}_{1} - \op{F}^{\mc{H}_{1}}_{\xi} \quad \textrm{ and }\quad 
 \op{C}_{2}= \op{X}_{2} - \op{F}^{\mc{H}_{2}}_{\xi}.
\end{equation}
Furthermore, \cref{eq:FxieC1C2,eq:C1C2X1X2} implies that the energy of the observer perturbs the horizons by 
\begin{equation}
\label{eq:eX1X2}
\op{\varepsilon} =- \op{X}_{1}-\op{X}_{2} \ .
\end{equation}
\Cref{eq:C1C2X1X2,eq:eX1X2} are stronger constraints than \cref{eq:FxieC1C2}. In particular, \cref{eq:C1C2X1X2} implies that, given any observable in $\mc{R}$, its corresponding initial data on $\mc{H}_{1}^{-}$ and $\mc{H}_{2}^{+}$ 
must commute with both $\op{X}_{1}-\op{F}^{\mc{H}_{1}}_{\xi}$ and $\op{X}_{2}-\op{F}^{\mc{H}_{2}}_{\xi}$. Furthermore, \cref{eq:eX1X2} implies that the observer's energy is not an independent degree of freedom and is determined by the charges $\op{X}_{1}$ and $\op{X}_{2}$. 

We now construct dressed quantum field observables which commute with $\op{C}_1$ and $\op{C}_2.$ 
We therefore extend the algebra to include the one-dimensional algebras $\Alg_{\textrm{X}_{1}}$ and $\Alg_{\textrm{X}_{2}}$ generated by $\op{X}_{1}$ and $\op{X}_{2}$. At this stage we will not impose \cref{eq:eX1X2} and so the Hilbert space representations of the algebra are $\Hilb_{\textrm{X}_{1}}=\Hilb_{\textrm{X}_{2}} \cong L^{2}(\bb{R})$. The conjugate variables will be denoted as $\op{t}_{1}$ and $\op{t}_{2}$ which satisfy 
\begin{equation}
[\op{X}_{1},\op{t}_{1}]=i \quad \textrm{and}\quad [\op{X}_{2},\op{t}_{2}]=i
\end{equation}
The extended Hilbert space is $\Fock\otimes \Hilb_{\textrm{X}_{1}}\otimes \Hilb_{\textrm{X}_{2}}$. The dressed quantum fields in $\mc{R}$ which simultaneously satisfies both constraints \cref{eq:C1C2X1X2} are 
\begin{equation}
\op{\phi}(f;\op{t}_{1},\op{t}_{2})\defn e^{i\op{F}^{\mc{H}_{1}}_{\xi}\op{t}_{1}}e^{i\op{F}^{\mc{H}_{2}}_{\xi}\op{t}_{2}}\op{\phi}(f)e^{-i\op{F}^{\mc{H}_{1}}_{\xi}\op{t}_{2}}e^{-i\op{F}^{\mc{H}_{1}}_{\xi}\op{t}_{1}}. 
\end{equation}
where the support of $f$ is in $\mc{R}_{1}$.
The von Neumann algebra of observables in $\mc{R}_{1}$ that satisfy the constraints \cref{eq:C1C2X1X2} is 
\begin{equation}
\label{eq:AextSdS}
\bar{\mf{A}}(\mc{R}_{1},\s_{0})\defn \{\op{\phi}(f;\op{t}_{1},\op{t}_{2}),\op{X}_{1},\op{X}_{2}\}^{\prime\prime}.
\end{equation}
where we note that we have not yet imposed \cref{eq:eX1X2} and so $\bar{\mf{A}}(\mc{R}_{1},\s_{0})$ is not the final algebra of dressed observables.

The modular flow of $\s_{0}$ is not geometric in $\mc{R}_{1}$ and therefore we cannot immediately conclude the type of $\bar{\mf{A}}(\mc{R}_{1},\s_{0})$ from the form of \cref{eq:AextSdS}. However, one can straightforwardly obtain the type of the algebra by decomposing the bulk field in terms of initial data 
\begin{equation}
\op{\phi}(f;\op{t}_{1},\op{t}_{2}) = \op{\Pi}_{1}(s_{1};\op{t}_{1})+\op{\Pi}_{2}(s_{2};\op{t}_{2})
\end{equation}
where 
\begin{equation}
\op{\Pi}_{1}(s_{1};\op{t}_{1}) \defn e^{i\op{F}_{\xi}^{\mc{H}_{1}}\op{t}_{1}}\op{\Pi}_{1}(s_{1})e^{-i\op{F}_{\xi}^{\mc{H}_{1}}\op{t}_{1}},\quad \op{\Pi}(s;\op{t}_{2})\defn e^{i\op{F}_{\xi}^{\mc{H}_{2}}\op{t}_{2}}\op{\Pi}_{2}(s_{2})e^{-i\op{F}_{\xi}^{\mc{H}_{2}}\op{t}_{2}}.
\end{equation}
The von Neumann algebra on $\mc{H}^{-}_{1,\textrm{R}}$ is given by 
\begin{equation}
\vNext(\mc{H}^{-}_{1,\textrm{R}},\s_{0})\defn \{\op{\Pi}_{1}(s_{1};\op{t}_{1}),\op{X}_{1}\}^{\prime\prime}
\end{equation}
and, by theorem \ref{thm:horizontheorem}, is Type II$_{\infty}$. The trace is 
\begin{equation}
\textrm{Tr}_{1}(a_{1})\defn \int_{\bb{R}}dX_{1}~e^{\beta_{1}X_{1}}\braket{\s^{\mc{H}_{1}}_{0},X_{1}|a_{1}|,\s^{\mc{H}_{1}}_{0},X_{1}}.
\end{equation}
for any $a_{1}\in \vNext(\mc{H}^{-}_{1,\textrm{R}},\s_{0})$. Similarly, the von Neumann algebra $\vNext(\mc{H}^{+}_{2,\textrm{L}},\s_{0})\defn \{\op{\Pi}_{2}(s_{2};\op{t}_{2}),\op{X}_{2}\}^{\prime\prime}$ is Type II$_{\infty}$ with trace 
\begin{equation}
\textrm{Tr}_{2}(a_{2})\defn \int_{\bb{R}}dX_{2}~e^{\beta_{2}X_{2}}\braket{\s^{\mc{H}_{2}}_{0},X_{2}|a_{2}|\s^{\mc{H}_{2}}_{0},X_{2}}.
\end{equation}
The algebra of dressed observables in $\mc{R}_{1}$ --- without imposing \cref{eq:eX1X2} --- is then 
\begin{equation}
\bar{\mf{A}}_{\textrm{ext.}}(\mc{R}_{1},\s_{0})=\vNext(\mc{H}^{-}_{\textrm{R}},\s_{0})
\otimes \vNext(\mc{H}^{+}_{\textrm{L}},\s_{0})
\end{equation}
Thus, the algebra $\bar{\mf{A}}_{\textrm{ext.}}(\mc{R}_{1},\s_{0})$ is Type II$_{\infty}$ with trace 
\begin{equation}
\textrm{Tr}(\bar{a})=\int_{\bb{R}^{2}}dX_{1}dX_{2}~e^{\beta_{1}X_{1}+\beta_{2}X_{2}}\braket{\s_{0},X_{1},X_{2}|a|\s_{0},X_{1},X_{2}}
\end{equation}
for any $\bar{a}\in \bar{\mf{A}}_{\textrm{ext.}}(\mc{R}_{1},\s_{0})$ and $\ket{\s_{0},X_{1},X_{2}}=\ket{\s_{0}}\otimes \ket{X_{1}}\otimes \ket{X_{2}}$. 
We now impose the final constraint \cref{eq:eX1X2} which implies, if the observer's energy is bounded from below by zero, the condition that $X_{2}+X_{1}<0$. Applying the projection operator $\op{P}_{X_{1}<-X_{2}}\defn \op{\Theta}(-X_{1}-X_{2})$ to the algebra yields 
\begin{equation}
\mf{A}_{\textrm{ext.}}(\mc{R}_{1},\s_{0})\defn \op{P}_{X_{1}<-X_{2}}\bar{\mf{A}}_{\textrm{ext.}}(\mc{R}_{1},\s_{0})\op{P}_{X_{1}<-X_{2}}
\end{equation}
which acts on the Hilbert space $\Fock \otimes \Hilb_{X_{1}<-X_{2}}$ where $\Hilb_{X_{1}<-X_{2}}\defn \op{P}_{X_{1}<-X_{2}}[\Hilb_{\textrm{X}_{1}}\otimes \Hilb_{\textrm{X}_{2}}]$. The corresponding trace is 
\begin{equation}
\textrm{Tr}(\hat{a})= \int_{-\infty}^{\infty}dX_{2}~\int_{-\infty}^{-X_{2}}dX_{1}~e^{\beta_{1}X_{1}+\beta_{2}X_{2}}\braket{\s_{0},X_{1},X_{2}|\hat{a}|\s_{0},X_{1},X_{2}}.
\end{equation}
for any $a\in \mf{A}_{\textrm{ext.}}(\mc{R}_{1},\s_{0})$.  
The projection does not change the algebra type,\footnote{A heuristic argument for the type of the algebra for Schwarzschild-de Sitter, consistent with our result, was given in \cite{2023arXiv230803663W}.} which can be easily seen by noting that $\textrm{Tr}(\hat{\op{1}})=\infty$.

We now consider a ``classical-quantum'' state of the form 
\begin{equation}
\ket{\hat{\s}}\defn \int_{-\infty}^{\infty}dX_{2}\int_{-\infty}^{-X_{2}}dX_{1}~f(X_{1},X_{2})\ket{\s,X_{1},X_{2}}
\end{equation}
and we choose, for simplicity, $\ket{\s}\in \Fock$ to be a product state. 
\begin{equation}
\ket{\s} = \ket{\s^{\mc{H}_{1}}}\otimes \ket{\s^{\mc{H}_{2}}}
\end{equation}

To simplify the following formulas, we will use the notation $\s^{1}$ to denote a state on $\Alg_{\mc{H}^{-}_{1}}$ and $\s^{2}$ to denote a state on $\Alg_{\mc{H}^{-}_{2}}$. Assuming $f(X_{1},X_{2})$ is slowly varying {in both $X_1$ and $X_2$}, the corresponding density matrix can then be expressed as 

\begin{align}
\begin{aligned}
 \rho_{\hat{\s}} =e^{-{i(\beta_1^{-1}{\op{t}_1\op{H}_{\s_{0}^1}}+\beta_2^{-1}{\op{t}_2\op{H}_{\s_{0}^2}})}}{f}(\op{X}_1+\beta_1^{-1}\op{H}_{\omega_0^{1}},\op{X}_2+\beta_2^{-1}\op{H}_{\omega_0^{2}})e^{-{ \beta_1\op{X}_1}-{ \beta_2\op{X}_2}}\op{\Delta}_{\s^1|\s_{0}^1}
 \\
 \times \op{\Delta}_{\s^2|\s_{0}^2}\bar{f}(\op{X}_1+\beta_1^{-1}\op{H}_{\omega_0^{1}},\op{X}_2+\beta_2^{-1}\op{H}_{\omega_0^{2}})e^{{i(\beta_1^{-1}{\op{t}_1\op{H}_{\s_{0}^1}}+\beta_2^{-1}{\op{t}_2\op{H}_{\s_{0}^2}})}},
\end{aligned}
\end{align}  
and the entropy, assuming $f(X_{1},X_{2})$ is slowly varying, is approximately 
\begin{equation}
S_{\textrm{vN.}}(\rho_{\hat{\s}})\simeq \hat{\s}(\beta_1\op{X}_{1}+\beta_2\op{X}_{2})-S_{\textrm{rel.}}(\s^{1}|\s_{0}^{1})
- S_{\textrm{rel.}}(\s^{2}|\s_{0}^{2}) + S(\rho_{f})
\end{equation}
where $\rho_{f}\defn |f(\op{X}_{1},\op{X}_{2})|$. Additionally, the von Neumann entropy $S_{\textrm{vN.}}(\rho_{\hat{\s}})$ can be formally expressed as 
\begin{equation}
S_{\textrm{vN.}}(\rho_{\hat{\s}}) \simeq S_{\textrm{gen.},\mc{B}_{1},\mc{B}_{2}}(\hat{\s}) + S(\rho_{f}) +C
\end{equation}
where $\mc{B}_{1}$ is the bifurcation surface of the black hole horizon and $\mc{B}_{2}$ is the bifurcation surface of the cosmological horizon,
\begin{equation}
S_{\textrm{gen.},\mc{B}_{1},\mc{B}_{2}}(\hat{\s}) \defn \hat{\s}\bigg(\frac{\op{A}_{\mc{B}_{1}}}{4G_{\textrm{N}}}\bigg)+\hat{\s}\bigg(\frac{\op{A}_{\mc{B}_{2}}}{4G_{\textrm{N}}}\bigg)+S_{\textrm{vN.}}(\s\vert_{\mc{R}})
\end{equation}
where $\op{A}_{\mc{B}_{1}}$ and $\op{A}_{\mc{B}_{2}}$ are the area of the black hole and cosmological horizon bifurcation surfaces up to second-order and are defined by \cref{eq:QVT,eq:ABdef}. $S_{\textrm{vN.}}(\s\vert_{\mc{R}})$ is the von Neumann entropy of $\s$ in $\mc{R}$. 

\section{Black Holes in Asymptotically Anti-de Sitter Spacetime}
\label{sec_AdS}

\subsection{Kerr-AdS}
Asymptotically AdS Kerr black holes are under more control than their asymptotically flat counterparts because they are essentially ``in a box.'' This allows large black holes to be in thermal equilibrium with the exterior region because Hawking radiation emitted from the black hole reflects off the asymptotic boundary and re-enters the black hole in finite time. For black holes rotating slower than the ``Hawking-Reall bound'' \cite{1999PhRvD..61b4014H} ($\Omega_{\textrm{H}} < 1$), the Killing vector field generating the horizon, $\xi^a = t^a + \Omega_{\textrm{H}} \psi^a$, is timelike throughout the exterior region and there exists a Hartle-Hawking state. Black holes above the Hawking-Reall bound cannot be in thermal equilibrium and we instead must consider an Unruh-like state.\footnote{We note that such states are also relevant for ``small black holes,'' which are black hole geometries that are never the dominant solution in the gravitational path integral \cite{hawking1983thermodynamics}. These black holes are hard to probe from the gravitational path integral or AdS/CFT, but are natural from the bulk algebraic perspective.} Our analysis is brief due to the similarity with \cref{sec_flat} and we will refrain from redefining quantities that were previously defined there.

We take $(\mc{M}_{\textrm{R}},g)$ to be the physically relevant region of the spacetime corresponding to an asymptotically AdS Kerr black hole, i.e. $\mc{M}_R$ is the union of the right exterior and future interior. We have not displayed the Penrose diagram from Kerr-AdS, but note that it looks like figure \ref{fig:AdS_tikz} in the exterior region $\mc{R}$ and figure \ref{fig:Kerr_tikz} in the interior region $\mc{F}$. Namely, it has a Cauchy horizon, past which the initial data cannot be uniquely propagated. $\mc{M}_{\textrm{R}}$ is 
globally hyperbolic (after imposing boundary conditions), so the construction of the classical phase space and $\ast$-algebra, $\mathscr{A}$, from sections \cref{subsec:scfieldth} and \cref{subsec:AlgQFTCS} apply with $V = 0$ (massive fields can be accounted for following \cref{app1:timequant}). The isometry group of $(\mc{R},g)$ is again $\mathbb{R}\times \textrm{U}(1)$.

For massless fields, the bulk observable can be specified on the past horizon
\begin{align}
 \op{\phi}(f) = \op{\Pi}(s),
\end{align}
where $s=[Ef]|_{\mathcal{H}^-}$ and the bulk algebra is isomorphic to the horizon algebra
\begin{align}
\label{eq:ads_iso}
 \ms{A} \cong \ms{A}_{\mc{H}^-}.
\end{align}
By assumption \ref{assump4}, there exists a (Hadamard) stationary state $\s_{0}$ in $\mc{M}_{\rm R}$. This state has vacuum initial data $\s_{0}^{\mc{H}}$ on $\mc{H}^{-}$ defined in \eqref{eq:vacH-} with Dirichlet boundary conditions on the timelike boundary\footnote{We expect that Neumann and more general Robin boundary conditions will pose no obstructions.}
\begin{align}
 \omega_{0} \defn \omega_0^{\mathcal{H}^-} \ .
\end{align}
Restricting the algebra to $\mc{H}_R^-$ and taking the double commutant leads to Type III$_1$ von Neumann algebra $\mf{A}(\mc{H}_R^-,\omega_0)$. \eqref{eq:ads_iso} implies that $\mf{A}(\mc{H}_R^-,\omega_0)$ is isomorphic to the von Neumann algebra for quantum field operators in $\mc{R}$, $\mf{A}(\mc{R},\omega_0)$. The commutant is the Type III$_1$ algebra $\mf{A}(\mc{H}_L^-,\s_0)$. 

Integrating \eqref{eq:BDYCharge}, for the two Killing vectors, on a spatial slice stretching from $i_L^+$ to the right asymptotic boundary yields the constraint relations 
\begin{align}
 \op{F}_{\psi} =-\delta^2 \op{J}_{i^0}+ \delta^2 \op{J}_+, \quad \op{F}_t = \delta^2 \op{M}_{i^0}-\frac{\delta^2 \op{\mathcal{Q}}_+}{4G_{\textrm{N}}\beta} - \Omega_{\textrm{H}} \delta^2 \op{J}_+.
\end{align}
We extend the algebra to include the charges in $\mc{R}$, namely $\delta^2 \op{J}_{i^0}\defn-\op{X}_{\psi}$ and $\delta^2 \op{M}_{i^0}\defn \op{X}_t$, with their quantization identical to \cref{sec_flat}. The other charges on the horizon are determined as
\begin{align}
 \frac{\delta^2 \op{\mathcal{Q}}_-}{4G_{\textrm{N}}\beta} = \delta^2 \op{M}_{i^0} - \Omega_{\textrm{H}} \delta^2 \op{J}_{i^0},\quad \delta^2 \op{J}_{-}= \delta^2 \op{J}_{i^0}.
\end{align}
The observables that commute with $\op{C}_{\psi} \defn -\delta^2 \op{J}_+ = \op{X}_{\psi} - \op{F}_{\psi}$ and $\op{C}_{t} \defn -\frac{\delta^2 \op{\mathcal{Q}}_+}{4G_{\textrm{N}}\beta} - \Omega_{\textrm{H}} \delta^2 \op{J}_+ = \op{X}_t - \op{F}_t$ are the charges $\op{X}_{\psi}$, $\op{X}_{t}$ and the dressed quantum fields 
\begin{equation}
\label{eq:dressKerrAdS}
\op{\phi}(f;\op{t},\op{\psi})\defn e^{i\op{F}_{t} \op{t}}e^{i\op{F}_{\psi}\op{\psi}}\op{\phi}(f)e^{-i\op{F}_{\psi}\op{\psi}}e^{-i\op{F}_{t}\op{t}},
\end{equation}
which together generate the algebra
\begin{equation}
\vNext(\mc{R},\s_{0})\cong \vNext(\mc{H}^{-}_{\textrm{R}},\s_{0})\defn \{\op{\phi}(f;\op{t},\op{\psi}),\op{X}_{\psi}, \op{X}_{t}\}^{\prime\prime}.
\end{equation}
This is once again a Type II$_{\infty}$ algebra. 

To consider entropies, we again consider the subalgebra of observables $\bar{{\mf{A}}}_{\textrm{ext.}}(\mc{R},\s_{0})$ dressed to the charge $\op{X}_{\xi}$. This algebra is isomorphic to $\bar{\mf{A}}_{\textrm{ext.}}(\mc{H}_{\textrm{R}}^{-},\s_{0})$ defined in \cref{eq:Ubar} and the trace, for any $a\in \bar{{\mf{A}}}_{\textrm{ext.}}(\mc{R},\s_{0})$, is
\begin{equation}
\label{eq:Tracekerrads}
{\textrm{Tr}(a)=\beta \int_{\bb{R}}dX_{\xi}~e^{\beta X_{\xi}}\braket{\s_{0}^{\mc{H}},X_{\xi}| a|\s_{0}^{\mc{H}},X_{\xi}}}
\end{equation}
which is identical to the trace in asymptotically flat Kerr \eqref{eq:Tracekerr}, just without the contribution from $\scri^-$.
The entropy of classical-quantum states $\ket{\hat{\s}}$ on $\Fock \otimes \Hilb_{\textrm{X}_{\xi}}$ of the form of \cref{eq:classqstatekerr} is given by
\begin{align}
 S_{\textrm{vN.}}(\rho_{\hat{\s}}) \simeq \hat{\s}(\beta \op{X}_{\xi})-S_{\textrm{rel.}}(\s|\s_0) + S(\rho_f) + \log \beta.
\end{align}
Following the heuristic arguments of \cref{subsec:dens_ent}, this is equivalent to the generalized entropy
\begin{align}
 S_{\textrm{vN.}}(\rho_{\hat{\s}})
\simeq S_{\textrm{gen.},\mc{B}}(\hat{\s})+S(\rho_{f})+C.
\end{align}
\subsection{AdS/CFT}
\label{subsec:adscft}
Throughout this work, we have focused on quantum physics in general spacetimes with black holes without appealing to any ultraviolet completions such as string theory, emphasizing the universality of the crossed product construction to spacetimes with Killing horizons. Nevertheless, it is important to return to the source of inspiration of this research direction \cite{2021arXiv211005497L,2021arXiv211212156L,2022JHEP...10..008W} where the algebra associated to the large-N limit of conformally invariant gauge theories was analyzed. The duality between these theories and string theory in AdS is our current best understanding of non-perturbative quantum gravity \cite{1999IJTP...38.1113M,1998AdTMP...2..253W,1998PhLB..428..105G}.

In the AdS/CFT duality, the bulk quantum gravity theory is equivalent to a conformal field theory that can be thought of as being defined on the asymptotic boundary of the asymptotically AdS space. The number of degrees of freedom in the boundary theory, $N$, is related to Newton's constant in the bulk as $G_{\textrm{N}} \sim N^{-2}$. The bulk theory is primarily well-understood in the semiclassical limit, where $G_{\textrm{N}}$ is small, corresponding to the large-$N$ limit of the CFT.

The eternal black hole in AdS (and corresponding Hartle-Hawking state for quantum field perturbations on this background) is described in the boundary theory by the thermofield double (TFD) state on two copies of the holographic CFT~\cite{Maldacena:2001kr}
\begin{align}
 \ket{\textrm{TFD}} \defn \frac{1}{\sqrt{\mathcal{Z}(\beta)}}\sum_i e^{-\beta E_i/2}\ket{E_i}_R \ket{E_i}_L,
\end{align}
where the sum is over energy eigenstates of the two copies, and $\mathcal{Z}(\beta)$ is the thermal partition function.
At finite values of $N$, each copy of the CFT acts on its own Hilbert space and the bounded CFT operators of each copy generate two von Neumann algebras of Type I that are each other's commutants. However, when one considers the $N \to \infty$ limit at a temperature above the Hawking-Page transition, there cease to be well-defined Hilbert spaces for the individual CFTs and the algebras of operators localized on one side of the thermofield double that survive the $N \to \infty$ limit are of Type III$_1$~\cite{2021arXiv211005497L, 2021arXiv211212156L}. In $\mathcal{N} = 4$ super Yang-Mills, these are the ``single-trace'' operators, which behave as generalized free fields (GFFs). We will denote the resulting right/left GFF algebra by $\mf{A}_{R/L}.$ These emergent Type III$_1$ algebras are holographically dual to the algebras of quantum field theory operators in the left and right exteriors of the black hole. 

Our goal is to understand the relevance of the crossed product and resulting Type II$_{\infty}$ algebras from the boundary perspective. 
To do so, we consider the Hamiltonian of either copy of the CFT, which does not survive the $N \to \infty$ limit because its thermal fluctuations are $O(N^2)$. However, the existence of a well-defined Hamiltonian at finite $N$ suggests that a limit of the CFT Hamiltonian (minus its expectation value) might still be defined on a subspace of states and included in the algebra of operators as $N \to \infty$ either by working perturbatively in a $1/N$ expansion~\cite{2022JHEP...10..008W} or by considering a microcanonical thermal state \cite{2022arXiv220910454C}. We consider the latter approach where the microcanonical TFD is defined at finite $N$ as
\begin{align}
\ket{\textrm{TFD}}_f \defn \frac{1}{\sqrt{\mathcal{Z}(\beta)}}\sum_i e^{-\beta E_i/2}f(E_i-E_0)\ket{E_i}_R \ket{E_i}_L, \quad f \in \mathcal{S},
\end{align}
where $\mathcal{S}$ is the Schwartz space of functions decaying faster than any power at infinity. This is a superposition of thermofield double states with different relative time shifts between the copies, which can be seen by writing $f(E_i-E_0)$ as the Fourier transform of some $F(t)$.
While expectation values of operators localized on one side of $\ket{\textrm{TFD}}_f$ equal that of $\ket{\textrm{TFD}}$ at leading order in $N$, the two-sided correlators are different \cite{2022arXiv220910454C}. The benefit of $\ket{\textrm{TFD}}_f$ is that the fluctuations of the subtracted (boundary) Hamiltonians, $\op{h}_{\textrm{R/L}} = \op{H}_{\textrm{R/L}}-E_0\op{1}$, are finite in the large-$N$ limit since all moments of $f(E)$ exist.

Unlike for $\ket{\textrm{TFD}}$, the one-sided modular Hamiltonian does not coincide with the physical Hamiltonian at finite $N$. This may be seen by the expression for the reduced density matrices,
\begin{align}
 \Tr_{\textrm{L/R}} \ket{\textrm{TFD}}\bra{\textrm{TFD}}_f \defn \frac{1}{\mathcal{Z}(\beta)} \sum_i |{f}(E_i-E_0)|^2 e^{-\beta E_i}\ket{E_i}\bra{E_i}_{\textrm{R/L}} ,
\end{align}
from which one may read off the total modular Hamiltonian\footnote{Recall that the modular Hamiltonian is $-\log \rho_R + \log \rho_L$ since we have a Type I algebra in this case.} as
\be \label{eq:finNModHamMicro}
 \op{H}_{\textrm{TFD}_f} = \beta\le(\op{h}_{\textrm{R}} - \op{h}_{\textrm{L}}\ri) - \log\le|f(\op{h}_{\textrm{R}})\ri|^2 + \log\le|f(\op{h}_{\textrm{L}})\ri|^2 \ .
\ee
Nevertheless, because the one-sided correlation functions agree as $N$ is taken to infinity, the large $N$ limit of the microcanonical TFD will be KMS with respect to time translations of the generalized free fields. Time translations are the modular automorphisms of the right GFFs in the large $N$ limit of the canonical TFD state. We denote the corresponding modular Hamiltonian by $\op{H}_{\textrm{TFD}}.$ Since the right GFF algebra is of Type III$_1$ its modular automorphisms are always outer, and thus no generator of time translations on the right GFF algebra can exist within the algebra itself. Thus, $\op{H}_{\textrm{R}}$ (as a generator of time translations on only the right GFFs) must be treated as an additional quantum mechanical degree of freedom. 

The limit of the right CFT Hamiltonian should be identified with an operator that acts as $\op{H}_{\textrm{TFD}}$ on $\mf{A}_{\textrm{R}}$ while acting trivially on $\mf{A}_L.$ Similarly, the limit of the left CFT Hamiltonian, $\op{h}_{\textrm{L}},$ must act as $-\op{H}_{\textrm{TFD}}$ on $\mf{A}_L$ while acting trivially on $\mf{A}_R.$ However, $\op{H}_{\textrm{TFD}}$ acts nontrivially on both algebras, so one cannot simply identify $\op{h}_{\textrm{R}}$ or $\op{h}_{\textrm{L}}$ with $\op{H}_{\textrm{TFD}}.$ Analogously to the case of the right CFT, $\op{h}_{\textrm{L}}$ is not affiliated to $\mf{A}_L$ and must be treated as an additional quantum mechanical degree of freedom in the $N \to \infty$ limit.

Including these additional degrees of freedom, the extended Hilbert space in the large-$N$ limit is $\sH_{\rm ext.} = \sH_0 \otimes L^2\le(\mathbb{R}_{\op{h}_{\textrm{L}}}\ri) \otimes L^2\le(\mathbb{R}_{\op{h}_{\textrm{R}}}\ri),$ where $\sH_0$ is the GNS Hilbert space of small excitations around the TFD. At this level the algebras of operators in the $N \to \infty$ limit of the two CFTs are
\be \label{eq:AdSnaiveAlgs}
 \mf{A}_{R,0} = \mf{A}_R \otimes \mathbf{1} \otimes \sB\le(L^2\le(\mathbb{R}_{\op{h}_{\textrm{R}}}\ri)\ri),~ \mf{A}_{L,0} = \mf{A}_L \otimes \sB\le(L^2\le(\mathbb{R}_{\op{h}_{\textrm{L}}}\ri)\ri) \otimes \mathbf{1} \ .
\ee
As written, $\op{h}_{\textrm{R}}$ and $\op{h}_{\textrm{L}}$ commute with the single-trace operators, so they do not implement the time translations that they are supposed to. We must recall that $\op{h}_{\textrm{L}/\textrm{R}}$ were introduced to act as $\mp \op{H}_{\textrm{TFD}}$ on the single-trace operators in their respective theories while commuting with operators in the other theory. This motivates the relation
\be \label{eq:AdSHLHRrelation}
 \beta(\op{h}_{\textrm{R}} - \op{h}_{\textrm{L}})= \op{H}_{TFD} \ ,
\ee
as this effectively sets ``$\beta \op{h}_{\textrm{R}} = \op{H}_{\textrm{TFD}}$'' for right operators, since $\op{h}_{\textrm{L}}$ commutes with right single-trace operators. Similarly, since $\op{h}_{\textrm{R}}$ commutes with left operators, this sets ``$\beta \op{h}_{\textrm{L}} = -\op{H}_{\textrm{TFD}}$'' for left operators.~\eqref{eq:AdSHLHRrelation} can be viewed as a constraint that tells us that the limits of left and right CFT Hamiltonians about the TFD state cannot truly be treated as independent degrees of freedom if they are to implement time-translations on their respective single-trace operators. 

By \textit{defining} the large $N$ limits, $\op{h}_{\textrm{L}/\textrm{R}}$, of the subtracted left/right CFT Hamiltonians to (i) implement time-translations on their respective single-trace algebras, (ii) commute with the single-trace operators of the other theory, we are forced to impose the condition~\eqref{eq:AdSHLHRrelation}. We must then restrict to the subalgebras of operators within $\mf{A}_{L/R,0}$ built from operators obeying~\eqref{eq:AdSHLHRrelation}. These are the subalgebras that commute with $\op{h}_{\textrm{R}} - \op{h}_{\textrm{L}} - \op{H}_{\textrm{TFD}},$ and they are generated as
\be \label{eq:AdSSymmDressedAlgs}
\begin{aligned}
 \mf{A}_R^{\rm (ext)} &= \le\{e^{-i t_R \op{H}_{\textrm{TFD}}} a_R e^{i t_R \op{H}_{\textrm{TFD}}} \otimes \mathbf{1}_{L^2(\mathbb{R}_{\op{h}_{\textrm{L}}})},~ \mathbf{1}_{\sH_0} \otimes \mathbf{1}_{L^2(\mathbb{R}_{\op{h}_{\textrm{L}}})} \otimes \op{h}_{\textrm{R}}\ri\}'',~ a_R \in \mf{A}_R \\
 \mf{A}_L^{\rm (ext)} &= \le\{e^{i t_L \op{H}_{\textrm{TFD}}} a_L e^{-i t_L \op{H}_{\textrm{TFD}}} \otimes \mathbf{1}_{L^2(\mathbb{R}_{\op{h}_{\textrm{R}}})},~ \mathbf{1}_{\sH_0} \otimes \op{h}_{\textrm{L}} \otimes \mathbf{1}_{L^2(\mathbb{R}_{\op{h}_{\textrm{R}}})}\ri\}'',~ a_L \in \mf{A}_L \ ,
\end{aligned}
\ee
where $t_{L/R}$ are the canonical conjugates to $\op{h}_{\textrm{L}/\textrm{R}}$.\footnote{Note that, if $t_{L/R}$ was a number rather than an operator, $\{e^{\mp i t_{L/R} \op{H}_{\textrm{TFD}}} a_{L/R} e^{\pm i t_{L/R} \op{H}_{\textrm{TFD}}}\}$ would just be an operator in $\mf{A}_{L/R}.$ This is the story that one expects for the CFT at finite $N,$ where the Hamiltonian is built from the CFT operators and is not an independent quantum mechanical degree of freedom with its own conjugate variable.} Reconciling the Type III$_1$ algebra obtained as the $N \to \infty$ limit of the CFT algebra with the fact that the CFT at finite $N$ is its own self-contained dynamical system forces the introduction of ``$\op{h}_{\textrm{R}}$'' as an independent quantum mechanical degree of freedom to reflect this finite $N$ fact in the large $N$ theory. We do not have a rigorous argument for the existence of the conjugates to $\op{h}_{\textrm{L}}$ and $\op{h}_{\textrm{R}}$ in the $N \to \infty$ limit from the CFT perspective; however, the fact that any state obtained by applying finite products of single-trace operators to the microcanonical TFD will have finite variance of $\op{h}_{\textrm{L}/\textrm{R}}$ in the $N \to \infty$ limit (see~\cite{Chandrasekaran:2022cip}) suggests that the allowed wave functions for $\op{h}_{\textrm{L}/\textrm{R}}$ should lie in $\mathcal{S}.$ Completion to a Hilbert space {using the standard inner product} will then give $L^2(\mathbb{R})$, a space on which one can define a canonical conjugate and obtain a representation of the canonical commutation relations. 

While the relation~\eqref{eq:AdSHLHRrelation} is puzzling from the finite $N$ CFT perspective, it does have a natural interpretation from the bulk. Recall that the formulation of the theory on the past horizon of the black hole required the introduction of certain zero modes of the second-order perturbed metric $\delta^2 \op{M}_{\textrm{L}/\textrm{R}}$ in order to properly analyze the integrated version of Raychaudhuri's equation at second-order. By identifying these perturbed ADM masses with the corresponding subtracted CFT energies in the large $N$ limit, i.e., $\delta^2 \op{M}_{\textrm{L}/\textrm{R}} = \op{h}_{\textrm{L}/\textrm{R}},$ one can understand the origin of~\eqref{eq:AdSHLHRrelation} from the bulk. Upon quantization, one obtains the extended bulk Hilbert space $\sH_{\rm ext.} = \sH_0 \otimes L^2\le(\mathbb{R}_{ \textrm{M}_\textrm{L}}\ri) \otimes L^2\le(\mathbb{R}_{ \textrm{M}_\textrm{R}}\ri),$ where now $\sH_0$ is interpreted as the Hilbert space of quantum field fluctuations on a fixed eternal AdS black hole geometry, while $\delta^2 \op{M}_{\textrm{L}/\textrm{R}}$ are the second-order perturbed ADM masses at the left and right boundaries, respectively. 

This extended Hilbert space is not the end of the story for dynamical gravity. The Hamiltonian formulation of general relativity is a parameterized theory whose constraints we have not solved in order to deparametrize the theory (see, e.g., \cite{Regge:1974zd} or Appendix E of \cite{Wald:1984rg}). We must therefore necessarily work with redundant degrees of freedom (whose quantization is included in the extended Hilbert space) and impose the constraints after the fact. For the bulk theory being formulated on the past horizon of a two-sided asymptotically AdS spacetime (i.e., a theory that will contain the eternal AdS black hole amongst its solutions), one of the constraints is precisely
\be \label{eq:AdSBulkHLHRrel}
 \delta^2 \op{M}_{\textrm{R}} - \delta^2 \op{M}_{\textrm{L}} = \op{F}_{\xi} \Rightarrow \beta(\op{h}_{\textrm{R}} - \op{h}_{\textrm{L}} )= \op{H}_{\textrm{TFD}} ,
\ee
i.e., the constraint imposes that the (total) ADM Hamiltonian implements Killing time translation on the bulk quantum fields everywhere. Identifying the ADM Hamiltonians with the large $N$ limits of the CFT Hamiltonians, this is precisely~\eqref{eq:AdSHLHRrelation} and we see that, from the bulk perspective, it arises from the constraints of the Hamiltonian formulation of general relativity.

\section{Discussion}
\label{sec_discussion}
In this paper, we have demonstrated how the algebras of observables and von Neumann entropies in perturbative quantum gravity can be deduced from initial data surfaces that are Killing horizons. We conclude by outlining some direct generalizations of the results of this paper as well as several comments that may deserve further investigation. 

\subsection{Higher dimensions and higher derivatives}

With the objective of optimal clarity, we have chosen to focus on four-dimensional spacetimes in Einstein gravity throughout this paper. This has been a matter of convenience more than a matter of necessity, and the derivations will go through immediately in higher spacetime dimensions under a straightforward generalization of our assumptions.\footnote{In higher dimensions, the vacuum initial data on $\mc{H}^{-}$ which yields a stationary state in $\mc{R}$ is a Gaussian state with vanishing one-point function and two-point function given by \cref{eq:vacH-} with $\delta_{\bb{S}^{2}}$ replaced by a delta-function $\delta_{\mc{B}}$ on the bifurcation surface \cite{Sanders_2023}.} In particular, in higher dimensions, the space of black hole solutions is much richer and the horizon topology is no longer fixed to be a sphere. Nevertheless, \cref{thm:horizontheorem} as well and our results in \cref{sec_flat,sec_dS,sec_AdS} directly generalize where now the black hole entropy is the area of the bifurcation surface. In lower spacetime dimensions, gravity does not have propagating degrees of freedom, so the gravitational constraints are more subtle and require separate consideration. For recent progress in two-dimensional Jackiw-Teitelboim gravity, see, e.g., \cite{2023arXiv230107257P,2023JHEP...06..067K}.

Higher derivative gravity also does not present any particular challenge in our approach.\footnote{See also \cite{2023arXiv230700241A} for a recent discussion.} As shown by Hollands, Wald and Zhang, for any diffeomorphism invariant Lagrangian any spacetime with a Killing horizon, one can obtain a second-order perturbed charge $\delta^{2}\mathcal{Q}^{\textrm{HWZ}}_{U}$ associated to the horizon Killing field for any constant $U$ cut of the horizon \cite{Hollands:2024vbe,Visser:2024pwz}. In particular, the Hollands-Wald-Zhang charge evaluated at the bifurcation surface $U=0$ is equivalent to the Wald entropy \cite{1993PhRvD..48.3427W}. Dressing quantum field operators to these generalized charges works in the exact same way because the gravitational constraint now relates the difference in these charges to the flux through the Killing horizon
\begin{align}
 \label{eq:HWZ}
 \delta^2\op{\mathcal{Q}}^{\textrm{HWZ}}_{+} - \delta^2\op{\mathcal{Q}}^{\textrm{HWZ}}_{-} = \beta \op{F}_{\xi} \quad \quad  \textrm{ (on $\mc{H}^{-}$)}
\end{align}
where $\delta^{2}\op{\mc{Q}}^{\textrm{HWZ}}_{\pm}$ are the limits of the perturbed Hollands-Wald-Zhang charge at asymptotically early and late times. While assumptions \ref{assump1}-\ref{assump3} can be directly generalized, the existence of a ``ground state'' $\omega_{0}$ in assumption \ref{assump4} will only be valid for alternative theories of gravity with positive ``canonical energy'' \cite{Hollands:2012sf}. For such theories, the right hand side of eq.~\ref{eq:HWZ} is proportional to the modular Hamiltonian $\s_{0}$ on $\mc{H}^{-}$.
In this case, \cref{thm:horizontheorem} as well as our results in sectons \ref{sec_flat}~--~\ref{sec_AdS} directly generalize with the appropriate replacement of the area with the Wald entropy in the generalized entropy. 

\subsection{Interactions}
\label{subsec:interactions}

{In this paper, we have considered the quantum theory of a linear scalar field on general curved spacetime satisfying assumptions \ref{assump1}---\ref{assump4}. These assumptions can be straightforwardly generalized to electromagnetic fields and linearized gravitational perturbations (see \cref{sec:grav_EM_class}) and so our results directly apply to these cases as well.}

{Our arguments crucially rely on the factorization of the algebra into an algebra of observables ``localized'' on the past Killing horizon and an algebra ``at infinity'' as well as the existence of a stationary state $\s_{0}$ in $\mc{M}_{\textrm{R}}$. The definition of a stationary state as one that is annihilated by $\op{F}_{\xi}$ is well-defined in the context of an interacting theory. However, for a general interacting theory, an algebra of observables cannot be strictly defined on a Cauchy surface. Nevertheless, one can formulate an initial value problem for the algebra on a suitably ``thickened'' Cauchy surface.\footnote{The physical expectation that the algebra of observables on a `thickened'' Cauchy surface $\Sigma$ is isomorphic to the algebra in the domain of dependence of $\Sigma$ is known as the ``time-slice axiom'' \cite{Haag:1963dh,Haag1962THEPO} which has been proven in flat spacetimes for certain super-renormalizable theories in $2$-dimensions \cite{1972AnPhy..70..412S,Glimm:1968kh} and perturbatively for interacting scalar fields on arbitrary curved, globally hyperbolic spacetimes \cite{Chilian:2008ye}.} Therefore, it would seem that the appropriate generalization of assumption \ref{assump3} is that the initial data can be independently specified on an algebra localized in a neighborhood of $\mc{H}^{-}$. To our knowledge, neither the existence and uniqueness of a stationary state in $\mc{M}_{\textrm{R}}$ nor the appropriate factorization properties of the algebra of observables have been investigated for the case of interacting theories. Our results indicate that progress in either direction would pave the way toward obtaining a generalized entropy for the algebra of observables of interacting quantum fields. }

\subsection{Generalized second law}

The generalized second law (GSL) of thermodynamics asserts that the generalized entropy of the exterior region of a black hole is non-decreasing in time \cite{bekenstein1974generalized}. A version of the GSL was proven by Wall in a setting where von Neumann entropies themselves were not well-defined by appealing to the monotonicity of relative entropy \cite{2012PhRvD..85j4049W}. In \cite{2022arXiv220910454C}, CPW built on this work in the case of asymptotically AdS black holes and by considering the properties of trace-preserving algebra inclusions. In this setting, the algebras are semi-finite and so the GSL could be proven using finite von Neumann entropies, though the setting was limited in the sense that the entropies were evaluated for discrete times; first, the generalized entropy of the bifurcation surface was proven to be smaller than that of the infinite time surface. It is this ``two-step'' GSL that we will discuss in the following. CPW also discussed a more subtle discrete GSL involving the free product algebras, though we will not comment on this approach.

The central property used is that the von Neumann entropy is monotonically increasing under trace-preserving algebra inclusions. Namely, consider a subalgebra $\mathscr{B} \subset \sA$ such that both algebras are equipped with traces that agree on the elements where they are both defined, i.e.
\begin{align}
 \Tr_{\sA} (b) = \Tr_{\mathscr{B}}(b) , ~~ \forall ~b \in \mathscr{B}.
\end{align}
Then, for all states in the Hilbert space, the von Neumann entropy of the state on $\mathscr{B}$ is greater than or equal to that on $\sA$ \cite{2022arXiv220203357L}. For asymptotically AdS black holes, the algebra inclusion was particularly simple in the two-step process. The subalgebra, $\mathscr{B}$, was taken to be the algebra at infinity which consists only of bounded functions of the conserved charges, namely the ADM Hamiltonian without any quantum field operators.

Let us first consider Schwarzschild-de Sitter. At asymptotically late times, the algebra of observables again only consists of the conserved charges. An identical discussion to the asymptotically AdS case goes through, so the GSL is satisfied. Importantly, by asymptotically late times, we mean first taking $G_{\textrm{N}} \rightarrow 0$ and then $t \rightarrow \infty$ such that the black hole does not evaporate and our perturbative analysis holds. If instead, we considered finite $G_{\textrm{N}}$ or took $t \sim G_{\textrm{N}}^{-1} \rightarrow \infty$ (where our perturbative analysis fails), the black hole would evaporate because, unlike AdS, Schwarzschild-de Sitter is not stable. Thus, at very late times (scaling with $G_{\textrm{N}}^{-1}$), all radiation will cross the cosmological horizon. The resulting state will approach the Bunch-Davies state in the static patch, which we have discussed is a maximum entropy state. As it is a maximum entropy state, the GSL will be satisfied.

We next consider asymptotically flat spacetimes where we now have a choice for the late-time subalgebra. We may take only the conserved charges at $i^+$ and the above arguments would be unchanged. However, it is more interesting to consider the intermediate of the algebra at $\scri^+$. The algebra, $\mf{A}_{\rm ext.}(\scri^+,\s_0) \subset \mf{A}_{\rm ext.}(\mc{R},\s_0)$, includes the quantum field theory operators localized at $\scri^+$ and dressed via the ADM Hamiltonian. 

We need to determine the algebra type of $\mf{A}_{\rm ext.}(\scri^+,\s_0)$. The ADM Hamiltonian generates an outer automorphism of quantum field theory operators, but it does not coincide with the modular Hamiltonian of the state because the state is not thermal at early retarded times. Note that if we had taken a finite time cut of the horizon, considering the quantum field theory algebra in the associated causal diamond (as in Wall's proof of the generalized second law \cite{2012PhRvD..85j4049W}), the ADM Hamiltonian would not generate an automorphism of the algebra, but instead generate all of $\mf{A}_{\rm ext.}(\mc{R},\s_0)$. This is an obstruction to proving a continuous version of the generalized second law in the present context.

The algebra is Type II if there exists a dense set of dressed operators in this crossed product algebra with finite trace. It is a logical possibility that no finite trace operators remain when taking the limit to $\scri^+$. In order to evaluate the trace, we must propagate back the operator to the initial value surface $\mc{H}_- \cup \scri^-$. This map is highly non-trivial due to transmission and reflection off of the potential {barrier}. This map is non-trivial even for observables at late retarded times. Back-propagating the corresponding ``mode-function'' for these observables will yield (highly blue-shifted) data in the asymptotic past with support on $\mc{H}_{\textrm{R}}^{-}$ as well as at late advanced times on $\scri^{-}$. To verify the validity of the GSL in this case, one would have to evaluate the trace \eqref{eq:Tracekerr} on these back-propagated observables.

\subsection{Nariai limit and discrete symmetries}
\label{sec:Nariai}

\begin{figure}
 \centering
 \includegraphics[width = \textwidth]{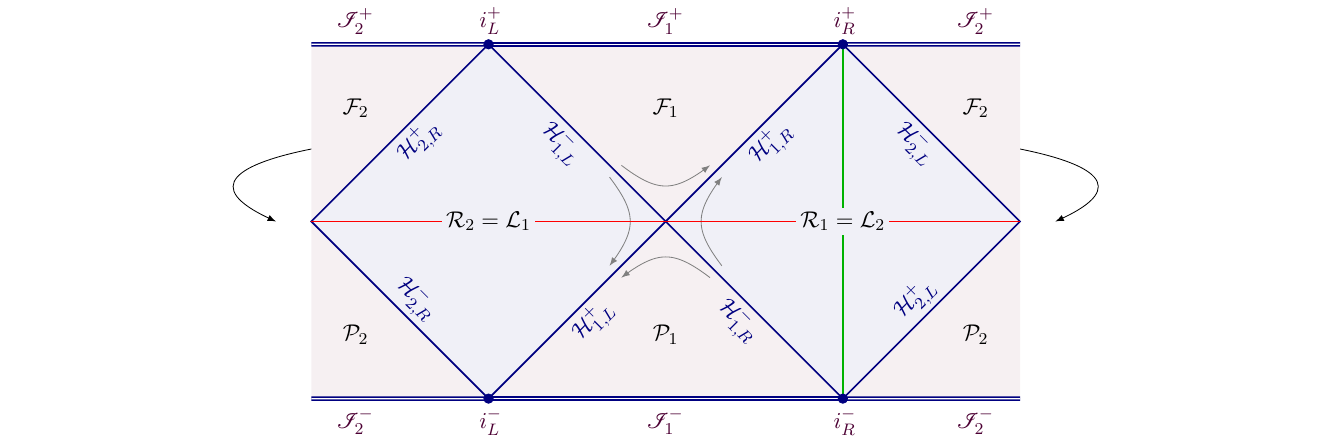}
 \caption{The Nariai solution has the Penrose diagram of dS$_2$ with every point an $\mathbb{S}_2$ of constant size. The left and right edges are identified. This is the same Penrose diagram as SdS except the singularities have been transformed to asymptotic spacelike boundaries.}
 \label{fig:nariai}
\end{figure}

There is a special limit of Schwarzschild-de Sitter where the surface gravities of the black hole and cosmological horizons become equal. These are the largest black holes in de Sitter. In this \textit{Nariai} limit, the curvature singularity disappears and the metric becomes that of dS$_2 \times \bb{S}_2$ \cite{nariai1950some,ginsparg1983semiclassical} with Penrose diagram shown in Figure \ref{fig:nariai}. The isometries of the Nariai geometry are $\textrm{SO}(1,2) \times \textrm{SO}(3)$. The isometries of $(\mc{R},g)$ are $\mathbb{R} \times \mathbb{Z}_2\times \textrm{SO}(3)$, where the $\mathbb{Z}_2$ symmetry is a spatial refection exchanging the two horizons. This facilitates the existence of a Hartle-Hawking state, which is the product of two (one Lorentzian and one Euclidean) two-dimensional Bunch-Davies state. Gauging the $\textrm{SO}(3)$ requires equipping the observer with an orthonormal frame, as discussed for the $\textrm{SO}(3)$ of empty de Sitter space by CLPW. 

To gauge discrete symmetries, which has been argued to be necessary in quantum gravity \cite{2018arXiv181005338H, 2023arXiv231109978H,2023arXiv230400589S}, it seems that one must endow the observer with a preferred orientation. 
{This amounts to adding an additional qubit to the clock and extending the action of the discrete symmetry operator, $\Theta$, such that it flips the spin of this qubit. All operators in the theory must commute with $\tilde{\Theta} \defn \Theta \otimes \sigma_x$. The operators in the crossed product algebra \cref{eq:xproductalg} can be made $\tilde{\Theta}$ invariant as
\begin{equation}
 \vNext = \{ \op{\Delta}_{\omega}^{i\op{t}}a \op{\Delta}_{\omega}^{-i\op{t}} \otimes \ket{\uparrow}\bra{\uparrow}+ \Theta\op{\Delta}_{\omega}^{i\op{t}}a \op{\Delta}_{\omega}^{-i\op{t}}\Theta\otimes \ket{\downarrow}\bra{\downarrow}, \op{X}; a\in \mf{A}, t\in \mathbb{R}\}.''
\end{equation}
The trace is
\begin{align} 
 \Tr(\hat{a}) = \frac{\beta}{2}\int_{\bb{R}}dX~e^{\beta X}\left(\bra{\s,X,\uparrow}\hat{a}\ket{\s,X,\uparrow} +\bra{\s,X,\downarrow}\hat{a}\ket{\s,X,\downarrow} \right),
\end{align}
so the discrete gauging does not change the type of the algebra. A particularly interesting, and ubiquitous, discrete symmetry is $\mc{CRT}$, which is anti-unitary. Gauging the anti-unitary symmetry in a closed universe curiously leads to a real Hilbert space \cite{2023arXiv231109978H} and thus a real von Neumann algebra. While this might initially seem disturbing, it has long been understood how to embed standard complex quantum mechanics in a modified quantum theory based on a real Hilbert space, with clear connections to the inclusion of an extra qubit for the observer as above \cite{stueckelberg1960quantum,2013PhRvA..87e2106A}. }

\subsection{Recovering a Type I algebra}

By adding in gravity, the algebra of observables was deformed from a Type III$_1$ algebra to a Type II algebra. In a sense, this was an ``improvement'' because for the Type II algebra, density matrices, traces, and entropies were well-defined. However, this structure still lacked information regarding individual black hole microstates in the Hilbert space of quantum gravity. If this was known, one could consider the algebra of all bounded operators on this Hilbert space, which is a Type I algebra.

There are certain scenarios where Type I algebras may be seen to arise in quantum gravity. The first is the case of AdS/CFT at finite-N. If N is large but finite, there is still a well-defined Hilbert space. The algebra of observables on this Hilbert space includes both the single-trace operators previously discussed and multi-trace operators. Together, these generate a Type I algebra. This is our best understanding of the microstates of quantum gravity and we subsequently have access to a Type I algebra.

There are two instances we may consider where we do not have the luxury of a dual theory to non-perturbatively define quantum gravity, yet still may construct a Type I algebra for the exterior region of a black hole. Neither are as powerful as the Type I algebra of CFT at finite-N because they appear unrelated to the quantum microstates of the black hole, but we hope they may serve as inspiration.

We first note that in this paper, we have considered the Unruh and Hartle-Hawking states, two of three frequently discussed vacuum states for black holes. The other common vacuum state is the so-called Boulware state \cite{boulware1975quantum}. The Boulware state is defined as the ground state with respect to the timelike Killing field $\partial_t$. It is Hadamard in the right exterior $\mc{R}$ but singular across the future horizon.\footnote{One may call this a firewall.} The right exterior is globally hyperbolic and the Boulware state is pure on this region. The algebra of observables constructed via closure in the GNS Hilbert space above the Boulware state is then a Type I von Neumann algebra.

The second way to construct a Type I algebra is to consider an evaporating black hole. In the semiclassical limit, the Hawking radiation continues to arbitrarily late retarded times at $\scri^+$. This build-up of thermal radiation creates a state with an infrared divergence, leading to a Type III$_1$ algebra at $\scri^+$. The commutant of this algebra contains the observables on $\mathcal{H}_R^+\cup \mathcal{H}_L^-$, which is also Type III$_1$ because it has the same IR divergence of Hawking modes in the black hole interior.

One may consider the backreaction of the Hawking radiation, causing the black hole to shrink and eventually entirely evaporate away. This is outside the regime of this paper because this time is $O(G_{\textrm{N}}^{-1})$, leading to $O(1)$ area changes of the horizon. The late-time spacetime approximates Minkowski space. The amount of Hawking radiation reaching $\scri^+$ is now finite, so the algebra of observables has the potential to be Type I$_{\infty}$. Note that even if a Type I algebra is recovered, there is still information loss in this picture because there is a large amount of entanglement between the Hawking modes at $\scri^+$ and $\mathcal{H}_R^+\cup \mathcal{H}_L^-$, causing there to be a mixed state at $\scri^+$. This is not a mysterious form of information loss because $\scri^+$ is not a Cauchy slice of the spacetime, a perspective emphasized in \cite{2017RPPh...80i2002U}. If one insisted on all information about the initial Unruh state reaching $\scri^+$, new physics would have to emerge in the black hole interior. One possibility, proposed by Horowitz and Maldacena \cite{2004JHEP...02..008H}, is for there to be postselection at the singularity, effectively teleporting the information about the interior into the exterior Hawking radiation. Such a projector will only make sense for interior algebras that are Type I and these will purify the state at $\scri^+$.

\subsection{General Diffeomorphisms}

In this paper, we considered perturbative quantum gravitational effects off a fixed background geometry $(\mc{M},g)$. In particular, we constructed the algebra of observables invariant under the isometries of an invariantly defined subregion $\mc{R}$. We restricted our attention to isometries since our algebra and perturbation theory were defined with respect to a fixed background metric. An obvious generalization would be to consider the algebra of observables invariant under all diffeomorphisms that preserve $\mc{R}$. These diffeomorphisms fall into two categories. The first are so-called ``small diffeomorphisms'' which are degeneracies of the symplectic form. As in the case of de Sitter spacetime, these diffeomorphisms have a trivial action on phase space and it seems that, as in the case of de Sitter, to gauge these diffeomorphisms one must include an additional gravitating body with now an infinite number of degrees of freedom. This could be achieved by dressing to a (quantum) field. The diffeomorphisms that are not degeneracies of the symplectic form are ``large diffeomorphisms'' which, in an open universe, are given by the asymptotic symmetry group at infinity as well as the symmetry group of the bifurcate Killing horizon. For $d$-dimensional asymptotically AdS spacetimes the asymptotic symmetry group is $O(d-1,2)$. For asymptotically flat spacetimes, the asymptotic symmetry group in $d=4$ is the infinite-dimensional BMS group \cite{Sachs:1962zza,Ashtekar:1978zz} and, in $d>4$, can be reduced down to the Poincaré group \cite{Hollands:2016oma}. Furthermore, the symmetry group of the bifurcate Killing horizon is also an infinite-dimensional group, analogous to that of the BMS group at $\scri$ \cite{Chandrasekaran:2018aop}. The isometries of $(\mc{R},g)$ are a subgroup of these 
large diffeomorphisms. The large diffeomorphisms act non-trivially on the phase space and correspond to ``conserved'' charges on the boundary of $\mc{R}$ \cite{Iyer:1994ys,2000PhRvD..61h4027W,Hollands:2005wt,Chandrasekaran:2018aop}. Constructing observables invariant under the full group of large diffeomorphisms corresponds to quantizing this (possibly infinite-dimensional) larger group. It is not presently clear to us how the inclusion of both large and small diffeomorphisms would affect the algebra of observables in $\mc{R}$.

\subsection{Near-Extremal Black Holes}
\label{sec:extremal}
The near extremal limit of black holes in asymptotically flat or AdS spacetime has provided an important tractable playground for understanding aspects of black hole physics, both classically and quantum mechanically. The simplifying feature of these spacetimes is that they develop a long AdS$_2$ throat with slowly varying internal space near the horizon. 

At extremality, black holes have zero temperature, violating the third law of black hole thermodynamics. This leads to a puzzle when raising the mass slightly to perturb away from extremality. Semiclassically, energy growth of the black hole scales as the square of the temperature, $\Delta E \sim T^2/M_{\textrm{gap}}$, where $M_{\textrm{gap}}$ is an energy scale where the semiclassical analysis of Hawking must break down because the black hole does not have enough mass to radiate a single quantum at typical energy $\sim T$. There are various proposed resolutions to this and related puzzles \cite{preskill1991limitations,1999JHEP...02..011M,2000hep.th...12020P,2021JHEP...05..145I}, all of which require significant modification of the physics at low energies. Recently, an analysis taking care of quantum fluctuations in the Gibbons-Hawking approach, resolved this puzzle by evaluating the density of states at low temperatures \cite{2021JHEP...05..145I} which has significant $\log (T/G)$ corrections due to a strongly-coupled zero mode. See \cite{2020arXiv201101953H} for a different resolution, involving a mass gap above the degenerate ground states, using similar techniques for supersymmetric black holes. 
 
This is a scenario where abandoning thermodynamic arguments in favor of the computation of an explicit von Neumann entropy is desirable. In the crossed product approach, our previous analysis goes through identically, leading to the areas of the black holes in both asymptotically flat and AdS spacetimes. This misses the $\log T/G$ corrections, just as the Noether charge approach misses such corrections. The culprit of the breakdown of the crossed product approach at subleading orders may be related to the strong coupling of the fields at low temperature. The near extremal limit of SdS (Nariai) also leads to a strong-coupling regime \cite{2019arXiv190401911M}, so may deserve further considerations.

\subsection{Von Neumann entropies in quantum field theory}

One of the most remarkable aspects of the crossed product construction for black hole entropy is that it provides a mathematically well-defined notion of von Neumann entropy for quantum fields, doing away with the universal ultraviolet divergence. While von Neumann entropies themselves are not well-defined in quantum field theory, one expects that entropy differences are better behaved, because their ultraviolet divergences cancel out. Entropy differences have played fundamental roles in recent progress on the deep connections between energy and information in quantum field theory, such as in the entropic c- and F-theorems \cite{2004PhLB..600..142C,2012PhRvD..85l5016C}, the Bekenstein bound \cite{bekenstein1981universal, 2004PhRvD..69f4006M,2008CQGra..25t5021C}, the generalized second law of thermodynamics \cite{2012PhRvD..85j4049W}, and the quantum null energy condition (QNEC) \cite{2016PhRvD..93f4044B,2016PhRvD..93b4017B}. While the derivations of these results are very convincing, they have not been made mathematically rigorous\footnote{The QNEC has been rigorously proven in \cite{2018arXiv181204683C} by reformulating the inequality purely in terms of relative entropies which are well-defined in quantum field theory.} due to the intermediate steps involving ultraviolet divergent von Neumann entropies.\footnote{{Some related discussions were given in \cite{2023arXiv230607323A,2023arXiv230609314K}.}} 

The crossed product appears to be a promising direction for providing rigorous proofs of these conjectures because of its well-defined renormalization of the von Neumann entropy. In CLPW and this work, it was physically motivated to do the crossed product due to the gravitational constraints, but there is nothing stopping one from leveraging this technique in non-gravitational quantum field theory. We elaborate on this idea, highlighting an application to the Bekenstein bound and the QNEC in \cite{2023arXiv231207646K}.

\acknowledgments
We would like to thank Venkatesa Chandrasekaran, Thomas Faulkner, Philipp Hohn, Stefan Hollands, Kristan Jensen, Hong Liu, Juan Maldacena, Geoff Penington, Massimo Porrati, Kartik Prabhu, Arvin Shahbazi-Moghaddam, Jonathan Sorce, Antony Speranza, Joaquin Turiaci, Robert Wald, and Edward Witten for discussions and comments. {We especially thank Antony Speranza for bringing to our attention an error in an earlier version of the manuscript.} Part of this work was carried out while G.S. was visiting IQOQI, Vienna and the Erwin Schrödinger Institute (ESI). It is a pleasure to thank IQOQI and the ESI for their hospitality and financial assistance. J.K.F.~is supported by the Marvin L.~Goldberger Member Fund at the Institute for Advanced Study and the National Science Foundation under Grant No. PHY-2207584. S.L.~acknowledges the support of NSF grant No. PHY-2209997. G.S.~and S.L.~are supported by the Princeton Gravity Initiative at Princeton University.

\appendix

\section{The Non-local Modular Flow of the Unruh State}
\label{app:Unruh}
While it was not essential for the construction of the exterior algebra nor the entropy, it is interesting to note that the modular flow may be exactly constructed in the Unruh state in the region $(\mc{R},g)$ of a black hole in asymptotically flat spacetime even though it is far from geometric. The key insight is that the modular flow acts, on the initial data, geometrically at $\sH_R^-$ and as a projector at $\scri^-$. The result is that we find the modular flow for a black hole formed from collapse in an asymptotically flat spacetime, which may be of use in future analysis.

We note that, given the decomposition of $\op{\phi}(f)$ as in \cref{eq:phiPiPi} in terms of initial data on $\mc{H}^{-}$ and $\scri^{-}$ the corresponding Weyl operators $\op{W}(f)$ defined by \cref{eq:Weyl} can be decomposed as
\begin{equation}
\label{eq:Weyl_decomp}
\op{W}(f)=e^{i\op{\Pi}(s)}e^{i\tilde{\op{\Pi}}(\tilde{s})}
\end{equation}
where we recall that $s\defn[Ef]\vert_{\mc{H}_{\textrm{R}}^{-}}$ and $\tilde{s}\defn [Ef]\vert_{\scri^{-}}$. 

We recall that the Unruh state is defined as the product state $\s_{0}\defn \s_{0}^{\mc{H}^{-}}\otimes \s_{0}^{\scri^{-}}$ and $\s_{0}^{\mc{H}^{-}}$ is a cyclic and separating state on $\mf{A}(\mc{H}^{-}_{\textrm{R}},\s_{0})$. However, $\s_{0}^{\scri^{-}}$ is a cyclic but {\em not} separating state on $\mf{A}(\scri^{-},\s_{0})$. Therefore, $\s_{0}$ is cyclic but not separating on the algebra $\mf{A}(\mc{R},\s_{0})$ of field observables in $\mc{R}$.

The modular flow of a cyclic but not separating state was considered in appendix A of \cite{2018arXiv181204683C}. To construct the modular flow we first define the projection operator on $\Fock^{\mc{H}}\otimes \Fock^{\scri}$
\begin{equation}
\op{P}_{\s_{0}}\defn \op{1}_{\mc{H}}\otimes \ket{\s_{0}^{\scri}}\bra{\s_{0}^{\scri}}.
\end{equation}
Since $\s_{0}$ is cyclic for $\mf{A}(\mc{R},\s_{0})$, the Tomita operator can be densely defined as 
\begin{equation}
 \op{S}_{\s_{0}}a \ket{\s_{0}} \defn \op{P}_{\s_{0}}a^{\dagger}\ket{\s_{0}}
\end{equation}
for any $a\in \mf{A}(\mc{R},\s_{0})$. The presence of the projector ensures that $\op{S}_{\s_{0}}$ is consistently defined as an antilinear operator. Namely, if $a\ket{\s_0} = 0$ but $a^{\dagger}\ket{\s_0} \neq 0$, $\op{P}_{\s_{0}}$ annihilates $a^{\dagger}\ket{\s_0}$.

The modular operator is then 
\begin{equation}
\op{\Delta}_{\s_{0}}\defn \op{S}_{\s_{0}}^{\dagger}\op{S}_{\s_{0}} = e^{-\op{H}_{\s_{0}}^{\mc{H}}}\otimes \ket{\s_{0}^{\scri}}\bra{\s_{0}^{\scri}}
\end{equation}
where $\op{H}^{\mc{H}}_{\s_{0}}$ is the modular Hamiltonian on $ \mf{A}(\mc{H}_{\textrm{R}}^{-},\s_{0})$. This defines the action of the modular flow on $ \mf{A}(\mc{R},\s_{0})$. We note that the modular flow only corresponds to a geometric flow on the past horizon $\mc{H}^{-}$.

The modular flow of a Weyl operator in the exterior is
\begin{equation}
 \op{\Delta}_{\s_0}^{it}\op{W}(f) \op{\Delta}_{\s_0}^{-it} =\s^{\scri}_0\big(e^{\frac{i}{2}\tilde{\op{\Pi}}(\tilde{s})}\big)\big[e^{\frac{i}{2}\op{\Pi}(s_{t})}\otimes \ket{\omega^{\scri}_0}\bra{\omega^{\scri}_0} \big], 
\end{equation}
where $s_{t}(x)= s(e^{-2\pi t}U,x^{A})$. The modular flow localizes operators to the past horizon. Propagating this data into the bulk yields a highly non-local action of the modular flow.

\section{Asymptotic Quantization at Timelike Infinity}
\label{app1:timequant}

In this appendix, we describe the quantization of fields at past timelike infinity in a spacetime that is asymptotically flat. As opposed to the asymptotic behavior and quantization of fields at spatial and null infinity, the classical and quantum behavior of fields at timelike infinity of a general asymptotically flat spacetime is generally less well-understood. We shall assume that the leading order decay of fields at $i^{-}$ is equivalent to their behavior in flat spacetime. This appears to be a reasonable assumption since, except for near isolated sources (e.g., a stationary black hole), the spacetime should be approaching a flat spacetime. We refer the reader to section 4.4 of \cite{Prabhu:2022zcr} for more details.

To approach $i^{-}$ it is useful to foliate the asymptotic spacetime with a family of hyperboloids $H_{\tau}$ and timelike infinity is approached by the limit $\tau \to -\infty$. 
The leading order asymptotic behavior of a scalar field of mass $m$ as $\tau\to -\infty$ is \cite{2015JHEP...07..115C,porrill1982structure}
\begin{align}
 \phi(p,\tau) \sim \frac{\sqrt{m}}{2(2\pi \tau)^{3/2}}\left[ a_{+}(p)e^{-im\tau} + a_{-}(p)e^{im\tau}\right],
\end{align}
where $p$ is a future-directed, unit normalized ``momentum,'' labeling a point on the unit hyperboloid. The initial data surface corresponds to the space of unit, timelike tangent directions at $i^{-}$ in a similar manner to the description of spatial infinity by Ashtekar and Hansen \cite{Ashtekar:1978zz}. The initial data for the asymptotic Klein-Gordon field corresponds to specifying the complex functions $a_{+}(p)$ and $a_{-}(p)$ on the unit hyperboloid which are respectively the positive and negative frequency part of $\phi$ with respect to $\tau$. Since $\phi$ is real, these functions satisfy $\bar{a}_{+}(p)=a_{-}(p)$. The symplectic form is 
\begin{equation}
\symp_{i^-}^{\KG}(\phi_1,\phi_2) =-\frac{im^2}{4(2\pi)^3}\int_{H^{-}} ~d^3p[a_{1,+}(p)a_{2,-}(p)-a_{2,+}(p)a_{1,-}(p)]
\end{equation}
where $d^{3}p$ is the measure on the unit hyperboloid $H^{-}$ at $i^{-}$. The non-trivial Poisson brackets are 
\begin{align}
 \left\{a_{+}(w_1),a_{-}(w_2) \right\} = -i \frac{4(2\pi)^3}{m^2} \langle w_1, w_2\rangle_{H^{-}} ~1
\end{align}
where
\begin{equation}
a_{+}(w)\defn \int_{H^{-}}d^{3}p~a_{+}(p)\bar{w}(p), \quad a_{-}(w)\defn \int_{H^{-}}d^{3}p~a_{-}(p)w(p)
\end{equation}
and $\braket{w_{1},w_{2}}_{H^{-}}$ is the ordinary $L^2$ inner product on the unit hyperboloid. The quantization proceeds in the obvious way, with a $\ast$-algebra $\Alg_{i^{-}}$ of smeared observables $\op{a}_{+}(w)$, $\op{a}_{-}(w)$, $\op{a}_{+}(w)^{\ast}$, $\op{a}_{-}(w)^{\ast}$ and the identity $\op{1}$ with non-trivial commutation relations
\begin{align}
 \left[\op{a}_{+}(w_1),\op{a}_{+}(w_2)^{\ast} \right] = \frac{4(2\pi)^3}{m^2} \langle w_1, w_2\rangle \op{1}. 
\end{align}
The standard Fock space $\Fock$ can be constructed from the GNS representation of the unique, Poincaré invariant vacuum state $\s_{0}^{i^{-}}$ with vanishing one-point function and the non-vanishing two-point function given by 
\begin{equation}
\s_{0}^{i^{-}}(\op{a}_{+}(p_{1}),\op{a}_{+}(p_{2})^{\ast}) = \frac{4(2\pi)^{3}}{m^{2}}\delta_{H}(p_{1},p_{2}).
\end{equation}
With this representation, we may take the double commutant of the $\ast$-algebra $\Alg_{i^{-}}$ to obtain a von Neumann algebra $\mf{A}(i^-,\s_0)$. The single commutant is just operators proportional to the identity, so $\mf{A}(i^-,\s_0)$ is all bounded operators on the GNS Hilbert space, thus a Type I$_{\infty}$ von Neumann algebra. In this sense, this algebra does not have a significant impact on the general lessons of the work because the tensor product of a Type II$_{\infty}$ algebra with a Type I$_{\infty}$ is still Type II$_{\infty}$. The trace is supplemented by the standard trace on the GNS Hilbert space for initial data at $i^-$. 

We conclude by giving the relevant fluxes and (perturbed) charges associated to a stationary and axisymmetric spacetime with asymptotic time like Killing vector $t^{a}$ and rotational Killing vector $\psi^{a}$.\footnote{See Section $4.2$ of \cite{Prabhu:2022zcr} for a general construction of charges associated to all asymptotic symmetries of $i^{-}$.} The action of $t^{a}$ changes the initial data by a phase so the infinitesimal transformation is 
\begin{equation}
\delta_{t}a_{+}(p) =-im a_{+}(p)\quad \delta_{t}a_{-}(p) = ima_{-}(p)
\end{equation}
and the observable which generates the corresponding infinitesimal transformation is 
\begin{equation}
\label{eq:Ft}
F_{t}^{i^{-}}\defn \symp^{\KG}_{i^{-}}(\phi,\pounds_{\xi}\phi) = \frac{2m^3}{4(2\pi)^3}\int_{H^{-}}d^{3}p~a_{+}(p)a_{-}(p) \ .
\end{equation}
The action of a general Lorentz transformation corresponds to an isometry of $H^{-}$. Therefore the action of the Killing vector $\psi^{a}$ is simply 
\begin{equation}
 \delta_{\psi}a_{+}(p) =\pounds_{\psi} a_{+}(p)\quad \delta_{\psi}a_{-}(p) =\pounds_{\psi} a_{-}(p)
\end{equation}
and the corresponding flux is given by 
\begin{equation}
\label{eq:Fpsi}
F_{\psi}^{i^{-}}\defn \symp^{\KG}_{i^{-}}(\phi,\pounds_{\psi}\phi)=\frac{im^2}{2(2\pi)^3}\int_{H^{-}} ~d^3p[a_{+}(p)\pounds_{\psi}a_{-}(p)-(\pounds_{\psi}a_{+}(p))a_{-}(p)] \ .
\end{equation}
Quite generally, the fluxes \cref{eq:Ft,eq:Fpsi} can be related to second-order perturbed charges which can be ``matched'' to limiting charges on $\scri^{-}$. If the spacetime has no internal boundaries then the second-order perturbed mass $\delta^{2}\op{M}_{i^{-}}$ and (minus) the second-order angular momentum $\delta^{2}\op{J}_{i^{-}}$ would simply be equal to the fluxes $\op{F}^{i^{-}}_{t}$ and $\op{F}_{\psi}^{i^{-}}$ respectively and so neither are independent degrees of freedom. However, following the same analysis which led to the constraint \cref{eq:i-match} at $i^{-}$ relating $\delta^{2}\op{A}_{-}$ to $\delta^{2}\op{M}_{i^{-}}$ and $\delta^{2}\op{J}_{i^{-}}$ gets modified by the flux through $i^{-}$
\begin{equation}
\label{eq:FMJAi-}
\op{F}_{\xi}^{i^{-}} = \delta^{2}\op{M}_{i^{-}}-\Omega_{\textrm{H}}\delta^{2}\op{J}_{i^{-}}-\frac{\delta^{2}\op{A}_{-}}{4G_{\textrm{N}}\beta}
\end{equation}
where $\op{F}_{\xi}^{i^{-}}=\op{F}_{t}^{i^{-}}+\Omega_{\textrm{H}}\op{F}_{\psi}$. 

\section{Linearized Gravitational Perturbations}
\label{sec:grav_EM_class}

In this appendix, we consider the phase space and quantization of gravitational perturbations off of a fixed background, globally hyperbolic spacetime $(\mc{M},g)$ (see \cite{Wald:1984rg} for a discussion of gravitational perturbations and notation). We recall that the linearized metric perturbation is denoted as 
\begin{equation}
\gamma_{ab}\defn \delta g_{ab}. 
\end{equation}
We denote the linearized Einstein tensor as $L_{ab}(\gamma)\defn \delta G_{ab}-\Lambda \gamma_{ab}$ and therefore the perturbation $\gamma$ satisfies 
\begin{equation}
\label{eq:linEE}
L_{ab}(\gamma) = -\frac{1}{2}\Box_{g}\gamma_{ab} -\frac{1}{2}\nabla_{a}\nabla_{b}\gamma^{c}{}_{c}+\frac{1}{2}\nabla_{c}\nabla_{a}\gamma^{c}{}_{b}+\frac{1}{2}\nabla_{c}\nabla_{b}\gamma^{c}{}_{a}-\Lambda \gamma_{ab}=0 \ ,
\end{equation}
where any two solutions $\gamma_{ab}$ and $\gamma^{\prime}_{ab}$ to \cref{eq:linEE} are physically equivalent if they are related by a linearized gauge transformation 
\begin{equation}
\label{eq:lindiff}
\gamma^{\prime}_{ab} - \gamma_{ab} = \nabla_{(a}\xi_{b)} \ ,
\end{equation}
for some vector field $\xi^{a}$ on $\mc{M}$. The Lagrangian of general relativity provides the space of metric perturbations $\gamma_{ab}$ with a symplectic structure. The (conserved) symplectic product of two solutions $\gamma_{1,ab}$ and $\gamma_{2,ab}$ is given by 
\begin{equation}
\label{eq:OmegaSigma}
\symp^{\GR}_{\Sigma}(\gamma_{1},\gamma_{2})= \frac{1}{16\pi G_{\textrm{N}}}\int_{\Sigma}\sqrt{h}d^{3}x~n_{a}P^{abcdef}[\gamma_{2,bc}\nabla_{d}\gamma_{1,ef}-\gamma_{1,bc}\nabla_{d}\gamma_{2,ef}]
\end{equation}
where 
\begin{equation}
P^{abcdef}\defn g^{ae}g^{fb}g^{cd}-\frac{1}{2}g^{ad}g^{be}g^{fc}-\frac{1}{2}g^{ab}g^{ae}g^{fd}+\frac{1}{2}g^{bc}g^{ad}g^{ef}.
\end{equation}
where $\Sigma$ is a Cauchy surface and $\sqrt{h}d^{3}x$ is the proper volume element on $\Sigma$. This symplectic product is conserved, i.e., it is independent of the choice of Cauchy surface $\Sigma$. If $h_{ab}(\lambda)$ corresponds to the induced metric on $\Sigma$, $p^{ab}(\lambda)$ is defined by 
\begin{equation}
p^{ab}\defn \sqrt{h}(K^{ab}-h^{ab}K),
\end{equation}
where $K_{ab}$ is the extrinsic curvature of $\Sigma$. Any perturbation $\gamma_{ab}$ corresponds to a pair $(\delta h_{ab},\delta p^{ab})$. Since $\gamma_{ab}$ satisfies the linearized equations of motion \cref{eq:linEE} then $(\delta h_{ab},\delta p^{ab})$ satisfies the linearized constraints and linearized Hamiltonian equations of motion. In terms of the pair $(\delta h_{ab},\delta p^{ab})$ the symplectic product of any two solutions can be simply expressed as 
\begin{equation}
\symp^{\GR}_{\Sigma}(\gamma_{1},\gamma_{2})=-\frac{1}{16\pi G_{\textrm{N}}}\int_{\Sigma}d^{3}x~[\delta h_{1, ab}\delta p_{2}^{ab} - \delta h_{2,ab}\delta p_{1}^{ab}].
\end{equation}
Therefore, the points in phase space $\mc{P}$ are given by specifying the pair $(\delta h_{ab},\delta p^{ab})$ on any spacelike Cauchy surface $\Sigma$ and we can view $\symp^{\GR}_{\Sigma}$ as a bilinear map on $\mc{P}$. 

The observable on $\mc{P}$ which is of central importance is the locally smeared field observable 
\begin{equation}
\label{eq:gammaf}
\gamma(f) \defn \int \sqrt{-g}d^{4}y~\gamma_{ab}(y)f^{ab}(y) \ ,
\end{equation}
where $f^{ab}$ is a symmetric, divergence-free test tensor field on $\mc{M}$ (i.e., the test tensor field $f^{ab}$ satisfies $f^{[ab]}=0=\nabla_{a}f^{ab}$). The restriction of the smearing tensors to symmetric, divergence-free test tensors is necessary and sufficient to eliminate the gauge freedom \eqref{eq:lindiff} in $\gamma_{ab}$
\begin{equation}
\int\sqrt{-g}d^{4}y~\nabla_{(a}\xi_{b)}f^{ab}=-2\int \sqrt{-g}d^{4}y~\xi_{b}\nabla_{a}f^{ab}=0.
\end{equation}
Therefore, the smeared fields $\gamma(f)$ are gauge invariant in the sense of \eqref{eq:lindiff}. The fields, however, are not invariant under the spacetime diffeomorphisms that change the background metric. The classical advanced and retarded Greens functions are well-defined when smeared with symmetric, divergence-free test tensors (i.e., $(Af)_{ab}$ and $(Rf)_{ab}$ are well-defined) so the propagator $(Ef)_{ab}=(Af)_{ab} - (Rf)_{ab}$ is well-defined \cite{Fewster:2012bj}. Therefore, one can straightforwardly repeat the constructions given in \cref{subsec:scfieldth}. In particular, we have that 
\begin{equation}
\gamma(f) = \symp^{\GR}_{\Sigma}(\gamma,Ef)
\end{equation}
and so, indeed, $\gamma(f)$ is an observable. As before, on a Cauchy surface $\Sigma$, one can again decompose $(Ef)_{ab}$ into initial data for the linearized Einstein equations \cref{eq:linEE} which thereby yields a decomposition of $\gamma(f)$ into a linear combination of three-smeared observables in an analogous manner to \eqref{eq:phivphipi}.

In exact parallel to the Klein-Gordon field considered in \cref{subsec:AlgQFTCS}, the quantization algebra $\Alg^{\GR}$ of local field observables is the unital $\ast$-algebra generated by $\op{\gamma}(f)$, $\op{\gamma}(f)^{\ast}$ and $\op{1}$ factored by the following relations: 
\begin{enumerate}[label=(B.{\Roman*})]
\label{GRalg}
\item $\op{\gamma}(c_{1}f_{1}+c_{2}f_{2})=c_{1}\op{\gamma}(f_{1})+c_{2}\op{\gamma}(f_{2})$ for any $f^{ab}_{1},f^{ab}_{2}$ and any $c_{1},c_{2}\in \mathbb{R}$, \label{B1}
\item $\op{\gamma}(L(f))=0$ for all $f^{ab}$\label{B2}
\item $\op{\gamma}(f)^{\ast}=\op{\gamma}(f)$ for all $f^{ab}$ \label{B3}
\item $[\op{\gamma}(f_{1}), \op{\gamma}(f_{2})]=iE(f_{1},f_{2})\op{1}$. \label{B4} 
\end{enumerate}

We now consider the case where the spacetime satisfies assumptions \ref{assump1}--\ref{assump4}. The description of the phase space of field observables on a bifurcate Killing horizon is analogous to the case of the phase space of a scalar field considered \cref{subsec:KillHor}. However, it will be convenient --- and necessary --- to choose a ``preferred gauge'' in a neighborhood of a Killing horizon. In particular, by Lemma $1$ of \cite{Hollands:2012sf}, the necessary and sufficient conditions to ensure that, in the (first-order) perturbed spacetime, the metric is in a gauge such that the horizon $\mc{H}^{-}$ remains null and coincides with the actual (past) event horizon to first-order in $\lambda$ are that the perturbation satisfies 
\begin{equation}
\label{eq:pertgauge}
\gamma_{ab}n^{b}=0 \quad \textrm{ and }\quad q^{ab}\gamma_{ab}=0
\end{equation}
on $\mc{H}^{-}$.\footnote{Since $q_{ab}$ is a degenerate metric on $\mc{H}^{-}$, the inverse metric $q^{ab}$ is defined up to multiples of $n^{a}$. However, since the first condition in \cref{eq:pertgauge} implies that $\gamma_{ab}$ is orthogonal to $n^{a}$, then the second condition is well-defined.} In this gauge, the perturbed expansion $\delta \theta$ vanishes on $\mc{H}^{-}$. Similar gauge conditions with $n^{a}$ replaced with $\ell^{a}$ in \cref{eq:pertgauge} can be imposed on the perturbation on the future horizon $\mc{H}^{+}$. The perturbed shear on $\mc{H}^{-}$ in terms of $\gamma_{ab}$ is
\begin{equation}
\delta \sigma_{ab} = \frac{1}{2}\bigg(q_{a}{}^{c}q_{b}{}^{d}-\frac{1}{2}q_{ab}q^{cd}\bigg)\pounds_{n}\gamma_{cd} \quad \quad \quad \textrm{(on $\mc{H}^{-}$)} \ .
\end{equation}
Since $\gamma_{ab}$ and $\delta \sigma_{ab}$ are orthogonal to $n^{a}$ we may write these quantities on $\mc{H}^{-}$ as $\gamma_{AB}$ and $\delta \sigma_{AB}$ respectively (see \cref{foot:AB}). Using the gauge conditions \cref{eq:pertgauge} and the coordinates $(U,x^{A})$ on $\mc{H}^{-}$ the shear can be simply expressed as 
\begin{equation}
\delta \sigma_{AB} = \frac{1}{2}\partial_{U}\gamma_{AB} \quad \quad \quad \textrm{(on $\mc{H}^{-}$)} \ .
\end{equation}
The symplectic product \cref{eq:OmegaSigma} on $\mc{H}^{-}$ is 
\begin{align}
\label{eq:OmegaHm}
\symp^{\GR}_{\mc{H}^{-}}(\gamma_{1},\gamma_{2}) = &-\frac{1}{16\pi G_{\textrm{N}}}\int_{\mc{H}^{-}}dUd\Omega_{2} ~ \big(\partial_{U}\gamma_{1}^{AB}\gamma_{2AB}-\gamma_{1}^{AB}\partial_{U}\gamma_{2AB}\big)\nonumber \\
=&-\frac{1}{8\pi G_{\textrm{N}}}\int_{\mc{H}^{-}}dUd\Omega_{2}~(\delta \sigma_{1}^{AB}\gamma_{2,AB}-\gamma_{1}^{AB}\delta \sigma_{2,AB}) \ .
\end{align}
As in the scalar case in \cref{subsec:scfieldth}, the observables on a Killing horizon $\mc{H}^{-}$ are given by the ``three-smeared'' observables $\symp^{\GR}_{\mc{H}^{-}}(\gamma,Ef)$. For any $s_{ab}=(Ef)_{ab}$ satisfying the gauge conditions \cref{eq:pertgauge}, the observable $\symp^{\GR}_{\mc{H}^{-}}(\gamma,Ef)$ is equivalent to the shear $\sigma_{AB}$ smeared with $s_{AB}$ on $\mc{H}^{-}$ 
\begin{equation}
\label{eq:shearsmear}
 \delta \sigma(s) =\int_{\mc{H}^{-}}dUd\Omega_{2}~\delta \sigma_{AB}s^{AB} = -4 \pi G_{\textrm{N}}\symp^{\GR}_{\mc{H}^{-}}(\gamma,Ef) \ .
\end{equation}
The Poisson brackets for the smeared perturbed shear $\delta \sigma(s)$ are 
\begin{equation}
\label{eq:sigmaPB}
\{\delta \sigma(s_{1}),\delta \sigma(s_{2})\} = 16\pi^{2}G_{\textrm{N}}\symp^{\GR}_{\mc{H}^{-}}(s_{1},s_{2})=- \int_{\mc{H}^{-}}dVd\Omega_{2}~[s_{1AB}\partial_{V}s_{2}^{AB}-s_{2AB}\partial_{V}s_{1}^{AB}] \ .
\end{equation}
Finally we define the flux observable on $\mc{P}$ as 
\begin{equation}
F_{\xi}\defn \symp^{\GR}_{\Sigma}(\gamma,\pounds_{\xi}\gamma)
\end{equation}
and the Poisson bracket of $F_{\xi}$ with $\gamma(f)$ is 
\begin{equation}
\{F_{\xi},\gamma(f)\}=\gamma(\pounds_{\xi}f). 
\end{equation}
We now evaluate $F_{\xi}$ on the past horizon $\mc{H}^{-}$. Using \cref{eq:OmegaHm} and the fact that $\xi^{a} = -\kappa U n^{a}$ on the horizon we obtain 
\begin{align}
\label{eq:symphor}
F_{\xi}=&~\symp^{\GR}_{\mc{H}^{-}}(\gamma,\pounds_{\xi}\gamma) = - \frac{\kappa}{4\pi G_{\textrm{N}}}\int_{\mc{H}^{-}}dUd\Omega_{2}~U\delta \sigma_{AB}\delta \sigma^{AB} 
\end{align}
The quantization of field observables on $\mc{H}^{-}$ is defined to be the unital $\ast$-algebra $\Alg^{\GR}_{\mc{H}^{-}}$ generated by the elements $\delta \op{\sigma}(s)$, $\delta \op{\sigma}(s)^{\ast}$ and $\op{1}$ and factored by relations analogous to that of \ref{B1}--\ref{B4}. Conditions \ref{B1} and \ref{B3} carry over straightforwardly to conditions on $\Alg^{\GR}_{\mc{H}^{-}}$. Condition \ref{B2} has no analog on $\Alg^{\GR}_{\mc{H}^{-}}$ since the test functions correspond to freely specifiable initial data on $\mc{H}^{-}$. The commutation relations \ref{B4} are now 
\begin{equation}
[\delta \op{\sigma}(s_{1}),\delta \op{\sigma}(s_{2})]=-64\pi^{2} G_{\textrm{N}} \symp^{\GR}_{\mc{H}^{-}}(s_{1},s_{2}) \ .
\end{equation}
The Hadamard condition on states is analogous to that of \cref{eq:2ptPI} and given by 
\begin{equation}
\label{eq:HadstateGR}
\s(\op{\sigma}_{AB}(x_{1})\op{\sigma}_{CD}(x_{2}))=\frac{1}{\pi}\frac{(q_{A(C}q_{D)B}-\frac{1}{2}q_{AB}q_{CD})\delta_{\bb{S}^{2}}(x_{1}^{A},x_{2}^{A})}{(U_{1}-U_{2}-i0^{+})^{2}}+ S_{ABCD}(x_{1},x_{2}) \ ,
\end{equation}
where $S_{ABCD}$ is a smooth bi-tensor on $\mc{H}^{-}$ which is symmetric in $AB$ and $CD$ and satisfies $q^{AB}S_{ABCD}=q^{CD}S_{ABCD}=0$ and symmetric under the interchange of $x_{1}$ and $x_{2}$ together with the interchange of $AB$ with $CD$. The connected $n$-point functions for $n\neq 2$ decay for any set of $|U_{i}|\to \infty$ as $O((\sum_{i}U_{i}^{2})^{-1/2-\epsilon})$ for some $\epsilon>0$. The unique Gaussian state $\s_{0}$ on $\Alg^{\GR}_{\mc{H}^{-}}$ invariant under affine time translations has a vanishing $1$-point function and the $2$-point function is given by \cref{eq:HadstateGR} with $S_{ABCD}=0$. Furthermore, it is straightforward to check that the state $\s_{0}$ is KMS with inverse temperature $\beta$ if restricted to $\mc{H}^{-}_{\textrm{R}}$. 

\section{Covariant Phase Space}
\label{app:cov_phase}

In this appendix, we will outline the derivation of \cref{eq:BDYCharge}. We refer the reader to \cite{Hollands:2012sf} for a complete discussion of the derivation, as well as the covariant phase space formalism. For simplicity, we will begin by considering vacuum general relativity with Lagrangian 
\begin{equation}
\mc{L}^{\GR} = \frac{1}{16\pi G_{\textrm{N}}} \left( R -2 \Lambda\right)
\end{equation}
and derive a formula for $F^{\GR}_{\xi}\defn \symp^{\GR}_{\Sigma}(\gamma,\pounds_{\xi}\gamma)$ in terms of ``boundary charges.'' We will then illustrate how the resulting formula is modified in the presence of other (first-order) matter fields such as the Klein-Gordon field considered in the main body of the paper. 

For ease of presentation, it will be convenient to present quantities as differential forms. For example, we shall take the Lagrangian density $L^{\GR}$ to be the four-form $(L^{\GR})_{abcd}\defn\mc{L}^{\GR} \epsilon_{abcd}$ where $\epsilon$ is the volume form on spacetime. For any\footnote{If the spacetime $(\mc{M},g)$ has any asymptotic conditions (e.g., asymptotic flatness) then we will take $X^{a}$ to preserve these conditions.} vector field $X$, the associated ``Noether current'' is a three-form defined as
\begin{align}
 \mc{J}^{\GR}_X \defn \theta^{\GR}(g,\pounds_X g) - i_{X}L^{\GR},
\end{align}
where $i_{X}$ denotes the contraction of $X$ into the first index of a differential form and $\theta$ is the pre-symplectic potential, which is a three-form given by
\begin{equation}
\theta^{\GR}_{efg} \defn \frac{1}{16\pi G_{\textrm{N}}}g^{ac}g^{bd}\bigg(\nabla_{d}\frac{d}{d\lambda}g_{bc} - \nabla_{c}\frac{d}{d\lambda}g_{bd}\bigg)\epsilon_{aefg}
\end{equation}
for any $g_{ab}(\lambda)$. The Noether current may also be expressed as \cite{1995PhRvD..52.4430I}
\begin{align}
 \mc{J}_X = C^{\GR}_X + dQ_X,
\end{align}
where
\begin{equation}
(C^{\GR}_X)_{cde} \defn \frac{1}{8\pi}X^{a}(G_{ab}+\Lambda g_{ab})\epsilon^{b}{}_{cde}
\end{equation}
are the gravitational constraints \cite{2007PhRvD..75h4029S} and $Q_X$ is the Noether charge, a two-form defined for all $g_{ab}(\lambda)$ as
\begin{equation}
(Q_{X})_{ab}\defn-\frac{1}{16\pi G_{\textrm{N}}}\nabla_{c}X_{d}\epsilon^{cd}{}_{ab} \ .
\end{equation}
The first variation of the Noether current is \cite{Iyer:1994ys}
\begin{align}
 \frac{d}{d\lambda} \mc{J}^{\GR}_X(g) = -i_{X}\left(E^{\GR}\cdot \frac{d}{d\lambda}g\right) +\mathscr{w}^{\GR}\left(\frac{d}{d\lambda}g, \pounds_X g\right)+ d\left[i_{X}\theta^{\GR}(g,\frac{d}{d\lambda} g) \right],
\end{align}
where $E^{\GR}_{abcdef}\defn(1/16\pi G_{\textrm{N}})\epsilon_{cdef}(\Lambda g_{ab}-G_{ab})$ are the equations of motion, ``$\cdot$'' denotes contraction over all indices and $\mathscr{w}^{\GR}$ is the symplectic current, defined as
\begin{align}
 \mathscr{w}^{\GR}(\partial_{\lambda_1}g,\partial_{\lambda_2}g) \defn \partial_{\lambda_1}\theta^{\GR}(g,\partial_{\lambda_2}g)-\partial_{\lambda_2}\theta^{\GR}(g,\partial_{\lambda_1}g) \ ,
\end{align}
which is related to the symplectic product by 
\begin{equation}
\symp^{\GR}_{\Sigma}(\partial_{\lambda_{1}}g,\partial_{\lambda_{2}}g) = \int_{\Sigma}\mathscr{w}^{\GR}(\partial_{\lambda_1}g,\partial_{\lambda_2}g).
\end{equation}
Therefore,
\begin{align}
\label{eq:symp_current}
 \mathscr{w}^{\GR}\left(\frac{d}{d\lambda}g,\pounds_X g\right) = X\cdot\left(E^{\GR}\cdot \frac{d}{d\lambda}g\right) + \frac{d}{d\lambda}C^{\GR}_X - d\left[\frac{d}{d\lambda} Q_X+X\cdot\theta^{\GR}(g,\frac{d}{d\lambda} g) \right].
\end{align}
Under the condition that the background metric and its perturbations satisfy Einstein's equations, the first two terms disappear. Integrating $\mathscr{w}$ over a Cauchy surface $\Sigma$ is thus a boundary term. 

The discussion up to this point has been completely general and applicable to any spacetime. We now consider the case where the spacetime $(\mc{M},g)$ contains a Killing horizon and is asymptotically flat at spatial infinity. We now integrate \cref{eq:symp_current} over a Cauchy surface $\Sigma$ extending from the bifurcation surface $\mc{B}$ of the Killing horizon to spatial infinity. The contribution at spatial infinity defines a conserved quantity \cite{Iyer:1994ys}
\begin{align}
 \frac{d}{d\lambda}H_X \defn \int_{\infty} \left[\frac{d}{d\lambda} Q_X-X\cdot\theta^{\GR}(g,\frac{d}{d\lambda} g)\right],
\end{align}
which is the ADM mass, $M$, when $X$ is asymptotically $\partial_t$ and (minus) the ADM angular momentum, $J_{[ij]}$, in the $ij$-plane when $X$ is asymptotically $x_i\partial_{x_j} - x_j\partial_{x_i}$.
On the bifurcation surface $\mc{B}$
, if $X=\xi$ generates the horizon Killing vector field, the contribution is given by the perturbed area
\begin{align}
 -\frac{\kappa}{8\pi G_{\textrm{N}}} \frac{d A_{\mc{B}} }{d\lambda}= \int_{\mc{B}} \left[\frac{d}{d\lambda} Q_\xi-\xi \cdot\theta^{\GR}(g,\frac{d}{d\lambda} g)\right].
\end{align}
In total, given a spacetime with a Killing horizon and a spatial infinity, integrating \cref{eq:symp_current} on a Cauchy surface $\Sigma$ which extends from the bifurcation surface to spatial infinity yields, for any $g_{ab}(\lambda)$ satisfying the equations of motion and the constraints, the following identity
\begin{align}
 \label{eq:firstordersymp}
 \symp^{\GR}\left(\frac{d}{d\lambda}g,\pounds_\xi g\right) = \frac{d}{d\lambda}M_{i^{0}} - \Omega_{\textrm{H}}\frac{d}{d\lambda} J_{i^{0}}- \frac{\kappa}{8\pi G_{\textrm{N}}} \frac{d A_{\mc{B}} }{d\lambda}.
\end{align}
Evaluated at $\lambda = 0$, \eqref{eq:firstordersymp} gives the first law of black hole mechanics \cite{bardeen1973four,Iyer:1994ys}.

We may now take a further $\lambda$ derivative of \eqref{eq:symp_current}, evaluating at $\lambda = 0$, in order to find our desired observable
\begin{equation}
\label{eq:FX_boundary}
F_X \defn \symp^{\GR}_{\Sigma}(\gamma,\pounds_{X}\gamma) = \int_{\Sigma}d\bigg[\frac{d^{2}}{d^{2}\lambda}Q_{X}(g)\bigg\vert_{\lambda=0}-X\cdot\frac{d}{d\lambda}\theta^{\GR}(g;\frac{d}{d\lambda}g)\bigg \vert_{\lambda=0}\bigg].
\end{equation}
From these general considerations, we can see the observable $F_{X}$ has a trivial action on phase space if the vector field $X^{a}$ is of compact support or if the spacetime manifold $(\mc{M},g)$ is closed.

Evaluating \cref{eq:FX_boundary} at $\lambda=0$ yields 
\begin{equation}
\label{eq:sympgamma}
\symp^{\GR}_{\Sigma}(\gamma,\pounds_{\xi}\gamma) = \delta^2M_{i^{0}} - \Omega_\textrm{H}\delta^2J_{i^{0}} - \frac{\kappa}{8\pi G_{\textrm{N}}} \delta^2 A_{\mc{B}} \ ,
\end{equation}
{where we note that, by \cref{eq:QU}, $\delta^{2}\mathcal{Q}_{U=0}=\delta^{2}A_{\mathcal{B}}$.} In the Einstein Klein-Gordon theory considered in \cref{subsec:charge}, the full symplectic form is simply the sum 
\begin{equation}
\symp_{\Sigma}((\gamma_{1},\phi_{1}),(\gamma_{2},\phi_{2})) = \symp^{\GR}_{\Sigma}(\gamma_{1},\gamma_{2}) + \symp^{\KG}_{\Sigma}(\phi_{1},\phi_{2}) .
\end{equation}
Nearly identical manipulations as outlined above yields that 
\begin{equation}
\label{eq:KGGRfluxcharge}
\symp^{\GR}_{\Sigma}(\gamma,\pounds_{\xi}\gamma)+\symp^{\KG}_{\Sigma}(\phi,\pounds_{\xi}\phi)= \int_{\Sigma}d\bigg[\frac{d^{2}}{d^{2}\lambda}Q_{X}(g)\bigg\vert_{\lambda=0}-X\cdot\frac{d}{d\lambda}\theta(g;\frac{d}{d\lambda}g)\bigg \vert_{\lambda=0}\bigg]
\end{equation}
where $\theta = \theta^{\GR}+\theta^{\KG}$ and 
\begin{equation}
\theta^{\KG}_{efg}\defn \epsilon_{efg}{}^{a}(\nabla_{a}\phi) \frac{d}{d\lambda}\phi.
\end{equation}
\Cref{eq:KGGRfluxcharge} with $\gamma_{ab}=0$ is precisely the form of \cref{eq:BDYCharge} as desired where $\mc{Q}_{X}$ in \cref{eq:BDYCharge} is given by quantity in brackets. 

Finally, we note that, indeed, integrating \cref{eq:KGGRfluxcharge} on a Cauchy surface from the bifurcation surface to infinity yields a similar relationship between the symplectic flux $F_{\xi}$ on $\mc{P}$ and the perturbed mass, angular momentum and area of the bifurcation surface obtained in \cite{Hollands:2012sf}. 
\begin{align}
\symp_{\Sigma}((\gamma,\phi),\pounds_{\xi}(\gamma,\phi))=\delta^2M_{i^{0}} - \Omega_{\textrm{H}}\delta^2J_{i^{0}}- \frac{\kappa}{8\pi G_{\textrm{N}}} \delta^2 A_{\mc{B}}.
\end{align}
This result together with integrating \cref{eq:pVQ} on the horizon from the bifurcation surface to asymptotically late affine times yields the global constraint \cref{eq:FxitotKerr}.

\bibliographystyle{JHEP}

\bibliography{main}

\end{document}